\documentclass[useAMS,usenatbib]{mn2e}

\usepackage{longtable}
\usepackage{morefloats}
\usepackage{supertabular,booktabs}
\usepackage{graphicx}
\usepackage{amssymb}
\usepackage{lscape}
\usepackage[usenames, dvipsnames]{color}
\usepackage{rotating}
\usepackage{amsmath}
 \usepackage{array}
\graphicspath{{images/}} 

\newcommand{\NHI}{$\rm N(H\,I)$}
\newcommand{\NHIgt}[1]{$\rm N(H\,I)>#1$}

\newcommand{\NHIbetween}[2]{$\rm #1\leq\log N(H\,I)<#2$}

\newcommand{\zabseq}[1]{$\rm z_{abs}=#1$}
\newcommand{\zabsgt}[1]{$\rm z_{abs}>#1$}
\newcommand{\zabslt}[1]{$\rm z_{abs}<#1$}
\newcommand{\zabsbetween}[2]{$\rm #1\leq z_{abs} \leq #2$}

\newcommand{\upperlimit}[2]{$\rm \log N(#1)<#2$}
\newcommand{\lowerlimit}[2]{$\rm \log N(#1)>#2$}
\newcommand{\abundance}[3]{$\rm \log N(#1)=#2\pm#3$}

\newcommand{\meanZ}{$\rm \mean{Z}$}
\newcommand{\aFe}{$\rm [\alpha/Fe]$}
\newcommand{\deltav}{$\rm \Delta V_{90}$}

\title[Sub-DLA Metallicity Measurements and the CGM]{The ESO UVES Advanced Data Products Quasar Sample -- VI. Sub-Damped Lyman-$\alpha$ Metallicity Measurements and the Circum-Galactic Medium\thanks{Includes observations collected during programme ESO 91.A-0300 at the European Southern Observatory (ESO) Very Large Telescope (VLT) with UVES on the 8.2 m telescopes operated at the Paranal Observatory, Chile.}
}
\author[S. Quiret et al.]{S. Quiret$^{1}$\thanks{E-mail:
samuel.quiret@lam.fr}, C. P\'eroux$^{1}$, T. Zafar$^{2}$, V. P. Kulkarni$^{3}$, E. B. Jenkins$^{4}$, B. Milliard$^{1}$,
 \newauthor
H. Rahmani$^{1}$, A. Popping$^{5}$,  S. M. Rao$^{6}$, D. A. Turnshek$^{6}$ and E. M. Monier$^{7}$\\\\
$^{1}$Aix Marseille Universit\'e, CNRS, LAM (Laboratoire d'Astrophysique de Marseille) UMR 7326, 13388, Marseille, France\\
$^{2}$European Southern Observatory, Karl-Schwarschild-Strasse 2, 85748, Garching, Germany\\
$^{3}$University of South Carolina, Dept. of Physics \& Astronomy, Columbia, USA\\
$^{4}$Princeton University Observatory, Princeton, NJ 08544-1001, USA\\
$^{5}$International Centre for Radio Astronomy Research (ICRAR), The University of Western Australia, 35 Stirling Hwy,\\~ 6009 Crawley WA, Australia\\
$^{6}$Department of Physics and Astronomy and PITTsburgh Particle physics, Astrophysics, and Cosmology Center (PITT PACC),\\ ~University of Pittsburgh, Pittsburgh, PA 15260, USA\\
$^{7}$Department of Physics, The College at Brockport, State University of New York, Brockport, NY 14420, USA}
\begin{document}

\pdfoutput=1
\def\mean#1{\left< #1 \right>}


\pagerange{\pageref{firstpage}--\pageref{lastpage}} \pubyear{2015}

\maketitle

\label{firstpage}

\begin{abstract}

The Circum-Galactic Medium (CGM) can be probed through the analysis of absorbing systems in the line-of-sight to bright background quasars. 
We present measurements of the metallicity of a new sample of 15 sub-damped Lyman-$\alpha$ absorbers (sub-DLAs, defined as absorbers with $\rm 19.0 < \log N(H\,I)<20.3$) with redshift \zabsbetween{0.584}{3.104} from the ESO Ultra-Violet Echelle Spectrograph (UVES) Advanced Data Products Quasar Sample (EUADP). 
We combine these results with other measurements from the literature to produce a compilation of metallicity measurements for 92 sub-DLAs as well as a sample of 362 DLAs. 
We apply a multi-element analysis to quantify the amount of dust in these two classes of systems. We find that either the element depletion patterns in these systems differ from the Galactic depletion patterns or they have a different nucleosynthetic history than our own Galaxy. 
We propose a new method to derive the velocity width of absorption profiles, using the modeled Voigt profile features. The correlation between the velocity width \deltav{} of the absorption profile and the metallicity is found to be tighter for DLAs than for sub-DLAs. 
We report hints of a bimodal distribution in the [Fe/H] metallicity of low redshift ($\rm z<1.25$) sub-DLAs, which is unseen at higher redshifts. This feature can be interpreted as a signature from the metal-poor, accreting gas and the metal-rich, outflowing gas, both being traced by sub-DLAs at low redshifts.

\end{abstract}


\section{Introduction}

In depth studies of galaxy evolution require an understanding of the complex processes occurring at the interface of the galaxy and its nearby environment, the Circum-Galactic Medium (CGM).

On the one hand, the star formation process is believed to be fed in galaxies via accretion mechanisms \citep{Rees1977, White1978, Prochaska2009, Bauermeister2010}. For galaxies with masses typically below $\rm \sim10^{11-12}M_{\odot}$, the accreting gas follows cold flows ($\rm T\sim 10^{4-5}K$) while for more massive galaxies, a second mode of accretion appears, the "hot mode", where the gas is shock heated near the virial temperature ($\rm T\sim 10^{6}K$) \citep{Rees1977, Silk1977a, White1978, Birnboim2003, Keres2005, Dekel2006, Ocvirk2008}. Simulations show that about 40\% of the accretion may be genuinely smooth \citep{Genel2010}. These modes also differ in metallicity \citep{Fumagalli2011, Shen2013}. Indeed, \cite{Ocvirk2008} showed that the "cold mode" accreting gas can reach metallicities up to tenth solar, while the hot mode accreting gas metallicities are usually lower and are highly dependent on the distance to the center of the galaxy and on how well the gas is mixed. These accreting streams may also provide the galaxy with additional angular momentum \citep{Fall1980}.
Observational evidences for accretion have been challenging to gather due to the low surface brightness and low filling factor of the infalling gas and its expected low metallicity. Nevertheless, cold accretion has been recently detected in a few objects \citep{Steidel2000, Martin2012, Rubin2012, Bouche2013a}.
Similarly, early evidences for cold accretion onto quasars have been recently reported by \cite{Cantalupo2014} and \cite{Martin2014a}.

On the other hand, galaxies release energy and material in their environment (up to $\rm \sim 125 kpc$) via supernovae (SNe), stellar winds or Active Galactic Nuclei (AGN) activity. These outflows tend to chemically enriched the IGM \citep{Songaila1996, Simcoe2004, Adelberger2005,Ryan-Weber2009, DOdorico2013,Shull2014, Shull2015}, and can regulate the star formation process of galaxies. Indeed, as the gas is released, it will starve the galaxy from fresh gas accreting along the galaxy major axis, quenching the star formation. It will also enhanced the star formation by cooling the gas via metal line emissions. Fountains can also be created if the gas does not leave the potential well of the galaxy. In this scenario, the fully metal enriched gas recycles and falls back onto the galactic disk and contribute directly to the star forming processes as it can cool efficiently. Simulations have shown that fountains dominate the global accretion mechanism for $\rm z\lesssim1$ galaxies \citep{Oppenheimer2010}. Even though outflows are ubiquitous at all redshifts around star forming galaxies \citep{Shapley2003, Martin2005, Rubin2014} and their existence is confirmed by signatures of OVI found within the CGM of low redshift star forming galaxies \citep{Tumlinson2011}, they remain poorly understood in the context of galaxy formation models.

In the context of emission line study, \cite{Bertone2010UV, Bertone2010X} argued that the ionization state of elements provides valuable insight on the physical state of the CGM (mainly its temperature but also its ionizing process) and can be used to study the different feedback processes taking place, including metal pollution of accreting gas via galactic fountains. \cite{Fumagalli2011} also argued that kinematic analysis of absorption lines can be used in addition to the metallicity analysis to distinguish metal-rich outflowing material from metal-poor ($\rm \lesssim0.01Z_{\odot}$) accreting gas.
Therefore, the study of metal lines (kinematics, line strengths, ionization states) might be the key diagnostic to observationally disentangle outflows from inflows and assess the level of metal enrichment of the CGM and thus galaxy evolution.

Absorbers observed in background quasar spectra are a tool to probe the low density gas and its metallicity. Indeed, simulations predict that cold accretion onto galaxies can be observed in absorption via dense H\,I absorbing systems with $\rm \log N(H\,I)>15.5$ \citep{Faucher-Giguere2011, VandeVoort2012b, Shen2013}. They predict that the cold streams could be traced with metal-poor H\,I absorption systems, mostly in the Lyman Limit System (LLS) range \NHIbetween{17.2}{19.0}. Recently, \cite{Lehner2013} showed observational evidence for low redshift LLS presenting a bimodal metallicity distribution, which they associated with infalls and outflows.
However, the metallicities of LLS depend sensitively on model-dependent ionization corrections, since the LLS gas is highly ionized. This makes it harder to reliably detect the difference between inflows and outflows using the LLS. A more robust way of detecting the metallicity distribution of the gas around galaxies is by using the damped Lyman-$\alpha$ (DLA; $\rm \log N(H\,I) \geq 20.3$) and sub-damped Lyman-$\alpha$ (sub-DLA; $\rm 19.0 \leq \log N(H\,I) < 20.3$) absorbers.
These systems are the primary neutral gas reservoir at $\rm 0 < z < 5$ \citep[][]{StorrieLombardi2000, Peroux2005, Prochaska2005, Rao2006,Zafar2013a} and offer the most precise element abundance measurements in distant galaxies. In particular, at $\rm z\leq2$, \cite{Fumagalli2011} anticipate that almost half of the cross-section in the sub-DLA H\,I column density range is due to streams, while at $\rm z\sim3$, \cite{VandeVoort2012b} anticipate that it is more than 80\%.

In an era of large quasar surveys, with samples of thousands of DLAs available \citep[e.g.,][]{Noterdaeme2012}, sub-DLAs remain little studied. 
Indeed, at low H\,I column densities, one requires a high spectral resolution and high signal to noise ratio (SNR) to derive element abundances. 
The large quasar samples observed with the high resolution spectrographs VLT/UVES \citep{Zafar2013} and Keck/HIRES \citep{Omeara2015} are therefore crucial tools for our understanding of sub-DLA properties. Here, we present a detailed study of the metallicity and kinematics of a large sample of DLAs and sub-DLAs observed at high resolution with UVES.

The paper is organised as follows. In \S 2 we present the data sample and in \S 3 we describe the abundance measurements. The results are discussed in \S 4 followed by conclusions in \S 5.

\section{The Data}
\label{sec:data}

\subsection{New absorbers}

In order to put together a significant sample of sub-DLAs observed at high spectral resolution, we make use of the ESO UVES Advanced Data Products (EUADP) sample from \cite{Zafar2013}. This sample consists of 250 high-resolution ($\rm R\sim42,000$) quasars spectra covering a total of 196 damped absorbers (with log \NHIgt{19.0}).

This dataset has motivated a number of studies including a report of new H\,I systems \citep{Zafar2013} and how they can be used to constrained the neutral gas mass density of sub-DLAs in particular \citep{Zafar2013a}, the nucleosynthetic history of Nitrogen \citep{Zafar2014} and the low Argon abundances observed in DLAs \citep{Zafar2014a}.

Most of the absorbers in the EUADP sample have their metallicity abundances published in the literature \citep[][and reference therein]{Peroux2006, Peroux2006a,Peroux2008,Zafar2013a}. We present here the analysis of 14 new EUADP sub-DLAs covering a redshift range \zabsbetween{0.584}{3.104}. We also include 6 new DLAs for completeness.
The measurements of HI column densities and redshifts of each system in the EUADP sample are reported in \cite{Zafar2013} and references therein.
 

In addition to these 14 new sub-DLAs from the EUADP sample, we present the UVES spectra of two other systems: one sub-DLA at \zabseq{0.584} and one DLA at \zabseq{0.647}. These two low-redshift absorbers have been observed with the HST ACS grism from which an estimate of their H\,I column densities has been derived \citep{Turnshek2015}. The quasars were subsequently observed with UVES on VLT under the programme 91.A-0300 (PI: C. P\'eroux) in Service Mode in August and September 2013. Each object was observed using a combined 346$+$564 nm setting with two different observations with exposure times lasting 4500\,+\,3600 sec (QSO J0018$-$0913) and 2 x 4500 sec (QSO J0132$-$0823). The data were reduced using the most recent version of the UVES pipeline in MIDAS (uves/5.4.3). Master bias and flat images were constructed using calibration frames taken closest in time to the science frames. The science frames were extracted with the ``optimal" option and corrected to the vacuum heliocentric reference. 
To combine the resulting spectra, we choose to weight them by the signal-to-noise ratio, as for the remaining of the EUADP sample \citep{Zafar2013}, in line with standard practice at this spectral resolution \citep{Omeara2015}.

The absorption redshifts, which are based on the \NHI{} or MgII features, are used to analyse the associated metal lines. Table \ref{table:subsample} summarises the properties of the quasars and absorbers in the sample studied here. The two additional objects which were not originally published by \cite{Zafar2013} are shown in bold.

\subsection{Literature sample}

In addition to these 15 new sub-DLA measurements (+7 DLAs), we gather metallicity estimates of sub-DLAs from the remaining part of the EUADP sample as well as other recently published samples \citep{Meiring2006, Meiring2009, Dessauges-Zavadsky2009, Battisti2012, Som15}. 
In order to compare the properties of sub-DLAs to that of DLAs, we add to the sample a collection of DLA metallicity measurements from the EUADP sample as well as from the literature (see earlier references and \citealt{Berg2015}). Altogether, this literature sample is the largest and most up-to-date sub-DLA sample published today.

\begin{figure}
\begin{center}
\includegraphics[width=.4\textwidth]{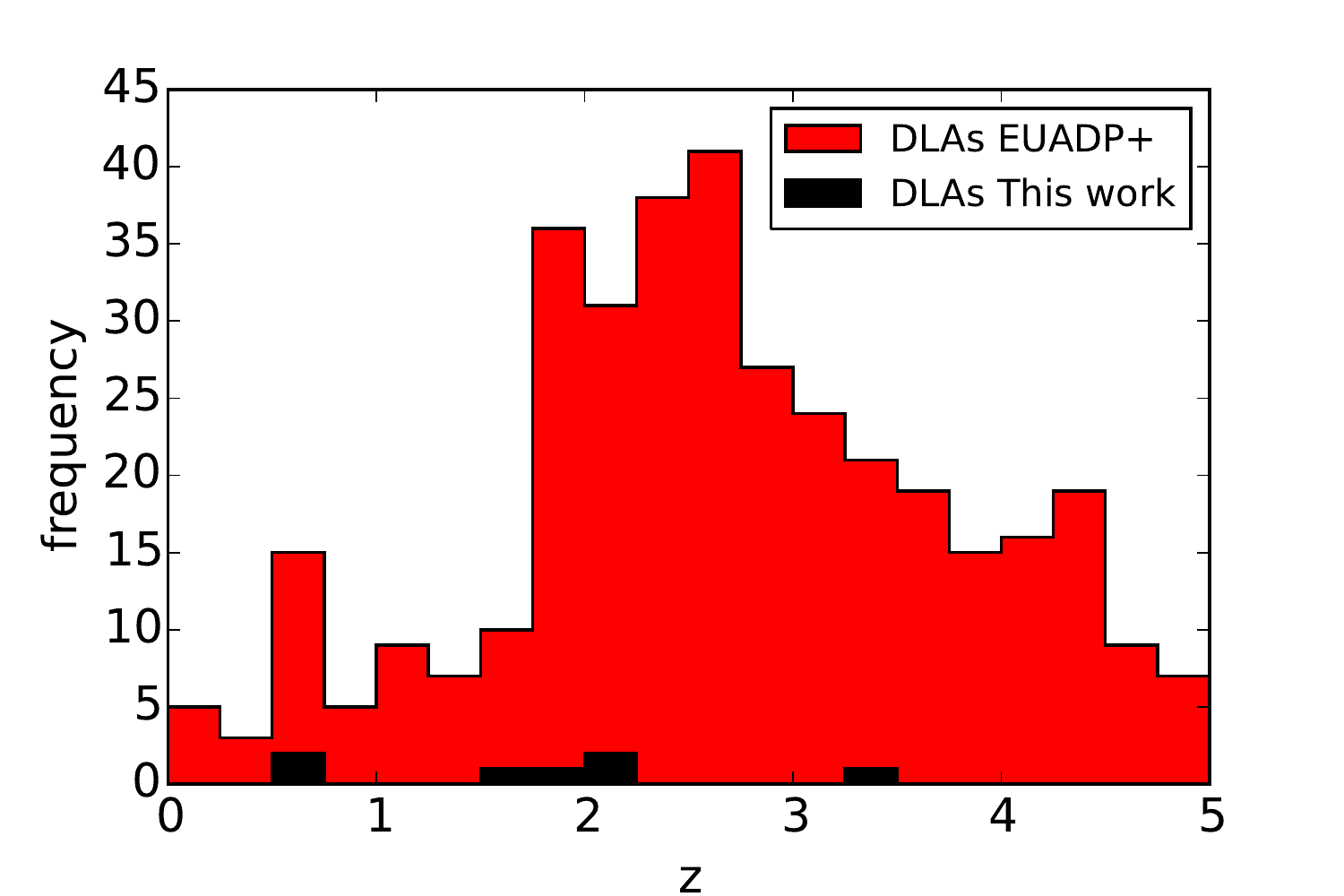}
\includegraphics[width=0.4\textwidth]{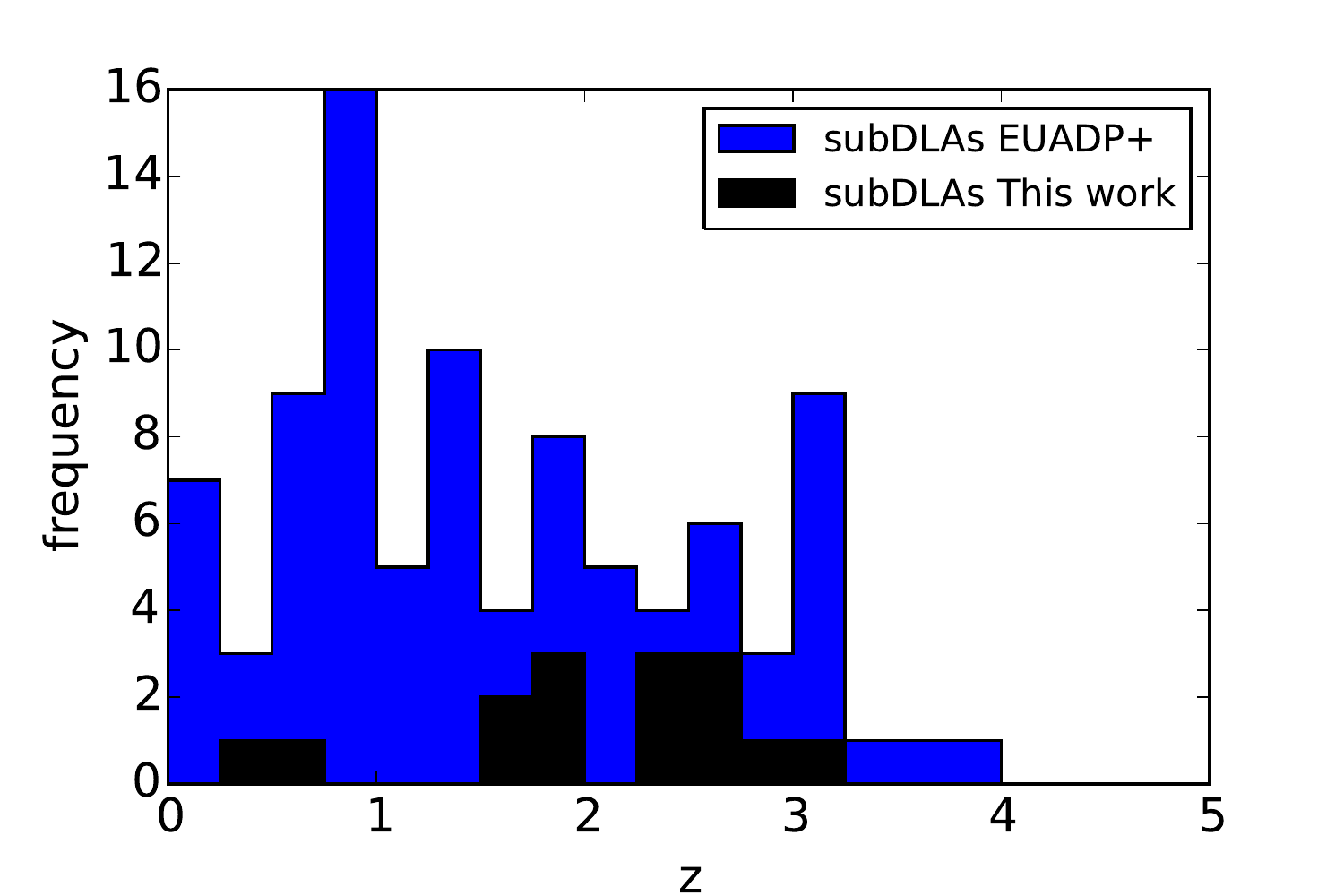}
\includegraphics[width=0.4\textwidth]{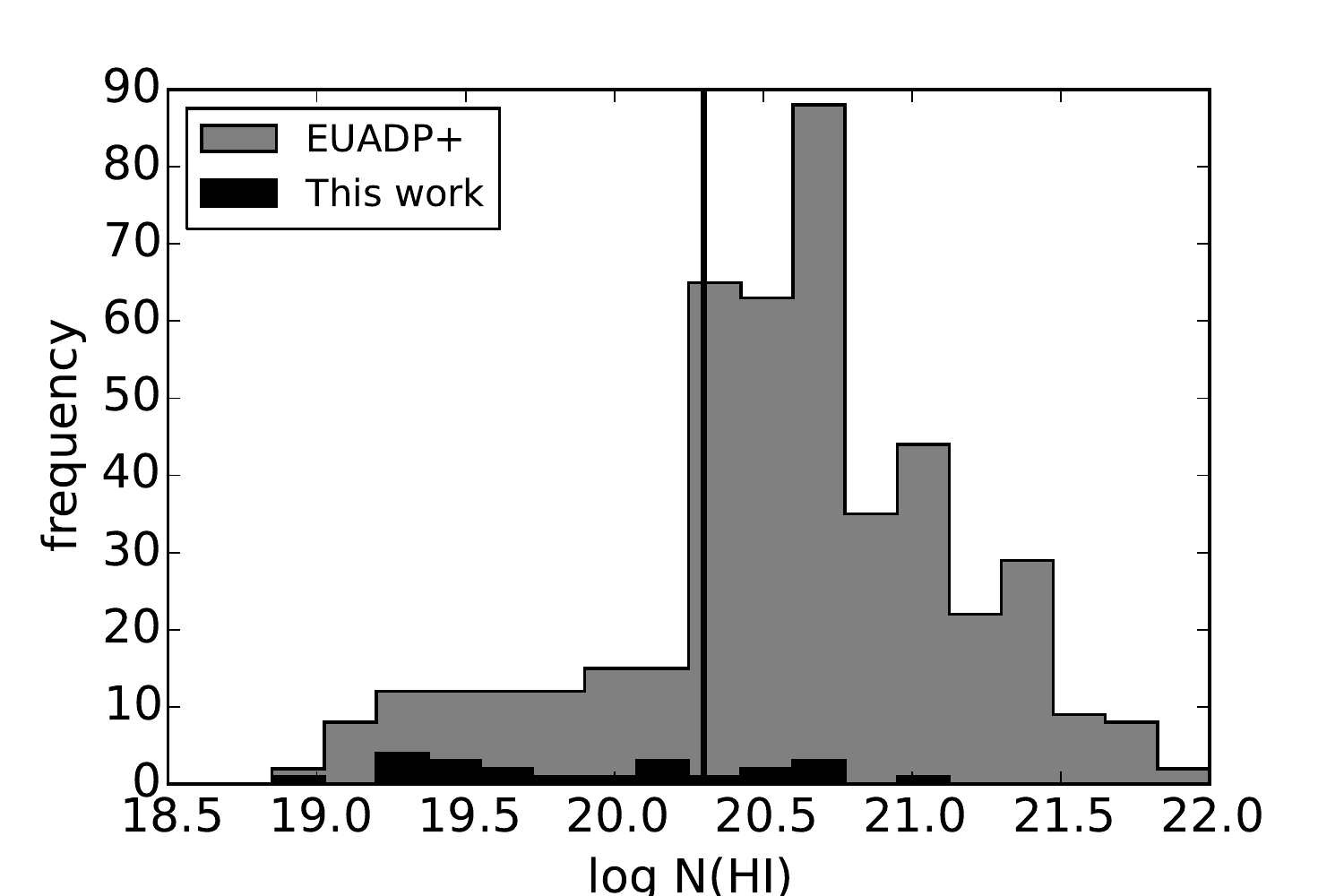}
\caption[Sample descriptive histograms]{Absorption redshifts and N(H\,I) distributions of the DLAs and sub-DLAs in our sample compared with the remaining absorbers covered by the EUADP survey and the literature (referred to as EUADP+ sample). The vertical line in the bottom panel indicates the canonical DLA definition. Clearly, the data presented here contribute most in the sub-DLA H\,I column density range at low redshift where few systems have been studied so far.}
\label{img:hist_samplelandfig}
\end{center}
\end{figure}

The table in Appendix \ref{ann:tableau} lists the metallicity estimates of the full sample of absorbers and associated references. Fig. \ref{img:hist_samplelandfig} illustrates the distribution in redshift of the absorber sample studied for both DLAs and sub-DLAs (top and middle panels respectively). The bottom panel presents the N(H\,I) distribution of the sample. 
We stress that the additional systems are consistent with the parent sample as they are not selected on their metal content or redshift but solely on their H\,I column density (see also Fig. \ref{img:M/HVSz}).

In conclusion, the final sample, referred to as the EUADP+ sample, contains 92 sub-DLAs (with 15 new measurements) and 362 DLAs (7 new measurements).
Clearly, the data presented here contribute most in the sub-DLA H\,I column density range.

\begin{table*}
\begin{minipage}{170mm}
\caption{Properties of the 22 quasar absorbers (15 subDLAs and 7 DLAs) studied here. The majority of these absorbers are from the EUADP sample \citep{Zafar2013}, but for the two systems shown in bold which have been observed recently with VLT/UVES by our group (see text for details).}
\label{table:subsample}
\begin{center}
\begin{tabular}{lccccc}
\hline\hline
QSO name & coordinates & z$_{\rm em}$ & z$_{\rm abs}$ & $\log$(N(H\,I))  & wavelength coverage (\AA) \\
\hline
QSO J0008-2900 & 2.219-29.012 & 2.645 & 2.254 & $20.22\pm0.10$ & 3300-4970,5730-10420 \\
QSO J0008-2901 & 2.24-29.024 & 2.607 & 2.491 & $19.94\pm0.11$ & 3300-4970,5730-10420 \\
\textbf{QSO J0018-0913} & 4.730-9.231 & 0.756 & 0.584 & $20.11\pm0.10$ & 3065-3875,4620-5602,5675-6650\\
QSO J0041-4936 & 10.381-49.603 & 3.24 & 2.248 & $20.46\pm0.13$ & 3290-4520,4620-5600,5675-6650 \\ 
QSO B0128-2150 & 22.773-21.58 & 1.9 & 1.857 & $20.21\pm0.09$ & 3045-3868,4785-5755,5830-6810 \\
\textbf{QSO J0132-0823} & 23.041-8.397 & 1.121 & 0.647 & $20.60\pm0.12$ & 3065-3875,4620-5602,5675-6650\\
QSO B0307-195B & 47.538-19.369 & 2.122 & 1.788 & $19.00\pm0.10$ & 3065-5758,5835-8520,8660-10420 \\
QSO J0427-1302 & 66.78-13.048 & 2.166 & 1.562 & $19.35\pm0.10$ & 3285-4515,4780-5760,5835-6810 \\
PKS 0454-220 & 74.037-21.986 & 0.534 & 0.474 & $19.45\pm0.03$ & 3050-3870,4170-5162,5230-6210 \\
J060008.1-504036 & 90.033-50.677 & 3.13 & 2.149 & $20.40\pm0.12$ & 3300-4520,4620-5600,5675-6650 \\
QSO B1036-2257 & 159.79-23.224 & 3.13 & 2.533 & $19.30\pm0.10$ & 3300-5758,5838-8525,8660-10420 \\
J115538.6+053050 & 178.911+5.514 & 3.475 & 3.327 & $21.00\pm0.10$ & 3300-5600,5675-7500,7665-9460 \\
LBQS 1232+0815 & 188.656+7.979 & 2.57 & 1.720 & $19.48\pm0.13$ & 3285-4520,4620-5600,5675-6650 \\
QSO J1330-2522 & 202.717-25.372 & 3.91 & 2.654 & $19.56\pm0.13$ & 3300-4515,4780-5757,5835-6810 \\ 
QSO J1356-1101 & 209.195-11.025 & 3.006 & 2.397 & $19.85\pm0.08$ & 3757-4985,6700-8520,8660-10420 \\
QSO J1621-0042 & 245.32-0.714 & 3.7 & 3.104 & $19.70\pm0.20$ & 3300-4515,4780-5757,5835-6810 \\
4C 12.59 & 247.938+11.934 & 1.792 & 0.531 & $20.70\pm0.09$ & 3060-3870,4780-5757,5835-6810 \\
LBQS 2114-4347 & 319.331-43.573 & 2.04 & 1.912 & $19.50\pm0.10$ & 3050-10420 \\
QSO B2126-15 & 322.3-15.645 & 3.268 & 2.638 & $19.25\pm0.15$ & 3300-5600,5675-6650,6695-8520, \\ 
... & ... & ... & 2.769 & $19.20\pm0.15$ & 8650-10420 \\ 
LBQS 2132-4321 & 324.025-43.138 & 2.42 & 1.916 & $20.74\pm0.09$ & 3290-4530,4620-5600,5675-6650 \\
QSO B2318-1107 & 350.369-10.856 & 2.96 & 1.629 & $20.52\pm0.14$ & 3050-4515,4780-5760,5840-6810\\
\hline
\end{tabular}
\end{center}
\end{minipage}
\end{table*}

\section{Analysis}

\subsection{Method}

The continua of the quasar spectra are fitted using a spline function connecting the regions of the spectrum free from absorption features as described in \citet{Zafar2013a}. The Voigt profile fits are performed with the FIT/LYMAN package within the MIDAS environment \citep{Fontana1995}. The routine calculates a $\chi^{2}$ Hessian minimization and enables fits of up to 50 free parameters including the central wavelength, the column density and the Doppler parameter of each component of the fit. This allows fitting several ions simultaneously as well as several transitions of the same species, thus making maximum use of the information available from the velocity profiles. The low-ionization species (OI, FeII, SiII, ...) are fitted as a separate group from the high-ionization species (CIV, SiIV, ...) \citep[e.g.,][]{Wolfe2005, Fox2007, Milutinovic2010, Crighton2013a}. The intermediate-ionization species AlIII are fitted either on its own, or with the low-ionization or high-ionization species, depending on the similarity in the absorption velocity profiles. 

This process allows us to identify possible blends of interloping absorbers at the positions of the features under study. In case of blending, the profiles are fitted using information on central wavelengths and Doppler parameters from other un-blended profiles, thus leading to upper limits in the column density determination. In addition, saturated transitions or components are avoided because the column density information cannot be recovered in that case. The quasar continuum solution is iteratively refined when necessary during the Voigt profile fitting process. The fits are performed minimizing the number of components. In cases where a transition is not detected, we derive a 3-$\sigma$ upper limit from an estimate of the SNR of the spectra at the expected position of the line. The laboratory wavelengths and oscillator strengths used throughout the fits are taken from \cite{Morton2003}\footnote{Recently, a new set of oscillator strengths for SII and ZnII lines has been derived for studies of the Inter-Stellar Medium (ISM), DLAs and sub-DLAs \citep{Kisielius2014, Kisielius2015}. A change from \cite{Morton2003} oscillator strengths to this new study would lower [Zn/H] by about 0.1 dex.}. 

We estimate the abundance for various elements of each absorbing system by summing the column densities of the different components found in the velocity profile described above. The metallicity $\rm [X/H]$ of an element X with respect to solar metallicity is derived from the following expression:

\begin{equation}
\rm [X/H] = \log \left(\frac{N(X)}{N(H)}\right) -  \log (X/H)_{\odot}
\end{equation}
where $\rm (X/H)_{\odot}$ is the photospheric solar abundance from \cite{Asplund2009} and $\rm N(X)$ is the column density of element X. The column density of each element is taken to be that of the dominant ion, and ionization correction is ignored here (see section \ref{sec:ionization} for further discussion on this point). The error estimate on the total column density $\rm \log N$ is calculated from the error on individual column density $\rm \log N$ of each component through the error propagation formula:

\begin{equation}
\rm \sigma_{\log (N(X))}=\frac{\sqrt{\sum_{i}(N(X)_{i}\sigma_{\log(N(X))_{i}})^{2}}}{N(X)}
\end{equation}

The global uncertainty on the abundance determination is then calculated from a quadratic sum of $\rm \sigma_{\log (N(X))}$ and $\rm \sigma_{\log (N(H))}$ since the errors in the solar abundances would introduce systematic effects which can be neglected in studies of relative abundances.

The resulting Voigt profile parameters and corresponding velocity plots for the low-, intermediate- and high-ionization species as well as a detailed description of the 22 individual systems mentioned earlier are provided in Appendix \ref{ann:individual}. The column densities and abundances derived for these systems are gathered in tables \ref{result:columndensities} (for total column densities) and \ref{result:metallicities} (for abundances).

For the different H\,I and metals column densities presented in this paper, the associated error on the abundances are based on $\chi^{2}$ minimization. The continuum placement error is not taken into account to be consistent with other measurements from the literature.

\begin{table*}
\begin{minipage}{180mm}
\caption{Total logarithmic column densities of the newly studied systems derived from the Voigt profile fits. In column N(X), (a) refers to ArI, (b) to OI, (c) to NI, (d) to TiII, (e) to CI, and (f) to CII. For PKS 0454-220, the abundances with the asterisk have been derived by Som et al. 2015.}
\label{result:columndensities}
\begin{center}
\hspace*{-.7cm}
\begin{tabular}{lcccccccc}
\hline\hline
\textbf{QSO} & \textbf{$z_{abs}$} & \textbf{N(HI)}  & \textbf{N(SII)} & \textbf{N(AlII)} & \textbf{N(SiII)} & \textbf{N(CrII)} & \textbf{N(MgI)} & \textbf{N(MgII)}   \\ \hline
QSO J0008-2900 & 2.254 & $20.22\pm0.1$ & - & - & $<14.40$ & $<12.37$ & - & $>15.01$ \\
QSO J0008-2901 & 2.491 & $19.94\pm0.11$ & $13.68\pm0.18$ & - & - & $<12.90$ & - & -   \\
\textbf{QSO J0018-0913} & 0.584 & $20.11\pm0.1$ & - & - & - & $<12.97$ & $<13.04$ & - \\
QSO J0041-4936 & 2.248 & $20.46\pm0.13$ & $<14.82$ & $>14.06$ & $14.78\pm0.03$ & $13.12\pm0.45$ & - & -  \\
QSO B0128-2150 & 1.857 & $20.21\pm0.09$ & $14.33\pm0.03$ & - & $14.82\pm0.02$ & - & $<13.21$ & -  \\
\textbf{QSO J0132-0823} & 0.647 & $20.60\pm0.12$ & - & - & - & $<13.17$ & $12.60\pm0.04$ & - \\
QSO B0307-195B & 1.788 & $19.00\pm0.10$ & - & - & $15.00\pm0.01$ & $<12.77$ & $12.54\pm0.00$ & -  \\
QSO J0427-1302 & 1.562 & $19.35\pm0.10$ & - & $11.78\pm0.10$ & - & $<12.39$ & $<12.38$ & -  \\
PKS 0454-220 & 0.474 & $19.45\pm0.03$ & $15.06\pm0.04^{*}$ & - & $>14.33^{*}$ & - & - & -  \\
J060008.1-504036 & 2.149 & $20.40\pm0.12$ & - & $>14.33$ & $15.08\pm0.01$ & $13.10\pm0.01$ & - & -  \\
QSO B1036-2257 & 2.533 & $19.30\pm0.1$ & - & $12.52\pm0.01$ & $13.64\pm0.01$ & $<12.54$ & - & $13.57\pm0.02$   \\
J115538.6+053050 & 3.327 & $21.00\pm0.1$ & $15.31\pm0.01$ & - & $15.93\pm0.01$ & - & $<13.33$ & -  \\
LBQS 1232+0815 & 1.720 & $19.48\pm0.13$ & $<14.19$ & - & $14.41\pm0.01$ & $<12.38$ & $<12.21$ & - \\
QSO J1330-2522 & 2.654 & $19.56\pm0.13$ & - & $12.18\pm0.02$ & - & - & - & -  \\
QSO J1356-1101 & 2.397 & $19.85\pm0.08$ & - & - & - & $<12.64$ & - & -   \\
QSO J1621-0042 & 3.104 & $19.70\pm0.2$ & - & - & $13.78\pm0.03$ & - & - & -  \\
4C 12.59 & 0.531 & $20.70\pm0.09$ & - & - & - & - & - & -  \\
LBQS 2114-4347 & 1.912 & $19.50\pm0.10$ & $<13.97$ & $13.00\pm0.01$ & $14.39\pm0.02$ & $<12.77$ & - & $14.40\pm0.01$   \\
QSO B2126-15 & 2.638 & $19.25\pm0.15$ & - & - & $14.67\pm0.02$ & - & - & -  \\
QSO B2126-15 & 2.769 & $19.20\pm0.15$ & - & $>14.04$ & $14.79\pm0.01$ & $<12.40$ & - & -   \\
LBQS 2132-4321 & 1.916 & $20.74\pm0.09$ & $>14.90$ & - & $15.55\pm0.01$ & $13.32\pm0.02$ & - & -  \\
QSO B2318-1107 & 1.629 & $20.52\pm0.14$ & $<14.54$ & $<14.93$ & - & $<12.47$ & $<12.37$ & -  \\
\hline
\end{tabular}
%
%

\hspace*{-.7cm}
\begin{tabular}{lcccccccc}
\hline\hline
\textbf{QSO} & \textbf{N(FeII)} & \textbf{N(NiII)} & \textbf{N(ZnII)} & \textbf{N(AlIII)} & \textbf{N(SiIV)} & \textbf{N(CIV)} & \textbf{N(MnII)} & \textbf{N(X)} \\ \hline
QSO J0008-2900 & $13.78\pm0.01$ & - & $<11.68$ & $12.39\pm0.04$ & $13.72\pm0.03$ & - & $<12.02$ & $<13.07^{(a)}$ \\ 
QSO J0008-2901 & $13.65\pm0.02$ & $<13.29$ & $<12.12$ & $<12.20$ & - & - & - & $15.31\pm0.24^{(b)}$  \\ 
\textbf{QSO J0018-0913} & $13.87\pm0.03$ & - & $<12.41$ & - & - & - & - & - \\ 
QSO J0041-4936 & $14.43\pm0.04$ & $13.07\pm0.07$ & $11.70\pm0.10$ & $12.90\pm0.01$ & - & $>14.56$ & - & $14.03\pm0.03^{(c)}$ \\ 
QSO B0128-2150 & $14.44\pm0.01$ & $13.26\pm0.05$ & $<12.26$ & $12.78\pm0.01$ & - & - & - & - \\ 
\textbf{QSO J0132-0823} & $14.96\pm0.07$ & - & - & - & - & - & - & $12.39\pm0.11^{(d)}$ \\ 
QSO B0307-195B & $14.48\pm0.00$ & $<13.22$ & $<12.18$ & - & $>14.55$ & $>15.13$ & $<12.13$ & - \\ 
QSO J0427-1302 & $12.23\pm0.04$ & $<13.23$ & $<11.75$ & - & $13.90\pm0.07$ & - & $<11.84$ & - \\ 
PKS 0454-220 & $14.71\pm0.01$ & $13.69\pm0.08^{*}$ & - & - & - & - & $12.58\pm0.01$ & - \\ 
J060008.1-504036 & $14.84\pm0.03$ & $13.62\pm0.02$ & $12.11\pm0.03$ & $12.78\pm0.01$ & - & - & - & $<12.5^{(e)}$ \\ 
QSO B1036-2257 & $12.93\pm0.01$ & $<12.93$ & $<11.74$ & - & $13.71\pm0.01$ & $>17.42$ & - & - \\
J115538.6+053050 & - & $13.74\pm0.01$ & - & $13.12\pm0.01$ & $13.56\pm0.01$ & $13.71\pm0.01$ & - & - \\ 
LBQS 1232+0815 & $13.50\pm0.01$ & $<13.05$ & $<11.58$ & $13.28\pm0.01$ & $>14.67$ & - & - & - \\ 
QSO J1330-2522 & - & $<13.22$ & - & $12.62\pm0.02$ & - & - & - & - \\ 
QSO J1356-1101 & $13.44\pm0.01$ & $<12.76$ & $<12.38$ & - & - & - & $<12.07$ & - \\ 
QSO J1621-0042 & $13.30\pm0.04$ & - & - & - & $14.24\pm0.03$ & $14.71\pm0.01$ & - & $<14.41^{(f)}$ \\ 
4C 12.59 & $14.26\pm0.08$ & - & - & - & - & - & - & - \\ 
LBQS 2114-4347 & $14.02\pm0.01$ & $<12.88$ & $<12.17$ & $<12.09$ & $13.43\pm0.01$ & $14.39\pm0.01$ & $<12.24$ & - \\ 
QSO B2126-15 & $14.05\pm0.01$ & $13.15\pm0.01$ & $<11.58$ & $13.24\pm0.02$ & - & - & - & - \\  
QSO B2126-15 & $14.17\pm0.00$ & - & $<11.95$ & $13.11\pm0.01$ & $13.84\pm0.13$ & - & $<12.28$ & - \\ 
LBQS 2132-4321 & $15.03\pm0.02$ & $13.77\pm0.02$ & $12.66\pm0.02$ & $13.25\pm0.01$ & $14.20\pm0.01$ & - & - & - \\  
QSO B2318-1107 & $14.14\pm0.02$ & - & $<11.74$ & $12.17\pm0.02$ & - & $<14.10$ & $11.78\pm0.04$ & - \\ 
\hline
\end{tabular}

\end{center}
\end{minipage}
\end{table*}
%
 

\begin{table*}
\begin{minipage}{170mm}
\caption{Abundances with respect to solar for the 22 systems studied in this work. In column [X/H], (a) refers to Ar, (b) refers to O, (c) refers to N and (d) refers to C. For PKS 0454-220, the metallicities with the asterisk have been derived by Som et al. 2015.}
\label{result:metallicities}
\begin{center}
\begin{tabular}{lcccccccc}
\hline\hline
\textbf{QSO} & \textbf{$z_{abs}$} & \textbf{log $\rm N(H\,I)$} & \textbf{[S/H]} & \textbf{[Al/H]} & \textbf{[Si/H]} & \textbf{[Cr/H]} \\  \hline
QSO J0008-2900 & 2.254 & $20.22\pm0.10$ & - & - & $<-1.33$ & $<-1.49$ \\ 
QSO J0008-2901 & 2.491 & $19.94\pm0.11$ & $-1.38\pm0.21$ & - & - & $<-0.68$ \\ 
\textbf{QSO J0018-0913} & 0.584 & $20.11\pm0.10$ & - & - & - & $<-0.78$ \\ 
QSO J0041-4936 & 2.248 & $20.46\pm0.13$ & $<-0.75$ & $>-0.85$ & $-1.19\pm0.16$ & $-0.98\pm0.58$ \\ 
QSO B0128-2150 & 1.857 & $20.21\pm0.09$ & $-1.00\pm0.09$ & - & $-0.90\pm0.09$ & - \\ 
\textbf{QSO J0132-0823} & 0.647 & $20.60\pm0.12$ & - & - & - & $<-1.07$ \\ 
QSO B0307-195B & 1.788 & $19.00\pm0.10$ & - & - & $0.49\pm0.10$ & $<0.13$ \\ 
QSO J0427-1302 & 1.562 & $19.35\pm0.10$ & - & $-2.02\pm0.14$ & - & $<-0.60$ \\ 
PKS 0454-220 & 0.474 & $19.45\pm0.03$ & $0.49\pm0.04^{*}$ & - & $>-0.78^{*}$ & - \\ 
J060008.1-504036 & 2.149 & $20.40\pm0.12$ & - & $>-0.52$ & $-0.83\pm0.12$ & $-0.94\pm0.12$ \\ 
QSO B1036-2257 & 2.533 & $19.30\pm0.10$ & - & $-1.24\pm0.10$ & $-1.17\pm0.10$ & $<-0.40$ \\ 
J115538.6+053050 & 3.327 & $21.00\pm0.10$ & $-0.81\pm0.10$ & - & $-0.58\pm0.10$ & - \\ 
LBQS 1232+0815 & 1.720 & $19.48\pm0.13$ & $<-0.41$ & - & $-0.58\pm0.13$ & $<-0.74$ \\ 
QSO J1330-2522 & 2.654 & $19.56\pm0.13$ & - & $-1.83\pm0.13$ & - & - \\ 
QSO J1356-1101 & 2.397 & $19.85\pm0.08$ & - & - & - & $<-0.85$ \\ 
QSO J1621-0042 & 3.104 & $19.70\pm0.20$ & - & - & $-1.43\pm0.20$ & - \\ 
4C 12.59 & 0.531 & $20.70\pm0.09$ & - & - & - & - \\ 
LBQS 2114-4347 & 1.912 & $19.50\pm0.10$ & $<-0.65$ & $-0.95\pm0.10$ & $-0.62\pm0.10$ & $<-0.37$ \\ 
QSO B2126-15 & 2.638 & $19.25\pm0.15$ & - & - & $-0.09\pm0.15$ & - \\ 
QSO B2126-15 & 2.769 & $19.20\pm0.15$ & - & $>0.39$ & $0.08\pm0.15$ & $<-0.44$ \\ 
LBQS 2132-4321 & 1.916 & $20.74\pm0.09$ & $>-0.96$ & - & $-0.70\pm0.10$ & $-1.06\pm0.11$ \\ 
QSO B2318-1107 & 1.629 & $20.52\pm0.14$ & $<-1.10$ & $<-0.04$ & - & $<-1.69$ \\ 
\hline
\end{tabular}
\begin{tabular}{lcccccc}
\hline\hline
\textbf{QSO} & \textbf{[Fe/H]} & \textbf{[Ni/H]} & \textbf{[Zn/H]} & \textbf{[Mg/H]}   & \textbf{[Mn/H]} & \textbf{[X/H]} \\ \hline
QSO J0008-2900 & $-1.94\pm0.10$ & - & $<-1.10$ & $>-0.81$ & $<-1.63$ & $<-1.55^{(a)}$ \\ 
QSO J0008-2901 & $-1.79\pm0.13$ & $<-0.87$ & $<-0.38$ & - & - & $-1.32\pm0.35^{(b)}$ \\ 
\textbf{QSO J0018-0913} & $-1.74\pm0.10$ & - & $<-0.26$ & - & - & - \\ 
QSO J0041-4936 & $-1.54\pm0.14$ & $-1.61\pm0.20$ & $-1.32\pm0.16$ & - & - & $-2.36\pm0.13^{(c)}$ \\ 
QSO B0128-2150 & $-1.27\pm0.09$ & $-1.17\pm0.10$ & $<-0.51$ & - & - & - \\ 
J013209-082349 & $-1.14\pm0.14$ & - & - & - & - & - \\ 
QSO B0307-195B & $-0.02\pm0.10$ & $<0.00$ & $<0.62$ & - & $<-0.30$ & - \\ 
QSO J0427-1302 & $-2.62\pm0.11$ & $<-0.34$ & $<-0.16$ & - & $<-0.94$ & - \\ 
PKS 0454-220 & $-0.24\pm0.03$ & $0.02\pm0.09^{*}$ & - & - & $-0.30\pm0.03$ & $-1.34\pm0.09^{*(c)}$ \\ 
\textbf{QSO J0132-0823} & $-1.06\pm0.12$ & $-1.00\pm0.12$ & $-0.85\pm0.12$ & - & - & - \\ 
QSO B1036-2257 & $-1.87\pm0.10$ & $<-0.59$ & $<-0.12$ & $-1.33\pm0.10$ & - & - \\ 
J115538.6+053050 & - & $-1.48\pm0.10$ & - & - & - & - \\ 
LBQS 1232+0815 & $-1.48\pm0.13$ & $<-0.65$ & $<-0.46$ & - & - & - \\ 
QSO J1330-2522 & - & $<-0.56$ & - & - & - & - \\ 
QSO J1356-1101 & $-1.91\pm0.08$ & $<-1.31$ & $<-0.03$ & - & $<-1.21$ & - \\ 
QSO J1621-0042 & $-1.90\pm0.20$ & - & - & - & - & $<-1.72^{(d)}$ \\ 
4C 12.59 & $-1.94\pm0.12$ & - & - & - & - &$-5.77^{(d)}$ \\ 
LBQS 2114-4347 & $-0.98\pm0.10$ & $<-0.84$ & $<0.11$ & $-0.70\pm0.10$ & $<-0.69$ & - \\ 
QSO B2126-15 & $-0.70\pm0.15$ & $-0.32\pm0.15$ & $<-0.23$ & - & - & - \\ 
QSO B2126-15 & $-0.53\pm0.15$ & - & $<0.19$ & - & $<-0.35$ & - \\ 
LBQS 2132-4321 & $-1.21\pm0.11$ & $-1.19\pm0.11$ & $-0.64\pm0.11$ & - & - & - \\ 
QSO B2318-1107 & $-1.88\pm0.14$ & - & $<-1.34$ & - & $-2.17\pm0.15$ & - \\ 
\hline
\end{tabular}
\end{center}
\end{minipage}
\end{table*}

\subsection{The Ionized Fraction of sub-DLAs}
\label{sec:ionization}

Given that observationally we are sensitive to the neutral gas in quasar absorbers, it is important to quantify the fraction of gas ionized in these systems. In the DLA column density range, the ionization corrections are below the typical abundance measurement errors \citep{Vladilo2001, Dessauges-Zavadsky2003}.

The situation might differ in the sub-DLA H\,I column density range given that the lower N(H\,I) might prevent complete self-shielding from the surrounding UV background. To address this issue, \cite{Dessauges-Zavadsky2003, Meiring2007, Meiring2009, Som2013, Som15} among others studied the ionized fraction of sub-DLAs based on photo-ionization CLOUDY modeling of individual systems. These studies show that the ionized fraction of hydrogen varies greatly within the sub-DLA H\,I column density range (see e.g. Fig 4 of
\citealt{Meiring2009} and Fig. 10 of \citealt{Lehner2014}). 
Nevertheless, while sub-DLAs might have an important fraction of their gas ionized in some cases, the ionization corrections to the measured abundances for sub-DLAs are often low. The large majority of elements require an ionization correction $\epsilon<0.3$ dex, while it is negligible for FeII but important for ZnII \citep{Dessauges-Zavadsky2003}. Based on these past results and in order to be in line with abundance measurements from the literature reported here, we choose not to apply ionization correction to the new abundances presented. 
A more statistical approach is now required. To this end, Fumagalli et al. (submitted) have recently built CLOUDY model grids to establish posterior probability distribution functions for different states of the gas with a Bayesian formalism and Markov Chain Monte Carlo algorithm.
%
 While such an analysis is beyond the scope of the current paper, we plan to address these issues in further publications.

\subsection{Assessing the Dust-Content of Quasar Absorbers: a Multi-Element Analysis}
\label{sec:obsbias}

Refractory elements are easily incorporated onto dust (e.g. Fe, Cr, Ni), while volatile elements are less prone to locking up into dust grains (e.g. Zn, S). To estimate the level of depletion of a given line of sight, it is possible to compare the abundance of a volatile element with that of a refractive element. The quantity $\rm [Zn/Fe]$ is therefore an excellent tool to probe the quantity of Fe atoms locked into dust \citep{Vladilo1998}. 
Indeed, Zn is thought to behave like Fe in different stages of chemical evolution, excluding the effects of dust depletion. From studies of low metallicity stars in our Galaxy, \citep{Saito2009, Barbuy2015}, [Zn/Fe] stays steady at $\rm [Zn/Fe]\sim 0$ down to metallicities $\rm[Fe/H] = -3$ and then increases for lower values of [Fe/H].
Hence, $\rm [Zn/H]$ provides a robust metallicity indicator. Unfortunately, its low cosmic abundance and long rest-frame wavelengths make it challenging to measure in sub-DLAs, preventing from a robust dust-metallicity derivation.

Here, we propose a different approach for the study of dust depletion based on the multi-element analysis proposed by \cite{Jenkins09} to assess the level of dust in a given line of sight. \cite{Jenkins09} proposed to use the abundances of different elements (namely C, N, O, Mg, Si, P, Cl, Ti, Cr, Mn, Fe, Ni, Cu, Zn, Ge, Kr and S) to compare the dust depletion of dense neutral hydrogen systems to that of the Interstellar Medium (ISM) of our Galaxy. Using a sample of 243 sight lines in our Galaxy, he established a connection between the line of sight depletion factor $\rm F_{*}$ and the different elements' abundances of each sight line. We refer the reader to Appendix \ref{ann:jenkins} for a mathematical description of the method.

\begin{figure*}
\begin{center}
\includegraphics[width=0.49\textwidth]{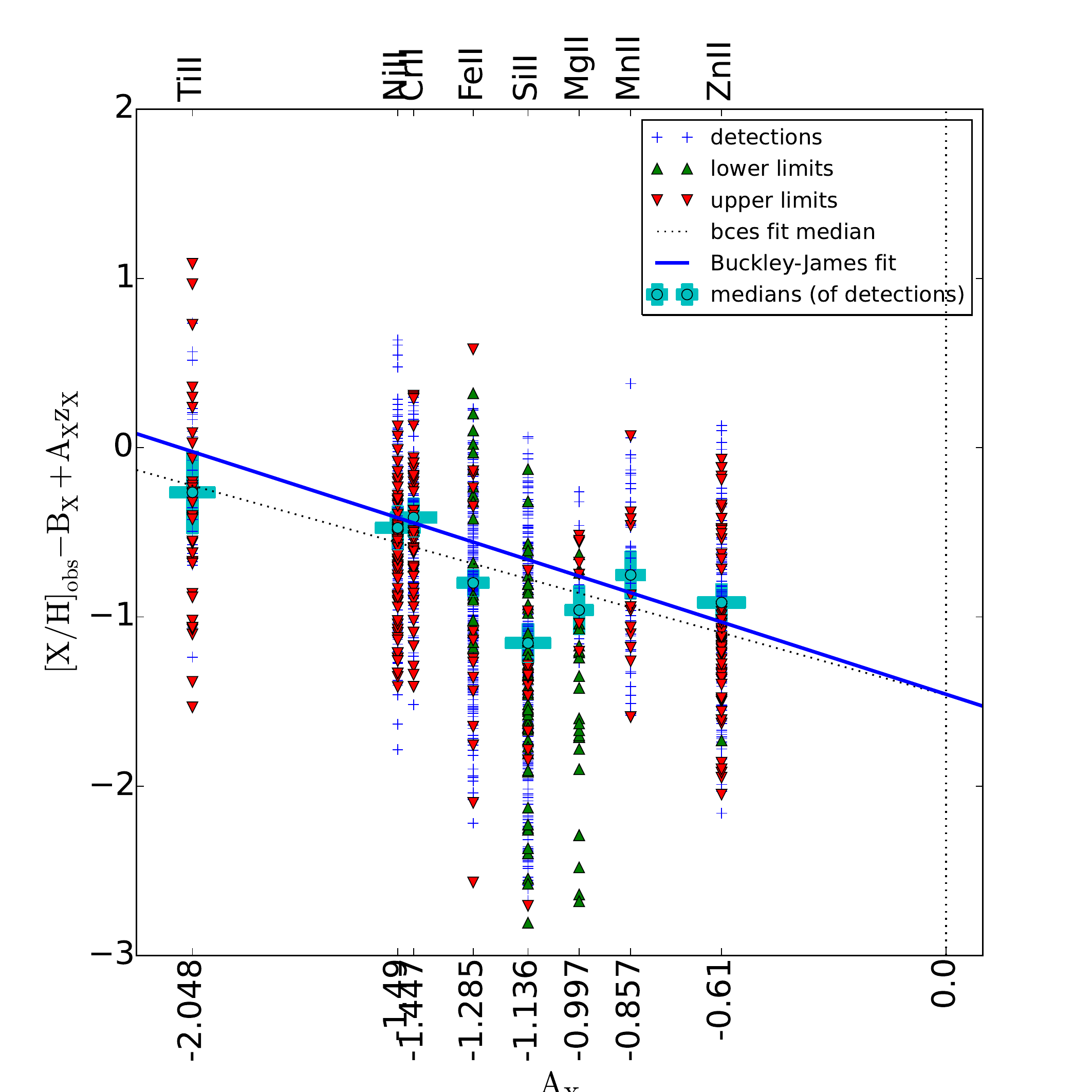}
\includegraphics[width=0.49\textwidth]{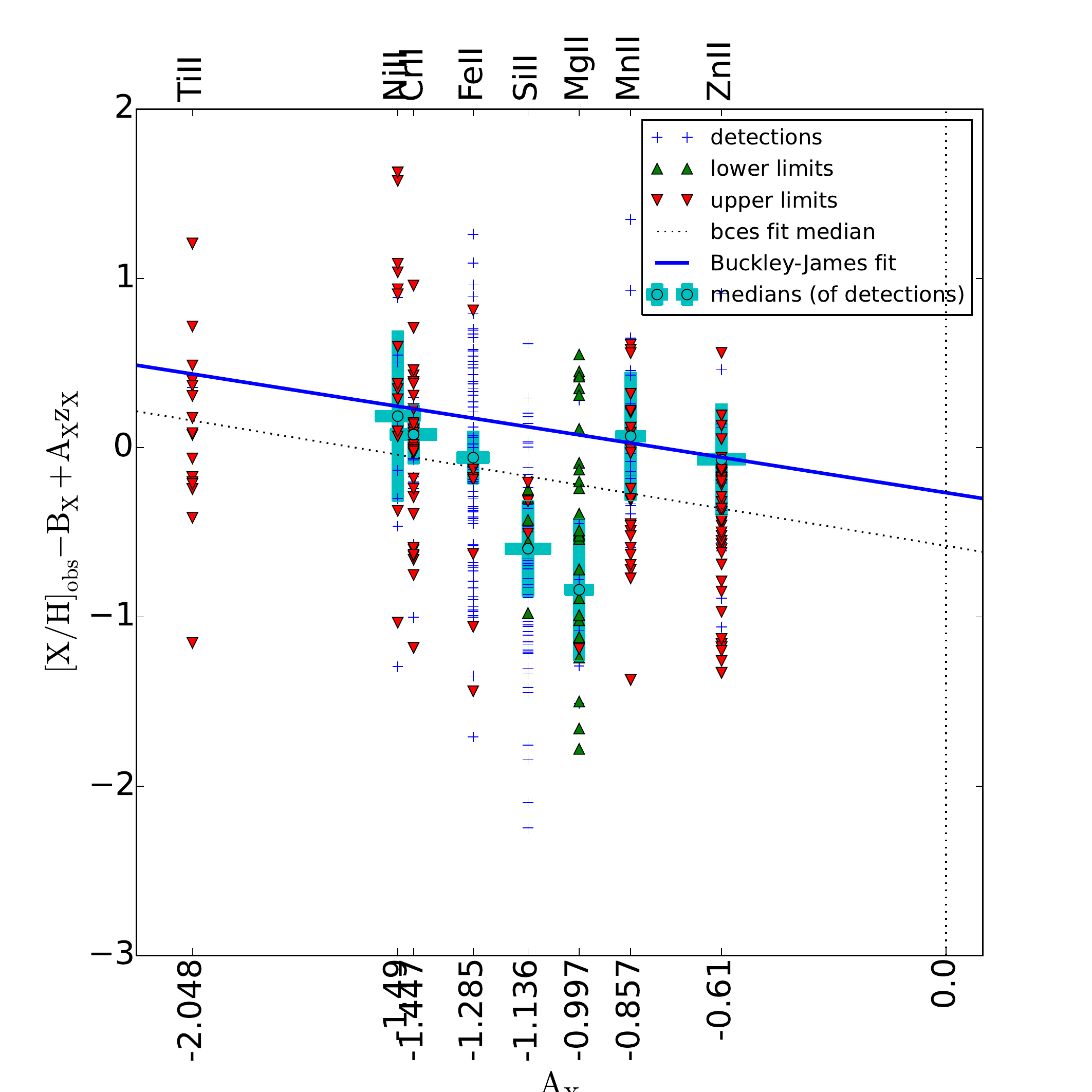}
\caption[Jenkins med]{Fits of $\rm F_{*}$ from equation \ref{eq:eq_dep_fit} for the DLAs (362 systems, left panel), and sub-DLAs (92 systems, right panel) in the EUADP+ sample. The blue crosses stand for the detections, the red triangles for the upper limits, the green triangles for lower lower limits, and the cyan points are the median of the detections for each element X. The fits are performed on the medians of the detections with a bisector fit (dashed line), and on the detections and the limits using a survival analysis technique, the Buckley-James method (solid line). We note that the $\rm \alpha$-elements (Mg and Si) are below the trend lines for both DLAs and sub-DLAs. We refer the reader to Appendix \ref{ann:jenkins} for a mathematical description of the fit.}
\label{img:exampleJenkins_med}
\end{center}
\end{figure*}

Fig. \ref{img:exampleJenkins_med} shows the fit for the line of sight depletion factor $\rm F_{*}$ (slope) for both populations of quasar absorbers from the EUADP+ sample. For each element, we plot in cyan the median of the detections if there is at least 4 systems measured. The vertical error bars represent the error on the median using a bootstrap technique with a confidence level of 95\%.

We are confronted with a large number of non-detections, creating a bias in the sample towards metal-rich systems. A large fraction of these upper (resp. lower) limits falls below (resp. above) the associated median. To address this issue, a survival analysis is considered. A Buckley-James linear regression, from the stsdas.statistics package in IRAF, results in $\rm F_{*}=-0.34\pm0.19$ for sub-DLAs and $\rm F_{*}=-0.70\pm0.06$ for DLAs.

On the one hand, both populations show negative values for $\rm F_{*}$, suggesting that sub-DLAs and DLAs arise in galaxies with a lower dust content than the Milky Way. On the other hand, the derived $\rm F_{*}$ values for both populations are different at the 1.8 $\rm \sigma$ level. The sub-DLA population is consistent with the Halo like ISM from our Galaxy\footnote{$\rm F_{*}=-0.28$ for \textit{Halo} like ISM, $\rm F_{*}=-0.08$ for \textit{Disk+Halo} like ISM, $\rm F_{*}=0.12$ for \textit{Warm Disk} like ISM and $\rm F_{*}=0.90$ for \textit{Cool Disk} like ISM} while the DLAs are described by an $\rm F_{*}$ value well below the ones measured in the Milky Way. This is counter intuitive as we expect DLAs to be self-shielded from the UV background towards the center of the galaxy. Indeed, numerous cosmological simulations predict DLAs to be closer to the center of the galaxy than sub-DLAs \citep{Fumagalli2011, Faucher-Giguere2015}. They should therefore exhibit an $\rm F_{*}$ value corresponding to regions within the halo. 

But DLAs and sub-DLAs might not be systematically associated with spiral galaxies. They might arise from a mixture of galaxy types, hence the non-physical values of $\rm F_{*}$.
In addition, the method described here is based on measurements in our Galaxy at log \NHIgt{19.5} to limit photo-ionization effects, while our quasar absorber sample goes down to $\rm \log N (H\,I) = 19.0$.
Furthermore, the ionization levels of the sub-DLAs and DLAs in our EUADP+ sample are higher than in the Milky Way ISM, as $\rm F_{*}$ is quite different between ionized and neutral gas \citep[$\rm F_{*}=-0.1$ for the warm ionized medium and $\rm F_{*}=0.1$ for the warm neutral medium, e.g.][]{Draine2011}. 
We derive $\rm F_{*}$ for log \NHIgt{19.5} sub-DLAs, and find similar results, suggesting that ionization effects do not affect the results much.
Moreover, the quasar absorbers trace gas at high redshifts, which may differ from the Milky Way properties as a local galaxy.
Overall, these results suggest that quasar absorbers differ from the Galactic depletion patterns or alternatively have a different nucleosynthetic history.

Also, the current QSO sample may suffer from dust selection bias. Indeed, it is possible that quasars in the background of dusty absorbers are not being accounted for in current selection techniques \citep{Boisse1998}. Programs to observe reddened quasars might bring valuable insights to the dust content of quasar absorbers \citep{Maddox2012, Krogager2015, Krogager2016}. Using the analysis from \cite{Vladilo2006}\footnote{see Appendix \ref{Av} for details of the calculation}, we recover estimates for the average extinction in our quasar absorber samples to be below 0.01, in line with results from \cite{Frank2010a} or \cite{Khare2012}. This suggests that the dust reddening is not observed in the current quasar selection.

%


Given these limitations, we do not apply dust corrections to the measured abundances. 
There is work underway \citep{Tchernyshyov2015} to derive the parameters $\rm A_{X}$, $\rm B_{X}$ and $\rm z_{X}$ for the Small Magellanic Cloud, which is more in line with the expected morphological type or $\rm H_{2}$ fraction of DLAs.

\subsection{$\alpha$-elements}

The production of $\alpha$-elements (O, N, Mg, Si, S, Ti, Ca...) and Fe-peak elements (V, Cr, Mn, Fe, Co, Ni...) has different origins in the history of star formation. $\alpha$-elements are mainly created during core-collapse Type II supernovae (SNe), whereas Fe-peak elements originate mainly from thermonuclear Type Ia SNe. These two processes have different time scales, as they originate from distinct stellar populations: the Type II SNe occur from short-lived massive stars while Type Ia SNe are thought to involve binary pairs containing a white dwarf exchanging material over longer periods of time. 
Observations of different objects suggest an excess of $\alpha$-elements with respect to Fe-peak elements (from \cite{Wallerstein1962} for G-dwarf stars, to \cite{Timmes1995} for QSO absorption line systems and \cite{Rafelski2012} for DLAs).

\begin{figure}
\begin{center}
\includegraphics[width=0.45\textwidth]{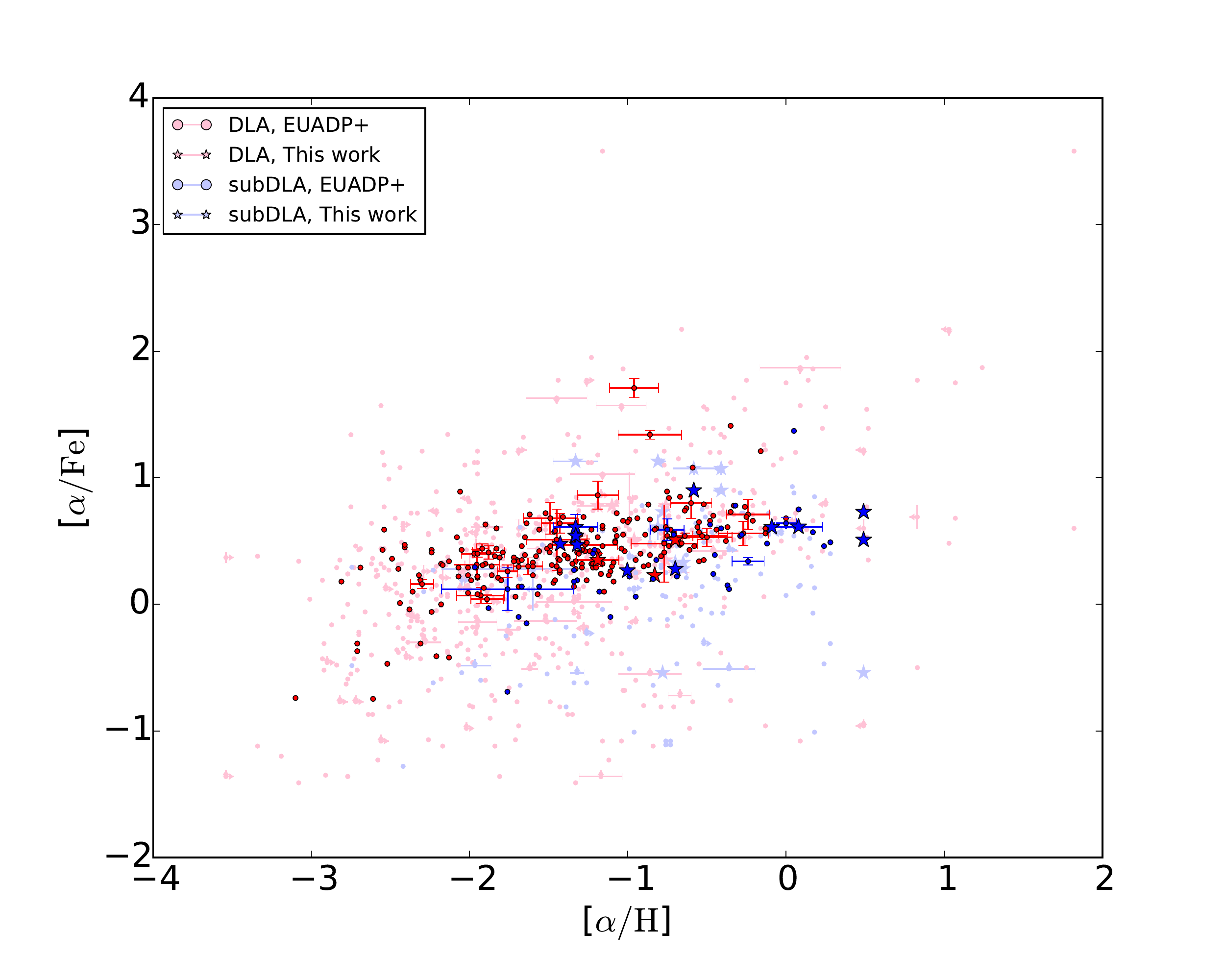}
\caption[]{The $\alpha$-enhancement of DLAs (red) and sub-DLAs (blue) versus metallicity, $\rm [\alpha/Fe]$ versus $\rm[\alpha/H]$, with $\alpha=$ OI, SII or SiII. For clarity, only one in ten data point displays error bars and the limits have faint colors.}
\label{img:alphaEnhancement_evolution}
\end{center}
\end{figure}

In Fig. \ref{img:alphaEnhancement_evolution}, we plot \aFe{} versus metallicity using $\alpha=$ OI, SII, MgII and SiII for the sub-DLA (blue) and DLA (red) populations. 

We observe a correlation between \aFe{} and $\rm [\alpha/H]$ for sub-DLAs. A Spearman test gives $\rm \rho_{sub-DLA}=0.69$ with a probability of no correlation $\rm P(\rho_{sub-DLA})<10^{-7}$. This correlation spans from low- to high-metallicity systems. 
The total number of DLA detections adds up to 227 systems. We do not see a flattening for DLAs with $\rm [\alpha/H]<-1$ as in \cite{Rafelski2012}, who attributed this flattening to the fact that the offsets in \aFe{} values for $\rm [\alpha/H]<-1$ are the effect of $\rm \alpha$-enhancement only. To avoid any dust extinction effects, we consider the 80 DLAs with $\rm [\alpha/H]<-1.5$ to derive a correction for $\rm \alpha$-enhancement for the EUADP+ DLAs \citep{DeCia2013}. We note that these systems might still be affected by dust depletion, as a trend is still visible between \aFe{} and $\rm [\alpha/H]$, even for $\rm [\alpha/H]$ down to $-3$ dex. 
We derive a mean value of $\rm [\alpha/Fe]=0.32\pm0.18$ using 80 DLAs with $\rm [\alpha/H]<-1.5$, which is consistent with the value found by \cite{Rafelski2012}. 

DLAs and sub-DLAs may have different nucleosynthetic histories, as they may originate from galaxies of different masses \citep{Khare2007,Kulkarni2010} and hence experience different star formation rates. Indeed, different [Mn/Fe] vs. [Zn/H] trends for DLAs and sub-DLAs suggest different nucleosynthetic histories for the two populations \citep{Meiring2007, Som15}. Therefore, one expects their $\rm \alpha$-enhancement to be statistically different and probably higher for sub-DLAs, which may experience higher star formation rates. In Fig. \ref{img:alphaEnhancement_evolution}, there is no apparent plateau for sub-DLAs, probably due to the small number of detections at low metallicities. More observations of sub-DLAs are needed to obtain more definitive conclusions in this H\,I column density regime. Nevertheless, to address the question of $\rm \alpha$-enhancement for sub-DLAs at least partly, we make use of the value derived for DLAs.
These corrections add $0.32\pm0.18$ dex to every metallicity derived using element Fe. This doesn't include a correction for dust extinction.



We emphasize that such a trend of \aFe{} versus $\rm [\alpha/H]$ in DLAs/sub-DLAs does not necessarily imply nucleosynthetic $\alpha$-enhancement. This is because of the increasing dust depletion of Fe with increasing metallicity, a trend that is seen to hold even at metallicities below -1 dex.
$\rm [\alpha/Zn]$ is indeed less prone to depletion than \aFe{}, but our current sub-DLA sample has only a limited number of Zn detections (19/92).
Additional Zn observations in the future will help address this question better.

\section{Results}

\subsection{Evolution of metals with redshift}

Together, DLA and sub-DLA populations contain the majority of the neutral gas mass in the Universe \citep{Zafar2013a}. Therefore, they present a valuable tool to estimate the cosmic metallicity throughout the ages.
Models of cosmic chemical evolution claim that the global interstellar metallicity would rise with decreasing redshift, to reach near solar metallicity values at present day \citep{Lanzetta1995,Pei1995,Malaney1996,Pei1999,Tissera2001}. Sub-DLAs in particular contribute substantially to the cosmic metal budget. Indeed, \cite{Kulkarni2007} show that the contribution of sub-DLAs to the metal budget increases with decreasing redshift considering a constant relative H\,I gas in DLAs and sub-DLAs at low and high redshifts. \cite{Bouche2007} anticipate that $\lesssim17$ per cent of the metals are in sub-DLAs at $\rm z\sim2.5$ but this estimate is highly dependent on the ionized fraction of the gas. It is therefore highly important to compare sub-DLA metallicities with those of DLAs.
Our study adds 15 new measurements of sub-DLA metallicity. We chose to use ZnII as our main metallicity indicator \citep{Pettini1994b} as it is nearly undepleted onto interstellar dust. Moreover, ZnII lines are usually unsaturated, and since ZnII is the dominant ionization state in neutral regions, it does not require strong ionization correction. 
However, Zn has an overall low cosmic abundance and the stronger lines $\lambda\lambda$ 2026 and 2062 can be blended with MgI $\lambda$ 2026 and CrII $\lambda$ 2062. When these ZnII lines are undetected, we use the dominant ions of other elements in the following order: OI, SII, SiII, MgII, FeII (corrected for the $\rm \alpha$ analysis) and NiII.

\begin{figure}
\begin{center}
\includegraphics[width=0.49\textwidth]{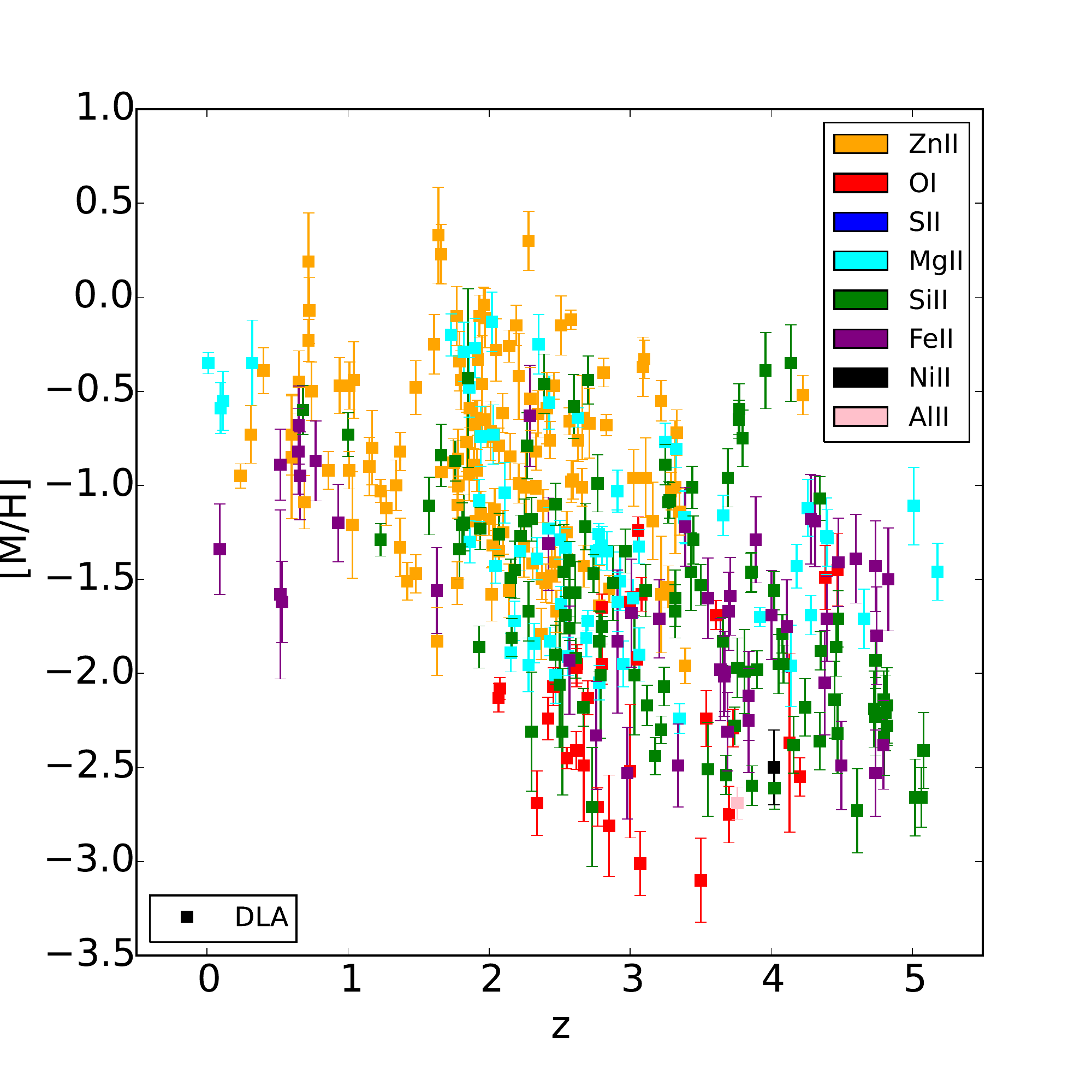}
\includegraphics[width=0.49\textwidth]{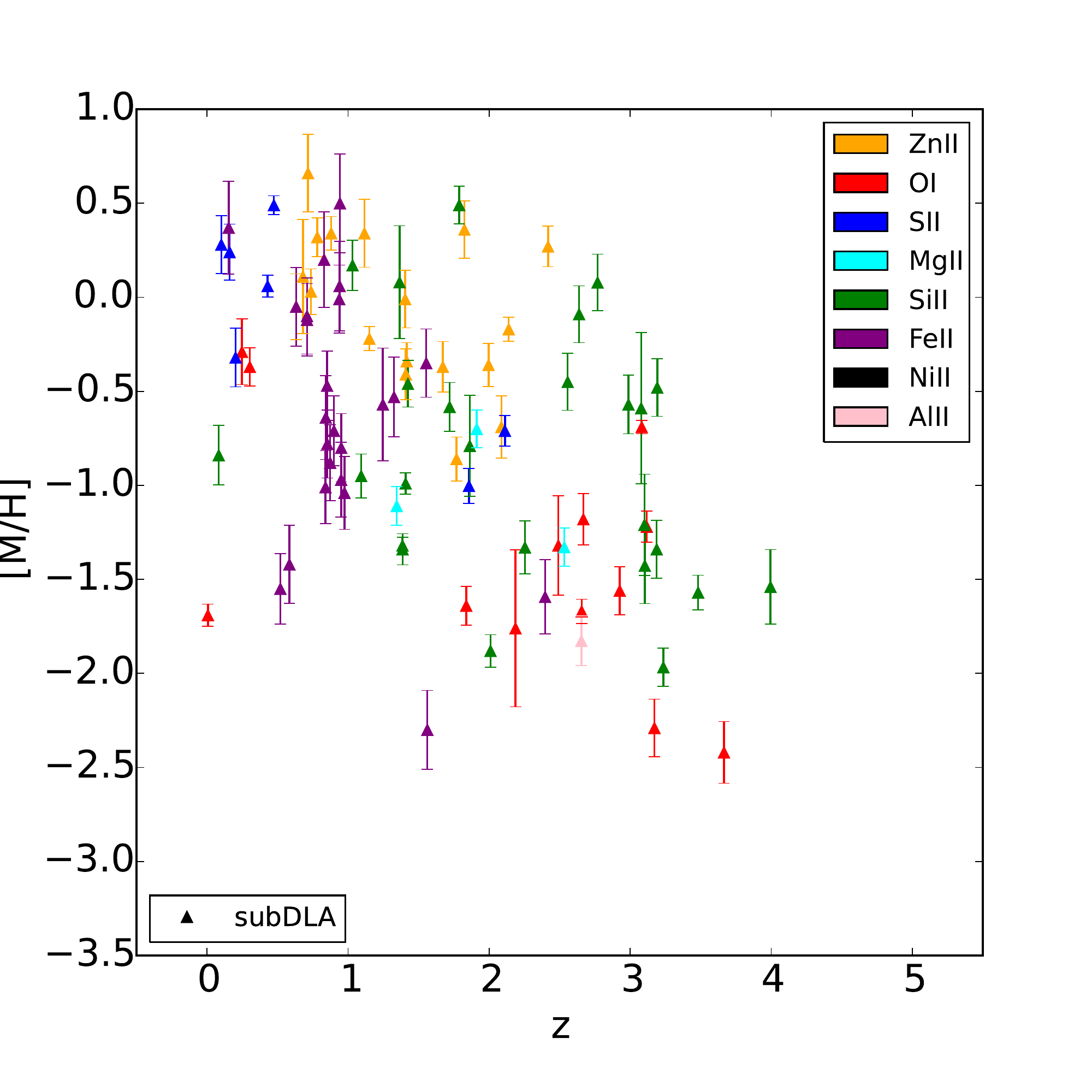}
\caption[Evolution of metallicity with redshift]{Evolution of [M/H] with redshift, color-coded with respect to the element used to derive the metallicity for DLAs (top panel) and sub-DLAs (bottom panel) from the EUADP+ sample. } 
\label{img:M/HVSzcolor}
\end{center}
\end{figure}

Fig. \ref{img:M/HVSzcolor} shows the evolution of the metallicity [M/H] with redshift of the systems for the 92 sub-DLAs (bottom panel) and the 362 DLAs (top panel) from the EUADP+ sample, color-coded with respect to the element used to derive the metallicity. We note that Zn is only detected up to $\rm z = 3$ (but for one DLA measured at $\rm z\sim4$), and O is only derived for metal-poor systems ($\rm [M/H] <-1$) because OI $\lambda$ 1302, the only OI line usually accessible to ground-based telescopes, is saturated otherwise.

\cite{Lanzetta1995} estimated the cosmic metallicity from the gas mass density $\Omega_{g}$ and metal mass density $\Omega_{m}$ via the H\,I-weighted mean metallicity \meanZ{}:
\begin{equation}
\rm <Z(z)>=\Omega_{m}(z)/\Omega_{g}(z)=\frac{\sum_{i}Z_{i}N(H\,I)_{i}}{\sum_{i}N(H\,I)_{i}}
\label{eq:HI-weight_Z}
\end{equation}

\begin{figure}
\begin{center}
\includegraphics[width=0.49\textwidth]{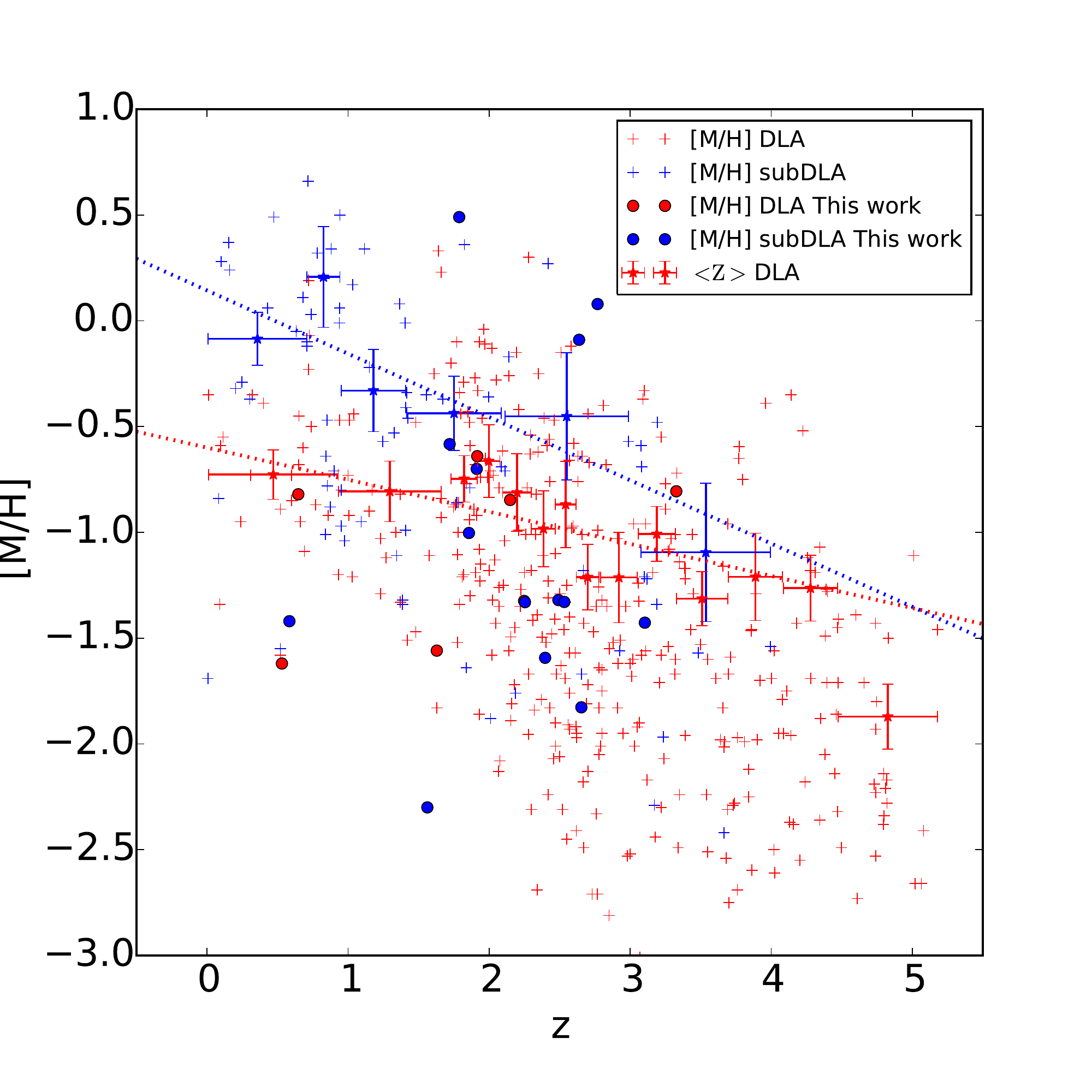}
\caption[Evolution of metallicity with redshift]{Evolution of the N(HI)-weighted mean metallicity $\rm \mean{Z}$ with redshift. Clearly, both DLA and sub-DLA populations show an increase of \meanZ{} with decreasing redshift, but sub-DLAs have a steeper evolution of \meanZ{} with redshift than DLAs. We also note a floor at $\rm [X/H]=-3$ below which no metals are detected.} 
\label{img:M/HVSz}
\end{center}
\end{figure}

Fig. \ref{img:M/HVSz} shows the metallicity derived in our sample, as well as the H\,I-weighted mean metallicity \meanZ{} for both populations (sub-DLAs in blue and DLAs in red).
 The bins for \meanZ{} are chosen such that there is an almost constant number of systems in each bin, that is 16 for the sub-DLAs and 26 for the DLAs.
The vertical error bars are derived from the consideration on sampling and measurements errors. The sampling errors are calculated from a bootstrap technique as described in \cite{Rafelski2012} and the measurement errors from the propagation formula. The total errors are the quadratic sums of these two quantities for each bin.
We note a large scatter for a third of the newly derived sub-DLA metallicities (blue dots). This points out to the need for a larger sample of sub-DLA measurements at all redshifts.

We measure an anti-correlation between redshift and metallicity for both populations. The Spearman coefficient for sub-DLAs is $\rm \rho=-0.49$ and $\rm \rho=-0.55$ for DLAs, with probabilities of no correlation $\rm P(\rho)<10^{-6}$ for both populations. The Kendall's $\rm \tau$ is $-0.34$ for sub-DLAs and $-0.38$ for DLAs, with a probability of no-correlation of below $10^{-5}$ for both populations. The dotted lines in Fig. \ref{img:M/HVSz} show the best bisector fits for the metallicity evolution with redshift of both populations. We measure\begin{equation}
\rm <Z>_{DLAs}=(-0.15\pm0.03)~ z  - (0.6\pm0.13)
\end{equation} 

\begin{equation}
\rm <Z>_{sub-DLAs}=(-0.30\pm0.07) ~z + (0.15\pm0.31)
\end{equation}
The fit has been performed shifting the y-axis to $\rm z=3$ to minimize the error on the intercept and ignoring the last DLA bin which presents a rapid decline in metallicity \citep{Rafelski2014}.
The evolution with redshift is steeper for sub-DLAs than for DLAs. 
Previous authors \citep{Khare2007, Kulkarni2010} argued that this effect might arise from the fact that sub-DLAs are more massive than DLAs.

%
%

Our results are in agreement with previous work \citep{Kulkarni2007,Kulkarni2010, Som15}, with a more significant result in the sub-DLA regime thanks to the larger sample presented here. The slope remains unchanged with respect to earlier studies. However, the significance of the result increases indicating a convergence towards a realistic value of the slope. 

\subsection{Kinematics}

In addition to the different abundances derived from Voigt profile fitting, information on the kinematics of the absorbers can be derived from the UVES high resolution spectra.

\subsubsection{Voigt Profile Optical Depth Method}
\label{sec:deltaVmethod}

We use the definition of the velocity interval \deltav{} as defined by \cite{Prochaska1997}, based on the integrated optical depth $\tau_{tot}=\int \tau(v)dv$ and considering the velocity interval from $5\%$ to $95\%$ of this quantity.

\begin{figure}
\begin{center}

\includegraphics[width=.45\textwidth]{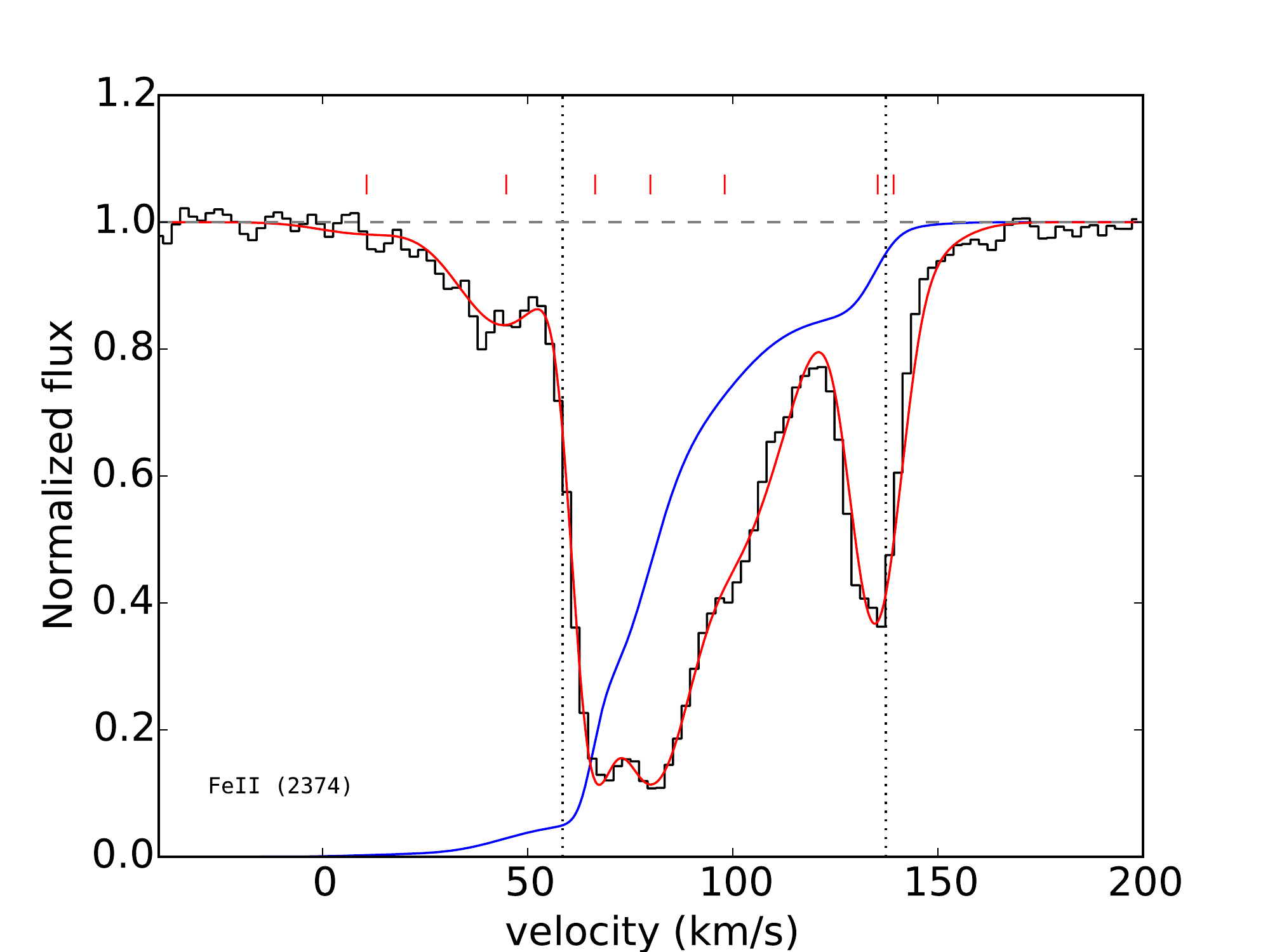}
\caption[]{An example illustrating the computation of the velocity interval \deltav{}. The black curve is the normalized spectrum of PKS 0454-220 centered on the FeII $\lambda$ 2374 line, the red curve is the Voigt profile fit of the absorption and the blue curve is the integrated optical depth derived from the Voigt profile. The vertical dotted lines indicates the 5\% and 95\% thresholds for the integrated optical depth, defining the velocity width \deltav{}.}
\label{img:VPOD}
\end{center}
\end{figure}

In this paper, we do not consider the apparent optical depth (AOD) $\tau_{app}=-\log (I/I_{c})$ to derive the velocity interval, as is usually done, but we use instead the optical depth derived from the Voigt profile fits (see appendix \ref{ann:individual} for a description of the fits for every system individually). 
This method, which we refer to as Voigt profile optical depth (VPOD) method, makes use of the information gathered from the fits. The saturation and contamination issues are then considered when deriving \deltav{}. This is the main difference with the AOD method, which might provide \deltav{} measurements affected by blends. In the VPOD method, we use simultaneously the information on several transitions to derive the velocity interval for any ion. Indeed, the only quantity that differs between transitions of the same ion is the oscillator strength, which has no impact on the velocity axis. Fig. \ref{img:VPOD} shows an example of the derivation of the velocity interval for an FeII line.
Table \ref{result:deltav90} summarizes the \deltav{} measurements for the 22 systems studied here. For 20 of them, we use the information from the FeII lines as it is the ion most detected in our sample.

One of the sub-DLA in our sample, towards PKS 0454-220, has already been studied by \cite{Som15}. They use SII $\lambda$ 1250 from an HST/COS spectrum and derive \deltav{}=155 km/s based on the AOD method. However, we find with the VPOD method described above a value almost twice smaller. We use the AOD method on the UVES spectrum with FeII $\lambda$ 2374 and derive \deltav{}$\sim$85.0 km/s, consistent with the result from the VPOD method. We note that the Line Spread Function (LSF) derived from the COS consortium is responsible for the reported large value. 
To overcome this problem, we exclude COS measurements from our analysis.

In conclusion, the VPOD \deltav{} values are not sensitive to blending and saturation effects, to the shape of the instrument's LSF, its resolution and to the SNR of the derived spectrum. We note that depletion of refractory elements contributes to the error in the \deltav{} because the different components can be affected differently by dust depletion. In the present study, FeII has been used because it is uniformly detected among the 22 new systems presented here.
In the remaining of the sample there is no object in common between the EUADP and the already derived \deltav{} found in the literature.

%
%
%
%
%
%


\subsubsection{\deltav{} versus Metallicity Relation}

\begin{figure}
\begin{center}

\includegraphics[width=.45\textwidth]{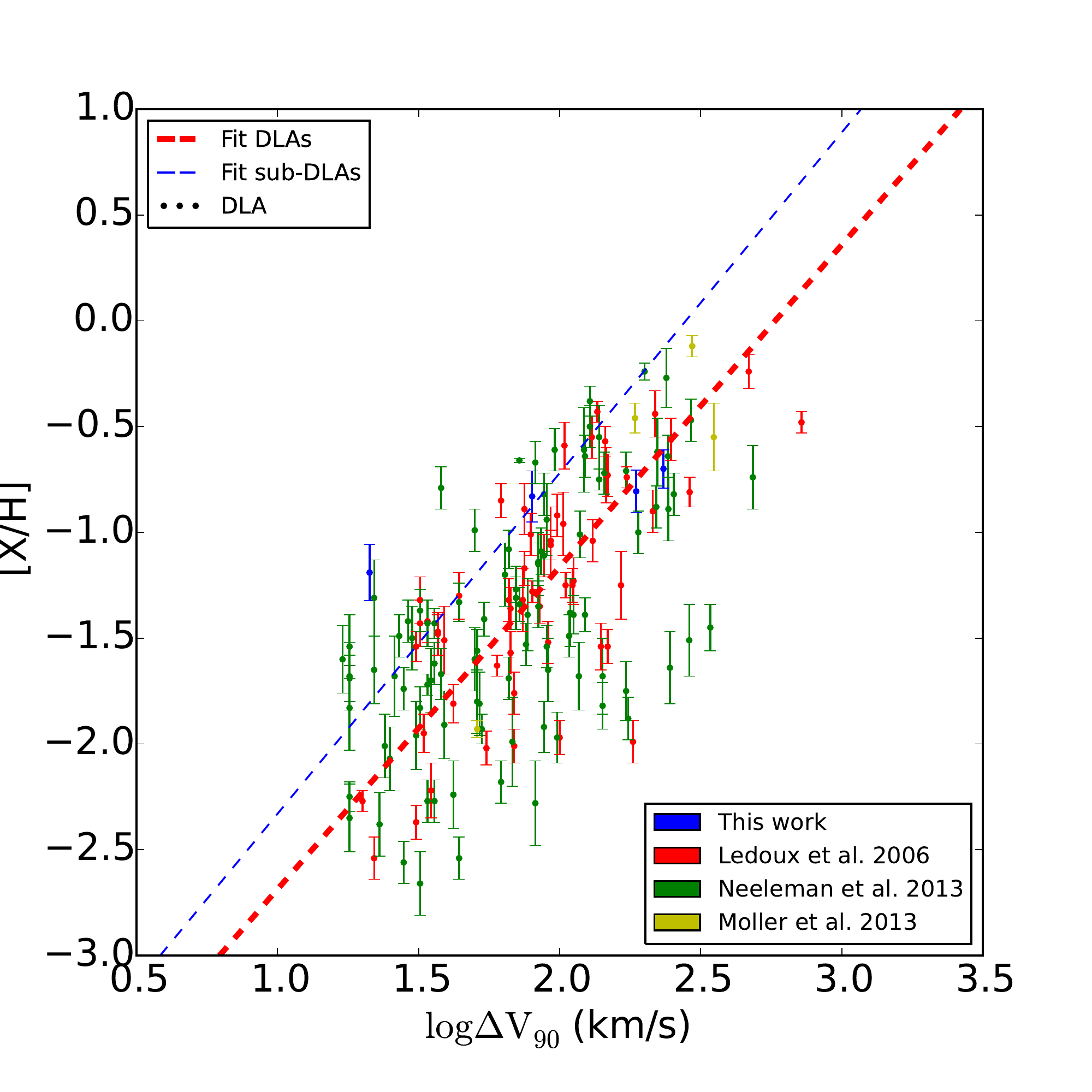}
\includegraphics[width=.45\textwidth]{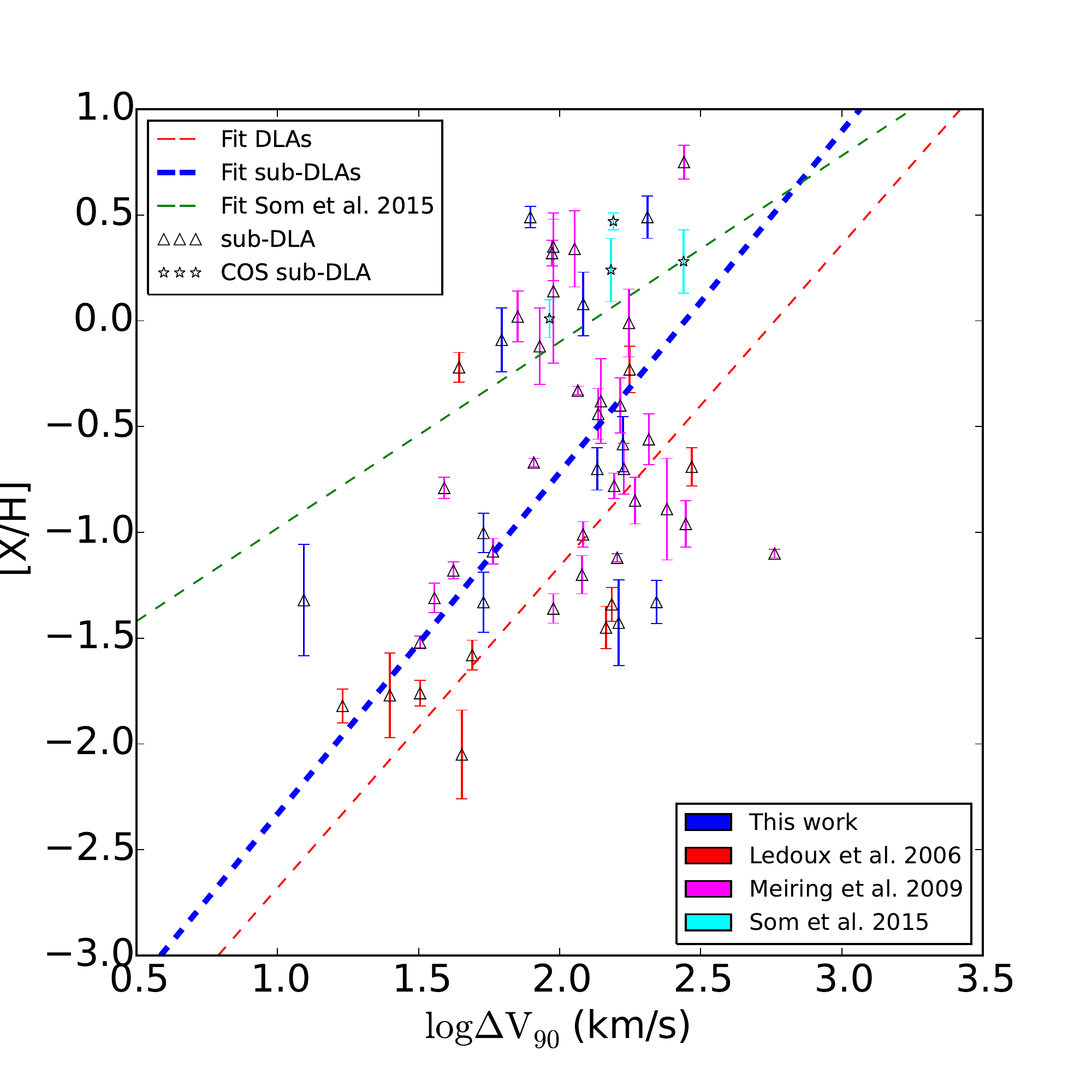}
\caption[]{[X/H] versus \deltav{} for the newly derived systems, the systems from \cite{Ledoux2006}, \cite{Neeleman2013}, \cite{Som15}, \cite{Meiring2009b}, and \cite{Moller2013} in blue, red, green, cyan, magenta and yellow, respectively. The upper panel shows the DLAs and the bottom panel the sub-DLAs. The data points measured with COS (stars) are not considered for the fit due to the discussion in section \ref{sec:deltaVmethod}. The open triangles represent the sub-DLAs and the dots the DLAs. The dashed blue line reproduces the bisector fit of the sub-DLAs, the dashed red line is the bisector fit for all the DLAs and the green dashed line in the lower panel represents the \cite{Som15} sub-DLA fit.}
\label{img:v90_M_HI}
\end{center}
\end{figure}

\begin{table}
\caption{Measures of \deltav{} in our sample derived from Voigt profile fits to the FeII lines (except for two systems with no Fe coverage, for which we used SiII (a) and AlII (b)).}
\label{result:deltav90}
\begin{center}
\begin{tabular}{lcccccccc}
\hline\hline
\textbf{QSO} & \textbf{$z_{abs}$} & log $\rm N(H\,I)$ [cm$^{-2}$] & $\Delta v_{90}$ [km/s] \\  \hline
QSO J0008-2900 & 2.254 & 20.22 & $53.7$ \\ 
QSO J0008-2901 & 2.491 & 19.94 & $12.4$ \\ 
\textbf{QSO J0018-0913} & 0.584 & 20.11 & $192.6$ \\ 
QSO J0041-4936 & 2.248 & 20.46 & $21.2$ \\ 
QSO B0128-2150 & 1.857 & 20.21 & $53.7$ \\ 
\textbf{QSO J0132-0823} & 0.647 & 20.60 & $76.5$ \\ 
QSO B0307-195B & 1.788 & 19.00 & $204.8$ \\ 
QSO J0427-1302 & 1.562 & 19.35 & $7.8$ \\ 
PKS 0454-220 & 0.474 & 19.45 & $78.7$\\ 
J060008.1-504036 & 2.149 & 20.40 & $79.9$ \\ 
QSO B1036-2257 & 2.533 & 19.30 & $220.4$ \\ 
J115538.6+053050 & 3.327 & 21.00 & $186.8^{a}$ \\ 
LBQS 1232+0815 & 1.72 & 19.48 & $167.6$ \\ 
QSO J1330-2522 & 2.654 & 19.56 & $21.0^{b}$ \\ 
QSO J1356-1101 & 2.397 & 19.85 & $337.7$ \\ 
QSO J1621-0042 & 3.104 & 19.70 & $161.9$ \\ 
4C 12.59 & 0.531 & 20.70 & $62.4$ \\ 
LBQS 2114-4347 & 1.912 & 19.50 & $135.9$ \\ 
QSO B2126-15 & 2.638 & 19.25 & $62.3$ \\ 
QSO B2126-15 & 2.769 & 19.20 & $121.2$ \\ 
LBQS 2132-4321 & 1.916 & 20.74 & $233.4$ \\ 
QSO B2318-1107 & 1.629 & 20.52 & $13.8$ \\ 
\hline
\end{tabular}
\end{center}
\end{table}

Recently, \cite{Som15} compared for the first time the sub-DLA metallicity versus velocity width trend over a statistically significant sample of 31 sub-DLAs at \zabsbetween{0.1}{3.1}. 
We propose here to extend their analysis to a wider sub-DLA sample using our new 15 sub-DLAs.
We consider a different sample than the one used in the remaining of the paper (EUADP+) as the velocity widths are not provided by all authors. We consider the data from \cite{Ledoux2006} (52 DLAs and 14 sub-DLAs at redshifts \zabsbetween{1.7}{4.3},  corrected for \cite{Asplund2009} photospheric solar abundances), 
observations from \cite{Meiring2009b} (29 sub-DLAs at redshifts \zabslt{1.5}, corrected for \cite{Asplund2009} photospheric solar abundances), 
observations from \cite{Neeleman2013} (98 DLAs at redshifts \zabsbetween{1.6613}{5.0647}), 
observations from \cite{Moller2013} (4 DLAs at redshifts \zabsbetween{1.9}{3.1}), 
as well as results from this study (see Table \ref{result:deltav90}). We only consider the systems with detected $\rm [\alpha/H]$ (11/15 sub-DLAs and 4/7 DLAs). 
The resulting sample gathers 54 sub-DLAs and 162 DLAs.

Fig. \ref{img:v90_M_HI} shows the trend between the metallicity and the velocity width \deltav{} for sub-DLAs (bottom panel) and DLAs (top panel) from this \deltav{} sample. 

We fit both populations with the best bisector fits:
\begin{equation}
\rm [X/H]_{DLA}=(1.52\pm0.08)~\log \Delta V_{90} -(4.20\pm0.16)
\end{equation}
\begin{equation}
\rm [X/H]_{sub-DLA}=(1.61\pm0.22)~\log \Delta V_{90} -(3.94\pm0.45)
\end{equation}
The fits are performed shifting the y-axis to $\rm \log \Delta V_{90}=2$ to minimize the error on the intercept.
%

%
\cite{Som15} find a higher intercept and a shallower slope for the sub-DLA population, using only Zn and S with ionization corrections. Although this result is free from dust depletion effect on the metallicity estimation, it might be biased towards higher metallicity sub-DLAs, where Zn or S can be measured. A larger Zn-based metallicity sub-DLAs samples is required to recover the metallicity-\deltav{} relation free from the effects of dust bias.

A larger Zn-based metallicity sub-DLAs samples is required to recover the metallicity-Delta v relation free from the effects of dust bias.

As in previous studies, the DLA sample is well correlated. The Spearman coefficient is  $\rm \rho=0.63$, with a probability of no correlation $<10^{-6}$. The Kendall's $\tau$ is $0.46$ with a probability of no correlation $<10^{-6}$.
The sub-DLA sample is less correlated than the DLA sample, in agreement with \cite{Som15}. The Spearman coefficient is $\rm \rho=0.39$, with a probability of no correlation of $\rm 0.004$. 
The Kendall's $\tau$ is $0.25$ with a probability of no correlation of $0.007$.
Adding more sub-DLAs to the \deltav{}-metallicity relation does not improve the correlation. \cite{Som15} showed that the ionization correction also does not improve this correlation.
This indicates a larger spread for the sub-DLAs, which may originate from more complex kinematic behaviors in sub-DLA clouds. 
However, the determination of \deltav{} appears to be sensitive to the resolution, the LSF and the SNR of the data. This might contribute to the observed scatter, although an intrinsic scatter is expected from CGM regions.

\subsubsection{Is \deltav{} a Reliable Tracer of Mass?}

A mass-metallicity relation (hereafter MZR) has been reported at low redshifts \citep{Lequeux1979, Tremonti2004}, intermediate redshifts \citep{Savaglio2005} and high redshifts \citep{Erb2006}. It relates the stellar mass of galaxies to the metallicity of their ISM. This relation is crucial in our understanding of galaxy evolution as it supports the theory of metal ejection from galactic outflows in low-mass (and hence low potential well) galaxies and their enrichment with accreting metal-poor IGM gas, diluting the galactic metallicity.

For quasar absorbers, some simulations indicate that the origin of the velocity width, \deltav{}, could be strongly related to the gravitational potential well of the absorption system's host galaxy \citep[e.g.][]{Prochaska1997, Haenelt1998, Pontzen2008a}. Similarly, assuming a scaling of the galaxies luminosity with dark matter haloes, \cite{Ledoux2006} and later \cite{Moller2013} proposed to interpret the \deltav{} versus metallicity relation of quasar absorbers as a MZR. Such a picture does not take into account the complex gas processes at play now known to take place in CGM regions. In other words, the \deltav{} may reflect bulk motions of the absorbing gas rather than motions governed by the gravitational potential well.

Observationally, a measurement of mass and \deltav{} has been possible in few individual systems. Infra-red IFU SINFONI observations of the galaxy hosts of 3 DLAs and 2 sub-DLAs in \cite{Peroux2011} and \cite{Peroux2014} allow one to determine the mass of the systems from a detailed kinematic study. 
In addition, \cite{Christensen2014} has used photometric information of the galaxy hosts and Spectral Energy Distribution (SED) fits to estimate the stellar mass of 13 DLAs. Combined together, these findings suggest that, individually, the absorption systems align well with the MZR reported at these redshifts. 
 
In addition to these measurements in a few specific systems, several authors have put constraints on the mass estimates of quasar absorbers in a statistical manner. Interestingly, the local analogues to DLAs, the 21cm $\rm z=0$ emitting galaxies studied with HIPASS by \cite{Zwaan2008} show that the quantity \deltav{} correlates little with mass. Similarly, \cite{Bouche2007} \citep[and later][]{Lundgren2009, Gauthier2014} have used the ratio of MgII systems auto-correlation with a correlation with Luminous Red Galaxies (LRG) at the same redshifts to derive an estimate of the overall mass of quasar absorbers. Their findings show an anti-correlation between equivalent width, a proxy for \deltav{}, and metallicity. Admittedly, the populations of absorbers do not completely overlap, the \cite{Ledoux2006} sample contains mostly DLAs, while the MgII sample of \cite{Bouche2007} might have at most 25\% of DLAs (according to the criterion of \cite{Rao2006}: 50\% meet the FeII/MgII criteria and 35-50\% of these are DLAs). In fact, \cite{Bouche2007} and \cite{Schroetter2015} argue that MgII absorbers can be used to trace superwinds as they are not virialized in the gaseous halo of the host-galaxies.
\cite{Bouche2012} also show that the inclination of the galaxy has a direct impact on the absorption profile and therefore on the velocity width. Put together, these many lines of evidence question the interpretation of the velocity width as a proxy for the mass of the host galaxy and the interpretation of the \deltav{}/metallicity correlation as a MZR for quasar absorbers.

\subsection{Tracing the Circum-Galactic Medium with sub-DLAs}

\begin{figure}
\begin{center}
\includegraphics[width=.4\textwidth]{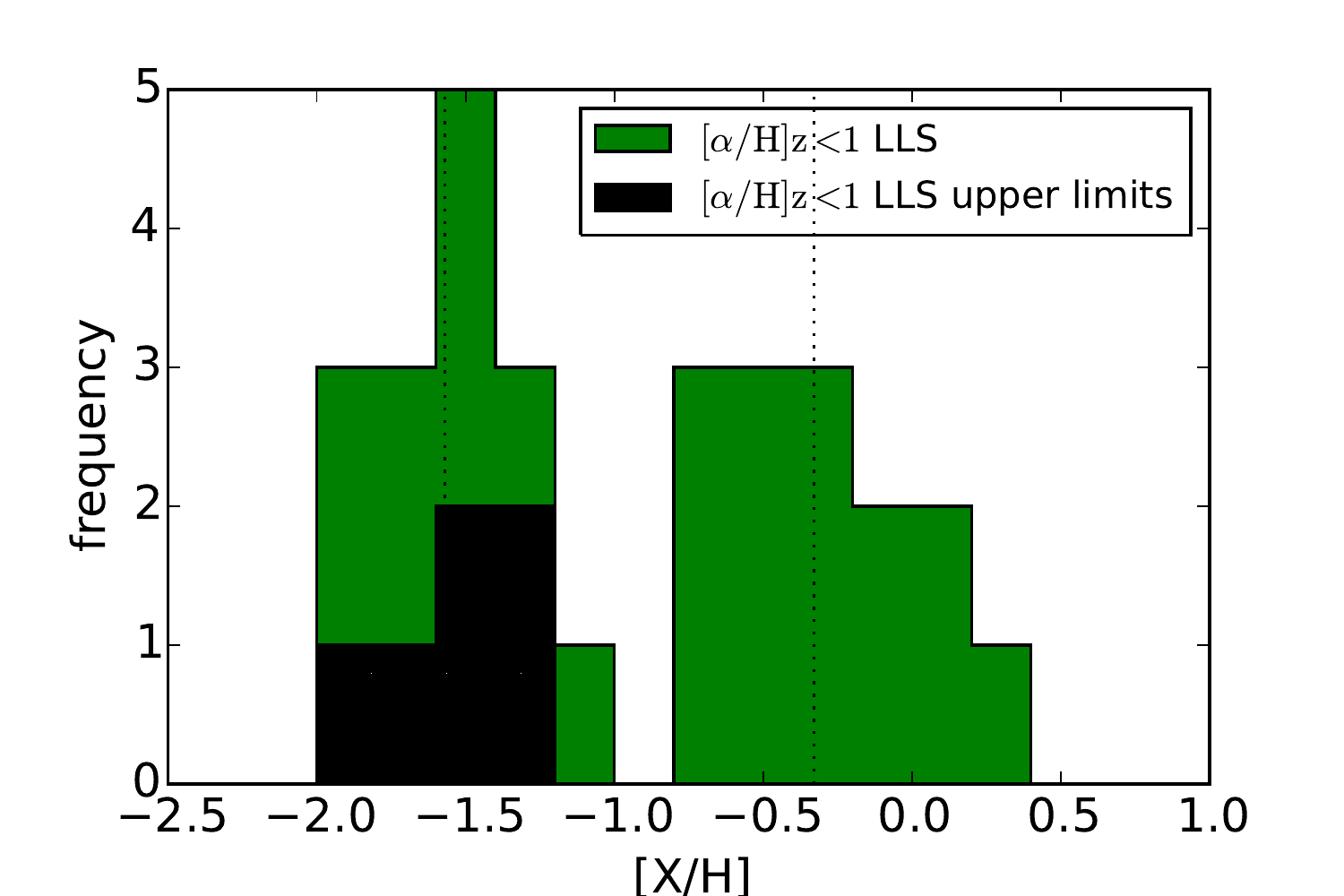}
\includegraphics[width=.4\textwidth]{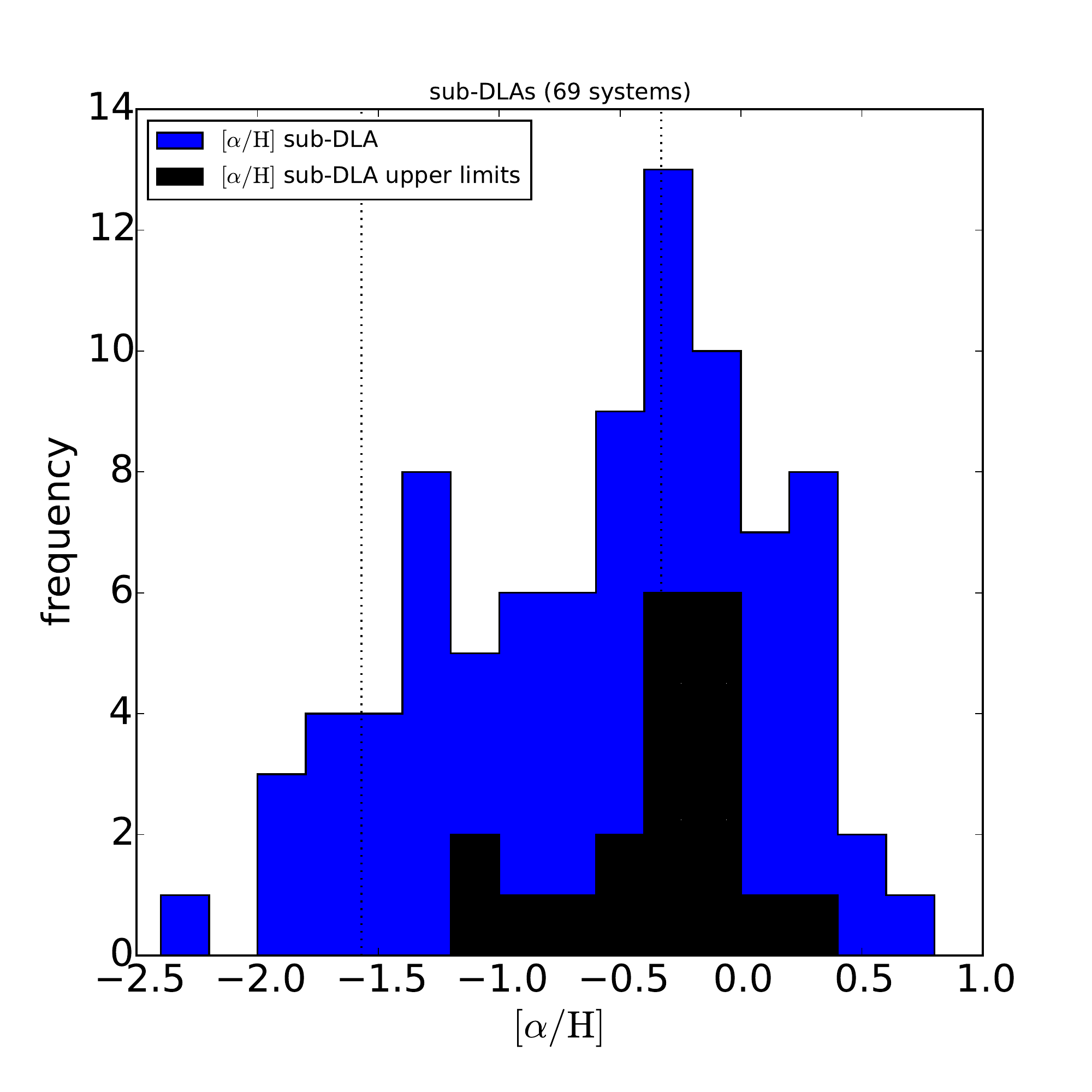}
\includegraphics[width=.4\textwidth]{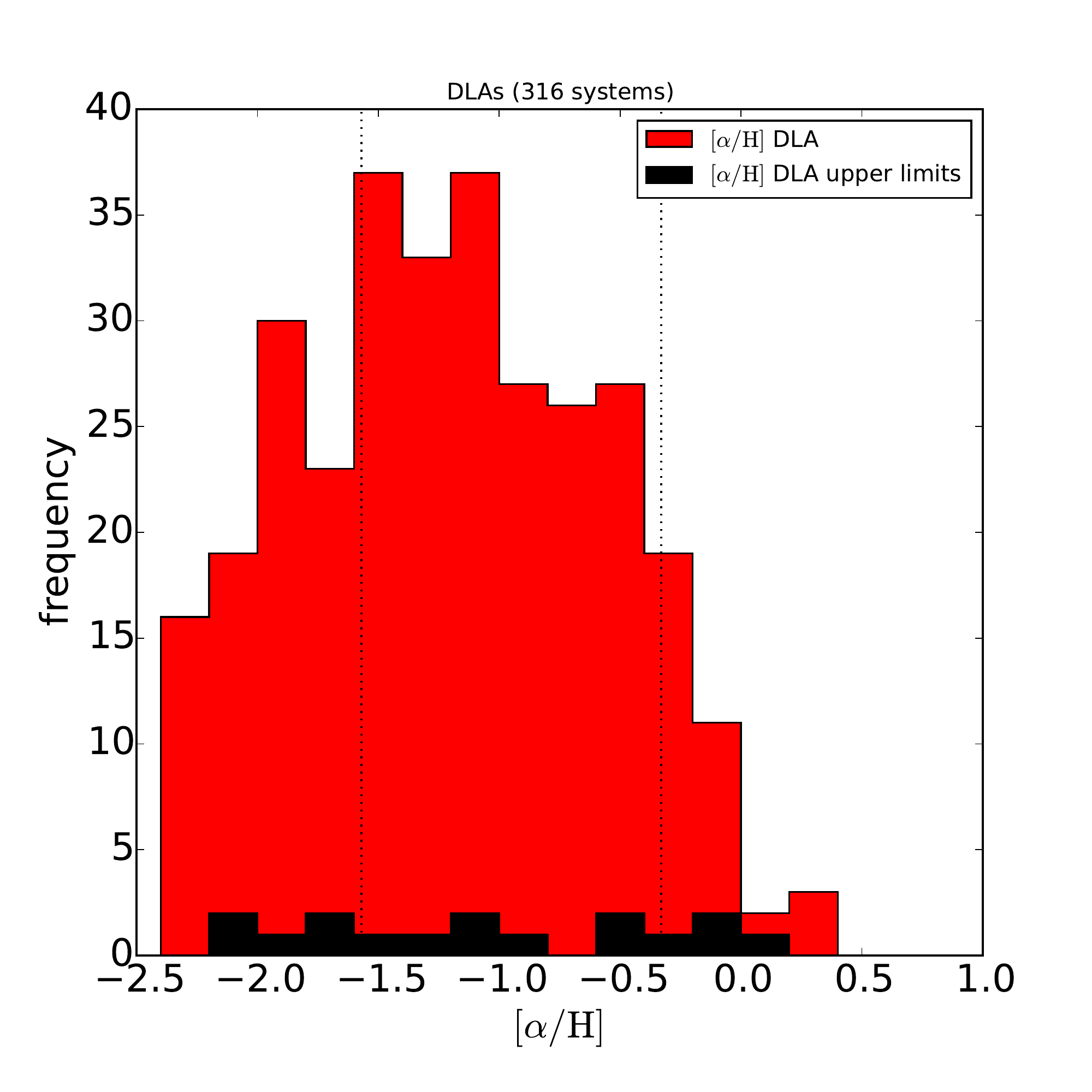}
\caption[Metallicity distribution in LLSs, sub-DLAs and DLAs]{Metallicity [$\alpha$/H] distribution of LLSs (top panel), sub-DLAs (middle panel) and DLAs (bottom panel). The histogram for LLSs has been taken from \cite{Lehner2013} and indicates a bimodality in the metallicity distribution for LLS at $\rm z<1$. The black vertical dashed lines represent the mean values derived from the \zabslt{1} LLS sub-groups by \cite{Lehner2013}.}
\label{img:bimodality_alpha}
\end{center}
\end{figure}

Our understanding of galaxy formation and evolution is tightly linked with the study of two opposite processes that take place within the CGM. Indeed, to create stars, the galaxy requires a continuous input of cold gas, that is believed to accrete along the filamentary structures from the cosmic web. In addition, cosmological simulations fail to reproduce the observed SFR without invoking feedback processes from star formation itself or AGN activity. 
These outflowing processes and their large scale impact have been confirmed observationnally \citep{Steidel2010,Bouche2012,Kacprzak2014}, but there is still little observational evidence for accretion of cool material \citep{Bouche2013a, Cantalupo2014, Martin2014a}.
Quasar absorbers with H\,I column densities in the range of LLS and sub-DLAs are believed to be good probes of this CGM \citep{Fumagalli2011, VandeVoort2012a}. \cite{Lehner2013, Lehner2014} report a bimodality in the metallicity distribution of 29 $z<1$ LLS, which they interpret as the signatures of outflows (metal-rich) and infalls (metal-poor).

\cite{Lehner2013} extended their analysis on 29 sub-DLAs and 26 DLAs, but do not report a bimodality distribution in the metallicity of these systems based on $\rm \alpha$-elements. Clearly, larger samples of quasar absorbers are required to perform such studies. 

Here, we perform similar analysis on a larger sample of sub-DLAs and DLAs with a broad redshift range.

\begin{figure*}
\begin{center}
\includegraphics[width=.4\textwidth]{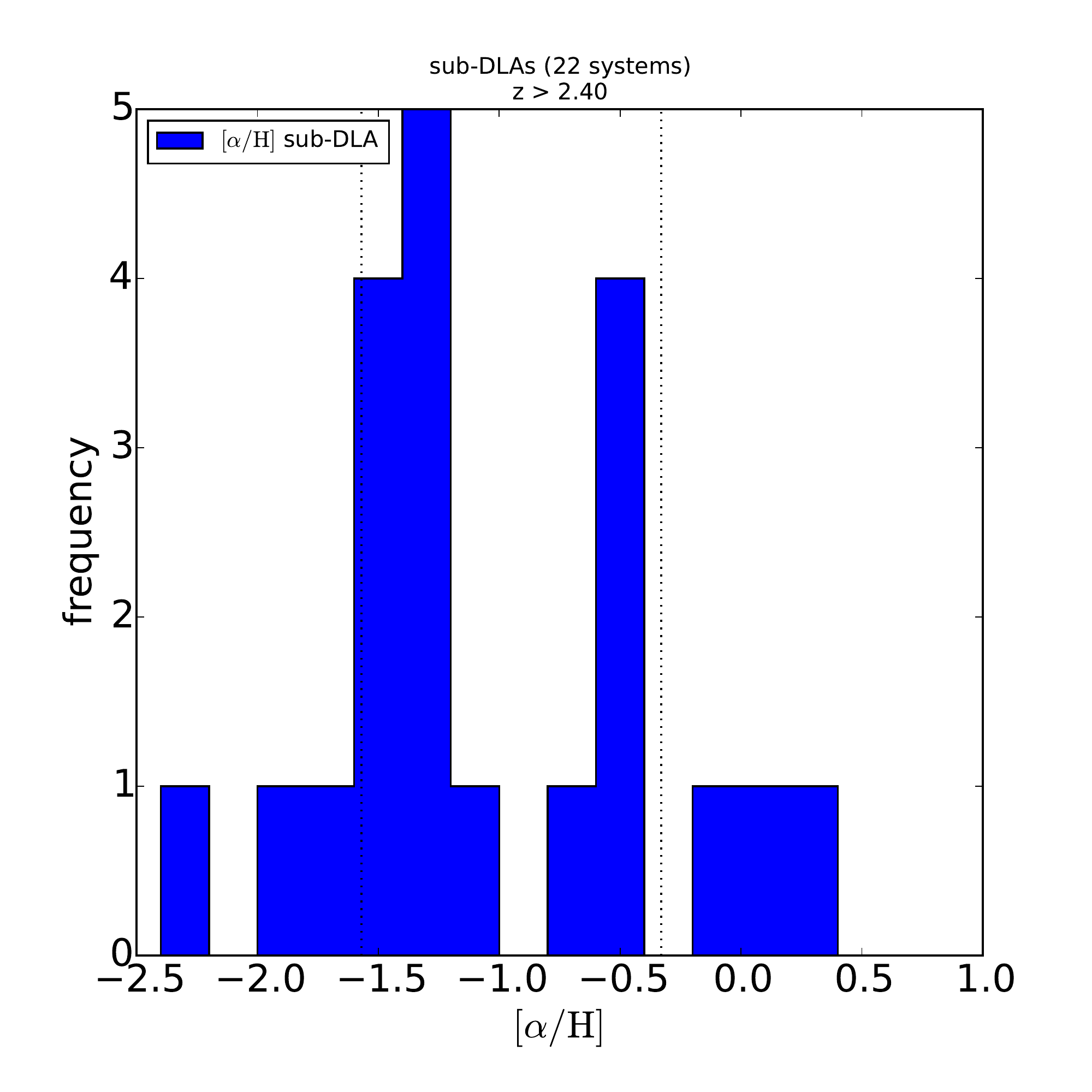}
\includegraphics[width=.4\textwidth]{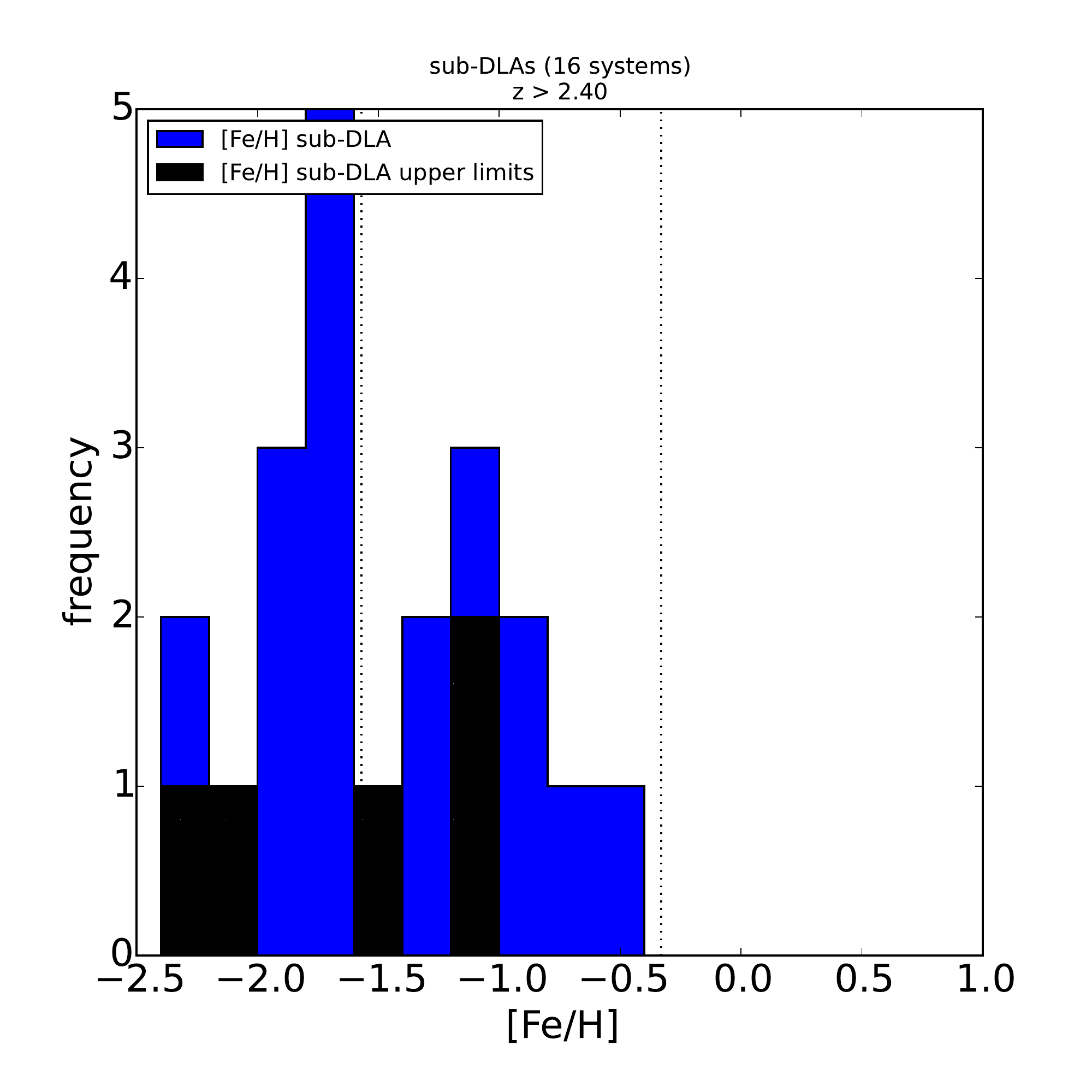}\\
\includegraphics[width=.4\textwidth]{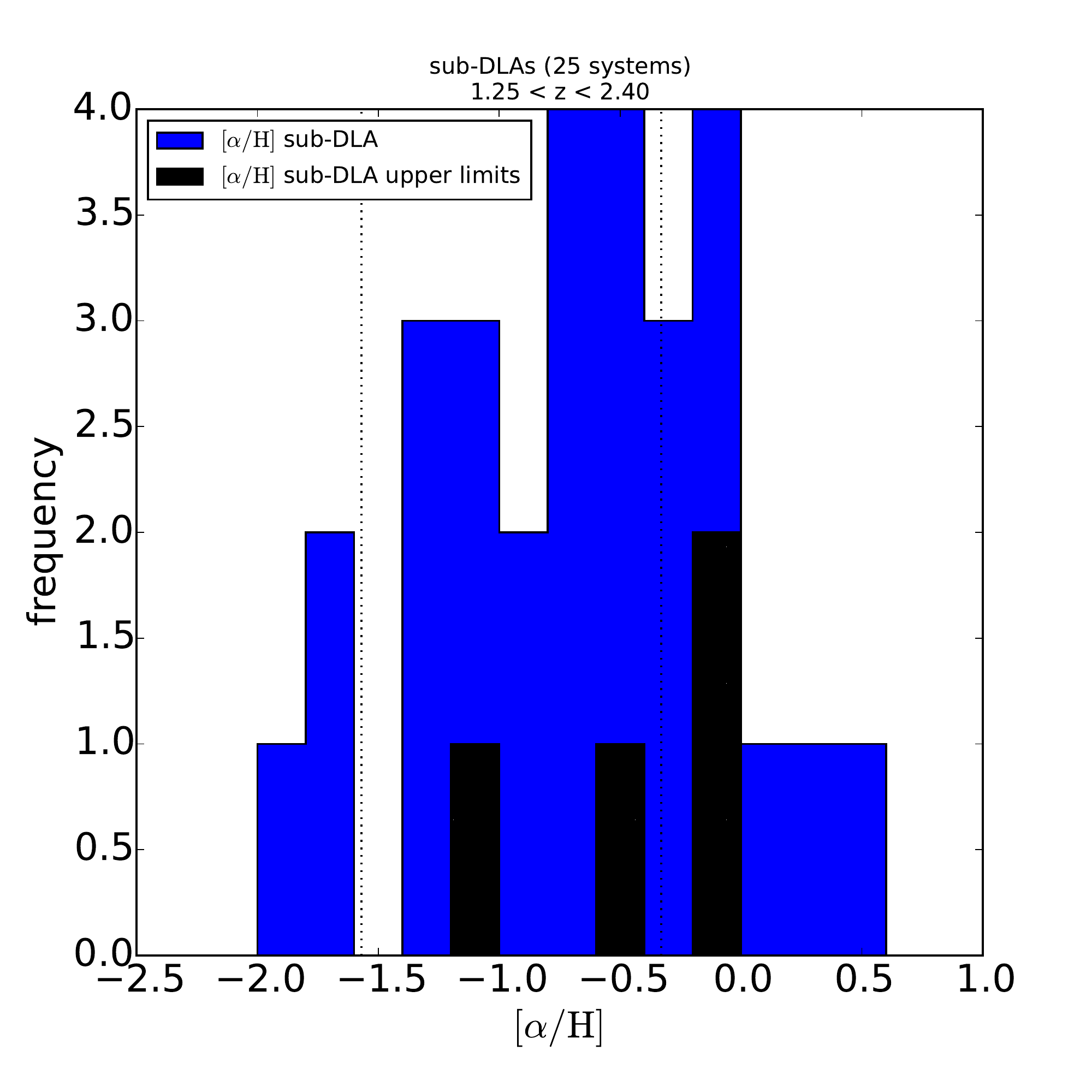}
\includegraphics[width=.4\textwidth]{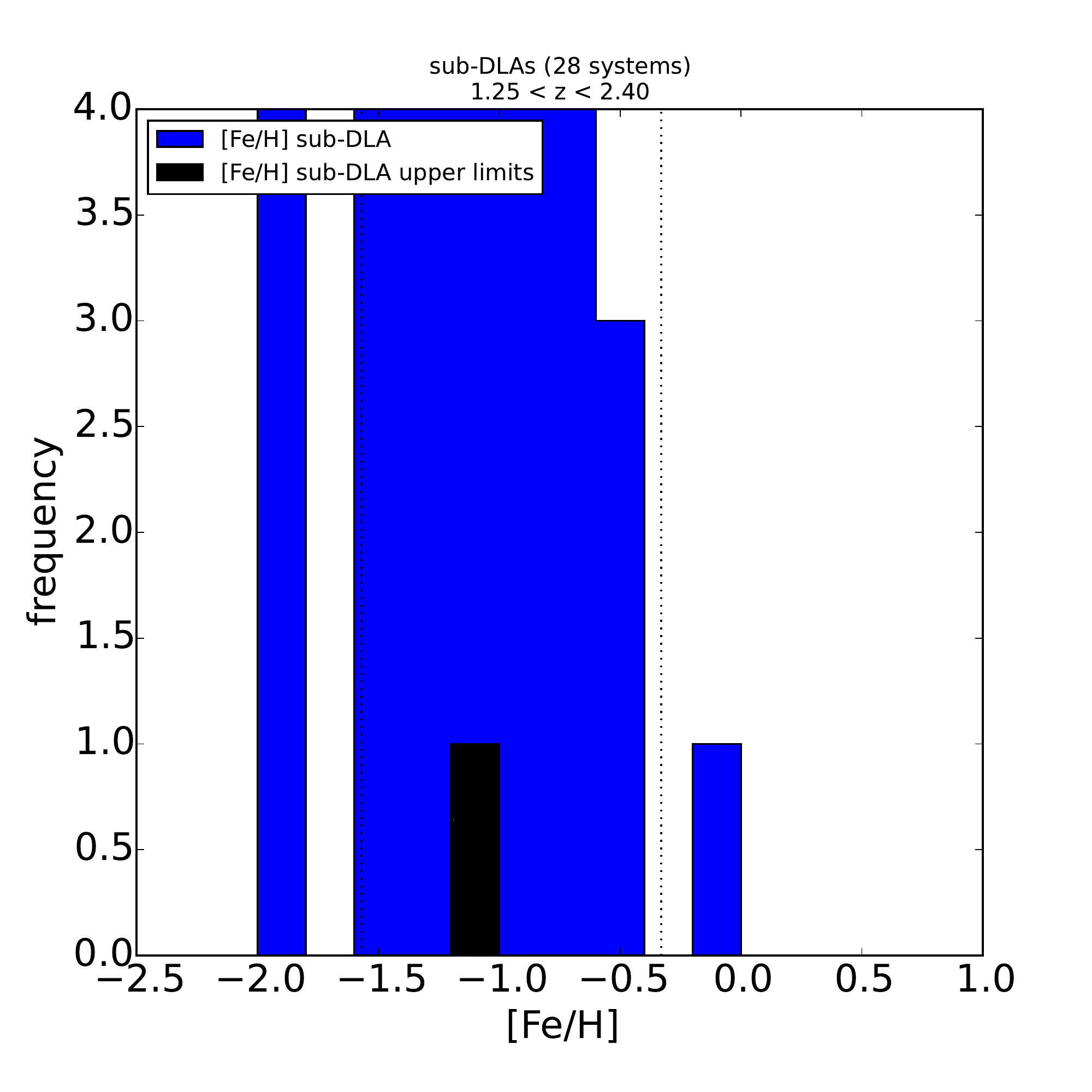}\\
\includegraphics[width=.4\textwidth]{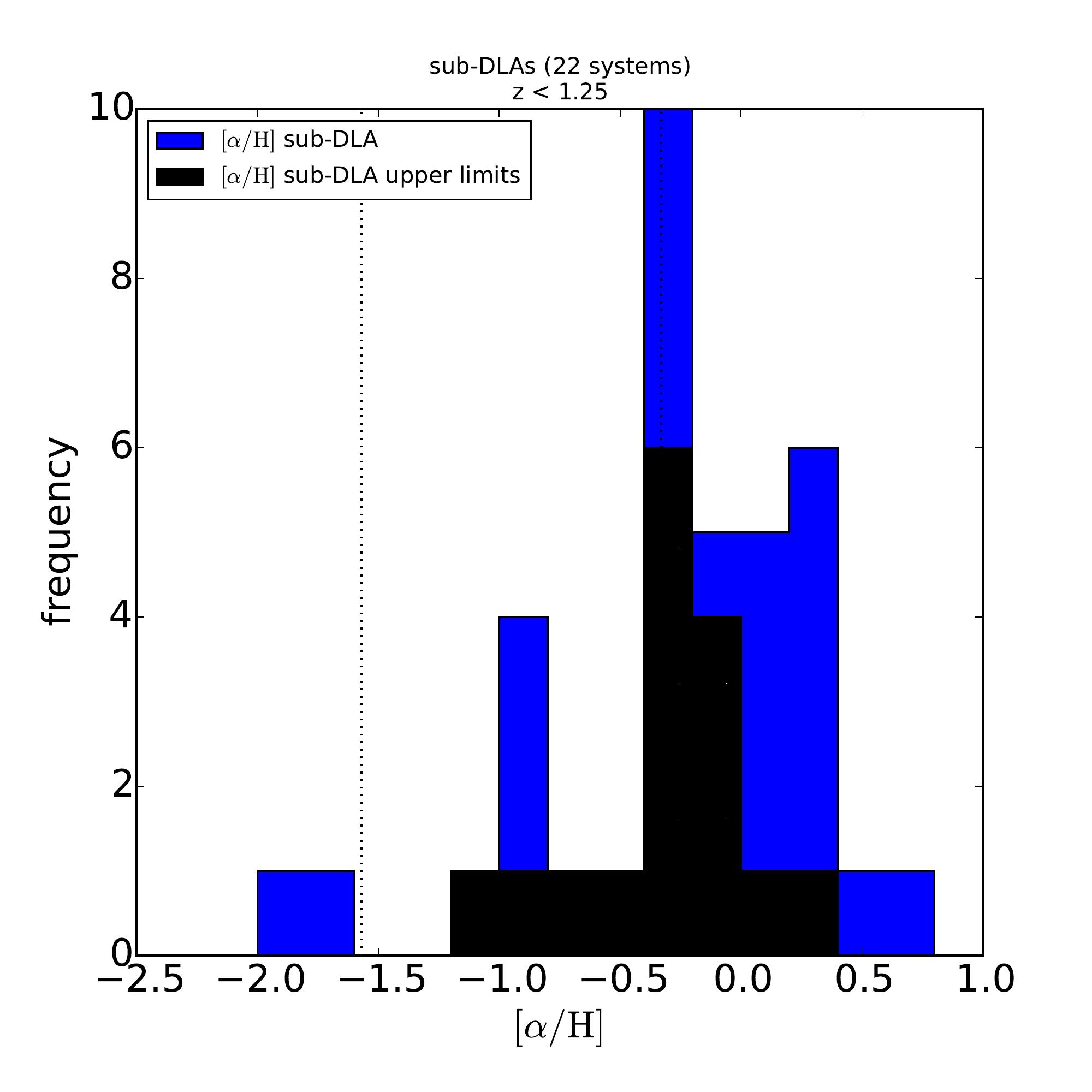}
\includegraphics[width=.4\textwidth]{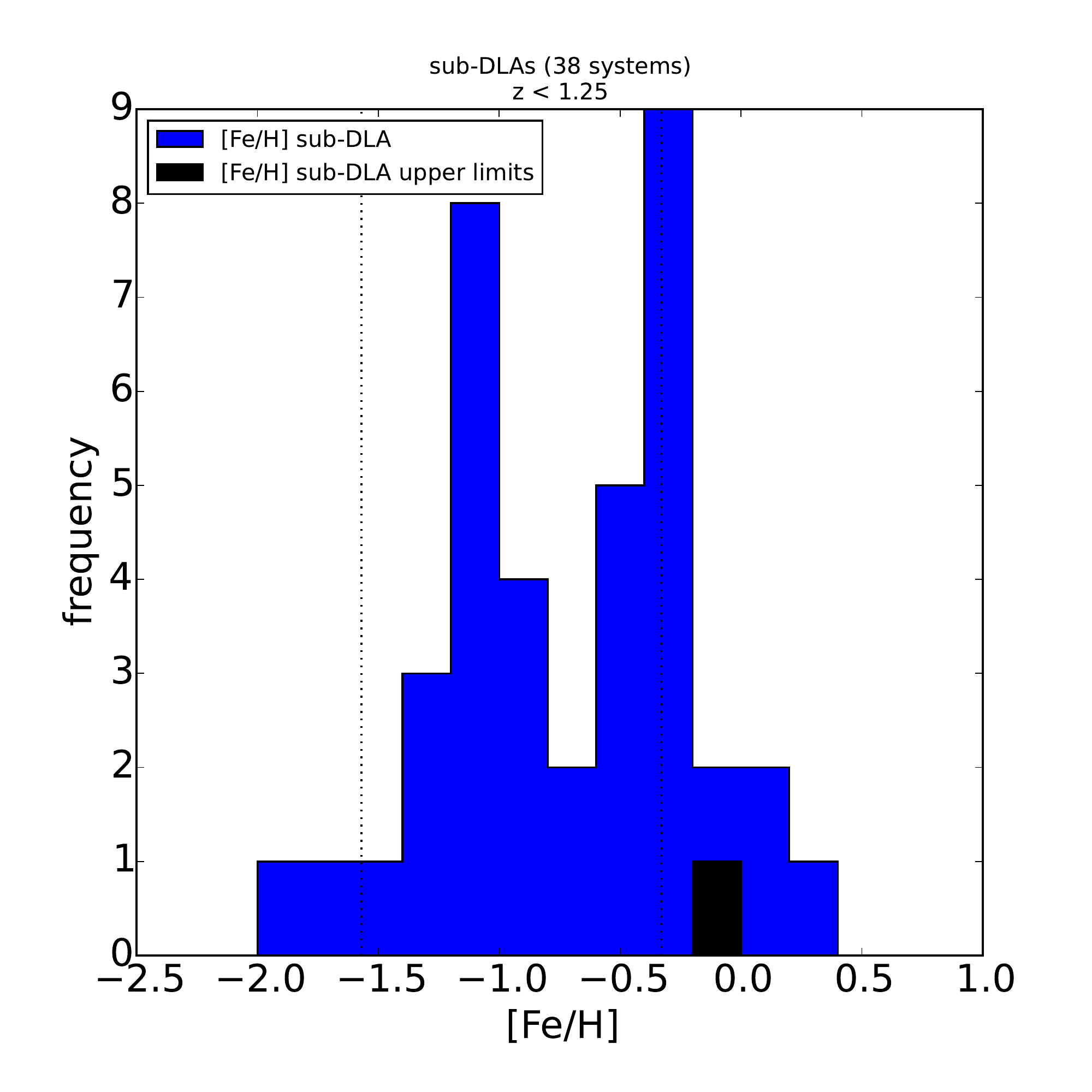}
\caption[Metallicity distribution in sub-DLAs at different redshift bins]{Metallicity (left panels: $\rm [\alpha/H]$, right panels: [Fe/H]) distribution of sub-DLAs for different redshift bins: $\rm z>2.4$ (top panels), $\rm 1.25<z<2.4$  (middle panels) and $\rm z<1.25$ (bottom panels). The black vertical dashed lines represent the mean values derived from the \zabslt{1} LLS sub-groups by \cite{Lehner2013}. The black areas represent upper limits. The metallicity distribution is a strong function of redshift and only the lowest redshift range presents hints of a bimodal distribution for the [Fe/H] metallicity.}
\label{img:bimodality}
\end{center}
\end{figure*}

Fig. \ref{img:bimodality_alpha} shows the bimodal metallicity distribution in $z<1$ LLS by \cite{Lehner2013} and the $\alpha$-element metallicity distribution for DLAs (316 systems) and sub-DLAs (68 systems) derived from our EUADP+ sample at all redshifts.
The sub-DLA $\rm [\alpha/H]$ distribution in the middle panel of Fig. \ref{img:bimodality_alpha} also suggests bimodality.

In Fig. \ref{img:bimodality}, we plot the distribution of the $\rm \alpha$ abundances (left panels) and Fe abundances (right panels) for the EUADP+ sub-DLAs in 3 redshift bins (\zabsgt{2.4} for the upper panels, \zabsbetween{1.25}{2.4} for the middle panels and \zabslt{1.25} for the bottom panels). 
We consider the metallicity traced by FeII as we have more detections with this ion and it is little affected by photo-ionization effect, even though Fe has an inclination to lock up onto dust grains. These histograms reveal the strong metallicity evolution with redshift for sub-DLAs.
The sub-DLAs in the high redshift bin, \zabsgt{2.4}, present an unimodal distribution centered around [M/H]$\sim-1.6$, similar to the metal-poor LLS population derived by \cite{Lehner2013}. 
At intermediate redshifts, \zabsbetween{1.25}{2.4}, a transition from low to higher metallicities appears. 
At low redshifts, \zabslt{1.25}, however, the distribution presents hints of a bimodal distribution. This trend is more pronounced for the [Fe/H] distribution. A DIP test rejects the unimodal distribution at a significance level of 83\%, taking the upper limit as a detection. The peaks of the distribution are located at $\rm [Fe/H]=$ -1.12 and $\rm [Fe/H]=$ -0.29, from a Gaussian Mixture Modeling. These values are compared with what is expected from simulations in terms of metallicity of accreting or outflowing gas. 
The prediction for the cold-mode accretion metallicity is above a hundredth solar, which is in line with the metal poor population in our distribution \citep{Ocvirk2008, Shen2013}. Therefore, the metal rich population should trace either outflowing gas or gas directly associated with the galaxy's ISM.
However, this is not seen in the $\rm [\alpha/H]$ distribution, where the DIP test rejects the unimodal distribution at a significance level of 31\%, still taking upper limits as detections. But these limits are located at the high metallicity end of the distribution, and therefore do not contradict the possible bimodal distribution seen in $\rm [Fe/H]$.

These results indicate a similar behavior for low redshift sub-DLAs as for low redshift LLS. However, we expect the position of the peaks to be higher than those derived for the $\rm z<1$ LLS, as the $\rm [Fe/H]$ metallicity is underestimated due to depletion of Fe onto dust grains. Lehner et al. (in prep.) show that the bimodal distribution for LLS disappears at higher redshifts, similarly to what we find with sub-DLAs at higher redshifts. We note however that the bimodality in LLS could perhaps be incorrect, given that there are larger uncertainties in the metallicity determination of LLS (due to ionization corrections).

We note that the effect of redshift plays an essential role in such an analysis, as illustrated in Fig. \ref{img:bimodality}. 

Altogether, larger samples of both LLS and sub-DLAs at low-redshifts are required to distinguish between metal-poor gas accreting onto the galaxy and metal-rich gas being expelled.

\section{Conclusion}

We present in this paper physical properties of 15 new sub-damped Lyman-$\rm \alpha$ absorbers seen in absorption in background quasar's high resolution UVES spectra. These systems cover a wide redshift range (\zabsbetween{0.584}{3.104}). The metallicity measurements were performed using Voigt profile fitting of the normalized high resolution UVES quasar spectra.
Our sub-DLA measurements add significantly to previous studies since high resolution spectroscopy is required to study these systems. 

We apply a multi-element based method to assess the level of dust depletion in the line of sight to the quasar. This study appears to be promising as it uses the combined information from several ions, and is relative to measurements of our Galaxy's ISM. With a survival analysis, we derive the best fit of the depletion factor $\rm F_{*}$ for the DLAs and sub-DLAs and find negative values, statistically different, for both groups: $\rm F^{sub-DLA}_{*}=-0.34\pm0.19$ and $\rm F^{DLA}_{*}=-0.70\pm0.06$. 
In comparison with values derived in our Galaxy, DLAs lie outside the halo and the sub-DLAs are associated with Halo like stars, in terms of depletion patterns. This is counter-intuitive as we expect DLAs to be more self-shielded to the UV background than sub-DLAs. We conclude that quasar absorbers differ from the Galactic depletion patterns or alternatively have a different nucleosynthetic history. Future analysis with the Small Magellanic Cloud will enable comparisons to a galaxy more in line with the morphology or the $\rm H_{2}$ fraction of DLAs. Moreover, we derive the averaged rest frame extinction $\rm A_{V}$ for both populations to be below 0.01, suggesting that dust reddening is not observed in the current quasar selection.

We then examine the relative abundances of Fe and $\rm \alpha$-elements. 
We derived an offset in \aFe{} for the DLAs in our sample of $0.32\pm0.18$, excluding systems with $\rm [\alpha/H]>-1.5$ to be less sensitive to dust depletion. This value is similar to that derived by \cite{Rafelski2012}.
However, we cannot derive a similar parameter for the sub-DLA population as we can not disentangle dust depletion effects from the $\alpha$-enhancement. We therefore apply the DLA corrections to the sub-DLAs.

We study the evolution of the cosmic metallicity $\Omega_{m}$, also described by the mean H\,I-weighted metallicity \meanZ{}. We confirm the steeper evolution of sub-DLAs than DLAs eventually reaching a solar metallicity at low redshifts as expected from chemical evolution models. We note that a third of the newly derived sub-DLA abundances appear as outliers from the previous data.

We measure the velocity width of the absorption systems in our new sample, \deltav{}, with a new method using the information from the Voigt profile fits. We confirm that there is a correlation between \deltav{} and metallicity for sub-DLAs. Indeed, sub-DLAs are potentially probing a different mass range than DLAs. Sub-DLAs could have a more important feedback mechanism than DLAs, thus increasing the scatter and weakening the possible velocity width/metallicity correlation.

Finally, we look at the metallicity distribution of sub-DLAs. At low redshifts, \zabslt{1.25}, we see a hint of a bimodal distribution which peaks at $\sim -1.1$ and $\sim -0.3$. This indicates that low-redshift sub-DLAs are tracing different mechanisms at play within the CGM, such as cold-mode accretion and outflows.
Larger samples of sub-DLAs and LLS abundances at low redshifts are required to better identify their connection to gas inflow/outflow processes.

\section*{Acknowledgements}
We thank N. Lehner, C. Howk, M. Fumagalli, M. Rafelski, M. Pieri, R. Bordoloi, J. O'Meara, D. Som and P. M\o ller for useful discussions. This work has been funded within the BINGO! (`history of Baryons: INtergalactic medium/Galaxies cO-evolution') project by the Agence Nationale de la Recherche (ANR) under the allocation ANR-08-BLAN-0316-01 as well as within the REGAL ('what REgulates the growth of GALaxies? ') project by the Labex (Laboratoire d'Excellence) OCEVU ('Origines, Constituants et Evolution de l'Univers'). This work has been carried out thanks to the support of the OCEVU Labex (ANR-11-LABX-0060) and the A*MIDEX project (ANR-11-IDEX-0001-02) funded by the "Investissements d'Avenir" French government program managed by the ANR. We thanks Pierre Mege for contributing in the development of the VPOD method. SQ acknowledges CNRS and CNES support for the funding of his PhD. CP thanks the ESO science visitor program for support. VPK acknowledges partial support from the NSF grant AST/1108830, with additional support from NASA grant NNX14AG74G and NASA/STScI grant for program GO 12536. EJ thanks Aix-Marseille University and C\'ecile Gry for a visit where part of this work was undertaken.

\bibliographystyle{mn2e}
\bibliography{/Users/squiret/Documents/work/Latex/library}  

\begin{thebibliography}{223}
\expandafter\ifx\csname natexlab\endcsname\relax\def\natexlab#1{#1}\fi

\bibitem[{Adelberger {et~al}\mbox{.}(2005)Adelberger, Shapley, Steidel,
  Pettini, Erb, \& Reddy}]{Adelberger2005}
Adelberger K.~L., Shapley a.~E., Steidel C.~C., Pettini M., Erb D.~K., Reddy
  N.~a., 2005, The Astrophysical Journal, 629, 636

\bibitem[{Akerman {et~al}\mbox{.}(2005)Akerman, Ellison, Pettini, \&
  Steidel}]{Akerman2005}
Akerman C.~J., Ellison S.~L., Pettini M., Steidel C.~C., 2005, Astronomy and
  Astrophysics, 440, 499

\bibitem[{Asplund {et~al}\mbox{.}(2009)Asplund, Grevesse, Sauval, \&
  Scott}]{Asplund2009}
Asplund M., Grevesse N., Sauval a.~J., Scott P., 2009, Annu. Rev. Astron.
  Astrophys, 47, 481

\bibitem[{Barbuy {et~al}\mbox{.}(2015)Barbuy, Hill, Zoccali, Minniti, \&
  Renzini}]{Barbuy2015}
Barbuy B., Hill V., Zoccali M., Minniti D., Renzini A., 2015

\bibitem[{Battisti {et~al}\mbox{.}(2012)Battisti, Meiring, Tripp, Prochaska,
  Werk, Jenkins, Lehner, Tumlinson, \& Thom}]{Battisti2012}
Battisti a.~J. {et~al.}, 2012, The Astrophysical Journal, 744, 93

\bibitem[{Bauermeister, Blitz \& Ma(2010)Bauermeister, Blitz, \&
  Ma}]{Bauermeister2010}
Bauermeister A., Blitz L., Ma C.-P., 2010, The Astrophysical Journal, 717, 323

\bibitem[{Berg {et~al}\mbox{.}(2015{\natexlab{a}})Berg, Ellison, Prochaska,
  Venn, \& Dessauges-Zavadsky}]{Berg2015}
Berg T. A.~M., Ellison S.~L., Prochaska J.~X., Venn K.~A., Dessauges-Zavadsky
  M., 2015{\natexlab{a}}, Monthly Notices of the Royal Astronomical Society,
  452, 4326

\bibitem[{Berg {et~al}\mbox{.}(2013)Berg, Ellison, Venn, \&
  Prochaska}]{Berg2013}
Berg T. A.~M., Ellison S.~L., Venn K.~A., Prochaska J.~X., 2013, Monthly
  Notices of the Royal Astronomical Society, 434, 2892

\bibitem[{Berg {et~al}\mbox{.}(2015{\natexlab{b}})Berg, Neeleman, Prochaska,
  Ellison, \& Wolfe}]{Berg2014}
Berg T. A.~M., Neeleman M., Prochaska J.~X., Ellison S.~L., Wolfe A.~M.,
  2015{\natexlab{b}}, Publications of the Astronomical Society of the Pacific,
  127, 167

\bibitem[{Bergeron \& Dodorico(1986)}]{Bergeron1986}
Bergeron J., Dodorico S., 1986, Monthly Notices of the Royal Astronomical
  Society, 220, 833

\bibitem[{Bertone {et~al}\mbox{.}(2010{\natexlab{a}})Bertone, Schaye, Booth,
  Vecchia, Theuns, \& Wiersma}]{Bertone2010UV}
Bertone S., Schaye J., Booth C.~M., Vecchia C.~D., Theuns T., Wiersma R. P.~C.,
  2010{\natexlab{a}}, Monthly Notices of the Royal Astronomical Society, 408,
  1120

\bibitem[{Bertone {et~al}\mbox{.}(2010{\natexlab{b}})Bertone, Schaye, Vecchia,
  Booth, Theuns, \& Wiersma}]{Bertone2010X}
Bertone S., Schaye J., Vecchia C.~D., Booth C.~M., Theuns T., Wiersma R. P.~C.,
  2010{\natexlab{b}}, Monthly Notices of the Royal Astronomical Society, 407,
  544

\bibitem[{Birnboim \& Dekel(2003)}]{Birnboim2003}
Birnboim Y., Dekel A., 2003, Monthly Notices of the Royal Astronomical Society,
  345, 349

\bibitem[{Blades {et~al}\mbox{.}(1982)Blades, Hunstead, Murdoch, \&
  Pettini}]{Blades1982}
Blades J.~C., Hunstead R.~W., Murdoch H.~S., Pettini M., 1982, Monthly Notices
  of the Royal Astronomical Society, 200, 1091

\bibitem[{Boiss{\'{e}} {et~al}\mbox{.}(1998)Boiss{\'{e}}, Brun, Bergeron, \&
  Deharveng}]{Boisse1998}
Boiss{\'{e}} P., Brun V.~L., Bergeron J., Deharveng J.-m., 1998, Astronomy and
  Astrophysics, 333, 23

\bibitem[{Bouch{\'{e}} {et~al}\mbox{.}(2012)Bouch{\'{e}}, Hohensee, Vargas,
  Kacprzak, Martin, Cooke, \& Churchill}]{Bouche2012}
Bouch{\'{e}} N., Hohensee W., Vargas R., Kacprzak G.~G., Martin C.~L., Cooke
  J., Churchill C.~W., 2012, Monthly Notices of the Royal Astronomical Society,
  426, 801

\bibitem[{Bouch{\'{e}} {et~al}\mbox{.}(2007)Bouch{\'{e}}, Murphy, P{\'{e}}roux,
  Csabai, \& Wild}]{Bouche2007}
Bouch{\'{e}} N., Murphy M., P{\'{e}}roux C., Csabai I., Wild V., 2007, New
  Astronomy Reviews, 51, 131

\bibitem[{Bouch{\'{e}} {et~al}\mbox{.}(2013)Bouch{\'{e}}, Murphy, Kacprzak,
  P{\'{e}}roux, Contini, Martin, \& Dessauges-Zavadsky}]{Bouche2013a}
Bouch{\'{e}} N., Murphy M.~T., Kacprzak G.~G., P{\'{e}}roux C., Contini T.,
  Martin C.~L., Dessauges-Zavadsky M., 2013, Science (New York, N.Y.), 341, 50

\bibitem[{Bowen {et~al}\mbox{.}(2005)Bowen, Jenkins, Pettini, \&
  Tripp}]{Bowen2005}
Bowen D.~V., Jenkins E.~B., Pettini M., Tripp T.~M., 2005, ApJ, 635, 880

\bibitem[{Cantalupo {et~al}\mbox{.}(2014)Cantalupo, Arrigoni-Battaia,
  Prochaska, Hennawi, \& Madau}]{Cantalupo2014}
Cantalupo S., Arrigoni-Battaia F., Prochaska J.~X., Hennawi J.~F., Madau P.,
  2014, Nature, 506, 63

\bibitem[{Carswell {et~al}\mbox{.}(1996)Carswell, Webb, Lanzetta, Baldwin,
  Cooke, Williger, Rauch, Irwin, Robertson, \& Shaver}]{Carswell1996}
Carswell R.~F. {et~al.}, 1996, Monthly Notices of the Royal Astronomical
  Society, 278, 506

\bibitem[{Centurion {et~al}\mbox{.}(2000)Centurion, Bonifacio, Molaro, \&
  Vladilo}]{Centurion2000}
Centurion M., Bonifacio P., Molaro P., Vladilo G., 2000, The Astrophysical
  Journal, 536, 540

\bibitem[{Centurion {et~al}\mbox{.}(2003)Centurion, Molaro, Vladilo, Peroux,
  Levshakov, \& D'Odorico}]{Centurion2003}
Centurion M., Molaro P., Vladilo G., Peroux C., Levshakov S.~a., D'Odorico V.,
  2003, Astronomy and Astrophysics, 403, 55

\bibitem[{Chen, Kennicutt \& Rauch(2005)Chen, Kennicutt, \& Rauch}]{Chen2005}
Chen H.-W., Kennicutt R., Rauch M., 2005, $\backslash$Apj, 620, 703

\bibitem[{Christensen {et~al}\mbox{.}(2014)Christensen, M{\o}ller, Fynbo, \&
  Zafar}]{Christensen2014}
Christensen L., M{\o}ller P., Fynbo J. P.~U., Zafar T., 2014, 000, 15

\bibitem[{Churchill {et~al}\mbox{.}(2000)Churchill, Mellon, Charlton, Jannuzi,
  Kirhakos, Steidel, \& Schneider}]{Churchill2000}
Churchill C.~W., Mellon R.~R., Charlton J.~C., Jannuzi B.~T., Kirhakos S.,
  Steidel C.~C., Schneider D.~P., 2000, The Astrophysical Journal, 543, 577

\bibitem[{Cooke, Pettini \& Murphy(2012)Cooke, Pettini, \& Murphy}]{Cooke2012}
Cooke R., Pettini M., Murphy M.~T., 2012, Monthly Notices of the Royal
  Astronomical Society, 425, 347

\bibitem[{Cooke {et~al}\mbox{.}(2010{\natexlab{a}})Cooke, Pettini, Steidel,
  King, Rudie, \& Rakic}]{Cooke2010a}
Cooke R., Pettini M., Steidel C.~C., King L.~J., Rudie G.~C., Rakic O.,
  2010{\natexlab{a}}, Monthly Notices of the Royal Astronomical Society, 409,
  679

\bibitem[{Cooke {et~al}\mbox{.}(2010{\natexlab{b}})Cooke, Pettini, Steidel,
  Rudie, \& Jorgenson}]{Cooke2011a}
Cooke R., Pettini M., Steidel C.~C., Rudie G.~C., Jorgenson R.~A.,
  2010{\natexlab{b}}, Monthly Notices of the Royal Astronomical Society, 412,
  no

\bibitem[{Cooke {et~al}\mbox{.}(2011)Cooke, Pettini, Steidel, Rudie, \&
  Nissen}]{Cooke2011}
Cooke R., Pettini M., Steidel C.~C., Rudie G.~C., Nissen P.~E., 2011, Monthly
  Notices of the Royal Astronomical Society, 417, 1534

\bibitem[{Crighton, Hennawi \& Prochaska(2013)Crighton, Hennawi, \&
  Prochaska}]{Crighton2013a}
Crighton N. H.~M., Hennawi J.~F., Prochaska J.~X., 2013, The Astrophysical
  Journal, 776, L18

\bibitem[{{De Cia} {et~al}\mbox{.}(2013){De Cia}, Ledoux, Savaglio, Schady, \&
  Vreeswijk}]{DeCia2013}
{De Cia} A., Ledoux C., Savaglio S., Schady P., Vreeswijk P.~M., 2013,
  Astronomy {\&} Astrophysics, 560, A88

\bibitem[{de~la Varga {et~al}\mbox{.}(2000)de~la Varga, Reimers, Tytler,
  Barlow, \& Burles}]{delaVarga2000}
de~la Varga A., Reimers D., Tytler D., Barlow T., Burles S., 2000, AA, 363, 69

\bibitem[{Dekel \& Birnboim(2006)}]{Dekel2006}
Dekel A., Birnboim Y., 2006, Monthly Notices of the Royal Astronomical Society,
  368, 2

\bibitem[{Dessauges-Zavadsky {et~al}\mbox{.}(2004)Dessauges-Zavadsky, Calura,
  Prochaska, D'Odorico, \& Matteucci}]{Dessauges-Zavadsky2004}
Dessauges-Zavadsky M., Calura F., Prochaska J.~X., D'Odorico S., Matteucci F.,
  2004, Astronomy and Astrophysics, 416, 79

\bibitem[{Dessauges-Zavadsky {et~al}\mbox{.}(2007)Dessauges-Zavadsky, Calura,
  Prochaska, D'Odorico, \& Matteucci}]{Dessauges-Zavadsky2007}
Dessauges-Zavadsky M., Calura F., Prochaska J.~X., D'Odorico S., Matteucci F.,
  2007, Astronomy and Astrophysics, 470, 431

\bibitem[{Dessauges-Zavadsky {et~al}\mbox{.}(2001)Dessauges-Zavadsky,
  D'Odorico, McMahon, Molaro, Ledoux, P�roux, \&
  Storrie-Lombardi}]{Dessauges-Zavadsky2001}
Dessauges-Zavadsky M., D'Odorico S., McMahon R.~G., Molaro P., Ledoux C.,
  P�roux C., Storrie-Lombardi L.~J., 2001, Astronomy and Astrophysics, 370,
  426

\bibitem[{Dessauges-Zavadsky, Ellison \& Murphy(2009)Dessauges-Zavadsky,
  Ellison, \& Murphy}]{Dessauges-Zavadsky2009}
Dessauges-Zavadsky M., Ellison S.~L., Murphy M.~T., 2009, {Revisiting the
  origin of the high metallicities of sub-damped Lyman-alpha systems}

\bibitem[{Dessauges-Zavadsky {et~al}\mbox{.}(2003)Dessauges-Zavadsky,
  P{\'{e}}roux, Kim, D'Odorico, \& McMahon}]{Dessauges-Zavadsky2003}
Dessauges-Zavadsky M., P{\'{e}}roux C., Kim T.-S., D'Odorico S., McMahon R.~G.,
  2003, Monthly Notices of the Royal Astronomical Society, 345, 447

\bibitem[{Dessauges-Zavadsky, Prochaska \& D'Odorico(2002)Dessauges-Zavadsky,
  Prochaska, \& D'Odorico}]{Dessauges-Zavadsky2002}
Dessauges-Zavadsky M., Prochaska J.~X., D'Odorico S., 2002, Astronomy and
  Astrophysics, 391, 801

\bibitem[{Dessauges-Zavadsky {et~al}\mbox{.}(2006)Dessauges-Zavadsky,
  Prochaska, D'Odorico, Calura, \& Matteucci}]{Dessauges-Zavadsky2006}
Dessauges-Zavadsky M., Prochaska J.~X., D'Odorico S., Calura F., Matteucci F.,
  2006, Astronomy and Astrophysics, 445, 93

\bibitem[{D'Odorico {et~al}\mbox{.}(2013)D'Odorico, Cupani, Cristiani,
  Maiolino, Molaro, Nonino, Centuri{\'{o}}n, Cimatti, {di Serego Alighieri},
  Fiore, Fontana, Gallerani, Giallongo, Mannucci, Marconi, Pentericci, Viel, \&
  Vladilo}]{DOdorico2013}
D'Odorico V. {et~al.}, 2013, Monthly Notices of the Royal Astronomical Society,
  435, 1198

\bibitem[{D'Odorico \& Molaro(2004)}]{Dodorico2004}
D'Odorico V., Molaro P., 2004, Astronomy and Astrophysics, 415, 879

\bibitem[{Draine(2011)}]{Draine2011}
Draine B., 2011, {Physics of the Interstellar and Intergalactic Medium}

\bibitem[{Dutta {et~al}\mbox{.}(2014)Dutta, Srianand, Rahmani, Petitjean,
  Noterdaeme, \& Ledoux}]{Dutta2014}
Dutta R., Srianand R., Rahmani H., Petitjean P., Noterdaeme P., Ledoux C.,
  2014, Monthly Notices of the Royal Astronomical Society, 440, 307

\bibitem[{Ellison {et~al}\mbox{.}(2007)Ellison, Hennawi, Martin, \&
  Sommer-Larsen}]{Ellison2007}
Ellison S.~L., Hennawi J.~F., Martin C.~L., Sommer-Larsen J., 2007, Monthly
  Notices of the Royal Astronomical Society, 378, 801

\bibitem[{Ellison {et~al}\mbox{.}(2012)Ellison, Kanekar, Prochaska, Momjian, \&
  Worseck}]{Ellison2012}
Ellison S.~L., Kanekar N., Prochaska J.~X., Momjian E., Worseck G., 2012,
  Monthly Notices of the Royal Astronomical Society, 424, 293

\bibitem[{Ellison \& Lopez(2001)}]{Ellison2001b}
Ellison S.~L., Lopez S., 2001, Astronomy and Astrophysics, 380, 117

\bibitem[{Ellison \& Lopez(2009)}]{Ellison2009}
Ellison S.~L., Lopez S., 2009, Monthly Notices of the Royal Astronomical
  Society, 397, 467

\bibitem[{Ellison {et~al}\mbox{.}(2001)Ellison, Pettini, Steidel, \&
  Shapley}]{Ellison2001}
Ellison S.~L., Pettini M., Steidel C.~C., Shapley A.~E., 2001, The
  Astrophysical Journal, 549, 770

\bibitem[{Ellison {et~al}\mbox{.}(2010)Ellison, Prochaska, Hennawi, Lopez,
  Usher, Wolfe, Russell, \& Benn}]{Ellison2010}
Ellison S.~L., Prochaska J.~X., Hennawi J., Lopez S., Usher C., Wolfe A.~M.,
  Russell D.~M., Benn C.~R., 2010, Monthly Notices of the Royal Astronomical
  Society, 406, no

\bibitem[{Ellison {et~al}\mbox{.}(2008)Ellison, York, Pettini, \&
  Kanekar}]{Ellison2008}
Ellison S.~L., York B.~a., Pettini M., Kanekar N., 2008, Monthly Notices of the
  Royal Astronomical Society, 388, 1349

\bibitem[{Erb {et~al}\mbox{.}(2006)Erb, Shapley, Pettini, Steidel, Reddy, \&
  Adelberger}]{Erb2006}
Erb D.~K., Shapley A.~E., Pettini M., Steidel C.~C., Reddy N.~a., Adelberger
  K.~L., 2006, The Astrophysical Journal, 644, 813

\bibitem[{Fall \& Efstathiou(1980)}]{Fall1980}
Fall S.~M., Efstathiou G., 1980, Royal Astronomical Society, 193, 189

\bibitem[{Faucher-Giguere {et~al}\mbox{.}(2015)Faucher-Giguere, Hopkins, Kere,
  Muratov, Quataert, \& Murray}]{Faucher-Giguere2015}
Faucher-Giguere C.-A., Hopkins P.~F., Kere D., Muratov A.~L., Quataert E.,
  Murray N., 2015, Monthly Notices of the Royal Astronomical Society, 449, 987

\bibitem[{Faucher-Giguere \& Kere{\v{s}}(2011)}]{Faucher-Giguere2011}
Faucher-Giguere C.~A., Kere{\v{s}} D., 2011, Monthly Notices of the Royal
  Astronomical Society: Letters, 412

\bibitem[{Fontana \& Ballester(1995)}]{Fontana1995}
Fontana A., Ballester P., 1995, The Messenger, 80, 37

\bibitem[{Fox {et~al}\mbox{.}(2007)Fox, Petitjean, Ledoux, \&
  Srianand}]{Fox2007}
Fox A.~J., Petitjean P., Ledoux C., Srianand R., 2007, 5

\bibitem[{Fox {et~al}\mbox{.}(2009)Fox, Prochaska, Ledoux, Petitjean, Wolfe, \&
  Srianand}]{Fox2009}
Fox A.~J., Prochaska J.~X., Ledoux C., Petitjean P., Wolfe A.~M., Srianand R.,
  2009, 746, 26

\bibitem[{Frank \& P{\'{e}}roux(2010)}]{Frank2010a}
Frank S., P{\'{e}}roux C., 2010, Monthly Notices of the Royal Astronomical
  Society, 406, 2235

\bibitem[{Fumagalli {et~al}\mbox{.}(2011)Fumagalli, Prochaska, Kasen, Dekel,
  Ceverino, \& Primack}]{Fumagalli2011}
Fumagalli M., Prochaska J.~X., Kasen D., Dekel A., Ceverino D., Primack J.~R.,
  2011, Monthly Notices of the Royal Astronomical Society, 418, 1796

\bibitem[{Fynbo {et~al}\mbox{.}(2013)Fynbo, Geier, Christensen, Gallazzi,
  Krogager, Kruhler, Ledoux, Maund, Moller, Noterdaeme, Rivera-Thorsen, \&
  Vestergaard}]{Fynbo2013}
Fynbo J. P.~U. {et~al.}, 2013, Monthly Notices of the Royal Astronomical
  Society, 436, 361

\bibitem[{Fynbo {et~al}\mbox{.}(2011)Fynbo, Ledoux, Noterdaeme, Christensen,
  M{\o}ller, Durgapal, Goldoni, Kaper, Krogager, Laursen, Maund,
  Milvang-Jensen, Okoshi, Rasmussen, Thorsen, Toft, \& Zafar}]{Fynbo2011}
Fynbo J. P.~U. {et~al.}, 2011, Monthly Notices of the Royal Astronomical
  Society, 413, 2481

\bibitem[{Gauthier {et~al}\mbox{.}(2014)Gauthier, Chen, Cooksey, Simcoe,
  Seyffert, \& O'Meara}]{Gauthier2014}
Gauthier J.-R., Chen H.-W., Cooksey K.~L., Simcoe R.~A., Seyffert E.~N.,
  O'Meara J.~M., 2014, Monthly Notices of the Royal Astronomical Society, 439,
  342

\bibitem[{Ge, Bechtold \& Kulkarni(2001)Ge, Bechtold, \& Kulkarni}]{Ge2001}
Ge J., Bechtold J., Kulkarni V.~P., 2001, The Astrophysical Journal, 547, L1

\bibitem[{Genel {et~al}\mbox{.}(2010)Genel, Bouch{\'{e}}, Naab, Sternberg, \&
  Genzel}]{Genel2010}
Genel S., Bouch{\'{e}} N., Naab T., Sternberg A., Genzel R., 2010, The
  Astrophysical Journal, 719, 229

\bibitem[{Guimar{\~{a}}es {et~al}\mbox{.}(2012)Guimar{\~{a}}es, Noterdaeme,
  Petitjean, Ledoux, Srianand, L{\'{o}}pez, \& Rahmani}]{Guimaraes2012}
Guimar{\~{a}}es R., Noterdaeme P., Petitjean P., Ledoux C., Srianand R.,
  L{\'{o}}pez S., Rahmani H., 2012, The Astronomical Journal, 143, 147

\bibitem[{Haehnelt, Steinmetz \& Rauch(1998)Haehnelt, Steinmetz, \&
  Rauch}]{Haenelt1998}
Haehnelt M.~G., Steinmetz M., Rauch M., 1998, The Astrophysical Journal, 495,
  647

\bibitem[{Heinmueller {et~al}\mbox{.}(2006)Heinmueller, Petitjean, Ledoux,
  Caucci, \& Srianand}]{Heinmueller2006}
Heinmueller J., Petitjean P., Ledoux C., Caucci S., Srianand R., 2006, 39, 33

\bibitem[{Henry \& Prochaska(2007)}]{Henry2007}
Henry R. B.~C., Prochaska J.~X., 2007, Publications of the Astronomical Society
  of the Pacific, 119, 962

\bibitem[{Herbert‐Fort {et~al}\mbox{.}(2006)Herbert‐Fort, Prochaska,
  Dessauges‐Zavadsky, Ellison, Howk, Wolfe, \& Prochter}]{Herbert-Fort2006}
Herbert‐Fort S., Prochaska J.~X., Dessauges‐Zavadsky M., Ellison S.~L.,
  Howk J.~C., Wolfe A.~M., Prochter G.~E., 2006, Publications of the
  Astronomical Society of the Pacific, 118, 1077

\bibitem[{Jenkins(2009)}]{Jenkins09}
Jenkins E.~B., 2009, The Astrophysical Journal, 700, 1299

\bibitem[{Kacprzak {et~al}\mbox{.}(2014)Kacprzak, Martin, Bouch{\'{e}},
  Churchill, Cooke, LeReun, Schroetter, Ho, \& Klimek}]{Kacprzak2014}
Kacprzak G.~G. {et~al.}, 2014, The Astrophysical Journal, 792, L12

\bibitem[{Kanekar {et~al}\mbox{.}(2014)Kanekar, Prochaska, Smette, Ellison,
  Ryan-Weber, Momjian, Briggs, Lane, Chengalur, Delafosse, Grave, Jacobsen, \&
  de~Bruyn}]{Kanekar2014}
Kanekar N. {et~al.}, 2014, Monthly Notices of the Royal Astronomical Society,
  438, 2131

\bibitem[{Kere{\v{s}} {et~al}\mbox{.}(2005)Kere{\v{s}}, Katz, Weinberg, \&
  Dav{\'{e}}}]{Keres2005}
Kere{\v{s}} D., Katz N., Weinberg D.~H., Dav{\'{e}} R., 2005, Monthly Notices
  of the Royal Astronomical Society, 363, 2

\bibitem[{Khare {et~al}\mbox{.}(2007)Khare, Kulkarni, P{\'{e}}roux, York,
  Lauroesch, \& Meiring}]{Khare2007}
Khare P., Kulkarni V.~P., P{\'{e}}roux C., York D.~G., Lauroesch J.~T., Meiring
  J.~D., 2007, Astronomy and Astrophysics, 464, 487

\bibitem[{Khare {et~al}\mbox{.}(2012)Khare, {Vanden Berk}, York, Lundgren, \&
  Kulkarni}]{Khare2012}
Khare P., {Vanden Berk} D., York D.~G., Lundgren B., Kulkarni V.~P., 2012,
  Monthly Notices of the Royal Astronomical Society, 419, 1028

\bibitem[{Kisielius {et~al}\mbox{.}(2014)Kisielius, Kulkarni, Ferland,
  Bogdanovich, \& Lykins}]{Kisielius2014}
Kisielius R., Kulkarni V.~P., Ferland G.~J., Bogdanovich P., Lykins M.~L.,
  2014, The Astrophysical Journal, 780, 76

\bibitem[{Kisielius {et~al}\mbox{.}(2015)Kisielius, Kulkarni, Ferland,
  Bogdanovich, Som, \& Lykins}]{Kisielius2015}
Kisielius R., Kulkarni V.~P., Ferland G.~J., Bogdanovich P., Som D., Lykins
  M.~L., 2015, The Astrophysical Journal, 804, 76

\bibitem[{Krogager {et~al}\mbox{.}(2013)Krogager, Fynbo, Ledoux, Christensen,
  Gallazzi, Laursen, Moller, Noterdaeme, Peroux, Pettini, \&
  Vestergaard}]{Krogager2013a}
Krogager J.-K. {et~al.}, 2013, Monthly Notices of the Royal Astronomical
  Society, 433, 3091

\bibitem[{Krogager {et~al}\mbox{.}(2016)Krogager, Fynbo, Noterdaeme, Zafar,
  M{\o}ller, Ledoux, Kr{\"{u}}hler, \& Stockton}]{Krogager2016}
Krogager J.-K., Fynbo J. P.~U., Noterdaeme P., Zafar T., M{\o}ller P., Ledoux
  C., Kr{\"{u}}hler T., Stockton A., 2016, Monthly Notices of the Royal
  Astronomical Society, 455, 2698

\bibitem[{Krogager {et~al}\mbox{.}(2015)Krogager, Geier, Fynbo, Venemans,
  Ledoux, M{\o}ller, Noterdaeme, Vestergaard, Kangas, Pursimo, Saturni, \&
  Smirnova}]{Krogager2015}
Krogager J.-K. {et~al.}, 2015, The Astrophysical Journal Supplement Series,
  217, 5

\bibitem[{Kulkarni {et~al}\mbox{.}(2005)Kulkarni, Fall, Lauroesch, York, Welty,
  Khare, \& Truran}]{Kulkarni2005}
Kulkarni V.~P., Fall S.~M., Lauroesch J.~T., York D.~G., Welty D.~E., Khare P.,
  Truran J.~W., 2005, The Astrophysical Journal, 618, 68

\bibitem[{Kulkarni {et~al}\mbox{.}(2007)Kulkarni, Khare, P{\'{e}}roux, York,
  Lauroesch, \& Meiring}]{Kulkarni2007}
Kulkarni V.~P., Khare P., P{\'{e}}roux C., York D.~G., Lauroesch J.~T., Meiring
  J.~D., 2007, The Astrophysical Journal, 661, 88

\bibitem[{Kulkarni {et~al}\mbox{.}(2010)Kulkarni, Khare, Som, Meiring, York,
  Peroux, \& Lauroesch}]{Kulkarni2010}
Kulkarni V.~P., Khare P., Som D., Meiring J., York D.~G., Peroux C., Lauroesch
  J.~T., 2010, 27

\bibitem[{Kulkarni {et~al}\mbox{.}(2012)Kulkarni, Meiring, Som, P{\'{e}}roux,
  York, Khare, \& Lauroesch}]{Kulkarni2012}
Kulkarni V.~P., Meiring J., Som D., P{\'{e}}roux C., York D.~G., Khare P.,
  Lauroesch J.~T., 2012, The Astrophysical Journal, 749, 176

\bibitem[{Lanzetta, Wolfe \& Turnshek(1995)Lanzetta, Wolfe, \&
  Turnshek}]{Lanzetta1995}
Lanzetta K.~M., Wolfe A.~M., Turnshek D.~A., 1995, The Astrophysical Journal,
  440, 435

\bibitem[{Ledoux, Bergeron \& Petitjean(2002)Ledoux, Bergeron, \&
  Petitjean}]{Ledoux2002b}
Ledoux C., Bergeron J., Petitjean P., 2002, European Space Agency, (Special
  Publication) ESA SP, 815, 17

\bibitem[{Ledoux {et~al}\mbox{.}(2006)Ledoux, Petitjean, Fynbo, M{\o}ller, \&
  Srianand}]{Ledoux2006}
Ledoux C., Petitjean P., Fynbo J. P.~U., M{\o}ller P., Srianand R., 2006,
  Astronomy and Astrophysics, 457, 71

\bibitem[{Ledoux, Petitjean \& Srianand(2003)Ledoux, Petitjean, \&
  Srianand}]{Ledoux2003}
Ledoux C., Petitjean P., Srianand R., 2003, Monthly Notices of the Royal
  Astronomical Society, 346, 209

\bibitem[{Ledoux, Srianand \& Petitjean(2002)Ledoux, Srianand, \&
  Petitjean}]{Ledoux2002}
Ledoux C., Srianand R., Petitjean P., 2002, 789, 9

\bibitem[{Lehner {et~al}\mbox{.}(2013)Lehner, Howk, Tripp, Tumlinson,
  Prochaska, O'Meara, Thom, Werk, Fox, \& Ribaudo}]{Lehner2013}
Lehner N. {et~al.}, 2013, The Astrophysical Journal, 770, 138

\bibitem[{Lehner {et~al}\mbox{.}(2014)Lehner, O'Meara, Fox, Howk, Prochaska,
  Burns, \& Armstrong}]{Lehner2014}
Lehner N., O'Meara J.~M., Fox A.~J., Howk J.~C., Prochaska J.~X., Burns V.,
  Armstrong A.~A., 2014, The Astrophysical Journal, 3

\bibitem[{Lequeux {et~al}\mbox{.}(1979)Lequeux, Peimbert, Rayo, Serrano, \&
  Torres-Peimbert}]{Lequeux1979}
Lequeux J., Peimbert M., Rayo J.~F., Serrano A., Torres-Peimbert S., 1979,
  Astronomy and Astrophysics, 80, 155

\bibitem[{Lopez \& Ellison(2003)}]{Lopez2003}
Lopez S., Ellison S.~L., 2003, Astronomy and Astrophysics, 403, 573

\bibitem[{Lopez {et~al}\mbox{.}(2002)Lopez, Reimers, D'Odorico, \&
  Prochaska}]{Lopez2002}
Lopez S., Reimers D., D'Odorico S., Prochaska J.~X., 2002, Astronomy and
  Astrophysics, 385, 778

\bibitem[{Lopez {et~al}\mbox{.}(2005)Lopez, Reimers, Gregg, Wisotzki, Wucknitz,
  \& Guzman}]{Lopez2005}
Lopez S., Reimers D., Gregg M.~D., Wisotzki L., Wucknitz O., Guzman A., 2005,
  The Astrophysical Journal, 626, 767

\bibitem[{Lopez {et~al}\mbox{.}(1999)Lopez, Reimers, Rauch, Sargent, \&
  Smette}]{Lopez1999}
Lopez S., Reimers D., Rauch M., Sargent W. L.~W., Smette A., 1999, 10, 49

\bibitem[{Lu {et~al}\mbox{.}(1996)Lu, Sargent, Barlow, Galaxies, Churchill, \&
  Vogt}]{Lu1996}
Lu L., Sargent W. L.~W., Barlow T.~A., Galaxies D.~L., Churchill C.~W., Vogt
  S.~S., 1996, The Astrophysical Journal Supplement Series, 107, 475

\bibitem[{Lundgren {et~al}\mbox{.}(2009)Lundgren, Brunner, York, Ross,
  Quashnock, Myers, Schneider, AlSayyad, \& Bahcall}]{Lundgren2009}
Lundgren B.~F. {et~al.}, 2009, The Astrophysical Journal, 698, 819

\bibitem[{Maddox {et~al}\mbox{.}(2012)Maddox, Hewett, P{\'{e}}roux, Nestor, \&
  Wisotzki}]{Maddox2012}
Maddox N., Hewett P.~C., P{\'{e}}roux C., Nestor D.~B., Wisotzki L., 2012,
  Monthly Notices of the Royal Astronomical Society, 424, 2876

\bibitem[{Malaney \& Chaboyer(1996)}]{Malaney1996}
Malaney R.~A., Chaboyer B., 1996, The Astrophysical Journal, 462, 57

\bibitem[{Martin(2005)}]{Martin2005}
Martin C.~L., 2005, The Astrophysical Journal, 621, 227

\bibitem[{Martin {et~al}\mbox{.}(2012)Martin, Shapley, Coil, Kornei, Bundy,
  Weiner, Noeske, \& Schiminovich}]{Martin2012}
Martin C.~L., Shapley A.~E., Coil A.~L., Kornei K.~a., Bundy K., Weiner B.~J.,
  Noeske K.~G., Schiminovich D., 2012, The Astrophysical Journal, 760, 127

\bibitem[{Martin {et~al}\mbox{.}(2014)Martin, Chang, Matuszewski, Morrissey,
  Rahman, Moore, \& Steidel}]{Martin2014a}
Martin D.~C., Chang D., Matuszewski M., Morrissey P., Rahman S., Moore A.,
  Steidel C.~C., 2014, The Astrophysical Journal, 786, 106

\bibitem[{Meiring {et~al}\mbox{.}(2006)Meiring, Kulkarni, Khare, Bechtold,
  York, Cui, Lauroesch, Crotts, \& Nakamura}]{Meiring2006}
Meiring J.~D. {et~al.}, 2006, Monthly Notices of the Royal Astronomical
  Society, 370, 43

\bibitem[{Meiring {et~al}\mbox{.}(2009{\natexlab{a}})Meiring, Lauroesch,
  Kulkarni, P{\'{e}}roux, Khare, \& York}]{Meiring2009}
Meiring J.~D., Lauroesch J.~T., Kulkarni V.~P., P{\'{e}}roux C., Khare P., York
  D.~G., 2009{\natexlab{a}}, Monthly Notices of the Royal Astronomical Society,
  397, 2037

\bibitem[{Meiring {et~al}\mbox{.}(2009{\natexlab{b}})Meiring, Lauroesch,
  Kulkarni, Peroux, Khare, \& York}]{Meiring2009b}
Meiring J.~D., Lauroesch J.~T., Kulkarni V.~P., Peroux C., Khare P., York
  D.~G., 2009{\natexlab{b}}, Monthly Notices of the Royal Astronomical Society,
  397, 2037

\bibitem[{Meiring {et~al}\mbox{.}(2007)Meiring, Lauroesch, Kulkarni,
  P{\'{e}}roux, Khare, York, \& Crotts}]{Meiring2007}
Meiring J.~D., Lauroesch J.~T., Kulkarni V.~P., P{\'{e}}roux C., Khare P., York
  D.~G., Crotts a. P.~S., 2007, Monthly Notices of the Royal Astronomical
  Society, 376, 557

\bibitem[{Meyer, Lanzetta \& Wolfe(1995)Meyer, Lanzetta, \& Wolfe}]{Meyer1995}
Meyer D.~M., Lanzetta K.~M., Wolfe A.~M., 1995, The Astrophysical Journal, 451,
  1

\bibitem[{Milutinovic {et~al}\mbox{.}(2010)Milutinovic, Ellison, Prochaska, \&
  Tumlinson}]{Milutinovic2010}
Milutinovic N., Ellison S.~L., Prochaska J.~X., Tumlinson J., 2010, Monthly
  Notices of the Royal Astronomical Society, 408, 2071

\bibitem[{Molaro {et~al}\mbox{.}(2000)Molaro, Bonifacio, Centurion,
  D’Odorico, Vladilo, Santin, \& {Di Marcantonio}}]{Molaro2000}
Molaro P., Bonifacio P., Centurion M., D’Odorico S., Vladilo G., Santin P.,
  {Di Marcantonio} P., 2000, The Astrophysical Journal, 541, 54

\bibitem[{Moller {et~al}\mbox{.}(2013)Moller, Fynbo, Ledoux, \&
  Nilsson}]{Moller2013}
Moller P., Fynbo J. P.~U., Ledoux C., Nilsson K.~K., 2013, Monthly Notices of
  the Royal Astronomical Society, 430, 2680

\bibitem[{Morton(2003)}]{Morton2003}
Morton D.~C., 2003, The Astrophysical Journal Supplement Series, 149, 205

\bibitem[{Neeleman {et~al}\mbox{.}(2013)Neeleman, Wolfe, Prochaska, \&
  Rafelski}]{Neeleman2013}
Neeleman M., Wolfe A.~M., Prochaska J.~X., Rafelski M., 2013, The Astrophysical
  Journal, 769, 54

\bibitem[{Nestor {et~al}\mbox{.}(2008)Nestor, Pettini, Hewett, Rao, \&
  Wild}]{Nestor2008}
Nestor D.~B., Pettini M., Hewett P.~C., Rao S., Wild V., 2008, Monthly Notices
  of the Royal Astronomical Society, 390, 1670

\bibitem[{Noterdaeme {et~al}\mbox{.}(2012{\natexlab{a}})Noterdaeme, Laursen,
  Petitjean, Vergani, Maureira, Ledoux, Fynbo, L{\'{o}}pez, \&
  Srianand}]{Noterdaeme2012b}
Noterdaeme P. {et~al.}, 2012{\natexlab{a}}, Astronomy {\&} Astrophysics, 540,
  A63

\bibitem[{Noterdaeme {et~al}\mbox{.}(2007{\natexlab{a}})Noterdaeme, Ledoux,
  Petitjean, {Le Petit}, Srianand, \& Smette}]{Noterdaeme2007}
Noterdaeme P., Ledoux C., Petitjean P., {Le Petit} F., Srianand R., Smette A.,
  2007{\natexlab{a}}, Astronomy and Astrophysics, 474, 393

\bibitem[{Noterdaeme {et~al}\mbox{.}(2008{\natexlab{a}})Noterdaeme, Ledoux,
  Petitjean, \& Srianand}]{Noterdaeme2008a}
Noterdaeme P., Ledoux C., Petitjean P., Srianand R., 2008{\natexlab{a}},
  Astronomy and Astrophysics, 481, 327

\bibitem[{Noterdaeme {et~al}\mbox{.}(2012{\natexlab{b}})Noterdaeme,
  L{\'{o}}pez, Dumont, Ledoux, Molaro, \& Petitjean}]{Noterdaeme2012a}
Noterdaeme P., L{\'{o}}pez S., Dumont V., Ledoux C., Molaro P., Petitjean P.,
  2012{\natexlab{b}}, Astronomy {\&} Astrophysics, 542, L33

\bibitem[{Noterdaeme {et~al}\mbox{.}(2012{\natexlab{c}})Noterdaeme, Petitjean,
  Carithers, P{\^{a}}ris, Font-Ribera, Bailey, Aubourg, Bizyaev, Ebelke,
  Finley, Ge, Malanushenko, Malanushenko, Miralda-Escud{\'{e}}, Myers, Oravetz,
  Pan, Pieri, Ross, Schneider, Simmons, \& York}]{Noterdaeme2012}
Noterdaeme P. {et~al.}, 2012{\natexlab{c}}, Astronomy {\&} Astrophysics, 547,
  L1

\bibitem[{Noterdaeme {et~al}\mbox{.}(2008{\natexlab{b}})Noterdaeme, Petitjean,
  Ledoux, Srianand, \& Ivanchik}]{Noterdaeme2008}
Noterdaeme P., Petitjean P., Ledoux C., Srianand R., Ivanchik a.,
  2008{\natexlab{b}}, Astronomy and Astrophysics, 491, 397

\bibitem[{Noterdaeme {et~al}\mbox{.}(2007{\natexlab{b}})Noterdaeme, Petitjean,
  Srianand, Ledoux, \& {Le Petit}}]{Noterdaeme2007a}
Noterdaeme P., Petitjean P., Srianand R., Ledoux C., {Le Petit} F.,
  2007{\natexlab{b}}, Astronomy and Astrophysics, 469, 425

\bibitem[{Ocvirk, Pichon \& Teyssier(2008)Ocvirk, Pichon, \&
  Teyssier}]{Ocvirk2008}
Ocvirk P., Pichon C., Teyssier R., 2008, Monthly Notices of the Royal
  Astronomical Society, 390, 1326

\bibitem[{O'Meara {et~al}\mbox{.}(2015)O'Meara, Lehner, Howk, Prochaska, Fox,
  Swain, Gelino, Berriman, \& Tran}]{Omeara2015}
O'Meara J.~M. {et~al.}, 2015, Astronomical Journal

\bibitem[{Oppenheimer {et~al}\mbox{.}(2010)Oppenheimer, Dav{\'{e}},
  Kere{\v{s}}, Fardal, Katz, Kollmeier, \& Weinberg}]{Oppenheimer2010}
Oppenheimer B.~D., Dav{\'{e}} R., Kere{\v{s}} D., Fardal M., Katz N., Kollmeier
  J.~a., Weinberg D.~H., 2010, Monthly Notices of the Royal Astronomical
  Society, 406, 2325

\bibitem[{Pei \& Fall(1995)}]{Pei1995}
Pei Y.~C., Fall S.~M., 1995, The Astrophysical Journal, 454, 69

\bibitem[{Pei, Fall \& Hauser(1999)Pei, Fall, \& Hauser}]{Pei1999}
Pei Y.~C., Fall S.~M., Hauser M.~G., 1999, The Astrophysical Journal, 522, 604

\bibitem[{Penprase {et~al}\mbox{.}(2010)Penprase, Prochaska, Sargent,
  Toro-Martinez, \& Beeler}]{Penprase2010}
Penprase B.~E., Prochaska J.~X., Sargent W. L.~W., Toro-Martinez I., Beeler
  D.~J., 2010, The Astrophysical Journal, 721, 1

\bibitem[{P{\'{e}}roux {et~al}\mbox{.}(2011)P{\'{e}}roux, Bouch{\'{e}},
  Kulkarni, York, \& Vladilo}]{Peroux2011}
P{\'{e}}roux C., Bouch{\'{e}} N., Kulkarni V.~P., York D.~G., Vladilo G., 2011,
  Monthly Notices of the Royal Astronomical Society, 410, 2251

\bibitem[{P{\'{e}}roux {et~al}\mbox{.}(2005)P{\'{e}}roux, Dessauges-Zavadsky,
  D'Odorico, Kim, \& McMahon}]{Peroux2005}
P{\'{e}}roux C., Dessauges-Zavadsky M., D'Odorico S., Kim T.~S., McMahon R.~G.,
  2005, Monthly Notices of the Royal Astronomical Society, 363, 479

\bibitem[{P{\'{e}}roux {et~al}\mbox{.}(2007)P{\'{e}}roux, Dessauges-Zavadsky,
  D'Odorico, Kim, \& McMahon}]{Peroux2007}
P{\'{e}}roux C., Dessauges-Zavadsky M., D'Odorico S., Kim T.~S., McMahon R.~G.,
  2007, Monthly Notices of the Royal Astronomical Society, 382, 177

\bibitem[{P{\'{e}}roux {et~al}\mbox{.}(2006{\natexlab{a}})P{\'{e}}roux,
  Kulkarni, Meiring, Ferlet, Khare, Lauroesch, Vladilo, \& York}]{Peroux2006}
P{\'{e}}roux C., Kulkarni V.~P., Meiring J., Ferlet R., Khare P., Lauroesch
  J.~T., Vladilo G., York D.~G., 2006{\natexlab{a}}, Astronomy {\&}
  Astrophysics, 450, 53

\bibitem[{P{\'{e}}roux, Kulkarni \& York(2014)P{\'{e}}roux, Kulkarni, \&
  York}]{Peroux2014}
P{\'{e}}roux C., Kulkarni V.~P., York D.~G., 2014, Monthly Notices of the Royal
  Astronomical Society, 437, 3144

\bibitem[{P{\'{e}}roux {et~al}\mbox{.}(2006{\natexlab{b}})P{\'{e}}roux,
  Meiring, Kulkarni, Ferlet, Khare, Lauroesch, Vladilo, \& York}]{Peroux2006a}
P{\'{e}}roux C., Meiring J.~D., Kulkarni V.~P., Ferlet R., Khare P., Lauroesch
  J.~T., Vladilo G., York D.~G., 2006{\natexlab{b}}, Monthly Notices of the
  Royal Astronomical Society, 372, 369

\bibitem[{P{\'{e}}roux {et~al}\mbox{.}(2008)P{\'{e}}roux, Meiring, Kulkarni,
  Khare, Lauroesch, Vladilo, \& York}]{Peroux2008}
P{\'{e}}roux C., Meiring J.~D., Kulkarni V.~P., Khare P., Lauroesch J.~T.,
  Vladilo G., York D.~G., 2008, Monthly Notices of the Royal Astronomical
  Society, 386, 2209

\bibitem[{Peroux {et~al}\mbox{.}(2002)Peroux, Petitjean, Aracil, \&
  Srianand}]{Peroux2002}
Peroux C., Petitjean P., Aracil B., Srianand R., 2002, New Astronomy, 7, 1

\bibitem[{Petitjean, Ledoux \& Srianand(2008)Petitjean, Ledoux, \&
  Srianand}]{Petitjean2008}
Petitjean P., Ledoux C., Srianand R., 2008, Astronomy {\&} Astrophysics, 480,
  349

\bibitem[{Petitjean, Srianand \& Ledoux(2000)Petitjean, Srianand, \&
  Ledoux}]{Petitjean2000}
Petitjean P., Srianand R., Ledoux C., 2000, AA, 364, 26

\bibitem[{Petitjean, Srianand \& Ledoux(2002)Petitjean, Srianand, \&
  Ledoux}]{Petitjean2002}
Petitjean P., Srianand R., Ledoux C., 2002, Monthly Notices of the Royal
  Astronomical Society, 332, 9

\bibitem[{Pettini \& Cooke(2012)}]{Pettini2012}
Pettini M., Cooke R., 2012, Monthly Notices of the Royal Astronomical Society,
  425, 2477

\bibitem[{Pettini {et~al}\mbox{.}(2002)Pettini, Ellison, Bergeron, \&
  Petitjean}]{Pettini2002}
Pettini M., Ellison S.~L., Bergeron J., Petitjean P., 2002, Astronomy and
  Astrophysics, 391, 21

\bibitem[{Pettini {et~al}\mbox{.}(1999)Pettini, Ellison, Steidel, \&
  Bowen}]{Pettini1999}
Pettini M., Ellison S.~L., Steidel C.~C., Bowen D.~V., 1999, The Astrophysical
  Journal, 510, 576

\bibitem[{Pettini {et~al}\mbox{.}(2000)Pettini, Ellison, Steidel, Shapley, \&
  Bowen}]{Pettini2000}
Pettini M., Ellison S.~L., Steidel C.~C., Shapley A.~E., Bowen D.~V., 2000, The
  Astrophysical Journal, 532, 65

\bibitem[{Pettini {et~al}\mbox{.}(1997{\natexlab{a}})Pettini, King, Smith, \&
  Hunstead}]{Pettini1997a}
Pettini M., King D.~L., Smith L.~J., Hunstead R.~W., 1997{\natexlab{a}}, The
  Astrophysical Journal, 478, 536

\bibitem[{Pettini {et~al}\mbox{.}(1994)Pettini, Smith, Hunstead, \&
  King}]{Pettini1994b}
Pettini M., Smith L.~J., Hunstead R.~W., King D.~L., 1994, {Metal enrichment,
  dust, and star formation in galaxies at high redshifts. 3: Zn and CR
  abundances for 17 damped Lyman-alpha systems}

\bibitem[{Pettini {et~al}\mbox{.}(1997{\natexlab{b}})Pettini, Smith, King, \&
  Hunstead}]{Pettini1997b}
Pettini M., Smith L.~J., King D.~L., Hunstead R.~W., 1997{\natexlab{b}}, The
  Astrophysical Journal, 486, 665

\bibitem[{Pettini {et~al}\mbox{.}(2008{\natexlab{a}})Pettini, Zych, Murphy,
  Lewis, \& Steidel}]{Pettini2008}
Pettini M., Zych B.~J., Murphy M.~T., Lewis A., Steidel C.~C.,
  2008{\natexlab{a}}, Monthly Notices of the Royal Astronomical Society, 391,
  1499

\bibitem[{Pettini {et~al}\mbox{.}(2008{\natexlab{b}})Pettini, Zych, Steidel, \&
  Chaffee}]{Pettini2008a}
Pettini M., Zych B.~J., Steidel C.~C., Chaffee F.~H., 2008{\natexlab{b}},
  Monthly Notices of the Royal Astronomical Society, 385, 2011

\bibitem[{Pontzen {et~al}\mbox{.}(2008)Pontzen, Governato, Pettini, Booth,
  Stinson, Wadsley, Brooks, Quinn, \& Haehnelt}]{Pontzen2008a}
Pontzen A. {et~al.}, 2008, Monthly Notices of the Royal Astronomical Society,
  390, 1349

\bibitem[{Prochaska, Castro \& Djorgovski(2003)Prochaska, Castro, \&
  Djorgovski}]{Prochaska2003a}
Prochaska J.~X., Castro S., Djorgovski S.~G., 2003, The Astrophysical Journal
  Supplement Series, 148, 317

\bibitem[{Prochaska, Gawiser \& Wolfe(2001)Prochaska, Gawiser, \&
  Wolfe}]{Prochaska2001b}
Prochaska J.~X., Gawiser E., Wolfe A.~M., 2001, The Astrophysical Journal, 552,
  99

\bibitem[{Prochaska {et~al}\mbox{.}(2003)Prochaska, Gawiser, Wolfe, Cooke, \&
  Gelino}]{Prochaska2003}
Prochaska J.~X., Gawiser E., Wolfe A.~M., Cooke J., Gelino D., 2003, The
  Astrophysical Journal Supplement Series, 147, 227

\bibitem[{Prochaska {et~al}\mbox{.}(2002)Prochaska, Henry, O’Meara, Tytler,
  Wolfe, Kirkman, Lubin, \& Suzuki}]{Prochaska2002a}
Prochaska J.~X., Henry R. B.~C., O’Meara J.~M., Tytler D., Wolfe A.~M.,
  Kirkman D., Lubin D., Suzuki N., 2002, Publications of the Astronomical
  Society of the Pacific, 114, 933

\bibitem[{Prochaska, Herbert‐Fort \& Wolfe(2005)Prochaska, Herbert‐Fort, \&
  Wolfe}]{Prochaska2005}
Prochaska J.~X., Herbert‐Fort S., Wolfe A.~M., 2005, The Astrophysical
  Journal, 635, 123

\bibitem[{Prochaska \& Wolfe(1996)}]{Prochaska1996}
Prochaska J.~X., Wolfe A.~M., 1996, The Astrophysical Journal, 470, 403

\bibitem[{Prochaska \& Wolfe(1997{\natexlab{a}})}]{Prochaska1997a}
Prochaska J.~X., Wolfe A.~M., 1997{\natexlab{a}}, $\backslash$Apj, 474, 140

\bibitem[{Prochaska \& Wolfe(1997{\natexlab{b}})}]{Prochaska1997}
Prochaska J.~X., Wolfe A.~M., 1997{\natexlab{b}}, The Astrophysical Journal,
  487, 73

\bibitem[{Prochaska \& Wolfe(1999)}]{Prochaska1999}
Prochaska J.~X., Wolfe A.~M., 1999, The Astrophysical Journal Supplement
  Series, 121, 369

\bibitem[{Prochaska \& Wolfe(2002)}]{Prochaska2001}
Prochaska J.~X., Wolfe A.~M., 2002, The Astrophysical Journal, 566, 68

\bibitem[{Prochaska \& Wolfe(2009)}]{Prochaska2009}
Prochaska J.~X., Wolfe A.~M., 2009, The Astrophysical Journal, 696, 1543

\bibitem[{Prochaska {et~al}\mbox{.}(2007)Prochaska, Wolfe, Howk, Gawiser,
  Burles, \& Cooke}]{Prochaska2007}
Prochaska J.~X., Wolfe A.~M., Howk J.~C., Gawiser E., Burles S.~M., Cooke J.,
  2007, The Astrophysical Journal Supplement Series, 171, 29

\bibitem[{Prochter {et~al}\mbox{.}(2010)Prochter, Prochaska, O'Meara, Burles,
  \& Bernstein}]{Prochter2010}
Prochter G.~E., Prochaska J.~X., O'Meara J.~M., Burles S., Bernstein R.~a.,
  2010, The Astrophysical Journal, 20

\bibitem[{Quast, Reimers \& Baade(2008)Quast, Reimers, \& Baade}]{Quast2008}
Quast R., Reimers D., Baade R., 2008, Astronomy {\&} Astrophysics, 457, 443

\bibitem[{Rafelski {et~al}\mbox{.}(2014)Rafelski, Neeleman, Fumagalli, Wolfe,
  \& Prochaska}]{Rafelski2014}
Rafelski M., Neeleman M., Fumagalli M., Wolfe A.~M., Prochaska J.~X., 2014, The
  Astrophysical Journal, 782, L29

\bibitem[{Rafelski {et~al}\mbox{.}(2012)Rafelski, Wolfe, Prochaska, Neeleman,
  \& Mendez}]{Rafelski2012}
Rafelski M., Wolfe A.~M., Prochaska J.~X., Neeleman M., Mendez A.~J., 2012, The
  Astrophysical Journal, 755, 89

\bibitem[{Rao {et~al}\mbox{.}(2005)Rao, Prochaska, Howk, \& Wolfe}]{Rao2005}
Rao S.~M., Prochaska J.~X., Howk J.~C., Wolfe A.~M., 2005, The Astronomical
  Journal, 129, 9

\bibitem[{Rao \& Turnshek(2000)}]{Rao2000}
Rao S.~M., Turnshek D.~a., 2000, The Astrophysical Journal Supplement Series,
  130, 1

\bibitem[{Rao, Turnshek \& Nestor(2006)Rao, Turnshek, \& Nestor}]{Rao2006}
Rao S.~M., Turnshek D.~a., Nestor D.~B., 2006, The Astrophysical Journal, 636,
  610

\bibitem[{Rees \& Ostriker(1977)}]{Rees1977}
Rees M., Ostriker J., 1977, Monthly Notices of the Royal Astronomical Society,
  179, 541

\bibitem[{Richter, Westmeier \& Bruens(2005)Richter, Westmeier, \&
  Bruens}]{Richter2005}
Richter P., Westmeier T., Bruens C., 2005, 52, 4

\bibitem[{Rix {et~al}\mbox{.}(2007)Rix, Pettini, Steidel, Reddy, Adelberger,
  Erb, \& Shapley}]{Rix2007}
Rix S.~a., Pettini M., Steidel C.~C., Reddy N.~a., Adelberger K.~L., Erb D.~K.,
  Shapley A.~E., 2007, 27

\bibitem[{Rubin {et~al}\mbox{.}(2014)Rubin, Prochaska, Koo, Phillips, Martin,
  \& Winstrom}]{Rubin2014}
Rubin K. H.~R., Prochaska J.~X., Koo D.~C., Phillips A.~C., Martin C.~L.,
  Winstrom L.~O., 2014, The Astrophysical Journal, 794, 156

\bibitem[{Rubin {et~al}\mbox{.}(2012)Rubin, {Xavier Prochaska}, Koo, \&
  Phillips}]{Rubin2012}
Rubin K. H.~R., {Xavier Prochaska} J., Koo D.~C., Phillips A.~C., 2012, The
  Astrophysical Journal, 747, L26

\bibitem[{Ryan-Weber {et~al}\mbox{.}(2009)Ryan-Weber, Pettini, Madau, \&
  Zych}]{Ryan-Weber2009}
Ryan-Weber E.~V., Pettini M., Madau P., Zych B.~J., 2009, Monthly Notices of
  the Royal Astronomical Society, 395, 1476

\bibitem[{Saito {et~al}\mbox{.}(2009)Saito, Takada-Hidai, Honda, \&
  Takeda}]{Saito2009}
Saito Y.-j., Takada-Hidai M., Honda S., Takeda Y., 2009, Publications of the
  Astronomical Society of Japan, 61, 549

\bibitem[{Savage \& Sembach(1996)}]{Savage1996}
Savage B.~D., Sembach K.~R., 1996, Annual Review of Astronomy and Astrophysics,
  34, 279

\bibitem[{Savaglio {et~al}\mbox{.}(2005)Savaglio, Glazebrook, Borgne, Juneau,
  Abraham, Chen, Crampton, McCarthy, Carlberg, Marzke, Roth, Jorgensen, \&
  Murowinski}]{Savaglio2005}
Savaglio S. {et~al.}, 2005, 1, 260

\bibitem[{Schroetter {et~al}\mbox{.}(2015)Schroetter, Bouch{\'{e}}, Peroux,
  Murphy, Contini, \& Finley}]{Schroetter2015}
Schroetter I., Bouch{\'{e}} N., Peroux C., Murphy M.~T., Contini T., Finley H.,
  2015

\bibitem[{Shapley {et~al}\mbox{.}(2003)Shapley, Steidel, Pettini, \&
  Adelberger}]{Shapley2003}
Shapley a.~E., Steidel C.~C., Pettini M., Adelberger K.~L., 2003, The
  Astrophysical Journal, 588, 65

\bibitem[{Shen {et~al}\mbox{.}(2013)Shen, Madau, Guedes, Mayer, Prochaska, \&
  Wadsley}]{Shen2013}
Shen S., Madau P., Guedes J., Mayer L., Prochaska J.~X., Wadsley J., 2013, The
  Astrophysical Journal, 765, 89

\bibitem[{Shull, Danforth \& Tilton(2014)Shull, Danforth, \&
  Tilton}]{Shull2014}
Shull J.~M., Danforth C.~W., Tilton E.~M., 2014, The Astrophysical Journal,
  796, 49

\bibitem[{Shull {et~al}\mbox{.}(2015)Shull, Moloney, Danforth, \&
  Tilton}]{Shull2015}
Shull J.~M., Moloney J., Danforth C.~W., Tilton E.~M., 2015, The Astrophysical
  Journal, 811, 3

\bibitem[{Silk(1977)}]{Silk1977a}
Silk J., 1977, The Astrophysical Journal, 211, 638

\bibitem[{Simcoe, Sargent \& Rauch(2004)Simcoe, Sargent, \& Rauch}]{Simcoe2004}
Simcoe R.~a., Sargent W. L.~W., Rauch M., 2004, The Astrophysical Journal, 606,
  92

\bibitem[{Som {et~al}\mbox{.}(2013)Som, Kulkarni, Meiring, York, P{\'{e}}roux,
  Khare, \& Lauroesch}]{Som2013}
Som D., Kulkarni V.~P., Meiring J., York D.~G., P{\'{e}}roux C., Khare P.,
  Lauroesch J.~T., 2013, Monthly Notices of the Royal Astronomical Society,
  435, 1469

\bibitem[{Som {et~al}\mbox{.}(2015)Som, Kulkarni, Meiring, York, P{\'{e}}roux,
  Lauroesch, Aller, \& Khare}]{Som15}
Som D., Kulkarni V.~P., Meiring J., York D.~G., P{\'{e}}roux C., Lauroesch
  J.~T., Aller M.~C., Khare P., 2015, The Astrophysical Journal, 806, 25

\bibitem[{Songaila \& Cowie(1996)}]{Songaila1996}
Songaila A., Cowie L.~L., 1996, The Astrophysical Journal, 112, 335

\bibitem[{Songaila \& Cowie(2002)}]{Songaila2002}
Songaila A., Cowie L.~L., 2002, The Astronomical Journal, 123, 2183

\bibitem[{Srianand, Gupta \& Petitjean(2007)Srianand, Gupta, \&
  Petitjean}]{Srianand2007}
Srianand R., Gupta N., Petitjean P., 2007, Monthly Notices of the Royal
  Astronomical Society, 375, 584

\bibitem[{Srianand {et~al}\mbox{.}(2012)Srianand, Gupta, Petitjean, Noterdaeme,
  Ledoux, Salter, \& Saikia}]{Srianand2012}
Srianand R., Gupta N., Petitjean P., Noterdaeme P., Ledoux C., Salter C.~J.,
  Saikia D.~J., 2012, Monthly Notices of the Royal Astronomical Society, 421,
  no

\bibitem[{Srianand \& Petitjean(1998)}]{Srianand1998}
Srianand R., Petitjean P., 1998, Astronomy {\&} Astrophysics, 40, 33

\bibitem[{Srianand \& Petitjean(2001)}]{Srianand2001}
Srianand R., Petitjean P., 2001, European Space Agency, (Special Publication)
  ESA SP, 826, 12

\bibitem[{Srianand, Petitjean \& Ledoux(2000)Srianand, Petitjean, \&
  Ledoux}]{Srianand2000}
Srianand R., Petitjean P., Ledoux C., 2000, Nature, 408, 1

\bibitem[{Srianand {et~al}\mbox{.}(2005)Srianand, Petitjean, Ledoux, Ferland,
  \& Shaw}]{Srianand2005}
Srianand R., Petitjean P., Ledoux C., Ferland G., Shaw G., 2005, {The VLT-UVES
  survey for molecular hydrogen in high-redshift damped Lyman a systems:
  Physical conditions in the neutral gas}

\bibitem[{Steidel {et~al}\mbox{.}(2000)Steidel, Adelberger, Shapley, Pettini,
  Dickinson, \& Giavalisco}]{Steidel2000}
Steidel C.~C., Adelberger K.~L., Shapley A.~E., Pettini M., Dickinson M.,
  Giavalisco M., 2000, The Astrophysical Journal, 532, 170

\bibitem[{Steidel {et~al}\mbox{.}(1997)Steidel, Dickinson, Meyer, Adelberger,
  \& Sembach}]{Steidel1997}
Steidel C.~C., Dickinson M., Meyer D.~M., Adelberger K.~L., Sembach K.~R.,
  1997, Astrophysical Journal, 480, 568

\bibitem[{Steidel {et~al}\mbox{.}(2010)Steidel, Erb, Shapley, Pettini, Reddy,
  Bogosavljevi{\'{c}}, Rudie, \& Rakic}]{Steidel2010}
Steidel C.~C., Erb D.~K., Shapley A.~E., Pettini M., Reddy N.,
  Bogosavljevi{\'{c}} M., Rudie G.~C., Rakic O., 2010, The Astrophysical
  Journal, 717, 289

\bibitem[{Storrie-Lombardi \& Wolfe(2000)}]{StorrieLombardi2000}
Storrie-Lombardi L.~J., Wolfe A.~M., 2000, The Astrophysical Journal, 543, 46

\bibitem[{Tchernyshyov {et~al}\mbox{.}(2015)Tchernyshyov, Meixner, Seale, Fox,
  Friedman, Sembach, \& Dwek}]{Tchernyshyov2015}
Tchernyshyov K., Meixner M., Seale J., Fox A., Friedman S.~D., Sembach K., Dwek
  E., 2015, The Astrophysical Journal, 811, 1

\bibitem[{Timmes, Lauroesch \& Truran(1995)Timmes, Lauroesch, \&
  Truran}]{Timmes1995}
Timmes F.~X., Lauroesch J.~T., Truran J.~W., 1995, 22

\bibitem[{Tissera, Mosconi \& Cora(2001)Tissera, Mosconi, \&
  Cora}]{Tissera2001}
Tissera P.~B., Mosconi D. G. L. M.~B., Cora S.~A., 2001, 20, 527

\bibitem[{Tremonti {et~al}\mbox{.}(2004)Tremonti, Heckman, Kauffmann,
  Brinchmann, Charlot, White, Seibert, Peng, Schlegel, Uomoto, Fukugita, \&
  Brinkmann}]{Tremonti2004}
Tremonti C.~A. {et~al.}, 2004, The Astrophysical Journal, 613, 15

\bibitem[{Tripp {et~al}\mbox{.}(2005)Tripp, Jenkins, Bowen, Prochaska, Aracil,
  \& Ganguly}]{Tripp2005}
Tripp T.~M., Jenkins E.~B., Bowen D.~V., Prochaska J.~X., Aracil B., Ganguly
  R., 2005, 714

\bibitem[{Tumlinson {et~al}\mbox{.}(2011)Tumlinson, Thom, Werk, Prochaska,
  Tripp, Weinberg, Peeples, O'Meara, Oppenheimer, Meiring, Katz, Dav{\'{e}},
  Ford, \& Sembach}]{Tumlinson2011}
Tumlinson J. {et~al.}, 2011, Science (New York, N.Y.), 334, 948

\bibitem[{Turnshek {et~al}\mbox{.}(2015)Turnshek, Monier, Rao, Hamilton,
  Sardane, \& Held}]{Turnshek2015}
Turnshek D.~A., Monier E.~M., Rao S.~M., Hamilton T.~S., Sardane G.~M., Held
  R., 2015, Monthly Notices of the Royal Astronomical Society, 449, 1536

\bibitem[{Turnshek {et~al}\mbox{.}(2004)Turnshek, Rao, Nestor, {Vanden Berk},
  Belfort-Mihalyi, \& Monier}]{Turnshek2004}
Turnshek D.~A., Rao S.~M., Nestor D.~B., {Vanden Berk} D., Belfort-Mihalyi M.,
  Monier E.~M., 2004, The Astrophysical Journal, 609, L53

\bibitem[{van~de Voort \& Schaye(2012)}]{VandeVoort2012a}
van~de Voort F., Schaye J., 2012, Monthly Notices of the Royal Astronomical
  Society, 423, 2991

\bibitem[{van~de Voort {et~al}\mbox{.}(2012)van~de Voort, Schaye, Altay, \&
  Theuns}]{VandeVoort2012b}
van~de Voort F., Schaye J., Altay G., Theuns T., 2012, Monthly Notices of the
  Royal Astronomical Society, 421, 2809

\bibitem[{Vladilo(1998)}]{Vladilo1998}
Vladilo G., 1998, The Astrophysical Journal, 493, 583

\bibitem[{Vladilo {et~al}\mbox{.}(2011)Vladilo, Abate, Yin, Cescutti, \&
  Matteucci}]{Vladilo2011}
Vladilo G., Abate C., Yin J., Cescutti G., Matteucci F., 2011, Astronomy {\&}
  Astrophysics, 530, A33

\bibitem[{Vladilo {et~al}\mbox{.}(2001)Vladilo, Centurion, Bonifacio, \&
  Howk}]{Vladilo2001}
Vladilo G., Centurion M., Bonifacio P., Howk J.~C., 2001, The Astrophysical
  Journal, 557, 1007

\bibitem[{Vladilo {et~al}\mbox{.}(2006)Vladilo, Centuri{\'{o}}n, Levshakov,
  P{\'{e}}roux, Khare, Kulkarni, \& York}]{Vladilo2006}
Vladilo G., Centuri{\'{o}}n M., Levshakov S.~a., P{\'{e}}roux C., Khare P.,
  Kulkarni V.~P., York D.~G., 2006, Astronomy and Astrophysics, 454, 151

\bibitem[{Wallerstein(1962)}]{Wallerstein1962}
Wallerstein G., 1962, The Astrophysical Journal Supplement Series, 6, 407

\bibitem[{White \& Rees(1978)}]{White1978}
White S. D.~M., Rees M.~J., 1978, Monthly Notices of the Royal Astronomical
  Society, 183, 341

\bibitem[{Wolfe, Gawiser \& Prochaska(2005)Wolfe, Gawiser, \&
  Prochaska}]{Wolfe2005}
Wolfe A.~M., Gawiser E., Prochaska J.~X., 2005, Annual Review of Astronomy and
  Astrophysics, 43, 861

\bibitem[{Zafar {et~al}\mbox{.}(2014{\natexlab{a}})Zafar, Centurion, Peroux,
  Molaro, D'Odorico, Vladilo, \& Popping}]{Zafar2014}
Zafar T., Centurion M., Peroux C., Molaro P., D'Odorico V., Vladilo G., Popping
  A., 2014{\natexlab{a}}, 16, 16

\bibitem[{Zafar {et~al}\mbox{.}(2013)Zafar, P{\'{e}}roux, Popping, Milliard,
  Deharveng, \& Frank}]{Zafar2013a}
Zafar T., P{\'{e}}roux C., Popping A., Milliard B., Deharveng J.-M., Frank S.,
  2013, Astronomy {\&} Astrophysics, 556, A141

\bibitem[{Zafar, Popping \& P{\'{e}}roux(2013)Zafar, Popping, \&
  P{\'{e}}roux}]{Zafar2013}
Zafar T., Popping A., P{\'{e}}roux C., 2013, Astronomy {\&} Astrophysics, 556,
  A140

\bibitem[{Zafar {et~al}\mbox{.}(2014{\natexlab{b}})Zafar, Vladilo, Peroux,
  Molaro, Centurion, D'Odorico, Abbas, \& Popping}]{Zafar2014a}
Zafar T., Vladilo G., Peroux C., Molaro P., Centurion M., D'Odorico V., Abbas
  K., Popping A., 2014{\natexlab{b}}, Monthly Notices of the Royal Astronomical
  Society, 14

\bibitem[{Zafar {et~al}\mbox{.}(2011)Zafar, Watson, Fynbo, Malesani, Jakobsson,
  \& {de Ugarte Postigo}}]{Zafar2011}
Zafar T., Watson D., Fynbo J. P.~U., Malesani D., Jakobsson P., {de Ugarte
  Postigo} A., 2011, Astronomy {\&} Astrophysics, 532, A143

\bibitem[{Zwaan {et~al}\mbox{.}(2008)Zwaan, Walter, Ryan-Weber, Brinks,
  de~Blok, \& Kennicutt}]{Zwaan2008}
Zwaan M., Walter F., Ryan-Weber E., Brinks E., de~Blok W. J.~G., Kennicutt
  R.~C., 2008, The Astronomical Journal, 136, 2886

\bibitem[{Zych {et~al}\mbox{.}(2009)Zych, Murphy, Hewett, \&
  Prochaska}]{Zych2009}
Zych B.~J., Murphy M.~T., Hewett P.~C., Prochaska J.~X., 2009, Monthly Notices
  of the Royal Astronomical Society, 392, 1429

\end{thebibliography}

\clearpage

\appendix

\section[]{Multi-Element Analysis}

\subsection{Derivation of $\rm F_{*}$}
\label{ann:jenkins}

\cite{Jenkins09} proposed to use the abundances of different elements (namely C, N, O, Mg, Si, P, Cl, Ti, Cr, Mn, Fe, Ni, Cu, Zn, Ge, Kr and S) to compare the dust depletion of dense neutral hydrogen systems to that of the Interstellar Medium (ISM) of our Galaxy. 
Specifically, the depletion $\rm [X_{gas}/H]$ described in \cite{Jenkins09} is the difference between the observed abundance of element X normalized to the total hydrogen abundance in both neutral and molecular phases, $\rm N(H)=N(H\,I)+2N(H_{2})$, and its intrinsic abundance (assumed to be solar, initially):

\begin{equation}
\rm [X_{gas}/H] = \log (X/H)_{obs} - log (X/H)_{\odot}
\label{eq:def_depl}
\end{equation}

Using $243$ sight lines in our Galaxy, \cite{Jenkins09} linearly fits the following formula:
\begin{equation}
\rm [X_{gas}/H]_{fit} = [X_{gas}/H]_{0} + A_{X}F_{*}
\end{equation}
where  $\rm F_{*}$ is defined as the line-of-sight depletion factor\footnote{We stress that this definition is based on calibration of $\rm F_{*}$ against the descriptive summary definitions reported in the review article by \cite{Savage1996}.}, $\rm A_{X}$ as the propensity of the element X to increase the absolute value of its particular depletion level as $\rm F_{*}$ becomes larger and $\rm[X_{gas}/H]_{0}$ as the depletion for element X when $\rm F_{*}=0$. This equation can linearly be rewritten as:
\begin{equation}
\rm [X_{gas}/H]_{fit} = B_{X} + A_{X}(F_{*}-z_{X})
\label{eq:depl_eq}
\end{equation}
\label{eq:jenkins}where $\rm F_{*}$ has its zero-point reference displaced to an intermediate value $\rm z_{X}$ (unique to element X), $\rm B_{X}$ being the depletion at $\rm F_{*}=z_{X}$. 
The three parameters, $\rm A_{X}$ $\rm B_{X}$ and $\rm z_{X}$, are then solved for each element \citep[Table 4 of][]{Jenkins09}.

In the case of DLAs and sub-DLAs\footnote{We note that the measured molecular hydrogen fraction for DLAs is rather low and that this fraction is not correlated with the HI column density \citep{Ledoux2003,Noterdaeme2008a}. Therefore, the definition $\rm N(H)=N(H\,I)+2N(H_{2})$ still holds for the present study.}, releasing the hypothesis of solar metallicity, equation \ref{eq:def_depl} can be rewritten as follows:
\begin{equation}
\rm [X_{gas}/H] = \log (X/H)_{obs} - log (X/H)_{intrinsic}
\end{equation}
where $\rm log (X/H)_{intrinsic}$ is the abundance of element X in the absence of depletion. 
Equation \ref{eq:depl_eq} then becomes:
\begin{equation}
\rm [X/H]_{obs} - B_{X}+A_{X}z_{X}=[X/H]_{intrinsic}+A_{X}F_{*}
\label{eq:eq_dep_fit}
\end{equation}
where $\rm [X/H]_{obs}$ is the metallicity compared to solar as we measure it (hence affected by depletion) and $\rm [X/H]_{intrinsic}$ is the intrinsic metallicity compared to solar of the system derived from element X (corrected for depletion). 
We derive $\rm F_{*}$ by linearly fitting the left hand side of equation \ref{eq:eq_dep_fit} as a function of $\rm A_{X}$, thus providing an estimate of the intrinsic metallicity.

\subsection{Derivation of $\rm A_{V}$}
\label{Av}

Our study relates to the rest frame extinction $\rm A_{V}$ through equation 3 of \cite{Vladilo2006}, which scales $\rm A_{V}$ to the dust-phase column density of iron $\rm \widehat{N_{Fe}}$. Assuming that zinc is completely undepleted, and that the intrinsic ratio $\rm Zn/Fe$ ratio in DLAs and sub-DLAs is solar, we can express $\rm \widehat{N_{Fe}}$ as the following:\begin{equation}
\rm \widehat{N_{Fe}} = f_{Fe}N_{Zn}\left(\frac{Fe}{Zn}\right)_{\odot}
\end{equation} 
where $\rm f_{Fe}=1-10^{\delta_{Fe}}$ is the fraction of iron in dust form. We can assume $\rm \delta_{Fe}=[Fe/Zn]_{obs}$. To recover $\rm N_{Zn}$, we use the assumption that $\rm [Zn/H]=[X/H]$, with $\rm [X/H]$ being the metallicity derived for each system using the ion X (when Zn is not detected). We obtain the following expression:\begin{equation}
\rm \log \widehat{N_{Fe}}=\log\left(1-10^{[Fe/Zn]_{obs}}\right)+\log(N_{X})-\log\left(\frac{N_{X}}{N_{Fe}}\right)_{\odot}
\end{equation}
where $\rm N_{X}$ is the column density of the ion used for the metallicity derivation for a given system.

We can use equation \ref{eq:eq_dep_fit} to derive an estimate of the quantity $\rm [Fe/Zn]_{J}$ from the Jenkins analysis:\begin{equation}
\rm [Fe/Zn]_{J} = ( -1.01\pm 0.07 ) +  (-0.68\pm0.08)F_{*}
\end{equation}
This results in $\rm \mean{[Fe/Zn]}_{J,\, sub-DLA} = -0.78\pm0.15$ and $\rm \mean{[Fe/Zn]}_{J, \,DLA} = -0.53\pm0.10$. 
Using a bootstrap method, we derive $\rm \mean{\log \widehat{N_{Fe}}}_{sub-DLA}=14.38\pm0.80$ and $\rm \mean{\log \widehat{N_{Fe}}}_{DLA}=14.73\pm0.85$. Based on Fig. 4 from \cite{Vladilo2006}, we can estimate the rest frame extinction $\rm \log A_{V}$ from $\rm \log \widehat{N_{Fe}}$. For both populations, the value of $\rm \mean{\log \widehat{N_{Fe}}}$ falls below the range of MW data used to derive the correlation, giving low values of $\rm \mean{\log A_{V}}\lesssim -2$ mag. This suggests that the dust reddening is not observed in the current quasar selection.

\section[]{Sample}
\label{ann:tableau}

This Appendix presents the full list of damped and sub-damped absorbers identified in the EUADP+. We compile estimates of abundances from various references in the literature, as specified in the last column. 
We list the column density of ZnII, FeII and SII, as well as estimates of their metallicity based on the ion specified. The online version of this table includes a more complete list of ions.

\onecolumn
\setlength\LTleft{0pt}
\setlength\LTright{0pt}
\begin{longtable}{cccccccc>{\centering\arraybackslash}p{1cm}}
\caption{Full list of damped and sub-damped absorbers identified in the EUADP+, with the column densities of ZnII, FeII and SII, as well as estimates of their metallicity based on the ion specified.}\\
\hline
\hline
QSO name &$z_{\rm abs}$ & log N(HI) & log N(Zn) & log N(Fe) & log N(S) & [X/H] & X & Ref\\ 
\hline
\endfirsthead
\multicolumn{9}{c}%
{\tablename\ \thetable\ -- \textit{Continued from previous page}} \\
\hline\hline
QSO name &$z_{\rm abs}$ & log N(HI) & log N(Zn) & log N(Fe) & log N(S) & [X/H] & X & Reference\\
\hline
\endhead
\hline \multicolumn{9}{r}{\textit{Continued on next page}} \\
\endfoot
\hline
\endlastfoot

CTQ418 & 2.5100 & $20.50\pm0.07$ & - & $14.07\pm0.03$ & $13.99\pm0.06$ & $-1.63\pm0.09$ & SII & 43\\ 
CTQ418 & 2.4300 & $20.68\pm0.07$ & - & $13.91\pm0.05$ & $13.97\pm0.03$ & $-1.83\pm0.08$ & SII & 43\\ 
QXO0001 & 3.0000 & $20.70\pm0.05$ & - & $<15.09$ & - & $-1.62\pm0.05$ & OI & 92\\ 
Q0000-262 & 3.3900 & $21.41\pm0.08$ & $12.01\pm0.05$ & $14.87\pm0.03$ & - & $-1.96\pm0.09$ & ZnII & 62, 90\\ 
QSO J0003-2323 & 2.1870 & $19.60\pm0.40$ & $<11.04$ & $13.22\pm0.12$ & $<13.12$ & $-1.76\pm0.42$ & OI & 103\\ 
Q0005+0524 & 0.8514 & $19.08\pm0.04$ & $<11.24$ & $13.79\pm0.01$ & - & $-0.47\pm0.18$ & FeII & 60\\ 
PSS0007+2417 & 3.5000 & $21.10\pm0.10$ & $<12.39$ & $>14.63$ & - & $-1.53\pm0.11$ & SiII & 94\\ 
PSS0007+2417 & 3.8400 & $20.85\pm0.15$ & - & $13.91\pm0.03$ & - & $-2.12\pm0.24$ & FeII & 94\\ 
J0007+0041 & 4.7300 & $20.65\pm0.20$ & - & - & - & $-2.19\pm0.20$ & SiII & 99\\ 
QSO J0008-2900 & 2.2540 & $20.22\pm0.10$ & $<11.68$ & $13.78\pm0.01$ & - & $-1.33\pm0.14$ & SiII & 114\\ 
QSO J0008-2901 & 2.4910 & $19.94\pm0.11$ & $<12.12$ & $13.65\pm0.02$ & $13.68\pm0.18$ & $-1.32\pm0.26$ & OI & 114\\ 
J0008-0958 & 1.7700 & $20.85\pm0.15$ & $13.31\pm0.05$ & $15.62\pm0.05$ & $15.84\pm0.05$ & $-0.10\pm0.16$ & ZnII & 44, 3, 4, 5\\ 
LBQS 0009-0138 & 1.3860 & $20.26\pm0.02$ & $<11.55$ & $14.25\pm0.01$ & - & $-1.32\pm0.04$ & SiII & 60\\ 
LBQS 0010-0012 & 2.0250 & $20.95\pm0.10$ & $12.19\pm0.05$ & $15.18\pm0.03$ & $14.98\pm0.05$ & $-1.32\pm0.11$ & ZnII & 52, 110\\ 
Q0012-0122 & 1.3862 & $20.26\pm0.02$ & $<11.55$ & $14.24\pm0.01$ & - & $-1.34\pm0.08$ & SiII & 60\\ 
LBQS 0013-0029 & 1.9730 & $20.83\pm0.05$ & $12.74\pm0.05$ & $14.81\pm0.03$ & $15.28\pm0.03$ & $-0.65\pm0.07$ & ZnII & 76, 110, 44\\ 
LBQS 0018+0026 & 0.5200 & $19.54\pm0.03$ & - & $13.17\pm0.04$ & - & $-1.55\pm0.19$ & FeII & 32\\ 
LBQS 0018+0026 & 0.9400 & $19.38\pm0.15$ & $<11.64$ & $14.62\pm0.04$ & - & $0.06\pm0.24$ & FeII & 32\\ 
J001855-091351 & 0.5840 & $20.11\pm0.10$ & $<12.41$ & $13.87\pm0.03$ & - & $-1.42\pm0.21$ & FeII & 114\\ 
Q0019-15 & 3.4400 & $20.92\pm0.10$ & - & $>14.79$ & - & $-1.01\pm0.11$ & SiII & 89, 90\\ 
Q0021+0104 & 1.3259 & $20.04\pm0.11$ & $<11.48$ & $14.69\pm0.01$ & - & $-0.53\pm0.21$ & FeII & 60\\ 
Q0021+0104 & 1.5756 & $20.48\pm0.15$ & $<11.95$ & $14.61\pm0.02$ & - & $-1.11\pm0.15$ & SiII & 60\\ 
J0021+0043 & 0.9424 & $19.38\pm0.13$ & $<11.64$ & $15.06\pm0.14$ & - & $0.50\pm0.26$ & FeII & 26\\ 
QSO B0027-1836 & 2.4020 & $21.75\pm0.10$ & $12.79\pm0.02$ & $14.97\pm0.02$ & $15.23\pm0.02$ & $-1.52\pm0.10$ & ZnII & 64\\ 
J0035-0918 & 2.3400 & $20.55\pm0.10$ & - & $13.07\pm0.04$ & $<13.13$ & $-2.69\pm0.17$ & OI & 17\\ 
QSO B0039-3354 & 2.2240 & $20.60\pm0.10$ & - & $14.41\pm0.03$ & - & $-1.27\pm0.11$ & SiII & 66\\ 
J004054.7-091526 & 4.7400 & $20.55\pm0.15$ & - & $14.05\pm0.06$ & - & $-1.93\pm0.15$ & SiII & 98\\ 
J0040-0915 & 4.7394 & $20.30\pm0.15$ & - & $14.05\pm0.06$ & - & $-1.43\pm0.24$ & FeII & 98\\ 
QSO J0041-4936 & 2.2480 & $20.46\pm0.13$ & $11.70\pm0.10$ & $14.42\pm0.04$ & $<14.82$ & $-1.32\pm0.16$ & ZnII & 114\\ 
LBQS 0042-2930 & 1.8090 & $20.40\pm0.10$ & - & - & - & $-1.21\pm0.12$ & SiII & 37\\ 
LBQS 0042-2930 & 1.9360 & $20.50\pm0.10$ & - & - & - & $-1.23\pm0.11$ & SiII & 37\\ 
J0044+0018 & 1.7300 & $20.35\pm0.10$ & $<12.61$ & $>14.77$ & $15.27\pm0.05$ & $-0.20\pm0.11$ & SII & 3, 4, 5\\ 
Q0049-2820 & 2.0700 & $20.45\pm0.10$ & - & $14.50\pm0.02$ & - & $-1.26\pm0.11$ & SiII & 67\\ 
QSO B0058-292 & 2.6710 & $21.10\pm0.10$ & $12.23\pm0.05$ & $14.75\pm0.03$ & $14.92\pm0.03$ & $-1.43\pm0.11$ & ZnII & 51, 110\\ 
J0058+0115 & 2.0100 & $21.10\pm0.15$ & $12.95\pm0.05$ & $15.18\pm0.05$ & $15.40\pm0.05$ & $-0.71\pm0.16$ & ZnII & 3, 4, 5\\ 
QSO B0100+1300 & 2.3090 & $21.35\pm0.08$ & $12.49\pm0.02$ & $15.10\pm0.04$ & $15.09\pm0.06$ & $-1.42\pm0.08$ & ZnII & 89, 22\\ 
QSO J0105-1846 & 2.3700 & $21.00\pm0.08$ & $11.77\pm0.11$ & $14.47\pm0.10$ & $14.30\pm0.04$ & $-1.79\pm0.14$ & ZnII & 51, 110\\ 
QSO J0105-1846 & 2.9260 & $20.00\pm0.10$ & - & $13.80\pm0.03$ & $13.82\pm0.03$ & $-1.56\pm0.13$ & OI & 66\\ 
B0105-008 & 1.3700 & $21.70\pm0.15$ & $12.93\pm0.04$ & $15.59\pm0.03$ & - & $-1.33\pm0.16$ & ZnII & 35\\ 
QSO B0112-30 & 2.4180 & $20.50\pm0.08$ & - & $13.33\pm0.04$ & $14.44\pm0.03$ & $-2.24\pm0.11$ & OI & 77\\ 
QSO B0112-30 & 2.7020 & $20.30\pm0.10$ & - & $14.77\pm0.07$ & - & $-0.44\pm0.13$ & SiII & 51, 110\\ 
Q0112+030 & 2.4200 & $20.90\pm0.10$ & - & $14.85\pm0.01$ & $14.79\pm0.05$ & $-1.23\pm0.11$ & SII & 110, 52, 67\\ 
QSO B0122-005 & 1.7610 & $20.78\pm0.07$ & - & $15.10\pm0.10$ & - & $-0.87\pm0.11$ & SiII & 33\\ 
QSO B0122-005 & 2.0100 & $20.04\pm0.07$ & $<11.40$ & $13.69\pm0.07$ & - & $-1.88\pm0.09$ & SiII & 33\\ 
QSO J0123-0058 & 1.4090 & $20.08\pm0.09$ & $12.23\pm0.10$ & $14.98\pm0.02$ & - & $-0.41\pm0.13$ & ZnII & 74\\ 
QSO J0124+0044 & 2.9880 & $19.18\pm0.10$ & - & $<13.55$ & $<14.27$ & $-0.57\pm0.16$ & SiII & 73\\ 
QSO J0124+0044 & 3.0780 & $20.21\pm0.10$ & - & $<14.13$ & - & $-0.59\pm0.40$ & SiII & 73\\ 
QSO B0128-2150 & 1.8570 & $20.21\pm0.09$ & $<12.26$ & $14.44\pm0.01$ & $14.33\pm0.03$ & $-1.00\pm0.09$ & SII & 114\\ 
J013209-082349 & 0.6470 & $20.60\pm0.12$ & - & $14.96\pm0.07$ & - & $-0.82\pm0.23$ & FeII & 114\\ 
QSO J0133+0400 & 3.6920 & $20.68\pm0.15$ & - & $13.51\pm0.07$ & - & $-0.96\pm0.16$ & SiII & 93\\ 
QSO J0133+0400 & 3.7730 & $20.55\pm0.13$ & $<13.10$ & $>14.87$ & - & $-0.59\pm0.13$ & SiII & 93\\ 
QSO J0133+0400 & 3.9950 & $19.94\pm0.15$ & - & $<13.43$ & - & $-1.54\pm0.20$ & SiII & 73\\ 
PSS0133+0400 & 3.6900 & $20.70\pm0.10$ & - & $13.57\pm0.04$ & $<13.35$ & $-2.31\pm0.21$ & FeII & 93, 77\\ 
PSS0133+0400 & 3.7700 & $20.60\pm0.10$ & $<13.10$ & $>14.87$ & - & $-0.65\pm0.10$ & SiII & 93, 67\\ 
QSO J0134+0051 & 0.8420 & $19.93\pm0.13$ & $<12.17$ & $14.47\pm0.01$ & - & $-0.64\pm0.22$ & FeII & 72\\ 
PSS0134+3317 & 3.7600 & $20.85\pm0.08$ & - & - & - & $-2.69\pm0.09$ & AlII & 93\\ 
QSO B0135-42 & 3.1010 & $19.81\pm0.10$ & - & $13.67\pm0.11$ & - & $-1.21\pm0.27$ & SiII & 73\\ 
QSO B0135-42 & 3.6650 & $19.11\pm0.10$ & - & $<13.47$ & - & $-2.42\pm0.16$ & OI & 73\\ 
Q0135-273 & 2.8000 & $21.00\pm0.10$ & - & $14.77\pm0.03$ & $14.80\pm0.02$ & $-1.32\pm0.10$ & SII & 110, 52, 67\\ 
Q0135-273 & 2.1100 & $20.30\pm0.15$ & - & - & $14.38\pm0.06$ & $-1.04\pm0.16$ & SII & 52\\ 
QSO J0138-0005 & 0.7820 & $19.81\pm0.09$ & $12.69\pm0.05$ & $<15.17$ & - & $0.32\pm0.10$ & ZnII & 74\\ 
J0140-0839 & 3.7000 & $20.75\pm0.15$ & - & $<12.73$ & $<13.33$ & $-2.75\pm0.15$ & OI & 34\\ 
UM673A & 1.6300 & $20.70\pm0.10$ & $11.43\pm0.15$ & $14.59\pm0.03$ & $14.53\pm0.00$ & $-1.83\pm0.18$ & ZnII & 15\\ 
J0142+0023 & 3.3500 & $20.38\pm0.05$ & $<11.50$ & $13.70\pm0.10$ & $13.26\pm0.06$ & $-2.24\pm0.08$ & SII & 34\\ 
Q0149+33 & 2.1400 & $20.50\pm0.10$ & $11.50\pm0.10$ & $14.20\pm0.02$ & $<14.80$ & $-1.56\pm0.14$ & ZnII & 89, 11, 90\\ 
Q0151+0448 & 1.9300 & $20.36\pm0.10$ & $<11.81$ & $13.70\pm0.01$ & $<13.47$ & $-1.86\pm0.11$ & SiII & 34, 118\\ 
QSO J0157-0048 & 1.4160 & $19.90\pm0.07$ & $12.12\pm0.07$ & $14.57\pm0.03$ & - & $-0.34\pm0.10$ & ZnII & 32\\ 
QSO B0201+113 & 3.3850 & $21.26\pm0.08$ & - & $15.35\pm0.05$ & $15.21\pm0.11$ & $-1.17\pm0.14$ & SII & 29\\ 
Q0201+365 & 2.4600 & $20.38\pm0.05$ & $12.47\pm0.05$ & $15.01\pm0.01$ & $15.29\pm0.01$ & $-0.47\pm0.07$ & ZnII & 87, 79, 90, 92, 3, 4, 5\\ 
Q0201+1120 & 3.3900 & $21.26\pm0.10$ & - & $15.35\pm0.10$ & $15.21\pm0.10$ & $-1.17\pm0.14$ & SII & 29\\ 
QSO J0209+0517 & 3.6660 & $20.47\pm0.10$ & - & $13.63\pm0.05$ & - & $-2.01\pm0.21$ & FeII & 93\\ 
QSO J0209+0517 & 3.8630 & $20.55\pm0.10$ & - & $<13.34$ & - & $-2.60\pm0.11$ & SiII & 93\\ 
PSS0209+0517 & 3.6700 & $20.45\pm0.10$ & - & $13.64\pm0.05$ & - & $-1.99\pm0.21$ & FeII & 93\\ 
J0211+1241 & 2.6000 & $20.60\pm0.15$ & - & $15.06\pm0.05$ & - & $-0.58\pm0.17$ & SiII & 3, 4, 5\\ 
QSO B0216+0803 & 1.7690 & $20.20\pm0.10$ & $11.90\pm0.06$ & $14.53\pm0.09$ & - & $-0.86\pm0.12$ & ZnII & 57\\ 
QSO B0216+0803 & 2.2930 & $20.45\pm0.16$ & $12.47\pm0.05$ & $14.88\pm0.02$ & $15.04\pm0.02$ & $-0.54\pm0.17$ & ZnII & 57, 117\\ 
QSO J0217+0144 & 1.3450 & $19.89\pm0.09$ & - & $14.38\pm0.10$ & - & $-1.11\pm0.10$ & MgII & 6, 7\\ 
SDSS0225+0054 & 2.7100 & $21.00\pm0.15$ & $12.89\pm0.11$ & $15.30\pm0.08$ & - & $-0.67\pm0.19$ & ZnII & 44\\ 
J0233+0103 & 1.7900 & $20.60\pm0.15$ & - & $14.62\pm0.05$ & - & $-1.34\pm0.16$ & SiII & 3, 4, 5\\ 
J0234-0751 & 2.3200 & $20.90\pm0.10$ & - & $14.18\pm0.03$ & $14.18\pm0.03$ & $-1.84\pm0.10$ & SII & 28\\ 
AO0235+164 & 0.5200 & $21.70\pm0.10$ & - & $15.30\pm0.40$ & - & $-1.58\pm0.45$ & FeII & 13\\ 
QSO B0237-2322 & 1.3650 & $19.30\pm0.30$ & - & $14.13\pm0.01$ & - & $0.08\pm0.30$ & SiII & 111\\ 
QSO B0237-2322 & 1.6720 & $19.65\pm0.10$ & $11.84\pm0.09$ & $14.57\pm0.02$ & - & $-0.37\pm0.13$ & ZnII & 35\\ 
Q0242-2917 & 2.5600 & $20.90\pm0.10$ & - & $14.36\pm0.03$ & $14.11\pm0.02$ & $-1.91\pm0.10$ & SII & 67\\ 
QSO B0244-1249 & 1.8630 & $19.48\pm0.18$ & $<11.50$ & $<13.90$ & - & $-0.79\pm0.27$ & SiII & 33\\ 
QSO B0253+0058 & 0.7250 & $20.70\pm0.17$ & $13.19\pm0.04$ & $15.13\pm0.30$ & - & $-0.07\pm0.17$ & ZnII & 72\\ 
QSO B0254-404 & 2.0460 & $20.45\pm0.08$ & - & $14.17\pm0.03$ & $14.14\pm0.04$ & $-1.43\pm0.09$ & SII & 66\\ 
J0255+00 & 3.9200 & $21.30\pm0.05$ & - & $14.75\pm0.09$ & $14.72\pm0.01$ & $-1.70\pm0.05$ & SII & 90\\ 
J0255+00 & 3.2500 & $20.70\pm0.10$ & - & $14.76\pm0.01$ & - & $-0.89\pm0.11$ & SiII & 90\\ 
Q0300-3152 & 2.1800 & $20.80\pm0.10$ & - & $14.21\pm0.02$ & $14.20\pm0.03$ & $-1.72\pm0.10$ & SII & 67\\ 
Q0302-223 & 1.0100 & $20.36\pm0.11$ & $12.45\pm0.06$ & $14.67\pm0.05$ & - & $-0.47\pm0.13$ & ZnII & 82\\ 
QSO B0307-195B & 1.7880 & $19.00\pm0.10$ & $<12.18$ & $14.48\pm0.00$ & - & $0.49\pm0.10$ & SiII & 114\\ 
J0307-4945 & 4.4700 & $20.67\pm0.09$ & - & $14.21\pm0.17$ & $<15.46$ & $-1.45\pm0.19$ & OI & 20\\ 
TXS0311+430 & 2.2900 & $20.30\pm0.00$ & $<12.50$ & $14.85\pm0.20$ & - & $-0.63\pm0.27$ & FeII & 32, 32\\ 
J0311-1722 & 3.7300 & $20.30\pm0.06$ & - & $<13.76$ & - & $-2.29\pm0.10$ & OI & 16\\ 
QSO J0332-4455 & 2.6560 & $19.82\pm0.05$ & - & $13.51\pm0.05$ & - & $-1.67\pm0.06$ & OI & 36\\ 
QSO B0335-122 & 3.1780 & $20.65\pm0.07$ & $<12.25$ & $13.70\pm0.04$ & - & $-2.44\pm0.10$ & SiII & 1, 67\\ 
QSO B0336-017 & 3.0620 & $21.20\pm0.09$ & - & $14.90\pm0.03$ & $14.99\pm0.01$ & $-1.33\pm0.09$ & SII & 90\\ 
0338-0005 & 2.9090 & $21.10\pm0.10$ & $<12.47$ & $15.02\pm0.06$ & $15.19\pm0.01$ & $-1.03\pm0.10$ & SII & 95\\ 
QSO B0347-383 & 3.0250 & $20.63\pm0.09$ & $12.23\pm0.12$ & $14.47\pm0.01$ & $14.73\pm0.01$ & $-0.96\pm0.15$ & ZnII & 90, 51\\ 
QSO J0354-2724 & 1.4050 & $20.18\pm0.15$ & $12.73\pm0.03$ & $15.10\pm0.03$ & - & $-0.01\pm0.15$ & ZnII & 59\\ 
B0405-331 & 2.5700 & $20.60\pm0.10$ & $<12.74$ & $14.31\pm0.00$ & - & $-1.40\pm0.10$ & SiII & 1\\ 
Q0405-443 & 2.5500 & $21.13\pm0.10$ & $12.44\pm0.05$ & $14.95\pm0.06$ & $14.82\pm0.06$ & $-1.25\pm0.11$ & ZnII & 55, 110\\ 
Q0405-443 & 2.6200 & $20.47\pm0.10$ & - & $13.60\pm0.02$ & $<14.34$ & $-1.97\pm0.10$ & OI & 55, 110, 68, 112\\ 
QSO J0407-4410 & 1.9130 & $20.80\pm0.10$ & $12.44\pm0.05$ & - & - & $-0.92\pm0.11$ & ZnII & 52\\ 
QSO J0407-4410 & 2.5510 & $21.13\pm0.10$ & $12.44\pm0.05$ & $14.95\pm0.06$ & $14.82\pm0.06$ & $-1.25\pm0.11$ & ZnII & 55, 110\\ 
QSO J0407-4410 & 2.5950 & $21.09\pm0.10$ & $12.68\pm0.02$ & $15.15\pm0.02$ & $15.19\pm0.05$ & $-0.97\pm0.10$ & ZnII & 55, 110\\ 
QSO J0407-4410 & 2.6210 & $20.45\pm0.10$ & - & $13.60\pm0.02$ & $<14.34$ & $-1.95\pm0.10$ & OI & 55, 110\\ 
Q0421-2624 & 2.1600 & $20.65\pm0.10$ & - & $13.97\pm0.01$ & - & $-1.81\pm0.10$ & SiII & 67\\ 
QSO J0422-3844 & 3.0820 & $19.37\pm0.02$ & - & $13.96\pm0.10$ & - & $-0.69\pm0.04$ & OI & 10\\ 
Q0425-5214 & 2.2200 & $20.30\pm0.10$ & - & $13.96\pm0.03$ & $14.07\pm0.03$ & $-1.35\pm0.10$ & SII & 67\\ 
BRJ0426-2202 & 2.9800 & $21.50\pm0.15$ & $<12.17$ & $14.15\pm0.07$ & - & $-2.53\pm0.24$ & FeII & 93\\ 
QSO J0427-1302 & 1.5620 & $19.35\pm0.10$ & $<11.75$ & $12.23\pm0.04$ & - & $-2.30\pm0.21$ & FeII & 114\\ 
QSO J0427-1302 & 1.4080 & $19.04\pm0.04$ & $<11.09$ & $13.33\pm0.02$ & - & $-0.99\pm0.06$ & SiII & 60\\ 
Q0432-4401 & 2.3000 & $20.95\pm0.10$ & $<12.20$ & $14.87\pm0.10$ & - & $-1.18\pm0.12$ & SiII & 1, 67, 119\\ 
QSO B0438-43 & 2.3470 & $20.78\pm0.12$ & $12.72\pm0.03$ & - & - & $-0.62\pm0.12$ & ZnII & 1\\ 
PKS 0439-433 & 0.1012 & $19.63\pm0.15$ & - & $14.92\pm0.03$ & $15.03\pm0.03$ & $0.28\pm0.15$ & SII & 105\\ 
QSO B0449-1645 & 1.0070 & $20.98\pm0.07$ & $12.62\pm0.07$ & $15.09\pm0.01$ & - & $-0.92\pm0.10$ & ZnII & 74\\ 
QSO B0450-1310B & 2.0670 & $20.50\pm0.07$ & - & $14.29\pm0.03$ & $14.18\pm0.06$ & $-2.13\pm0.08$ & OI & 24, 119\\ 
PKS 0454-220 & 0.4740 & $19.45\pm0.03$ & - & $14.71\pm0.01$ & $15.06\pm0.04$ & $0.49\pm0.05$ & SII & 114\\ 
Q0454+039 & 0.8600 & $20.69\pm0.06$ & $12.33\pm0.08$ & - & - & $-0.92\pm0.10$ & ZnII & 82\\ 
4C-02.19 & 2.0400 & $21.70\pm0.10$ & $13.13\pm0.05$ & $15.38\pm0.05$ & - & $-1.13\pm0.11$ & ZnII & 42\\ 
QSO B0512-3329 & 0.9310 & $20.49\pm0.08$ & - & $14.47\pm0.06$ & - & $-1.20\pm0.21$ & FeII & 56\\ 
QSO B0515-4414 & 1.1510 & $19.88\pm0.05$ & $12.22\pm0.04$ & $14.31\pm0.03$ & - & $-0.22\pm0.06$ & ZnII & 97, 122\\ 
HE0512-3329A & 0.9300 & $20.49\pm0.08$ & - & $14.47\pm0.06$ & - & $-1.20\pm0.21$ & FeII & 56\\ 
HE0515-4414 & 1.1500 & $20.45\pm0.15$ & $12.11\pm0.04$ & $14.31\pm0.20$ & - & $-0.90\pm0.16$ & ZnII & 123\\ 
QSO B0528-2505 & 2.1410 & $20.70\pm0.08$ & $13.00\pm0.03$ & $14.94\pm0.26$ & $14.83\pm0.04$ & $-0.26\pm0.09$ & ZnII & 57, 12\\ 
QSO B0528-2505 & 2.8110 & $21.11\pm0.07$ & $13.27\pm0.03$ & $15.47\pm0.02$ & $15.56\pm0.02$ & $-0.40\pm0.08$ & ZnII & 107, 57, 12\\ 
QSO B0551-36 & 1.9620 & $20.50\pm0.08$ & $13.02\pm0.05$ & $15.05\pm0.05$ & $15.38\pm0.11$ & $-0.04\pm0.09$ & ZnII & 49\\ 
J060008.1-504036 & 2.1490 & $20.40\pm0.12$ & $12.11\pm0.03$ & $14.84\pm0.03$ & - & $-0.85\pm0.12$ & ZnII & 114\\ 
QSO B0642-5038 & 2.6590 & $20.95\pm0.08$ & $12.50\pm0.06$ & $15.10\pm0.04$ & - & $-1.01\pm0.10$ & ZnII & 66, 119\\ 
Q0738+313 & 0.0900 & $21.18\pm0.06$ & $<12.66$ & $15.02\pm0.15$ & - & $-1.34\pm0.24$ & FeII & 58, 47\\ 
HS0741+4741 & 3.0200 & $20.48\pm0.10$ & - & $14.05\pm0.01$ & $14.00\pm0.02$ & $-1.60\pm0.10$ & SII & 90, 92\\ 
J0747+4434 & 4.0196 & $20.95\pm0.15$ & - & $>14.32$ & - & $-2.50\pm0.20$ & NiII & 98\\ 
FJ0747+2739 & 3.9000 & $20.50\pm0.10$ & $<12.40$ & $<13.80$ & $<14.36$ & $-1.98\pm0.10$ & SiII & 93\\ 
J0759+1800 & 4.6577 & $20.85\pm0.15$ & - & $<15.16$ & $14.26\pm0.05$ & $-1.71\pm0.16$ & SII & 98\\ 
SDSS0759+3129 & 3.0300 & $20.60\pm0.10$ & - & $13.80\pm0.20$ & - & $-2.01\pm0.32$ & SiII & 70\\ 
PSSJ0808+52 & 3.1100 & $20.65\pm0.07$ & $<12.13$ & $14.17\pm0.04$ & - & $-1.56\pm0.14$ & SiII & 91, 93\\ 
FJ0812+32 & 2.0700 & $21.00\pm0.10$ & $12.21\pm0.02$ & $14.89\pm0.02$ & - & $-1.35\pm0.10$ & ZnII & 95, 43\\ 
FJ0812+32 & 2.6300 & $21.35\pm0.10$ & $13.15\pm0.05$ & $15.09\pm0.05$ & $15.63\pm0.07$ & $-0.76\pm0.11$ & ZnII & 93, 95, 3, 4, 5\\ 
J0815+1037 & 1.8500 & $20.30\pm0.15$ & - & $>14.87$ & - & $-0.43\pm0.47$ & SiII & 3, 4, 5\\ 
J0816+1446 & 3.2900 & $22.00\pm0.10$ & $13.53\pm0.00$ & $15.89\pm0.00$ & - & $-1.03\pm0.10$ & ZnII & 41\\ 
J0817+1351 & 4.2584 & $21.30\pm0.15$ & - & $15.45\pm0.06$ & $15.30\pm0.02$ & $-1.12\pm0.15$ & SII & 98\\ 
J0824+1302 & 4.4700 & $20.65\pm0.20$ & - & $13.60\pm0.08$ & - & $-2.32\pm0.21$ & SiII & 99\\ 
J0825+3544 & 3.2073 & $20.30\pm0.10$ & - & $13.77\pm0.03$ & - & $-1.71\pm0.21$ & FeII & 98\\ 
J0825+3544 & 3.6567 & $21.25\pm0.10$ & - & $>14.65$ & - & $-1.83\pm0.13$ & SiII & 98\\ 
J0825+5127 & 3.3180 & $20.85\pm0.10$ & - & $14.22\pm0.01$ & - & $-1.67\pm0.14$ & SiII & 98\\ 
Q0826-2230 & 0.9110 & $19.04\pm0.04$ & $12.71\pm0.08$ & $13.57\pm0.06$ & - & $1.11\pm0.09$ & ZnII & 60\\ 
Q0827+243 & 0.5200 & $20.30\pm0.05$ & $<12.80$ & $14.59\pm0.02$ & - & $-0.89\pm0.19$ & FeII & 58, 45\\ 
J0831+4046 & 4.3440 & $20.75\pm0.15$ & - & $13.79\pm0.07$ & - & $-2.36\pm0.15$ & SiII & 98\\ 
J0834+2140 & 3.7102 & $20.85\pm0.10$ & - & $14.44\pm0.02$ & - & $-1.59\pm0.21$ & FeII & 98\\ 
J0834+2140 & 4.3900 & $21.00\pm0.20$ & - & $14.76\pm0.02$ & $14.85\pm0.04$ & $-1.27\pm0.20$ & SII & 98\\ 
J0834+2140 & 4.4610 & $20.30\pm0.15$ & - & $13.71\pm0.07$ & $<14.13$ & $-1.86\pm0.16$ & SiII & 98\\ 
Q0836+11 & 2.4700 & $20.58\pm0.10$ & $<12.12$ & $14.68\pm0.01$ & $<14.66$ & $-1.10\pm0.11$ & SiII & 90, 92\\ 
J0839+3524 & 4.2800 & $20.30\pm0.15$ & - & $14.30\pm0.04$ & - & $-1.18\pm0.24$ & FeII & 98\\ 
QSO B0841+129 & 1.8640 & $21.00\pm0.10$ & - & - & $14.82\pm0.05$ & $-1.30\pm0.11$ & SII & 52\\ 
QSO B0841+129 & 2.3750 & $21.05\pm0.10$ & $12.12\pm0.05$ & $14.76\pm0.11$ & $14.69\pm0.15$ & $-1.50\pm0.11$ & ZnII & 89, 119\\ 
QSO B0841+129 & 2.4760 & $20.80\pm0.10$ & $11.69\pm0.05$ & $14.43\pm0.03$ & $14.48\pm0.12$ & $-1.67\pm0.11$ & ZnII & 89, 24\\ 
SDSS0844+5153 & 2.7700 & $21.45\pm0.15$ & - & $15.29\pm0.06$ & - & $-0.99\pm0.15$ & SiII & 44\\ 
J0900+42 & 3.2500 & $20.30\pm0.10$ & - & $14.54\pm0.01$ & $14.65\pm0.01$ & $-0.77\pm0.10$ & SII & 95, 43\\ 
J0909+3303 & 3.6584 & $20.55\pm0.10$ & - & $14.43\pm0.01$ & $14.51\pm0.04$ & $-1.16\pm0.11$ & SII & 98\\ 
QSO B0913+0715 & 2.6180 & $20.35\pm0.10$ & $<11.90$ & $12.99\pm0.01$ & $13.88\pm0.03$ & $-2.41\pm0.10$ & OI & 84, 77\\ 
B0913+003 & 2.7400 & $20.74\pm0.10$ & $<12.82$ & $14.60\pm0.00$ & - & $-1.47\pm0.10$ & SiII & 1\\ 
Q0918+1636 & 2.4100 & $21.26\pm0.06$ & $13.23\pm0.18$ & $15.51\pm0.23$ & - & $-0.59\pm0.19$ & ZnII & 39\\ 
Q0918+1636 & 2.5800 & $20.96\pm0.05$ & $13.40\pm0.01$ & $15.43\pm0.01$ & $15.82\pm0.01$ & $-0.12\pm0.05$ & ZnII & 38\\ 
J0925+4004 & 0.2477 & $19.55\pm0.15$ & - & $14.22\pm0.09$ & $<14.72$ & $-0.29\pm0.17$ & OI & 2\\ 
J0927+5823 & 1.6400 & $20.40\pm0.25$ & $13.29\pm0.05$ & $>15.27$ & $15.61\pm0.05$ & $0.33\pm0.25$ & ZnII & 3, 4, 5\\ 
J0928+6025 & 0.1538 & $19.35\pm0.15$ & - & $14.90\pm0.08$ & $<14.65$ & $0.37\pm0.25$ & FeII & 2\\ 
SDSS0928+0939 & 2.9100 & $20.75\pm0.15$ & - & $14.10\pm0.30$ & - & $-1.83\pm0.38$ & FeII & 70\\ 
Q0930+28 & 3.2400 & $20.35\pm0.10$ & - & $13.49\pm0.03$ & - & $-2.07\pm0.10$ & SiII & 92, 93\\ 
QSO B0933-333 & 2.6820 & $20.50\pm0.10$ & $<11.99$ & $14.46\pm0.08$ & - & $-1.22\pm0.12$ & SiII & 1, 66\\ 
Q0933+733 & 1.4800 & $21.62\pm0.10$ & $12.71\pm0.02$ & $15.19\pm0.01$ & - & $-1.47\pm0.10$ & ZnII & 101\\ 
Q0935+417 & 1.3700 & $20.52\pm0.10$ & $12.26\pm0.02$ & $14.82\pm0.10$ & - & $-0.82\pm0.10$ & ZnII & 61, 79, 100\\ 
Q0948+433 & 1.2300 & $21.62\pm0.06$ & $13.15\pm0.01$ & $15.56\pm0.01$ & - & $-1.03\pm0.06$ & ZnII & 101\\ 
QSO B0951-0450 & 3.2350 & $20.25\pm0.10$ & - & $13.49\pm0.03$ & - & $-1.97\pm0.10$ & SiII & 93\\ 
QSO B0951-0450 & 3.8580 & $20.60\pm0.10$ & - & $14.06\pm0.06$ & - & $-1.47\pm0.10$ & SiII & 89\\ 
QSO B0951-0450 & 4.2030 & $20.55\pm0.10$ & - & $13.07\pm0.19$ & $<13.89$ & $-2.55\pm0.10$ & OI & 89\\ 
BR0951-04 & 3.8600 & $20.60\pm0.10$ & - & $14.06\pm0.06$ & - & $-1.46\pm0.10$ & SiII & 89, 90\\ 
QSO B0952+179 & 0.2380 & $21.32\pm0.05$ & $12.93\pm0.04$ & - & - & $-0.95\pm0.06$ & ZnII & 47\\ 
QSO B0952-0115 & 4.0240 & $20.55\pm0.10$ & - & $14.19\pm0.08$ & - & $-2.61\pm0.11$ & SiII & 90\\ 
PC0953+4749 & 4.2400 & $20.90\pm0.15$ & - & $13.90\pm0.07$ & - & $-2.18\pm0.15$ & SiII & 106, 93\\ 
PC0953+4749 & 3.8900 & $21.20\pm0.10$ & - & $15.09\pm0.10$ & - & $-1.29\pm0.23$ & FeII & 106, 93\\ 
PSSJ0957+33 & 3.2800 & $20.45\pm0.08$ & $<12.13$ & $14.37\pm0.02$ & $<14.58$ & $-1.08\pm0.09$ & SiII & 90, 93\\ 
PSSJ0957+33 & 4.1800 & $20.70\pm0.10$ & - & $14.13\pm0.05$ & $14.39\pm0.06$ & $-1.43\pm0.12$ & SII & 90, 93\\ 
J0958+0145 & 1.9300 & $20.40\pm0.10$ & $<12.00$ & $14.23\pm0.05$ & $14.44\pm0.05$ & $-1.08\pm0.11$ & SII & 3, 4, 5\\ 
J1001+5944 & 0.3035 & $19.32\pm0.10$ & - & $14.30\pm0.04$ & $<14.53$ & $-0.37\pm0.10$ & OI & 2\\ 
SDSS1003+5520 & 2.5000 & $20.35\pm0.15$ & - & $12.90\pm0.30$ & - & $-2.06\pm0.34$ & SiII & 70\\ 
J1004+0018 & 2.6900 & $21.39\pm0.10$ & - & $14.71\pm0.04$ & $14.70\pm0.02$ & $-1.81\pm0.10$ & SII & 28\\ 
J1004+0018 & 2.5400 & $21.30\pm0.10$ & - & $15.13\pm0.02$ & $15.09\pm0.01$ & $-1.33\pm0.10$ & SII & 28\\ 
Q1007+0042 & 1.0400 & $21.15\pm0.20$ & $13.27\pm0.04$ & - & - & $-0.44\pm0.20$ & ZnII & 63\\ 
Q1008+36 & 2.8000 & $20.70\pm0.05$ & - & $<15.11$ & - & $-1.75\pm0.05$ & SiII & 43\\ 
QSO J1009-0026 & 0.8400 & $20.20\pm0.07$ & $<11.85$ & $14.37\pm0.03$ & - & $-1.01\pm0.19$ & FeII & 59\\ 
QSO J1009-0026 & 0.8800 & $19.48\pm0.08$ & $12.38\pm0.04$ & $15.33\pm0.06$ & - & $0.34\pm0.09$ & ZnII & 59\\ 
J1009+0713 & 0.1140 & $20.68\pm0.10$ & - & $15.29\pm0.17$ & $15.25\pm0.12$ & $-0.55\pm0.16$ & SII & 2\\ 
Q1010+0003 & 1.2700 & $21.52\pm0.07$ & $12.96\pm0.06$ & $15.26\pm0.05$ & - & $-1.12\pm0.09$ & ZnII & 58, 63, 3, 4, 5\\ 
J1013+4240 & 4.7979 & $20.60\pm0.15$ & - & - & - & $-2.14\pm0.15$ & SiII & 98\\ 
J1013+5615 & 2.2800 & $20.70\pm0.15$ & $13.56\pm0.05$ & $>15.45$ & - & $0.30\pm0.16$ & ZnII & 3, 4, 5\\ 
BRI1013+0035 & 3.1000 & $21.10\pm0.10$ & $13.33\pm0.02$ & $15.18\pm0.05$ & - & $-0.33\pm0.10$ & ZnII & 95\\ 
J1017+6116 & 2.7684 & $20.60\pm0.10$ & - & $13.76\pm0.05$ & - & $-2.71\pm0.10$ & OI & 98\\ 
Q1021+30 & 2.9500 & $20.70\pm0.10$ & $<12.23$ & $14.04\pm0.01$ & $13.87\pm0.07$ & $-1.95\pm0.12$ & SII & 93, 95\\ 
J1024+0600 & 1.9000 & $20.60\pm0.15$ & - & $15.27\pm0.08$ & $15.45\pm0.05$ & $-0.27\pm0.16$ & SII & 3, 4, 5\\ 
LBQS 1026-0045B & 0.6320 & $19.95\pm0.07$ & $12.46\pm0.16$ & $15.11\pm0.06$ & - & $-0.05\pm0.17$ & ZnII & 32\\ 
LBQS 1026-0045B & 0.7090 & $20.04\pm0.06$ & $<12.51$ & $15.10\pm0.03$ & - & $-0.12\pm0.19$ & FeII & 32\\ 
J1028-0100 & 0.6321 & $19.95\pm0.07$ & $<12.38$ & $15.08\pm0.08$ & - & $-0.05\pm0.21$ & FeII & 26\\ 
J1028-0100 & 0.7089 & $20.04\pm0.06$ & $<12.49$ & $15.12\pm0.07$ & - & $-0.10\pm0.20$ & FeII & 26\\ 
SDSS1031+4055 & 2.5700 & $20.55\pm0.10$ & - & $13.80\pm0.20$ & - & $-1.93\pm0.29$ & FeII & 70\\ 
QSO B1036-2257 & 2.5330 & $19.30\pm0.10$ & $<11.74$ & $12.93\pm0.01$ & - & $-1.33\pm0.10$ & MgII & 114\\ 
QSO B1036-2257 & 2.7770 & $20.93\pm0.05$ & $<12.36$ & $14.68\pm0.01$ & $14.79\pm0.02$ & $-1.26\pm0.05$ & SII & 120, 52, 67\\ 
Q1037+0028 & 1.4244 & $20.04\pm0.12$ & $<12.04$ & $14.84\pm0.02$ & - & $-0.46\pm0.12$ & SiII & 60\\ 
J1037+0139 & 2.7000 & $20.50\pm0.08$ & - & $13.53\pm0.02$ & - & $-2.13\pm0.09$ & OI & 16, 70\\ 
QSO J1039-2719 & 2.1390 & $19.70\pm0.05$ & $12.09\pm0.04$ & $14.56\pm0.02$ & $14.82\pm0.04$ & $-0.17\pm0.06$ & ZnII & 109\\ 
J1042+3107 & 4.0865 & $20.75\pm0.10$ & - & $14.22\pm0.03$ & - & $-1.95\pm0.10$ & SiII & 98\\ 
J1042+0628 & 1.9400 & $20.70\pm0.15$ & - & $15.00\pm0.15$ & $15.08\pm0.05$ & $-0.74\pm0.16$ & SII & 3, 4, 5\\ 
SDSS1042+0117 & 2.2700 & $20.75\pm0.15$ & $<12.74$ & $15.08\pm0.13$ & - & $-0.79\pm0.17$ & SiII & 44\\ 
SDSS1043+6151 & 2.7900 & $20.60\pm0.15$ & - & $14.00\pm0.20$ & - & $-2.01\pm0.34$ & SiII & 70\\ 
QSO B1045+056 & 0.9510 & $19.28\pm0.02$ & $<11.70$ & $13.49\pm0.08$ & - & $-0.97\pm0.20$ & FeII & 58\\ 
SDSS1048+3911 & 2.3000 & $20.70\pm0.10$ & - & $13.70\pm0.20$ & - & $-2.31\pm0.32$ & SiII & 70\\ 
J1049-0110 & 1.6600 & $20.35\pm0.15$ & $13.14\pm0.05$ & $15.17\pm0.05$ & $15.47\pm0.05$ & $0.23\pm0.16$ & ZnII & 3, 4, 5\\ 
J1051+3107 & 4.1392 & $20.70\pm0.20$ & - & $13.95\pm0.03$ & $13.86\pm0.08$ & $-1.96\pm0.22$ & SII & 98\\ 
J1051+3545 & 4.3498 & $20.45\pm0.10$ & - & $13.66\pm0.05$ & - & $-1.88\pm0.10$ & SiII & 98\\ 
J1051+3545 & 4.8206 & $20.35\pm0.10$ & - & - & - & $-2.28\pm0.10$ & SiII & 98\\ 
Q1054-0020 & 0.8301 & $18.95\pm0.18$ & $<11.76$ & $14.33\pm0.01$ & - & $0.20\pm0.25$ & FeII & 60\\ 
Q1054-0020 & 0.9514 & $19.28\pm0.02$ & $<11.70$ & $13.66\pm0.01$ & - & $-0.80\pm0.18$ & FeII & 60\\ 
J1054+1633 & 3.8400 & $20.65\pm0.20$ & - & $13.58\pm0.07$ & - & $-2.25\pm0.28$ & FeII & 99\\ 
J1054+1633 & 4.8200 & $20.65\pm0.20$ & - & - & - & $-2.17\pm0.20$ & SiII & 99\\ 
J1054+1633 & 4.1400 & $20.65\pm0.20$ & - & - & - & $-0.35\pm0.20$ & SiII & 99\\ 
QSO B1055-301 & 1.9040 & $21.54\pm0.10$ & $12.91\pm0.03$ & - & - & $-1.19\pm0.10$ & ZnII & 1\\ 
Q1055+46 & 3.3200 & $20.34\pm0.10$ & - & $13.94\pm0.06$ & - & $-1.60\pm0.15$ & SiII & 91, 43\\ 
J1056+1208 & 1.6100 & $21.45\pm0.15$ & $13.76\pm0.05$ & $15.81\pm0.05$ & $>16.15$ & $-0.25\pm0.16$ & ZnII & 3, 4, 5\\ 
J1100+1122 & 4.3947 & $21.74\pm0.10$ & - & $15.21\pm0.09$ & - & $-1.71\pm0.22$ & FeII & 98\\ 
QSO B1101-26 & 1.8380 & $19.50\pm0.05$ & $<11.27$ & $13.51\pm0.02$ & $13.66\pm0.11$ & $-1.64\pm0.10$ & OI & 22\\ 
J1101+0531 & 4.3446 & $21.30\pm0.10$ & - & $15.19\pm0.14$ & - & $-1.07\pm0.12$ & SiII & 98\\ 
QSO B1104-181 & 1.6610 & $20.85\pm0.01$ & $12.48\pm0.01$ & $14.77\pm0.02$ & - & $-0.93\pm0.01$ & ZnII & 53\\ 
J1106+1044 & 1.8200 & $20.50\pm0.15$ & - & $>15.15$ & $15.33\pm0.05$ & $-0.29\pm0.16$ & SII & 3, 4, 5\\ 
QSO J1107+0048 & 0.7400 & $21.00\pm0.04$ & $13.06\pm0.15$ & $15.53\pm0.02$ & - & $-0.50\pm0.16$ & ZnII & 72, 122\\ 
QSO B1108-07 & 3.4820 & $19.95\pm0.07$ & - & - & - & $-1.57\pm0.09$ & SiII & 52\\ 
QSO B1108-07 & 3.6080 & $20.37\pm0.07$ & - & $13.88\pm0.01$ & - & $-1.69\pm0.08$ & OI & 90, 77\\ 
J1111+3509 & 4.0520 & $20.80\pm0.15$ & - & $14.13\pm0.05$ & $<14.34$ & $-1.95\pm0.16$ & SiII & 98\\ 
Q1111-152 & 3.2700 & $21.30\pm0.05$ & $12.32\pm0.10$ & $14.81\pm0.01$ & $14.62\pm0.04$ & $-1.54\pm0.11$ & ZnII & 52, 67, 120, 119\\ 
SDSS1116+4118A & 2.6600 & $20.48\pm0.10$ & $12.40\pm0.20$ & $14.36\pm0.10$ & - & $-0.64\pm0.22$ & ZnII & 31\\ 
BR1117-1329 & 3.3500 & $20.84\pm0.12$ & $12.26\pm0.03$ & $14.83\pm0.03$ & - & $-1.14\pm0.12$ & ZnII & 71, 110\\ 
HE1122-1649 & 0.6800 & $20.45\pm0.05$ & $<11.76$ & $14.55\pm0.01$ & - & $-0.60\pm0.13$ & SiII & 123, 50\\ 
Q1127-145 & 0.3100 & $21.70\pm0.08$ & $13.53\pm0.13$ & $>15.16$ & - & $-0.73\pm0.15$ & ZnII & 45\\ 
J1131+6044 & 2.8800 & $20.50\pm0.15$ & - & $13.76\pm0.03$ & $<13.29$ & $-1.52\pm0.20$ & SiII & 34\\ 
HS1132+2243 & 2.7800 & $21.00\pm0.07$ & $<11.99$ & $14.02\pm0.01$ & $14.07\pm0.06$ & $-2.05\pm0.09$ & SII & 93\\ 
J1132+1209 & 4.3800 & $20.65\pm0.20$ & - & $13.78\pm0.07$ & - & $-2.05\pm0.28$ & FeII & 99\\ 
J1132+1209 & 5.0200 & $20.65\pm0.20$ & - & $<13.55$ & - & $-2.66\pm0.20$ & SiII & 99\\ 
J1135-0010 & 2.2100 & $22.05\pm0.10$ & $13.62\pm0.03$ & $15.76\pm0.03$ & $>16.19$ & $-0.99\pm0.10$ & ZnII & 48, 69\\ 
Q1137+3907 & 0.7200 & $21.10\pm0.10$ & $13.43\pm0.05$ & $15.45\pm0.05$ & - & $-0.23\pm0.11$ & ZnII & 58\\ 
J1142+0701 & 1.8400 & $21.50\pm0.15$ & $13.29\pm0.05$ & $15.47\pm0.05$ & - & $-0.77\pm0.16$ & ZnII & 3, 4, 5\\ 
QSO B1151+068 & 1.7750 & $21.30\pm0.08$ & $12.34\pm0.08$ & - & - & $-1.52\pm0.11$ & ZnII & 80\\ 
J115538.6+053050 & 3.3270 & $21.00\pm0.10$ & - & - & $15.31\pm0.00$ & $-0.81\pm0.10$ & SII & 114\\ 
J1155+3510 & 2.7582 & $21.00\pm0.10$ & - & $<14.73$ & $14.77\pm0.01$ & $-1.35\pm0.10$ & SII & 98\\ 
J1155+0530 & 2.6100 & $20.37\pm0.11$ & - & - & - & $-1.57\pm0.16$ & SiII & 119\\ 
J1155+0530 & 3.3300 & $21.05\pm0.10$ & $12.89\pm0.07$ & $15.37\pm0.05$ & $15.40\pm0.05$ & $-0.72\pm0.12$ & ZnII & 3, 4, 5\\ 
Q1157+014 & 1.9400 & $21.70\pm0.10$ & $13.11\pm0.06$ & $15.49\pm0.05$ & $>15.16$ & $-1.15\pm0.12$ & ZnII & 75, 24, 25, 3, 4, 5\\ 
J1200+4015 & 3.2200 & $20.85\pm0.10$ & $12.86\pm0.04$ & $15.31\pm0.04$ & $15.36\pm0.01$ & $-0.55\pm0.11$ & ZnII & 98\\ 
J1200+4618 & 4.4765 & $20.50\pm0.15$ & - & $14.27\pm0.02$ & - & $-1.41\pm0.24$ & FeII & 98\\ 
J1201+2117 & 3.7975 & $21.35\pm0.15$ & - & $15.56\pm0.04$ & - & $-0.75\pm0.15$ & SiII & 98\\ 
J1201+2117 & 4.1578 & $20.60\pm0.15$ & - & $13.76\pm0.03$ & - & $-2.38\pm0.15$ & SiII & 98\\ 
QSO B1202-074 & 4.3830 & $20.55\pm0.16$ & - & $13.88\pm0.11$ & - & $-1.49\pm0.17$ & OI & 57, 106, 27\\ 
J1202+3235 & 4.7955 & $21.10\pm0.15$ & - & $13.90\pm0.03$ & - & $-2.38\pm0.24$ & FeII & 98\\ 
J1202+3235 & 5.0647 & $20.30\pm0.15$ & - & - & - & $-2.66\pm0.16$ & SiII & 98\\ 
J1204-0021 & 3.6400 & $20.65\pm0.20$ & - & $13.85\pm0.04$ & - & $-1.98\pm0.27$ & FeII & 99\\ 
J120550.2+020131 & 1.7470 & $20.40\pm0.10$ & $12.08\pm0.08$ & - & - & $-0.88\pm0.13$ & ZnII & 37\\ 
J1208+0010 & 5.0800 & $20.65\pm0.20$ & - & $13.27\pm0.08$ & - & $-2.41\pm0.20$ & SiII & 99\\ 
QSO B1209+0919 & 2.5840 & $21.40\pm0.10$ & $12.98\pm0.05$ & $15.25\pm0.03$ & - & $-0.98\pm0.11$ & ZnII & 95\\ 
LBQS 1210+1731 & 1.8920 & $20.70\pm0.08$ & $12.37\pm0.03$ & $14.95\pm0.06$ & $14.96\pm0.03$ & $-0.89\pm0.09$ & ZnII & 90, 24\\ 
Q1215-0034 & 1.5543 & $19.56\pm0.02$ & $<11.63$ & $14.39\pm0.01$ & - & $-0.35\pm0.18$ & FeII & 60\\ 
Q1215+33 & 2.0000 & $20.95\pm0.07$ & $12.33\pm0.05$ & $14.75\pm0.05$ & $<15.36$ & $-1.18\pm0.09$ & ZnII & 89, 11, 90, 91\\ 
PG1216+069 & 0.0063 & $19.32\pm0.03$ & - & $13.23\pm0.14$ & - & $-1.69\pm0.06$ & OI & 115\\ 
J1219+1603 & 3.0000 & $20.35\pm0.10$ & - & $13.80\pm0.10$ & - & $-2.52\pm0.35$ & OI & 70\\ 
QSO B1220-1800 & 2.1120 & $20.12\pm0.07$ & - & $14.36\pm0.04$ & $14.53\pm0.04$ & $-0.71\pm0.08$ & SII & 66\\ 
Q1220-0040 & 0.9746 & $20.20\pm0.07$ & $<11.69$ & $14.34\pm0.02$ & - & $-1.04\pm0.19$ & FeII & 60\\ 
J1221+4445 & 4.8100 & $20.65\pm0.20$ & - & $14.35\pm0.06$ & - & $-2.21\pm0.20$ & SiII & 99\\ 
LBQS 1223+1753 & 2.4660 & $21.40\pm0.10$ & $12.55\pm0.03$ & $15.16\pm0.02$ & $15.14\pm0.04$ & $-1.41\pm0.10$ & ZnII & 90, 110\\ 
LBQS 1223+1753 & 2.5570 & $19.32\pm0.15$ & $<11.51$ & $13.98\pm0.03$ & - & $-0.45\pm0.15$ & SiII & 22\\ 
Q1224+0037 & 1.2300 & $20.88\pm0.05$ & $<11.89$ & $>15.11$ & - & $-1.29\pm0.09$ & SiII & 59\\ 
Q1225+0035 & 0.7700 & $21.38\pm0.11$ & $<13.01$ & $15.69\pm0.03$ & - & $-0.87\pm0.21$ & FeII & 58, 63\\ 
PHL 1226 & 0.1602 & $19.48\pm0.10$ & - & $14.76\pm0.18$ & $14.84\pm0.11$ & $0.24\pm0.15$ & SII & 105\\ 
QSO B1228-113 & 2.1930 & $20.60\pm0.10$ & $13.01\pm0.04$ & - & - & $-0.15\pm0.11$ & ZnII & 1\\ 
Q1228+1018 & 0.9376 & $19.41\pm0.02$ & $<11.67$ & $14.58\pm0.01$ & - & $-0.01\pm0.18$ & FeII & 60\\ 
PKS1229-021 & 0.4000 & $20.75\pm0.07$ & $12.92\pm0.10$ & $<14.95$ & - & $-0.39\pm0.12$ & ZnII & 8\\ 
QSO B1230-101 & 1.9310 & $20.48\pm0.10$ & $12.94\pm0.05$ & - & - & $-0.10\pm0.11$ & ZnII & 1\\ 
LBQS 1232+0815 & 1.7200 & $19.48\pm0.13$ & $<11.58$ & $13.50\pm0.01$ & $<14.19$ & $-0.58\pm0.13$ & SiII & 114\\ 
LBQS 1232+0815 & 2.3340 & $20.90\pm0.04$ & $12.64\pm0.09$ & $14.68\pm0.08$ & $14.83\pm0.10$ & $-0.82\pm0.10$ & ZnII & 108, 40, 12, 120\\ 
J1238+3437 & 2.4714 & $20.80\pm0.10$ & - & $14.06\pm0.03$ & $13.91\pm0.11$ & $-2.01\pm0.15$ & SII & 98\\ 
J1240+1455 & 3.1100 & $21.30\pm0.20$ & $12.90\pm0.07$ & $14.60\pm0.03$ & $15.56\pm0.02$ & $-0.96\pm0.21$ & ZnII & 34\\ 
J1241+4617 & 2.6674 & $20.70\pm0.10$ & - & $14.02\pm0.04$ & - & $-2.18\pm0.10$ & SiII & 98\\ 
LBQS1242+0006 & 1.8200 & $20.45\pm0.10$ & - & - & - & $-1.20\pm0.15$ & SiII & 119\\ 
J1245+3822 & 4.4500 & $20.65\pm0.20$ & - & $<13.93$ & - & $-2.14\pm0.20$ & SiII & 99\\ 
LBQS 1246-0217 & 1.7810 & $21.45\pm0.00$ & $13.01\pm0.05$ & $15.47\pm0.02$ & - & $-1.00\pm0.05$ & ZnII & 44\\ 
J1248+3110 & 3.6973 & $20.60\pm0.10$ & - & $14.11\pm0.03$ & - & $-1.67\pm0.21$ & FeII & 98\\ 
SDSS1249-0233 & 1.7800 & $21.45\pm0.15$ & $13.15\pm0.05$ & $15.47\pm0.02$ & $15.53\pm0.05$ & $-0.86\pm0.16$ & ZnII & 44, 3, 4, 5\\ 
SDSS1251+4120 & 2.7300 & $21.10\pm0.10$ & - & $14.20\pm0.30$ & - & $-2.71\pm0.32$ & SiII & 70\\ 
J1253+1046 & 4.6001 & $20.30\pm0.15$ & - & $14.09\pm0.03$ & - & $-1.39\pm0.24$ & FeII & 98\\ 
PSS1253-0228 & 2.7800 & $21.85\pm0.20$ & $12.77\pm0.07$ & $15.36\pm0.04$ & - & $-1.64\pm0.21$ & ZnII & 93\\ 
J1257-0111 & 4.0208 & $20.30\pm0.10$ & - & $13.65\pm0.07$ & $<13.90$ & $-1.56\pm0.10$ & SiII & 98\\ 
J1304+1202 & 2.9131 & $20.55\pm0.15$ & $<11.83$ & $13.72\pm0.04$ & $14.05\pm0.05$ & $-1.62\pm0.16$ & SII & 98\\ 
J1304+1202 & 2.9289 & $20.30\pm0.15$ & $<11.95$ & $13.85\pm0.03$ & $13.91\pm0.04$ & $-1.51\pm0.16$ & SII & 98\\ 
J1305+0924 & 2.0200 & $20.40\pm0.15$ & - & $15.21\pm0.14$ & $15.39\pm0.05$ & $-0.13\pm0.16$ & SII & 3, 4, 5\\ 
J1310+5424 & 1.8000 & $21.45\pm0.15$ & $13.57\pm0.05$ & $15.64\pm0.05$ & $>16.05$ & $-0.44\pm0.16$ & ZnII & 3, 4, 5\\ 
J1323-0021 & 0.7160 & $20.21\pm0.20$ & $13.43\pm0.05$ & $15.15\pm0.03$ & - & $0.66\pm0.21$ & ZnII & 72\\ 
Q1323-0021 & 0.7200 & $20.54\pm0.15$ & $13.29\pm0.21$ & - & - & $0.19\pm0.26$ & ZnII & 63\\ 
SDSS1325+1255 & 3.5500 & $20.50\pm0.15$ & - & $<13.69$ & - & $-2.51\pm0.25$ & SiII & 70\\ 
Q1328+307 & 0.6900 & $21.25\pm0.10$ & $12.72\pm0.10$ & $14.98\pm0.10$ & - & $-1.09\pm0.14$ & ZnII & 79, 8, 50\\ 
QSO J1330-2522 & 2.6540 & $19.56\pm0.13$ & - & - & - & $-1.83\pm0.13$ & AlII & 114\\ 
Q1330-2056 & 0.8526 & $19.40\pm0.02$ & $<11.96$ & $13.80\pm0.01$ & - & $-0.78\pm0.18$ & FeII & 60\\ 
QSO B1331+170 & 1.7760 & $21.15\pm0.07$ & $12.61\pm0.01$ & $14.60\pm0.00$ & $15.08\pm0.11$ & $-1.10\pm0.07$ & ZnII & 89, 22\\ 
J1335+0824 & 1.8600 & $20.65\pm0.15$ & - & $>15.17$ & $15.29\pm0.05$ & $-0.48\pm0.16$ & SII & 3, 4, 5\\ 
Q1337+113 & 2.8000 & $21.00\pm0.08$ & $<12.25$ & $14.33\pm0.01$ & $14.33\pm0.02$ & $-1.95\pm0.11$ & OI & 93, 110, 52, 95, 77, 67\\ 
J1340+1106 & 2.8000 & $21.00\pm0.06$ & - & $14.32\pm0.01$ & $14.30\pm0.02$ & $-1.65\pm0.07$ & OI & 16\\ 
J1340+3926 & 4.8300 & $20.65\pm0.20$ & - & $14.33\pm0.05$ & - & $-1.50\pm0.27$ & FeII & 99\\ 
QSO J1342-1355 & 3.1180 & $20.05\pm0.08$ & - & $13.93\pm0.03$ & $13.83\pm0.03$ & $-1.22\pm0.08$ & OI & 77\\ 
J1345+2329 & 5.0100 & $20.65\pm0.20$ & - & - & $14.66\pm0.05$ & $-1.11\pm0.21$ & SII & 99\\ 
BRI1346-03 & 3.7400 & $20.72\pm0.10$ & - & $<14.13$ & - & $-2.28\pm0.10$ & SiII & 89, 90\\ 
SDSS1350+5952 & 2.7600 & $20.65\pm0.10$ & - & $13.50\pm0.20$ & - & $-2.33\pm0.29$ & FeII & 70\\ 
J1353+5328 & 2.8349 & $20.80\pm0.10$ & - & $>14.46$ & $14.57\pm0.02$ & $-1.35\pm0.10$ & SII & 98\\ 
Q1354+258 & 1.4200 & $21.54\pm0.06$ & $12.59\pm0.08$ & $15.01\pm0.04$ & - & $-1.51\pm0.10$ & ZnII & 81\\ 
PKS1354-17 & 2.7800 & $20.30\pm0.15$ & - & $13.37\pm0.08$ & - & $-1.83\pm0.16$ & SiII & 93\\ 
QSO J1356-1101 & 2.3970 & $19.85\pm0.08$ & $<12.38$ & $13.44\pm0.01$ & - & $-1.59\pm0.20$ & FeII & 114\\ 
QSO J1356-1101 & 2.5010 & $20.44\pm0.05$ & $<11.70$ & $14.36\pm0.08$ & $14.27\pm0.09$ & $-1.29\pm0.10$ & SII & 1, 66\\ 
QSO J1356-1101 & 2.9670 & $20.80\pm0.10$ & $<11.93$ & $14.63\pm0.05$ & - & $-1.35\pm0.12$ & SiII & 1, 66\\ 
J1358+0349 & 2.8500 & $20.50\pm0.10$ & - & $13.01\pm0.05$ & - & $-2.81\pm0.27$ & OI & 70\\ 
J1358+6522 & 3.0700 & $20.35\pm0.15$ & - & $<12.80$ & - & $-3.01\pm0.17$ & OI & 70, 18\\ 
QSO B1409+0930 & 2.0190 & $20.65\pm0.10$ & $11.63\pm0.10$ & - & - & $-1.58\pm0.14$ & ZnII & 52\\ 
QSO B1409+0930 & 2.4560 & $20.53\pm0.08$ & - & $13.74\pm0.02$ & - & $-2.07\pm0.10$ & OI & 83\\ 
QSO B1409+0930 & 2.6680 & $19.80\pm0.08$ & $<11.22$ & $14.02\pm0.13$ & $13.54\pm0.06$ & $-1.18\pm0.14$ & OI & 83, 22\\ 
J1412+0624 & 4.1095 & $20.40\pm0.15$ & - & $13.83\pm0.08$ & - & $-1.75\pm0.25$ & FeII & 98\\ 
J1417+4132 & 1.9500 & $21.45\pm0.25$ & $13.55\pm0.05$ & $15.58\pm0.05$ & $>15.80$ & $-0.46\pm0.25$ & ZnII & 3, 4, 5\\ 
J1418+3142 & 3.9600 & $20.65\pm0.20$ & - & $<15.78$ & - & $-0.39\pm0.20$ & SiII & 99\\ 
J1419+0829 & 3.0500 & $20.40\pm0.03$ & - & $13.54\pm0.03$ & - & $-1.92\pm0.04$ & OI & 16, 86\\ 
QSO J1421-0643 & 3.4480 & $20.40\pm0.10$ & $<11.98$ & $14.18\pm0.08$ & - & $-1.29\pm0.13$ & SiII & 1, 66\\ 
Q1425+6039 & 2.8300 & $20.30\pm0.04$ & $12.18\pm0.04$ & $14.48\pm0.01$ & - & $-0.68\pm0.06$ & ZnII & 57, 91, 92, 95\\ 
J1431+3952 & 0.6000 & $21.20\pm0.10$ & $13.03\pm0.19$ & $15.15\pm0.11$ & - & $-0.73\pm0.21$ & ZnII & 35\\ 
PSS1432+39 & 3.2700 & $21.25\pm0.10$ & $<12.65$ & $>14.93$ & - & $-1.09\pm0.11$ & SiII & 93\\ 
J1435+3604 & 0.2026 & $19.80\pm0.10$ & - & $14.20\pm0.08$ & $14.60\pm0.12$ & $-0.32\pm0.16$ & SII & 2\\ 
SDSS1435+0420 & 1.6600 & $21.25\pm0.15$ & $<13.21$ & $15.70\pm0.07$ & - & $-0.84\pm0.17$ & SiII & 44\\ 
J1435+5359 & 2.3400 & $21.05\pm0.10$ & - & - & $14.78\pm0.05$ & $-1.39\pm0.11$ & SII & 43\\ 
Q1436-0051 & 0.7377 & $20.08\pm0.11$ & $12.67\pm0.05$ & $14.94\pm0.02$ & - & $0.03\pm0.12$ & ZnII & 60\\ 
J1437+2323 & 4.8000 & $20.65\pm0.20$ & - & - & - & $-2.34\pm0.20$ & SiII & 99\\ 
J1438+4314 & 4.3990 & $20.89\pm0.15$ & - & $14.42\pm0.01$ & $14.73\pm0.01$ & $-1.28\pm0.15$ & SII & 98\\ 
QSO J1439+1117 & 2.4180 & $20.10\pm0.10$ & $12.93\pm0.04$ & $14.28\pm0.05$ & $15.27\pm0.06$ & $0.27\pm0.11$ & ZnII & 66\\ 
SDSS1440+0637 & 2.5200 & $21.00\pm0.15$ & - & $14.50\pm0.30$ & - & $-2.31\pm0.34$ & SiII & 70\\ 
QSO J1443+2724 & 4.2240 & $20.95\pm0.10$ & $12.99\pm0.03$ & $15.33\pm0.03$ & $15.52\pm0.01$ & $-0.52\pm0.10$ & ZnII & 90, 66, 52\\ 
LBQS 1444+0126 & 2.0870 & $20.25\pm0.07$ & $12.12\pm0.15$ & $14.41\pm0.03$ & $14.62\pm0.08$ & $-0.69\pm0.17$ & ZnII & 51, 22\\ 
Q1451+123 & 2.4700 & $20.39\pm0.10$ & - & $13.36\pm0.07$ & $<13.55$ & $-1.90\pm0.16$ & SiII & 75, 110\\ 
Q1451+123 & 2.2600 & $20.30\pm0.15$ & $11.85\pm0.11$ & $14.33\pm0.07$ & - & $-1.01\pm0.19$ & ZnII & 22\\ 
J1454+0941 & 1.7900 & $20.50\pm0.15$ & $12.72\pm0.05$ & $15.02\pm0.12$ & $15.25\pm0.06$ & $-0.34\pm0.16$ & ZnII & 3, 4, 5\\ 
Q1455-0045 & 1.0929 & $20.08\pm0.06$ & $<11.91$ & $14.57\pm0.01$ & - & $-0.95\pm0.12$ & SiII & 60\\ 
J1456+0407 & 2.6700 & $20.35\pm0.10$ & - & $13.00\pm0.10$ & - & $-2.49\pm0.30$ & OI & 70\\ 
Q1501+0019 & 1.4800 & $20.85\pm0.13$ & $12.93\pm0.06$ & - & - & $-0.48\pm0.14$ & ZnII & 58\\ 
Q1502+4837 & 2.5700 & $20.30\pm0.15$ & - & $14.15\pm0.12$ & - & $-1.57\pm0.17$ & SiII & 93\\ 
PSS1506+5220 & 3.2200 & $20.67\pm0.07$ & $<12.11$ & $13.71\pm0.03$ & - & $-2.30\pm0.07$ & SiII & 93\\ 
J1507+4406 & 3.0644 & $20.75\pm0.10$ & - & $14.03\pm0.03$ & $13.97\pm0.10$ & $-1.90\pm0.14$ & SII & 98\\ 
J1509+1113 & 2.0300 & $21.30\pm0.15$ & - & $15.48\pm0.07$ & $15.69\pm0.05$ & $-0.73\pm0.16$ & SII & 3, 4, 5\\ 
PSS1535+2943 & 3.7600 & $20.40\pm0.15$ & - & - & - & $-1.97\pm0.16$ & SiII & 94\\ 
J1541+3153 & 2.4435 & $20.95\pm0.10$ & $12.03\pm0.11$ & $14.50\pm0.11$ & - & $-1.48\pm0.15$ & ZnII & 98\\ 
SBS1543+393 & 0.0100 & $20.42\pm0.04$ & - & - & $15.19\pm0.04$ & $-0.35\pm0.06$ & SII & 9\\ 
J1553+3548 & 0.0830 & $19.55\pm0.15$ & - & $14.01\pm0.07$ & $<14.24$ & $-0.84\pm0.16$ & SiII & 2\\ 
J1555+4800 & 2.3900 & $21.50\pm0.15$ & $<13.95$ & $15.84\pm0.05$ & $>15.88$ & $-0.46\pm0.16$ & SiII & 3, 4, 5\\ 
SDSS1557+2320 & 3.5400 & $20.65\pm0.10$ & - & $13.50\pm0.30$ & - & $-2.24\pm0.15$ & OI & 70\\ 
SDSSJ1558+4053 & 2.5500 & $20.30\pm0.04$ & - & $13.07\pm0.06$ & - & $-2.45\pm0.06$ & OI & 85\\ 
J1558-0031 & 2.7000 & $20.67\pm0.05$ & - & $14.11\pm0.03$ & $14.07\pm0.02$ & $-1.72\pm0.05$ & SII & 43\\ 
PHL 1598 & 0.4297 & $19.18\pm0.03$ & - & - & $14.36\pm0.05$ & $0.06\pm0.06$ & SII & 105\\ 
J1604+3951 & 3.1600 & $21.75\pm0.20$ & $13.12\pm0.05$ & $15.47\pm0.05$ & $15.71\pm0.05$ & $-1.19\pm0.21$ & ZnII & 34, 3, 4, 5\\ 
J1607+1604 & 4.4741 & $20.30\pm0.15$ & - & $14.03\pm0.06$ & - & $-1.71\pm0.15$ & SiII & 98\\ 
SDSS1610+4724 & 2.5100 & $21.15\pm0.15$ & $13.56\pm0.05$ & $15.62\pm0.05$ & $>16.01$ & $-0.15\pm0.16$ & ZnII & 44, 3, 4, 5\\ 
J1616+4154 & 0.3211 & $20.60\pm0.20$ & - & $15.02\pm0.05$ & $15.37\pm0.11$ & $-0.35\pm0.23$ & SII & 2\\ 
J1619+3342 & 0.0963 & $20.55\pm0.10$ & - & $14.38\pm0.15$ & $15.08\pm0.09$ & $-0.59\pm0.13$ & SII & 2\\ 
QSO J1621-0042 & 3.1040 & $19.70\pm0.20$ & - & $13.30\pm0.04$ & - & $-1.43\pm0.20$ & SiII & 114\\ 
3C336 & 0.6600 & $20.36\pm0.10$ & - & $14.59\pm0.11$ & - & $-0.95\pm0.23$ & FeII & 113, 14, 50\\ 
J1623+0718 & 1.3400 & $21.35\pm0.10$ & $12.91\pm0.09$ & $15.28\pm0.05$ & - & $-1.00\pm0.13$ & ZnII & 35\\ 
J1626+2751 & 4.3110 & $21.34\pm0.15$ & - & $15.33\pm0.06$ & - & $-1.19\pm0.24$ & FeII & 98\\ 
J1626+2751 & 4.4975 & $21.39\pm0.15$ & - & $14.08\pm0.02$ & - & $-2.49\pm0.24$ & FeII & 98\\ 
J1626+2751 & 5.1791 & $20.94\pm0.15$ & - & $>14.59$ & $14.60\pm0.02$ & $-1.46\pm0.15$ & SII & 98\\ 
J1626+2858 & 4.6100 & $20.65\pm0.20$ & - & - & - & $-2.73\pm0.22$ & SiII & 99\\ 
J1629+0913 & 1.9000 & $20.80\pm0.10$ & $12.68\pm0.08$ & $>14.93$ & $15.24\pm0.05$ & $-0.68\pm0.13$ & ZnII & 3, 4, 5\\ 
4C 12.59 & 0.5310 & $20.70\pm0.09$ & - & $14.26\pm0.08$ & - & $-1.62\pm0.22$ & FeII & 114\\ 
4C 12.59 & 0.9000 & $19.70\pm0.04$ & $<12.18$ & $14.17\pm0.03$ & - & $-0.71\pm0.19$ & FeII & 60\\ 
J1637+2901 & 3.5000 & $20.70\pm0.10$ & - & $13.84\pm0.10$ & - & $-3.10\pm0.22$ & OI & 70\\ 
J1654+2227 & 4.0022 & $20.60\pm0.15$ & - & $14.09\pm0.03$ & - & $-1.69\pm0.24$ & FeII & 98\\ 
SDSS1709+3417 & 2.5300 & $20.45\pm0.15$ & - & $14.30\pm0.20$ & - & $-1.46\pm0.25$ & SiII & 70\\ 
SDSS1709+3417 & 3.0100 & $20.40\pm0.10$ & - & $13.90\pm0.20$ & - & $-1.68\pm0.29$ & FeII & 70\\ 
J1712+5755 & 2.2500 & $20.60\pm0.10$ & - & $14.49\pm0.02$ & - & $-1.19\pm0.12$ & SiII & 43\\ 
Q1715+4606 & 0.6500 & $20.44\pm0.10$ & $<12.87$ & $14.94\pm0.03$ & - & $-0.68\pm0.21$ & FeII & 58\\ 
PSS1715+3809 & 3.3400 & $21.05\pm0.12$ & $<12.11$ & $13.74\pm0.04$ & - & $-2.49\pm0.22$ & FeII & 94\\ 
Q1727+5302 & 1.0300 & $21.41\pm0.15$ & $12.76\pm0.24$ & $14.81\pm0.01$ & - & $-1.21\pm0.28$ & ZnII & 116, 63\\ 
Q1727+5302 & 0.9400 & $21.16\pm0.10$ & $13.25\pm0.11$ & $15.29\pm0.01$ & - & $-0.47\pm0.15$ & ZnII & 116, 63\\ 
Q1733+5533 & 1.0000 & $20.70\pm0.10$ & $<12.11$ & - & - & $-0.73\pm0.12$ & SiII & 58, 63\\ 
SDSS1737+5828 & 4.7400 & $20.65\pm0.10$ & - & $13.30\pm0.10$ & - & $-2.53\pm0.23$ & FeII & 106\\ 
J1737+5828 & 4.7400 & $20.65\pm0.20$ & - & - & - & $-2.23\pm0.21$ & SiII & 99\\ 
Q1755+578 & 1.9700 & $21.40\pm0.15$ & $13.85\pm0.05$ & $15.79\pm0.05$ & $>16.12$ & $-0.11\pm0.16$ & ZnII & 3, 4, 5\\ 
Q1759+75 & 2.6300 & $20.76\pm0.05$ & $>11.65$ & $15.08\pm0.02$ & $15.24\pm0.01$ & $-0.64\pm0.05$ & SII & 89, 90, 43\\ 
PSS1802+5616 & 3.8100 & $20.35\pm0.20$ & - & $13.67\pm0.10$ & - & $-1.99\pm0.22$ & SiII & 94\\ 
PSS1802+5616 & 3.5500 & $20.50\pm0.10$ & $<12.63$ & $14.08\pm0.06$ & - & $-1.60\pm0.21$ & FeII & 94\\ 
PSS1802+5616 & 3.3900 & $20.30\pm0.10$ & $<12.41$ & $14.26\pm0.04$ & - & $-1.22\pm0.21$ & FeII & 94\\ 
QSO B2000-330 & 3.1720 & $19.75\pm0.15$ & - & $<12.86$ & - & $-2.29\pm0.15$ & OI & 96\\ 
QSO B2000-330 & 3.1880 & $19.80\pm0.15$ & - & $13.69\pm0.04$ & - & $-1.34\pm0.15$ & SiII & 96\\ 
QSO B2000-330 & 3.1920 & $19.10\pm0.15$ & - & $13.49\pm0.07$ & - & $-0.48\pm0.15$ & SiII & 96\\ 
J2036-0553 & 2.2800 & $21.20\pm0.15$ & - & $14.68\pm0.11$ & - & $-1.67\pm0.16$ & SiII & 43\\ 
Q2051+1950 & 1.1157 & $20.00\pm0.15$ & $12.90\pm0.10$ & $15.02\pm0.02$ & - & $0.34\pm0.18$ & ZnII & 60\\ 
SDSS2059-0529 & 2.2100 & $20.80\pm0.20$ & $12.94\pm0.11$ & $15.00\pm0.11$ & - & $-0.42\pm0.23$ & ZnII & 44\\ 
Q2059-360 & 3.0800 & $20.98\pm0.08$ & - & $14.52\pm0.07$ & $14.41\pm0.04$ & $-1.58\pm0.09$ & OI & 75, 110, 77\\ 
SDSS2100-0641 & 3.0900 & $21.05\pm0.15$ & $13.24\pm0.05$ & $15.37\pm0.05$ & $15.49\pm0.05$ & $-0.37\pm0.16$ & ZnII & 44, 3, 4, 5\\ 
LBQS 2114-4347 & 1.9120 & $19.50\pm0.10$ & $<12.17$ & $14.02\pm0.01$ & $<13.97$ & $-0.70\pm0.10$ & MgII & 114\\ 
QSO J2119-3536 & 1.9960 & $20.10\pm0.07$ & $12.30\pm0.09$ & $14.77\pm0.09$ & $<14.95$ & $-0.36\pm0.11$ & ZnII & 22\\ 
QSO B2126-15 & 2.6380 & $19.25\pm0.15$ & $<11.58$ & $14.05\pm0.01$ & - & $-0.09\pm0.15$ & SiII & 114\\ 
QSO B2126-15 & 2.7690 & $19.20\pm0.15$ & $<11.95$ & $14.17\pm0.00$ & - & $0.08\pm0.15$ & SiII & 114\\ 
LBQS 2132-4321 & 1.9160 & $20.74\pm0.09$ & $12.66\pm0.02$ & $15.03\pm0.02$ & $>14.90$ & $-0.64\pm0.09$ & ZnII & 114\\ 
LBQS 2138-4427 & 2.3830 & $20.60\pm0.05$ & $12.05\pm0.07$ & - & - & $-1.11\pm0.09$ & ZnII & 52\\ 
LBQS 2138-4427 & 2.8520 & $20.98\pm0.05$ & $11.99\pm0.05$ & $14.65\pm0.02$ & $14.50\pm0.02$ & $-1.55\pm0.07$ & ZnII & 51, 110\\ 
J2144-0632 & 4.1300 & $20.40\pm0.15$ & - & $<13.51$ & - & $-2.37\pm0.47$ & OI & 70\\ 
PSSJ2155+1358 & 3.3200 & $20.50\pm0.15$ & $12.05\pm0.32$ & $14.51\pm0.13$ & - & $-1.01\pm0.35$ & ZnII & 19, 93\\ 
LBQS 2206-1958A & 2.0760 & $20.44\pm0.05$ & $<11.20$ & $13.33\pm0.01$ & - & $-2.08\pm0.06$ & OI & 84\\ 
Q2206-199 & 1.9200 & $20.68\pm0.03$ & $12.91\pm0.01$ & $15.30\pm0.02$ & $15.42\pm0.02$ & $-0.33\pm0.03$ & ZnII & 88, 90, 91, 117\\ 
QSO B2222-396 & 2.1540 & $20.85\pm0.10$ & - & $14.42\pm0.03$ & $14.08\pm0.02$ & $-1.89\pm0.10$ & SII & 66\\ 
SDSS2222-0946 & 2.3500 & $20.50\pm0.15$ & $<12.78$ & $15.06\pm0.08$ & $15.37\pm0.05$ & $-0.25\pm0.16$ & SII & 44, 3, 4, 46, 5\\ 
Q2223+20 & 3.1200 & $20.30\pm0.10$ & - & $13.32\pm0.06$ & - & $-2.17\pm0.11$ & SiII & 93\\ 
Q2228-3954 & 2.1000 & $21.20\pm0.10$ & $12.51\pm0.06$ & $15.17\pm0.02$ & - & $-1.25\pm0.12$ & ZnII & 67\\ 
LBQS 2230+0232 & 1.8640 & $20.83\pm0.10$ & $12.80\pm0.03$ & $15.19\pm0.02$ & $15.29\pm0.10$ & $-0.59\pm0.10$ & ZnII & 89, 90, 91, 24, 25, 3, 4, 5\\ 
Q2231-00 & 2.0700 & $20.53\pm0.08$ & $12.30\pm0.05$ & $14.83\pm0.03$ & $15.10\pm0.15$ & $-0.79\pm0.09$ & ZnII & 89, 90, 21, 23\\ 
QSO B2237-0607 & 4.0790 & $20.55\pm0.10$ & - & $13.85\pm0.11$ & - & $-1.79\pm0.10$ & SiII & 57, 106, 43\\ 
J223941.8-294955 & 1.8250 & $19.84\pm0.14$ & $12.76\pm0.06$ & $14.33\pm0.04$ & - & $0.36\pm0.15$ & ZnII & 24\\ 
J2241+1225 & 2.4200 & $21.15\pm0.15$ & - & $15.02\pm0.08$ & $>15.01$ & $-1.31\pm0.25$ & FeII & 3, 4, 5\\ 
PSS2241+1352 & 4.2800 & $21.15\pm0.10$ & - & $>14.65$ & $14.58\pm0.03$ & $-1.69\pm0.10$ & SII & 93\\ 
HE2243-6031 & 2.3300 & $20.67\pm0.02$ & $12.22\pm0.03$ & $14.92\pm0.01$ & $14.88\pm0.01$ & $-1.01\pm0.04$ & ZnII & 54\\ 
J2252+1425 & 4.7475 & $20.60\pm0.15$ & - & $13.98\pm0.11$ & $<14.41$ & $-1.80\pm0.26$ & FeII & 98\\ 
QSO B2311-373 & 2.1820 & $20.48\pm0.13$ & $<11.82$ & $14.23\pm0.04$ & - & $-1.45\pm0.15$ & SiII & 1, 66\\ 
B2314-409 & 1.8600 & $20.90\pm0.10$ & $12.52\pm0.10$ & $15.08\pm0.10$ & $15.10\pm0.15$ & $-0.94\pm0.14$ & ZnII & 30\\ 
PSS2315+0921 & 3.4300 & $21.10\pm0.20$ & - & $>14.63$ & - & $-1.46\pm0.21$ & SiII & 94\\ 
QSO B2318-1107 & 1.6290 & $20.52\pm0.14$ & $<11.74$ & $14.14\pm0.02$ & $<14.54$ & $-1.56\pm0.23$ & FeII & 114\\ 
QSO B2318-1107 & 1.9890 & $20.68\pm0.05$ & $12.50\pm0.03$ & $14.91\pm0.01$ & $15.09\pm0.02$ & $-0.74\pm0.06$ & ZnII & 64\\ 
J2321+1421 & 2.5700 & $20.70\pm0.05$ & $<11.84$ & $14.18\pm0.03$ & $<13.60$ & $-1.76\pm0.06$ & SiII & 34\\ 
PSS2323+2758 & 3.6800 & $20.95\pm0.10$ & - & $13.32\pm0.13$ & - & $-2.54\pm0.10$ & SiII & 93\\ 
QSO J2328+0022 & 0.6520 & $20.32\pm0.07$ & $12.43\pm0.15$ & $14.84\pm0.01$ & - & $-0.45\pm0.17$ & ZnII & 72\\ 
QSO B2332-094 & 3.0570 & $20.50\pm0.07$ & $<12.17$ & $14.34\pm0.03$ & $14.34\pm0.18$ & $-1.24\pm0.07$ & OI & 51, 93, 77, 119\\ 
J233544.2+150118 & 0.6800 & $19.70\pm0.30$ & $12.37\pm0.04$ & $14.83\pm0.03$ & - & $0.11\pm0.30$ & ZnII & 74\\ 
J2340-00 & 2.0500 & $20.35\pm0.15$ & $12.63\pm0.07$ & $14.98\pm0.05$ & $14.95\pm0.05$ & $-0.28\pm0.17$ & ZnII & 95, 3, 4, 5\\ 
Q2342+34 & 2.9100 & $21.10\pm0.10$ & $<12.60$ & $14.91\pm0.07$ & $15.19\pm0.05$ & $-1.03\pm0.11$ & SII & 93, 95, 3, 4, 120, 5\\ 
QSO B2343+125 & 2.4310 & $20.40\pm0.07$ & $12.20\pm0.07$ & $14.52\pm0.02$ & $14.66\pm0.02$ & $-0.76\pm0.10$ & ZnII & 64\\ 
Q2344+12 & 2.5400 & $20.36\pm0.10$ & - & $14.03\pm0.03$ & $<14.20$ & $-1.69\pm0.10$ & SiII & 90, 92\\ 
PSSJ2344+0342 & 3.2200 & $21.25\pm0.08$ & $12.23\pm0.30$ & $15.06\pm0.15$ & - & $-1.58\pm0.31$ & ZnII & 19, 93\\ 
QSO J2346+1247 & 2.5690 & $20.98\pm0.04$ & $12.88\pm0.06$ & $15.24\pm0.04$ & $15.38\pm0.05$ & $-0.66\pm0.07$ & ZnII & 104\\ 
QSO B2348-0180 & 2.4260 & $20.50\pm0.10$ & $<11.20$ & $14.83\pm0.07$ & $15.06\pm0.10$ & $-0.56\pm0.14$ & SII & 65\\ 
QSO B2348-0180 & 2.6150 & $21.30\pm0.08$ & $<11.87$ & $14.57\pm0.09$ & - & $-1.92\pm0.11$ & SiII & 90\\ 
QSO B2348-147 & 2.2790 & $20.56\pm0.08$ & $<11.28$ & $13.79\pm0.02$ & $13.72\pm0.12$ & $-1.95\pm0.14$ & SII & 89, 24\\ 
Q2352-0028 & 0.8730 & $19.18\pm0.09$ & $<11.67$ & $13.48\pm0.02$ & - & $-0.88\pm0.20$ & FeII & 60\\ 
Q2352-0028 & 1.0318 & $19.81\pm0.13$ & $<11.93$ & $14.91\pm0.01$ & - & $0.17\pm0.13$ & SiII & 60\\ 
Q2352-0028 & 1.2467 & $19.60\pm0.24$ & $<11.53$ & $14.21\pm0.01$ & - & $-0.57\pm0.30$ & FeII & 60\\ 
Q2353-0028 & 0.6000 & $21.54\pm0.15$ & $13.25\pm0.29$ & - & - & $-0.85\pm0.33$ & ZnII & 63\\ 
B2355-106 & 1.1700 & $21.00\pm0.10$ & $12.76\pm0.17$ & $15.08\pm0.10$ & - & $-0.80\pm0.20$ & ZnII & 35\\ 
LBQS 2359-0216 & 2.0950 & $20.65\pm0.10$ & $12.60\pm0.03$ & $14.51\pm0.03$ & - & $-0.61\pm0.10$ & ZnII & 89\\ 
LBQS 2359-0216 & 2.1540 & $20.30\pm0.10$ & $<11.90$ & $13.89\pm0.03$ & - & $-1.49\pm0.10$ & SiII & 89\\

\end{longtable}

\textbf{References: }1: \citealt{Akerman2005}, 2: \citealt{Battisti2012}, 3: \citealt{Berg2013}, 4: \citealt{Berg2014}, 5: \citealt{Berg2015}, 6: \citealt{Bergeron1986}, 7: \citealt{Blades1982}, 8: \citealt{Boisse1998}, 9: \citealt{Bowen2005}, 10: \citealt{Carswell1996}, 11: \citealt{Centurion2000}, 12: \citealt{Centurion2003}, 13: \citealt{Chen2005}, 14: \citealt{Churchill2000}, 15: \citealt{Cooke2010a}, 16: \citealt{Cooke2011}, 17: \citealt{Cooke2011a}, 18: \citealt{Cooke2012}, 19: Dessauge-Zavadsky (unpublished), 20: \citealt{Dessauges-Zavadsky2001}, 21: \citealt{Dessauges-Zavadsky2002}, 22: \citealt{Dessauges-Zavadsky2003}, 23: \citealt{Dessauges-Zavadsky2004}, 24: \citealt{Dessauges-Zavadsky2006}, 25: \citealt{Dessauges-Zavadsky2007}, 26: \citealt{Dessauges-Zavadsky2009}, 27: \citealt{Dodorico2004}, 28: \citealt{Dutta2014}, 29: \citealt{Ellison2001}, 30: \citealt{Ellison2001b}, 31: \citealt{Ellison2007}, 32: \citealt{Ellison2008}, 33: \citealt{Ellison2009}, 34: \citealt{Ellison2010}, 35: \citealt{Ellison2012}, 36: \citealt{Fox2007}, 37: \citealt{Fox2009}, 38: \citealt{Fynbo2011}, 39: \citealt{Fynbo2013}, 40: \citealt{Ge2001}, 41: \citealt{Guimaraes2012}, 42: \citealt{Heinmueller2006}, 43: \citealt{Henry2007}, 44: \citealt{Herbert-Fort2006}, 45: \citealt{Kanekar2014}, 46: \citealt{Krogager2013a}, 47: \citealt{Kulkarni2005}, 48: \citealt{Kulkarni2012}, 49: \citealt{Ledoux2002}, 50: \citealt{Ledoux2002b}, 51: \citealt{Ledoux2003}, 52: \citealt{Ledoux2006}, 53: \citealt{Lopez1999}, 54: \citealt{Lopez2002}, 55: \citealt{Lopez2003}, 56: \citealt{Lopez2005}, 57: \citealt{Lu1996}, 58: \citealt{Meiring2006}, 59: \citealt{Meiring2007}, 60: \citealt{Meiring2009}, 61: \citealt{Meyer1995}, 62: \citealt{Molaro2000}, 63: \citealt{Nestor2008}, 64: \citealt{Noterdaeme2007}, 65: \citealt{Noterdaeme2007a}, 66: \citealt{Noterdaeme2008}, 67: \citealt{Noterdaeme2008a}, 68: \citealt{Noterdaeme2012a}, 69: \citealt{Noterdaeme2012b}, 70: \citealt{Penprase2010}, 71: \citealt{Peroux2002}, 72: \citealt{Peroux2006}, 73: \citealt{Peroux2007}, 74: \citealt{Peroux2008}, 75: \citealt{Petitjean2000}, 76: \citealt{Petitjean2002}, 77: \citealt{Petitjean2008}, 78: \citealt{Pettini1994b}, 79: \citealt{Pettini1997a}, 80: \citealt{Pettini1997b}, 81: \citealt{Pettini1999}, 82: \citealt{Pettini2000}, 83: \citealt{Pettini2002}, 84: \citealt{Pettini2008}, 85: \citealt{Pettini2008a}, 86: \citealt{Pettini2012}, 87: \citealt{Prochaska1996}, 88: \citealt{Prochaska1997a}, 89: \citealt{Prochaska1999}, 90: \citealt{Prochaska2001}, 91: \citealt{Prochaska2001b}, 92: \citealt{Prochaska2002a}, 93: \citealt{Prochaska2003}, 94: \citealt{Prochaska2003a}, 95: \citealt{Prochaska2007}, 96: \citealt{Prochter2010}, 97: \citealt{Quast2008}, 98: \citealt{Rafelski2012}, 99: \citealt{Rafelski2014}, 100: \citealt{Rao2000}, 101: \citealt{Rao2005}, 102: \citealt{Rao2006}, 103: \citealt{Richter2005}, 104: \citealt{Rix2007}, 105: \citealt{Som15}, 106: \citealt{Songaila2002}, 107: \citealt{Srianand1998}, 108: \citealt{Srianand2000}, 109: \citealt{Srianand2001}, 110: \citealt{Srianand2005}, 111: \citealt{Srianand2007}, 112: \citealt{Srianand2012}, 113: \citealt{Steidel1997}, 114: This work, 115: \citealt{Tripp2005}, 116: \citealt{Turnshek2004}, 117: \citealt{Vladilo2011}, 118: \citealt{Zafar2011}, 119: \citealt{Zafar2014}, 120: \citealt{Zafar2014a}, 121: Zafar et al. (in prep), 122: \citealt{Zych2009}, 123: \citealt{delaVarga2000}

\twocolumn
\section[]{Individual objects}
\label{ann:individual}

This Appendix summarizes a description of the individual systems as well as figures and tables providing the Voigt profile parameters for the low, intermediate and high-ionization species when available.
For the following figures, we use $\rm z_{abs}$ from the literature as the zero velocity component.

\subsection{QSOJ0008-2900 z$_{\rm em}=2.645$, z$_{\rm abs}=2.254$, $\rm \log N(HI)=20.22\pm0.10$}

The EUADP spectrum for this absorber covers multiple low-ionization transitions including FeII $\lambda\lambda\lambda\lambda$ 2374, 2382, 2586, 2600, SiII $\lambda$ 1526, and the saturated MgII doublet $\lambda\lambda$ 2796, 2803. The velocity profiles are well fitted with 5 components. The fit to the Fe lines is found to be consistent with the non-detection of FeII $\lambda$ 2260. The resulting total column density of Fe is \abundance{FeII}{13.78}{0.01}. By comparing the detected SiII $\lambda$ 1526 line with the weak SiII $\lambda$ 1808 transition, we deduce that the former is slightly contaminated. An estimation on the column density based on SiII $\lambda$ 1808 gives \abundance{SiII}{14.40}{0.1}. This total column density is confirmed with the apparent optical depth method applied on SiII $\lambda$ 1808 from $\rm v=-72$ km/s and $\rm v=+10$ km/s.
Because it is saturated, the MgII doublet is fitted fixing the number of components and parameters to the low-ionization lines. The result provides a lower limit to the Mg column density of \lowerlimit{MgII}{15.1}. The non-detection of CrII $\lambda$ 2062, ZnII $\lambda$ 2026 and MnII $\lambda$ 2576 leads to the determination of the following upper limits: \upperlimit{CrII}{12.37}, \upperlimit{ZnII}{11.68} and \upperlimit{MnII}{12.02}.

In addition to these low-ionization ions, the quasar spectrum covers high-ionization species including the SiIV doublet $\lambda\lambda$ 1393 and 1402. However, the bluest SiIV line lies on the Ly$\alpha$ emission line of the quasar thus complicating the quasar continuum placement. Therefore the reddest component of the SiIV doublet is used to model the Voigt profile. In addition, the intermediate-ionization lines of AlIII $\lambda\lambda$ 1854 and 1862 present a similar velocity profile to the SiIV doublet. Therefore, the fit is performed using these transitions simultaneously. A satisfactory 3-component fit leads to the following column densities: \abundance{SiIV}{13.72}{0.03} and \abundance{AlIII}{12.39}{0.04}.  

\begin{figure}
\begin{center}
\hspace*{-.8in}
\includegraphics[width=1.2\textwidth]{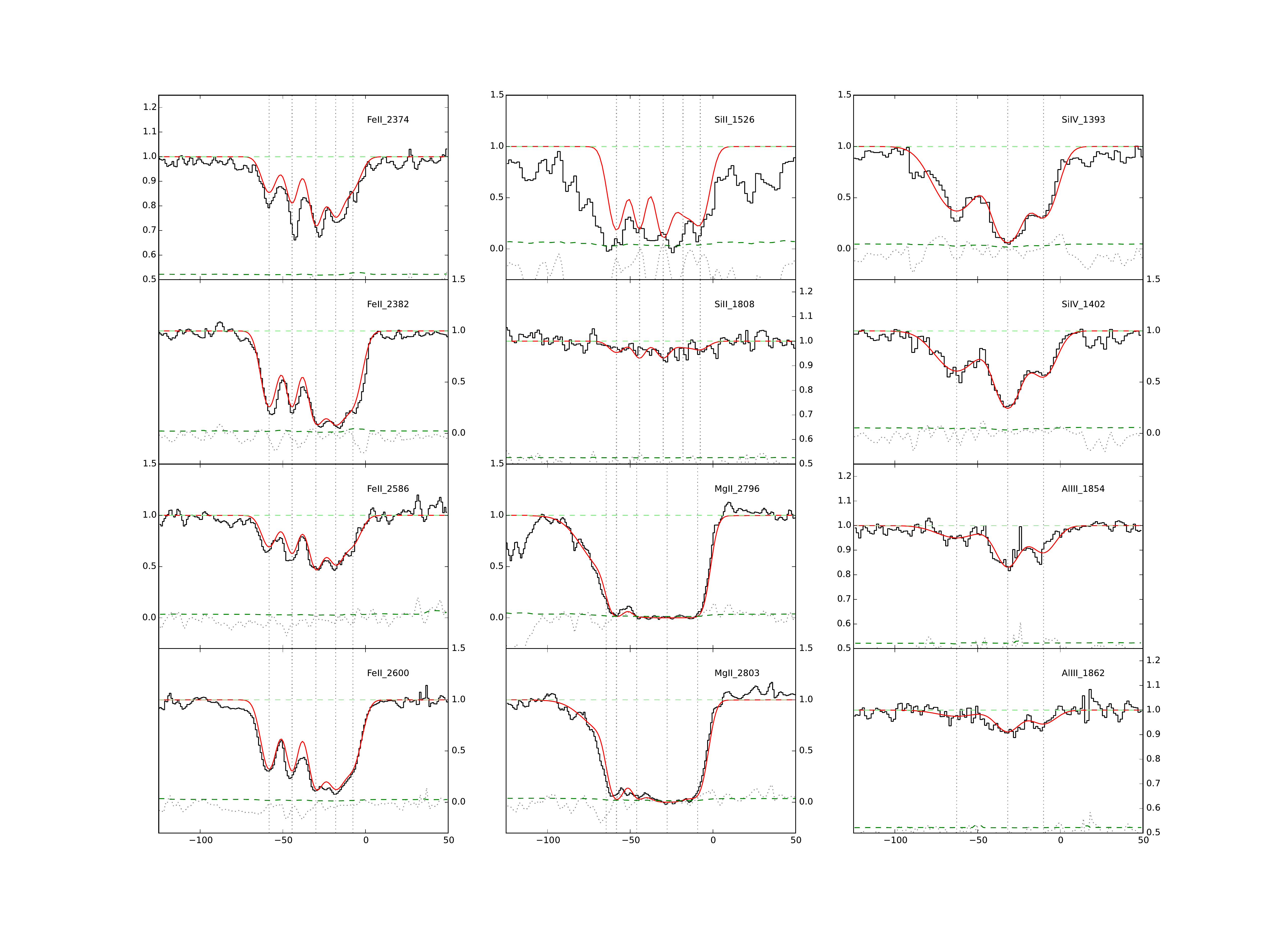}
\caption[QSOJ0008-2900]{Fit to the low-ionization transitions of the z$_{\rm abs} = 2.254$, log N(H\,I) = $20.22\pm0.10$ absorber towards QSOJ0008-2900 (see
Table \ref{ind:0008-2900}). In this and the following figures, the Voigt profile fits are overlaid in red above the observed quasar spectrum (black) and the green horizontal line indicates the normalised flux level to one. The zero velocity corresponds to the absorption redshift listed in Table \ref{table:subsample} and the vertical dotted lines correspond to the redshift of the fitted components. We warn the reader that the y-axis varies from one panel to another in order to optimise for each transitions so that weaker lines can be readily seen. }
\label{img:0008-2900}
\end{center}
\end{figure}

The parameter fits are summarised in Table \ref{ind:0008-2900} and Voigt profile fits are shown in Fig. \ref{img:0008-2900}.

\begin{table}
\begin{center}
\caption{Voigt profile fit parameters to the low- and high-ionisation species for the z$_{\rm abs}$=2.645 log N(H\,I)=$20.22\pm0.10$ absorber towards QSO J0008-2900. In this table and in the following ones, the values with no uncertainties have been manually fixed to improve the fitting process.}
\label{ind:0008-2900}
\begin{tabular}{ccccc}
\hline
\hline
Comp. & $z_{abs}$ & b & Ion & log $N$ \\
 & & km $s^{-1}$ & & cm$^{-2}$ \\
 \hline
 1 & $2.25337$ & $4.5\pm0.2$ & FeII & $12.95\pm0.01$\\
   &   &   & SiII & $13.70$\\
 2 & $2.25352$ & $2.8\pm0.2$ & FeII & $13.00\pm0.01$\\
   &   &   & SiII & $13.80$\\
 3 & $2.25367$ & $4.0\pm0.2$ & FeII & $13.24\pm0.01$\\
   &   &   & SiII & $13.85$\\
 4 & $2.25380$ & $5.0\pm0.3$ & FeII & $13.20\pm0.02$\\
   &   &   & SiII & $13.40$\\
 5 & $2.25392$ & $5.3\pm0.3$ & FeII & $12.91\pm0.02$\\
   &   &   & SiII & $13.60$\\
 \hline
 1 & $2.25332$ & $16.1\pm2.7$ & SiIV & $13.19\pm0.05$\\
   &   &   & AlIII & $11.74\pm0.10$\\
 2 & $2.25366$ & $9.0\pm1.1$ & SiIV & $13.42\pm0.05$\\
   &   &   & AlIII & $12.07\pm0.06$\\
 3 & $2.25389$ & $9.2\pm1.9$ & SiIV & $13.05\pm0.08$\\
   &   &   & AlIII & $11.88\pm0.08$\\
   \hline
\end{tabular}
\end{center}
\end{table}

\onecolumn

\twocolumn

\subsection{QSOJ0008-2901z$_{\rm em}=2.607$, z$_{\rm abs}=2.491$, $\rm \log N(HI)=19.94\pm0.11$}
In this case, the quasar spectrum is of modest SNR. The wavelength coverage includes low-ionization lines of FeII $\lambda\lambda\lambda\lambda$ 2260, 2344, 2382, 2586, SII $\lambda$ 1259 and OI $\lambda$ 1039. A two-component fit is used given the asymmetrical shape of the FeII and SII lines. 
The resulting column densities are $\log$ N(FeII)$=13.65\pm0.02$, $\log$ N(OI)$=15.31\pm0.24$ and $\log$ N(SII)$=13.68\pm0.18$.
Finally, we used the non-detection of the following transitions to derive upper limits on the column densities: CrII $\lambda$ 2062 $\log$ N(CrII)$<12.9$, NiII $\lambda$ 1741 $\log$ N(NiII)$<13.29$, ZnII $\lambda$ 2026 $\log$ N(ZnII)$<12.12$ and AlIII $\lambda$ 1854 $\log$ N(AlII)$<12.2$.

The EUADP quasar spectrum does not cover the CIV doublet for this absorber. In addition, the high-ionization SiIV doublet $\lambda\lambda$ 1393 and 1402 is covered but no satisfactory fit could be determined given the low SNR of the quasar spectrum in this region.

\begin{figure}
\begin{center}
\includegraphics[width=.5\textwidth, height = 12cm]{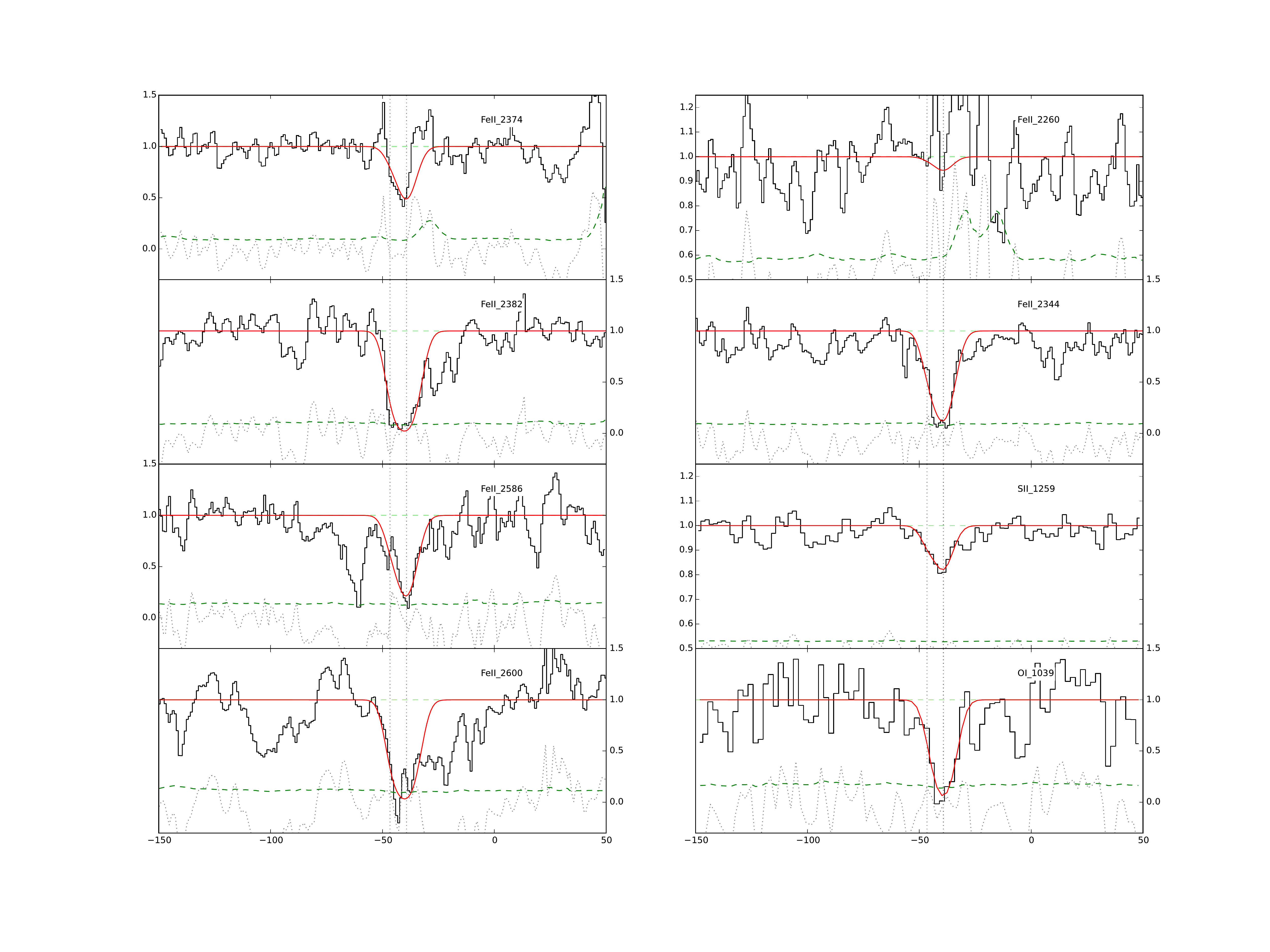}
\caption[QSOJ0008-2901]{QSOJ0008-2901}
\label{img:0008-2901}
\end{center}
\end{figure}

The resulting parameter fits are summarised in Table \ref{tab:0008-2901} and Voigt profile fits are shown in Fig. \ref{img:0008-2901}.

\begin{table}
\begin{center}
\caption{Voigt profile fit parameters to the low-ionization species for the z$_{\rm abs}$=2.491 log N(H\,I)=$19.94\pm0.11$ absorber towards QSO J0008-2901.}
\label{tab:0008-2901}
\begin{tabular}{ccccc}
\hline
\hline
Comp. & $z_{abs}$ & b & Ion & log $N$ \\
 & & km $s^{-1}$ & & cm$^{-2}$ \\
 \hline
 1 & $2.49046$ & $1.6\pm0.2$ & FeII & $12.60\pm0.07$\\
   &   &   & SII & $12.98\pm0.91$\\
   &   &   & OI & $13.32\pm1.54$\\
 2 & $2.49054$ & $4.1\pm0.1$ & FeII & $13.61\pm0.02$\\
   &   &   & SII & $13.58\pm0.02$\\
   &   &   & OI & $15.31\pm0.24$\\
   
   \hline
\end{tabular}

\end{center}
\end{table}

\subsection{QSO J0018-0913 z$_{\rm em}=0.75593$, z$_{\rm abs}=0.584$, $\rm \log N(HI)=20.11\pm0.10$}
We detect in this low-redshift absorber's spectrum the following low-ionization ions: FeII $\lambda\lambda\lambda$ 2344 2374 and 2382. The profile is fitted using 11 components, spread over a velocity range of about $\rm 200 km/s$, resulting in an abundance for FeII of $\log \rm N(FeII)=13.87\pm0.03$. The main component is located on the blue edge of the profile and is accounting for about 40\% of the total abundance.
Also, we derive upper limits from non detection of ZnII $\lambda$ 2026, $\rm \log N(ZnII)<12.41$, CrII $\lambda$ 2056, $\rm \log N(CrII)<12.97$, TiII $\lambda$ 3384, $\rm \log N(TiII)<11.57$, MgI $\lambda$ 2026, $\rm \log N(MgI)<13.04$, NaI $\lambda$ 3303.3 and  $\rm \log N(NaI)<13.15$.

The metallicity for this low-redshift sub-DLA is surprisingly low (even considering the $\alpha$-enhancement correction of $\sim.4$ dex), $\rm [Fe/H] = -1.70\pm0.13$. It may be an effect of dust depletion, but we are unable to conclude on this particular issue due to the low number of detected ions. The metallicity derived here is therefore to be considered as a lower limit.

There is no coverage of the high-ionization ions due to the low redshift of the absorber.

\begin{figure}
\begin{center}
\includegraphics[width=.5\textwidth, height=8cm]{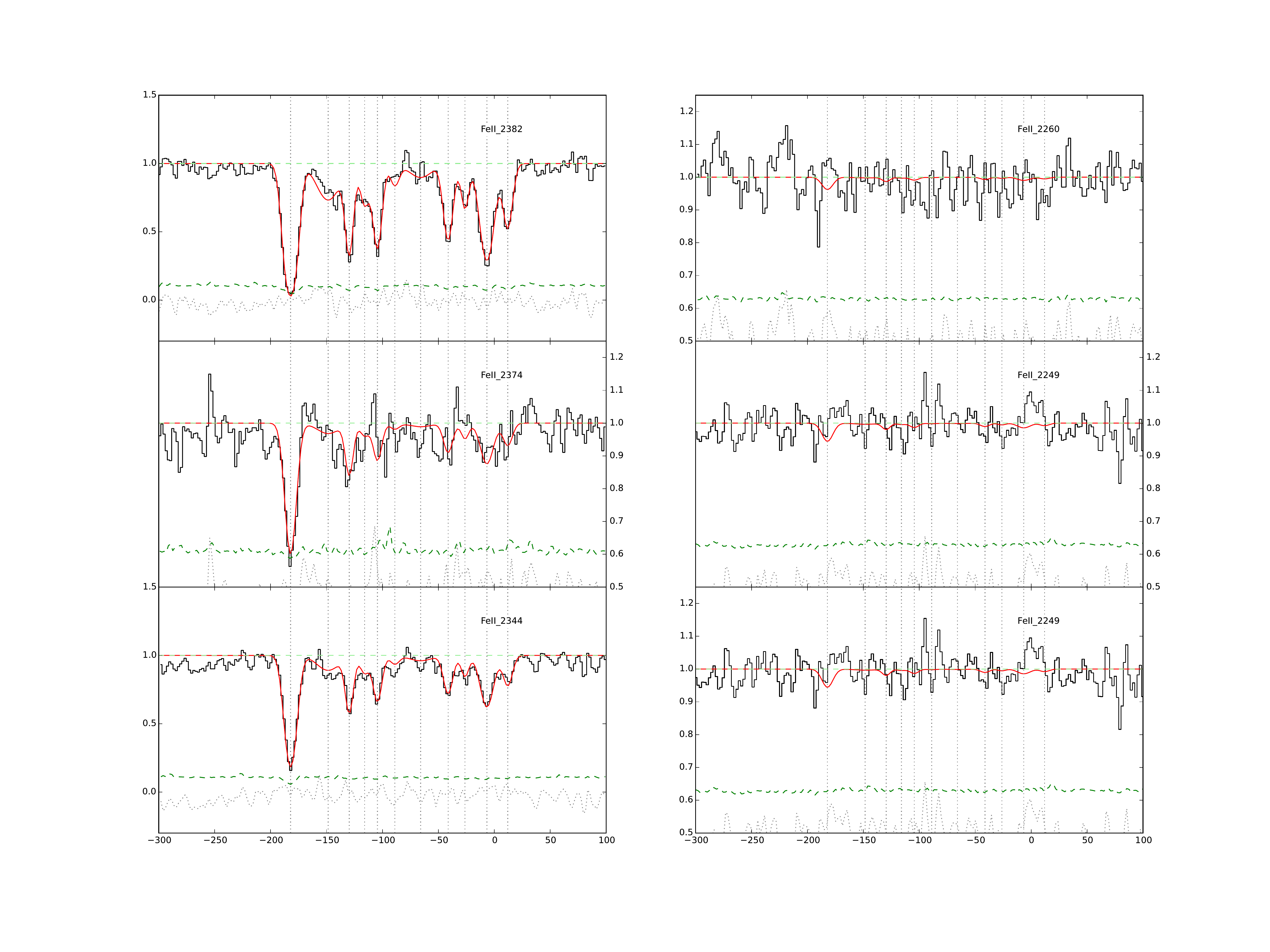}
\caption[QSOJ0018-0913]{QSOJ0018-0913}
\label{img:0018-0913}
\end{center}
\end{figure}

The parameter fits of the individual components are listed in Table \ref{tab:0018-0913} and the corresponding Voigt profile fits are shown in Fig. \ref{img:0018-0913}.

\begin{table}
\begin{center}
\caption{Voigt profile fit parameters to the low-ionization species for the z$_{\rm abs}$=0.584 log N(H\,I)=$20.11\pm0.1$ absorber towards QSO J0018-0913.}
\label{tab:0018-0913}
\begin{tabular}{ccccc}
\hline
\hline
Comp. & $z_{abs}$ & b & Ion & log $N$ \\
 & & km $s^{-1}$ & & cm$^{-2}$ \\
 \hline
1 & $0.58304$ & $5.4\pm0.3$ & FeII & $13.51\pm0.04$\\
2 & $0.58322$ & $13.0\pm5.4$ & FeII & $12.60\pm0.14$\\
3 & $0.58332$ & $2.5\pm0.7$ & FeII & $12.90\pm0.09$\\
4 & $0.58339$ & $3.1\pm5.3$ & FeII & $12.28\pm0.13$\\
5 & $0.58346$ & $3.3\pm1.2$ & FeII & $12.78\pm0.06$\\
6 & $0.58353$ & $3.5\pm3.5$ & FeII & $11.96\pm0.17$\\
7 & $0.58365$ & $13.4\pm14.9$ & FeII & $12.15\pm0.43$\\
8 & $0.58378$ & $3.8\pm2.1$ & FeII & $12.69\pm0.10$\\
9 & $0.58386$ & $2.0\pm3.1$ & FeII & $12.33\pm0.19$\\
10 & $0.58397$ & $6.7\pm0.8$ & FeII & $12.98\pm0.03$\\
11 & $0.58406$ & $3.8\pm1.2$ & FeII & $12.57\pm0.05$\\
\hline
\end{tabular}
\end{center}
\end{table}

\subsection{QSOJ0041-4936 z$_{\rm em}=3.24$, z$_{\rm abs}=2.248$, $\rm \log N(HI)=20.46\pm0.13$}

This spectrum covers many low-ionization transitions associated with the absorber:  SII $\lambda$ 1259, ZnII $\lambda$ 2026, FeII $\lambda$ 1608, SiII $\lambda\lambda$ 1808, 1526 and AlII $\lambda$ 1670. The velocity profile can be conveniently separated into a group of red and blue components. The overall profile is well fitted with 4 components. Interestingly, the blue group which is weaker than its red counterpart is only detected in SII $\lambda$ 1259, SiII $\lambda$ 1526 and AlII $\lambda$ 1670. Conversely, the red group of components is saturated in the case of the last two transitions. Therefore, these two groups are fitted separately: the blue-component group is fitted with two components using the lines of SiII $\lambda$ 1526 and AlII $\lambda$ 1670. The SII $\lambda$ 1259 line is too blended to provide useful information and we derive an upper limit of $\rm \log N(SII)<14.82$. The red-component group includes ZnII $\lambda$ 2026, FeII $\lambda$ 1608 and SiII $\lambda$ 1808. It is also fitted with two components given the asymmetrical profile of FeII $\lambda$ 1608. A fifth component redward of the profile is considered for SII $\lambda$ 1526 and AlII $\lambda$ 1670. The resulting column densities are: $\log$ N(FeII)=$14.43\pm0.04$, $\log$ N(ZnII)=$11.70\pm0.10$, $\log$ N(SiII)=$14.78\pm0.03$, $\log$ N(NiII)=$13.07\pm0.07$ and $\log$ N(CrII)=$13.12\pm0.45$. We detect four NI absorption lines NI $\lambda\lambda\lambda\lambda\lambda$ 1199.5, 1134.1, 1134.4 and 1134.9. The profile is fitted with the low-ionization profile in this case. In spite of the modest SNR, a satisfactory fit is found for NI $\lambda\lambda\lambda$ 1134.1, 1134.9 and 1199.5, while the remaining line (NI $\lambda$ 1134.4) appears to be blended. The resulting column density is $\log$ N(NI)=$14.03\pm0.03$. Finally, the AlII $\lambda$ 1670 line is saturated thus providing a lower limit on the column density: $\rm \log N(AlII)>14.06$. 

Regarding the high-ionization ions, the CIV doublet $\lambda\lambda$ 1548 and 1550 is covered by the EUADP spectrum. In addition, the intermediate-ionization AlIII transitions $\lambda\lambda$ 1854 and 1862 are present and match well the high-ionization profile. In the case of the CIV doublet, the CIV $\lambda$ 1550 is blended with a broad line in the blue part ($\rm v<-120km/s$), whereas CIV $\lambda$ 1548 seems blended in the red part ($\rm v>-120km/s$). In addition, the AlIII transition shows absorption features in the red part of the profile only. A 6-component profile is used to fit the red part of the profile (using transitions from CIV $\lambda$ 1550, AlIII $\lambda\lambda$ 1862 and 1854) and one-component is used to fit the blue part of the profile (using CIV $\lambda$ 1548 at $\rm v\sim-240km/s$). This results in an estimated column density of AlIII $\rm \log N(AlIII)=12.90\pm0.01$ and a lower limit (blending) for CIV $\rm \log N(CIV)>14.56$. 

\begin{figure}
\begin{center}
\hspace*{-.8in}
\includegraphics[width=1.2\textwidth]{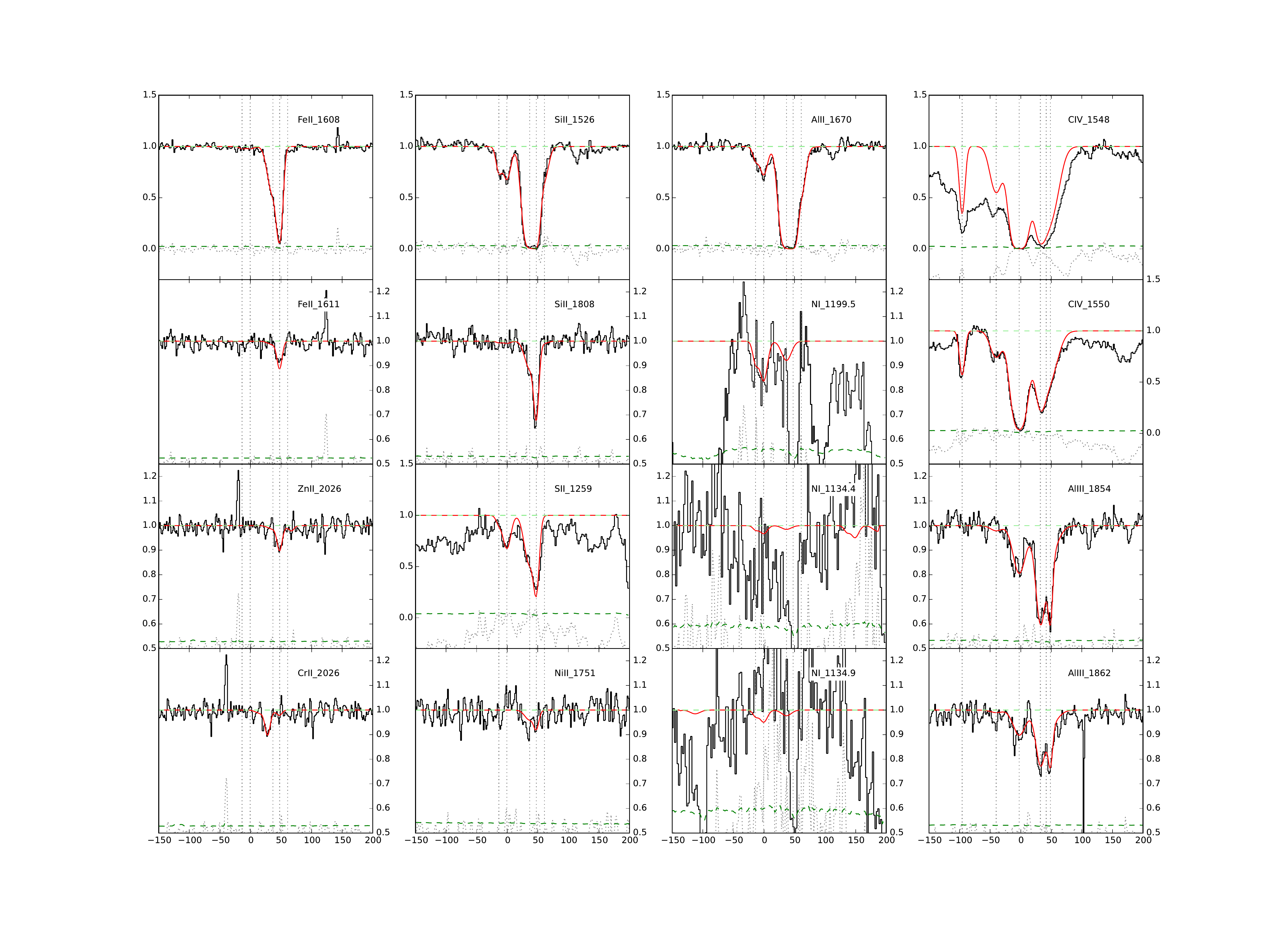}
\caption[QSOJ0041-4936]{QSOJ0041-4936}
\label{img:0041-4936}
\end{center}
\end{figure}

The parameter fits are summarised in Table \ref{tab:0041-4936} and Voigt profile fits are shown in Fig. \ref{img:0041-4936}.

\begin{table}
\begin{center}
\caption{Voigt profile fit parameters to the low-, intermediate- and high-ionization species for the z$_{\rm abs}$=2.248 log N(H\,I)=$20.46\pm0.13$ absorber towards QSO J0041-4936.}
\label{tab:0041-4936}
\begin{tabular}{ccccc}
\hline
\hline
Comp. & $z_{abs}$ & b & Ion & log $N$ \\
 & & km $s^{-1}$ & & cm$^{-2}$ \\
 \hline
1 & $2.24785$ & $4.6\pm0.8$ & FeII & $12.01\pm0.35$\\
   &   &   & ZnII & $-$\\
   &   &   & SiII & $12.85\pm0.04$\\
   &   &   & NiII & $-$\\
   &   &   & CrII & $-$\\
   &   &   & AlII & $11.28\pm0.05$\\
   &   &   & SII & $<13.24$\\
   &   &   & NI & $12.38\pm0.17$\\
2 & $2.24799$ & $7.3\pm0.4$ & FeII & $12.01\pm0.35$\\
   &   &   & ZnII & $-$\\
   &   &   & SiII & $13.12\pm0.02$\\
   &   &   & NiII & $-$\\
   &   &   & CrII & $-$\\
   &   &   & AlII & $11.78\pm0.02$\\
   &   &   & SII & $<14.04$\\
   &   &   & NI & $12.80\pm0.07$\\
3 & $2.24840$ & $10.2\pm0.9$ & FeII & $13.71\pm0.05$\\
   &   &   & ZnII & $11.03\pm0.36$\\
   &   &   & SiII & $14.35\pm0.06$\\
   &   &   & NiII & $12.80\pm0.10$\\
   &   &   & CrII & $12.73\pm0.91$\\
   &   &   & AlII & $>13.11$\\
   &   &   & SII & $<14.38$\\
   &   &   & NI & $12.58\pm0.15$\\
4& $2.24852$ & $3.8\pm0.2$ & FeII & $14.33\pm0.05$\\
   &   &   & ZnII & $11.59\pm0.09$\\
   &   &   & SiII & $14.54\pm0.04$\\
   &   &   & NiII & $12.74\pm0.09$\\
   &   &   & CrII & $12.89\pm0.44$\\
   &   &   & AlII & $>15.13$\\
   &   &   & SII & $<14.48$\\
   &   &   & NI & $14.25\pm0.04$\\
5& $2.24866$ & $8.0\pm0.3$ & FeII & $-$\\
   &   &   & ZnII & $-$\\
   &   &   & SiII & $13.13\pm0.02$\\
   &   &   & NiII & $-$\\
   &   &   & CrII & $-$\\
   &   &   & AlII & $12.09\pm0.02$\\
   &   &   & SII & $-$\\
   &   &   & NI & $-$\\

 \hline
 1 & $2.24547$ & $3.5\pm0.5$ & CIV & $12.84\pm0.02$\\
   &   &   & AlIII &$ -$\\
 2 & $2.24696$ & $4.1\pm0.4$ & CIV & $13.21\pm0.01$\\
   &   &   & AlIII & $-$\\
 3 & $2.24757$ & $12.1\pm0.7$ & CIV & $13.24\pm0.01$\\
   &   &   & AlIII & $11.25\pm0.06$\\
 4 & $2.24797$ & $13.1\pm0.1$ & CIV & $14.32\pm0.01$\\
   &   &   & AlIII & $12.28\pm0.01$\\
 5 & $2.24835$ & $8.7\pm0.3$ & CIV & $13.58\pm0.10$\\
   &   &   & AlIII & $12.37\pm0.01$\\
 6 & $2.24845$ & $19.6\pm0.4$ & CIV & $13.87\pm0.01$\\
   &   &   & AlIII & $12.36\pm0.01$\\
 7 & $2.24853$ & $2.6\pm0.7$ & CIV & $-$\\
   &   &   & AlIII & $12.08\pm0.01$\\
   
   \hline
\end{tabular}

\end{center}
\end{table}

\onecolumn

\twocolumn

\subsection{QSO B0128-2150 z$_{\rm em}=1.9$, z$_{\rm abs}=1.857$, $\rm \log N(HI)=20.21\pm0.09$}

For this low-redshift absorber, the EUADP spectrum covers a number of the low-ionization ions including FeII $\lambda\lambda\lambda$ 2374 2260 2249, NiII $\lambda\lambda\lambda$ 1751 1741 1709, SII $\lambda\lambda\lambda$ 1259 1253 1250 and SiII $\lambda$ 1808. The detected intermediate-ionization transitions, AlIII $\lambda\lambda$ 1854 and 1862 show the same velocity profile as the low-ionization ions. A 5-component Voigt profile is used to fit FeII $\lambda\lambda$ 2374 2249, SiII $\lambda$ 1808 SII $\lambda$1259, AlIII $\lambda$ 1854 and NiII $\lambda$ 1709. The full absorption profile extends to a velocity range of about 100 km/s. The resulting total column densities are $\log$ N(FeII)=$14.44\pm0.01$, $\log$ N(SiII)=$14.82\pm0.02$, $\log$ N(SII)=$14.33\pm0.03$, $\log$ N(NiII)=$13.26\pm0.05$ and $\log$ N(AlIII)=$12.78\pm0.01$. In addition, CII $\lambda$ 1334 is detected but, as often in DLAs, heavily saturated. Finally, the non-detection of both ZnII $\lambda$ 2062 and MgI $\lambda$ 1827 provides robust column density upper limits: $\log N(ZnII)<12.26$, and $\rm \log N(MgI)<13.21$. 

We note that no high-ionization species are covered by this spectrum.

\begin{figure}
\begin{center}
\hspace*{-.8in}
\includegraphics[width=1.2\textwidth]{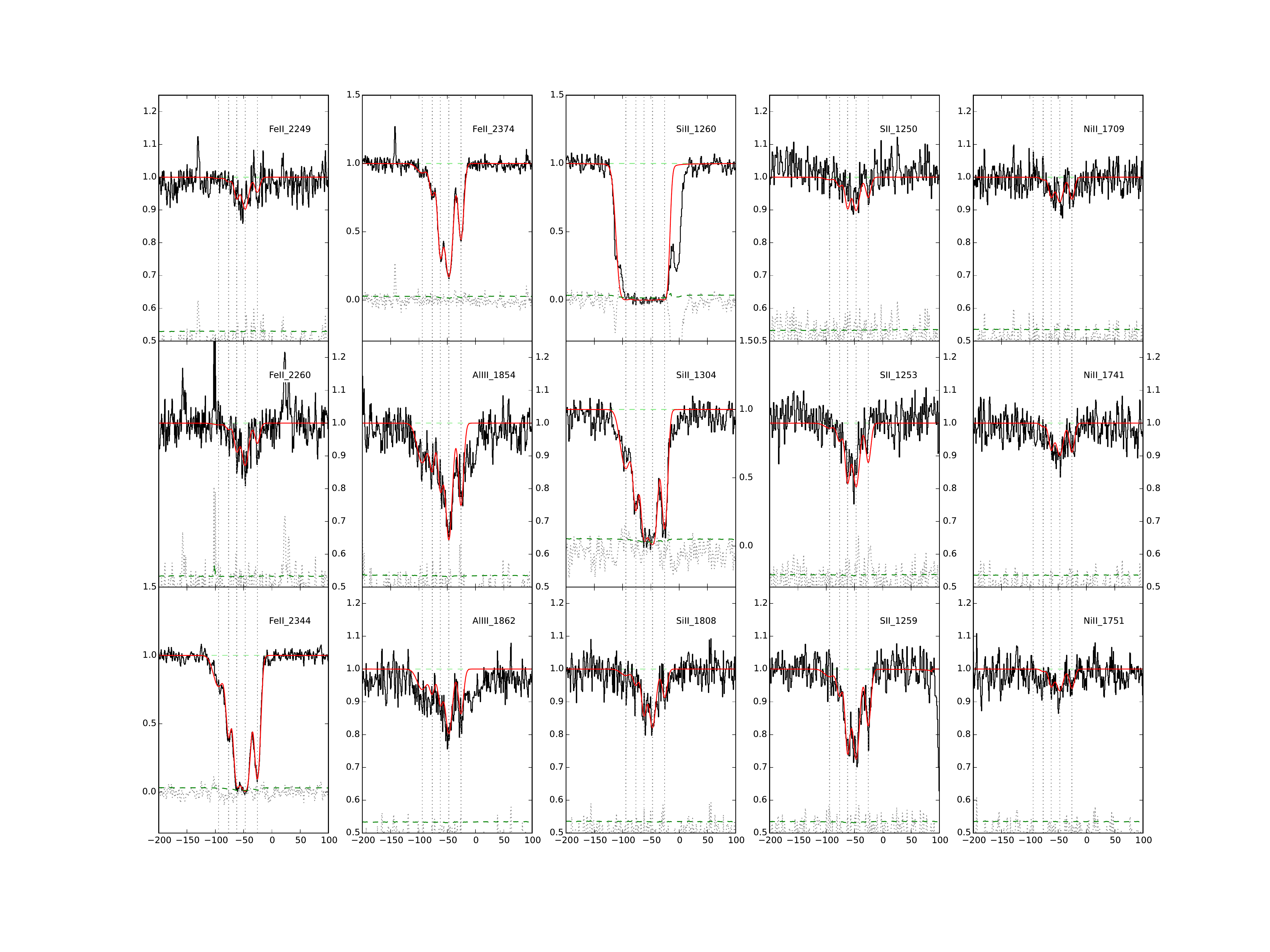}
\caption[QSOB0128-2150]{QSOB0128-2150}
\label{img:0128-2150}
\end{center}
\end{figure}

The parameter fits are summarised in Table \ref{tab:0128-2150} and Voigt profile fits are shown in Fig. \ref{img:0128-2150}.

\begin{table}
\begin{center}
\caption{Voigt profile fit parameters to the low-ionization species for the z$_{\rm abs}$=1.857 log N(H\,I)=$20.21\pm0.09$ absorber towards QSO B0128-2150.}
\label{tab:0128-2150}
\begin{tabular}{ccccc}
\hline
\hline
Comp. & $z_{abs}$ & b & Ion & log $N$ \\
 & & km $s^{-1}$ & & cm$^{-2}$ \\
 \hline
1 & $1.85610$ & $11.8\pm2.1$ & FeII & $12.91\pm0.06$\\
   &   &   & SiII & $13.93\pm0.10$\\
   &   &   & SII & $12.98\pm0.29$\\
   &   &   & AlIII & $12.02\pm0.05$\\
   &   &   & NiII & $-$\\
2 & $1.85627$ & $4.5\pm0.7$ & FeII & $13.16\pm0.04$\\
   &   &   & SiII & $13.71\pm0.12$\\
   &   &   & SII & $13.23\pm0.12$\\
   &   &   & AlIII & $11.76\pm0.07$\\
   &   &   & NiII & $11.80\pm0.67$\\
3 & $1.85641$ & $4.3\pm0.3$ & FeII & $13.82\pm0.01$\\
   &   &   & SiII & $14.19\pm0.04$\\
   &   &   & SII & $13.77\pm0.04$\\
   &   &   & AlIII & $11.94\pm0.03$\\
   &   &   & NiII & $12.66\pm0.1$\\
4 & $1.85655$ & $7.0\pm0.2$ & FeII & $14.14\pm0.01$\\
   &   &   & SiII & $14.44\pm0.01$\\
   &   &   & SII & $13.94\pm0.03$\\
   &   &   & AlIII & $12.37\pm0.01$\\
   &   &   & NiII & $12.89\pm0.06$\\
5 & $1.85676$ & $4.6\pm0.2$ & FeII & $13.69\pm0.01$\\
   &   &   & SiII & $13.99\pm0.05$\\
   &   &   & SII & $13.60\pm0.05$\\
   &   &   & AlIII & $12.08\pm0.02$\\
   &   &   & NiII & $12.73\pm0.08$\\
   \hline
\end{tabular}

\end{center}
\end{table}

\onecolumn

\twocolumn

\subsection{QSO J0132-0823 z$_{\rm em}=1.121$, z$_{\rm abs}=0.6467$, $\rm \log N(HI)=20.60\pm0.12$}
A few ions are covered and detected in this low-redshift absorber's spectrum: FeII $\lambda\lambda$ 2249 2260, MgI $\lambda$ 2582 and TiII $\lambda\lambda$ 3242 and 3384. The absorption is weak, we use the MgI profile to derive the two main components. The saturated FeII $\lambda$ 2344 unveils two weak components on either side of the profile. This results in abundances of FeII $\log \rm N(FeII)=14.96\pm0.07$, of TiII $\log \rm N(TiII)=12.39\pm0.11$ and of MgI $\log \rm N(MgI)=12.60\pm0.04$.

The overall SNR ($<10$) gives a reasonable upper limit for CrII, using CrII $\lambda$ 2056, of $\rm \log N(CrII)<13.17$. ZnII and CII are also covered, but the best upper limits we can derive are above $16.30$.

There is no coverage of the high-ionization ions due to the low redshift of the absorber.

\begin{figure}
\begin{center}
\includegraphics[width=.5\textwidth, height=8cm]{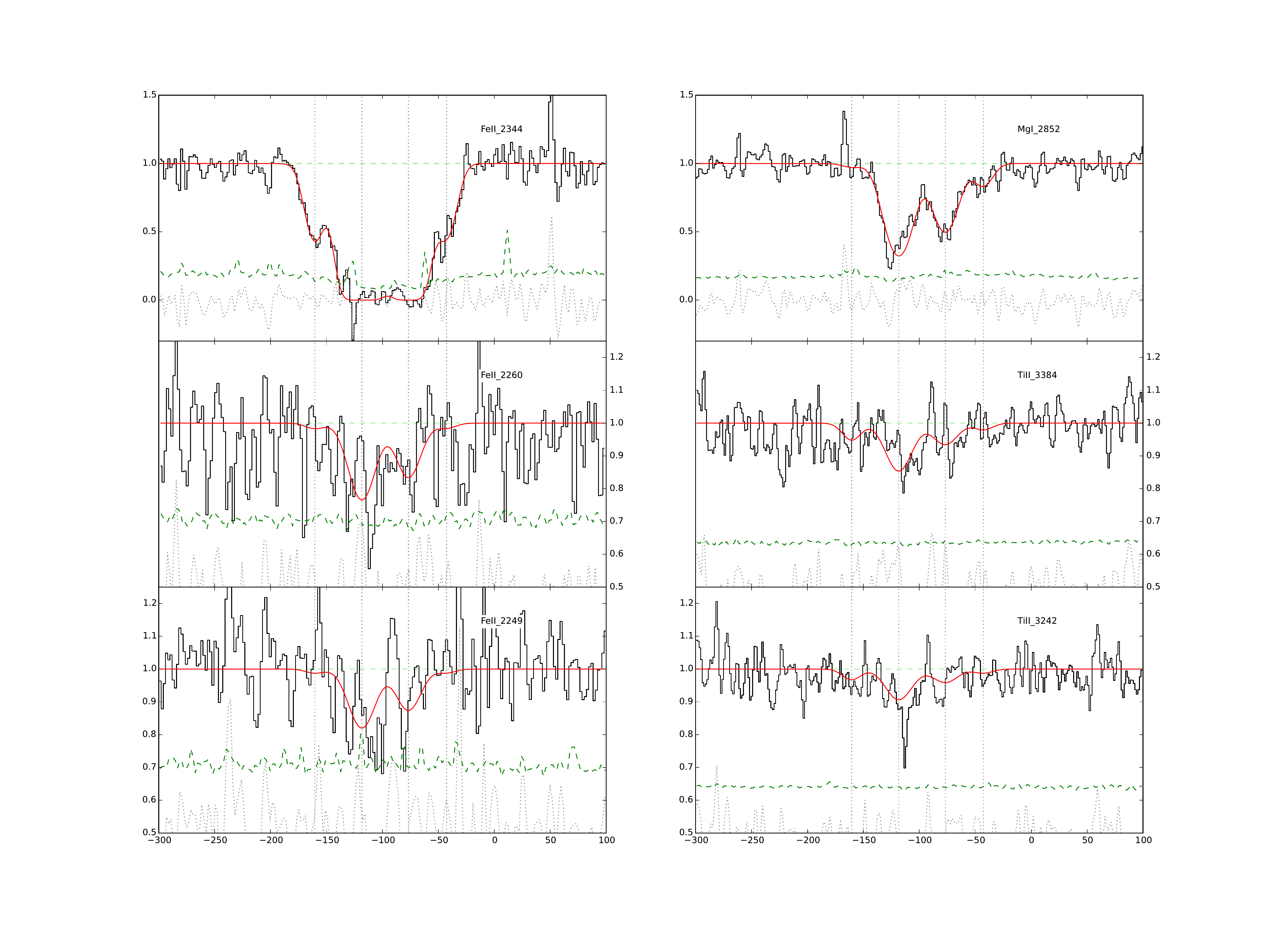}
\caption[QSOJ0132-0823]{QSOJ0132-0823}
\label{img:0132-0823}
\end{center}
\end{figure}

The parameter fits of the individual components are listed in Table \ref{tab:0132-0823} and the corresponding Voigt profile fits are shown in Fig. \ref{img:0132-0823}.

\begin{table}
\begin{center}
\caption{Voigt profile fit parameters to the low-ionization species for the z$_{\rm abs}$=0.647 log N(H\,I)=$20.60\pm0.12$ absorber towards QSO J0132-0823.}
\label{tab:0132-0823}
\begin{tabular}{ccccc}
\hline
\hline
Comp. & $z_{abs}$ & b & Ion & log $N$ \\
 & & km $s^{-1}$ & & cm$^{-2}$ \\
 \hline
1 & $0.64612$ & $10.7\pm2.3$ & FeII & $13.38\pm0.08$\\
  &   &  & TiII & $11.53\pm0.44$\\
 &  &  & MgI & $10.62\pm1.76$\\
2 & $0.64635$ & $15.6\pm1.2$ & FeII & $14.72\pm0.09$\\
  &   &  & TiII & $12.15\pm0.10$\\
 &  &  & MgI & $12.37\pm0.04$\\
3 & $0.64658$ & $14.4\pm2.1$ & FeII & $14.52\pm0.11$\\
  &   &  & TiII & $11.75\pm0.21$\\
 &  &  & MgI & $12.13\pm0.06$\\
4 & $0.64677$ & $10.8\pm3.0$ & FeII & $13.35\pm0.13$\\
 &  &  & TiII & $11.12\pm0.77$\\
 &  &  & MgI & $11.43\pm0.19$\\
\hline
\end{tabular}
\end{center}
\end{table}

\subsection{QSO B0307-195B z$_{\rm em}=2.122$, z$_{\rm abs}=1.788$, $\rm \log N(HI)=19.0\pm0.10$}
Many low-ionization ions are detected in this low-redshift absorber including MgI $\lambda$ 2852, FeII $\lambda\lambda\lambda\lambda\lambda\lambda$ 2374, 1608, 2586, 2344, 2382, 2600, SiII $\lambda\lambda\lambda$ 1808, 1304, 1526, MgII $\lambda\lambda$ 2803, 2796, AlII $\lambda$ 1670, AlIII $\lambda\lambda$ 1862, 1854, CIV $\lambda\lambda$ 1550, 1548 and SiIV $\lambda\lambda$ 1402, 1393. The absorption profile presents two distinct parts, one in the red (5 components), the other one in the blue (3 components), with a total velocity ranging about 300 km/s. The fit is performed using the transitions which are free from any saturation i.e. FeII $\lambda\lambda\lambda$ 2374 1608 2586, SiII $\lambda$ 1808 and MgI $\lambda$ 2852. The FeII $\lambda$ 2586 line in particular is not considered for the final fits given the medium quality of the EUADP spectrum around $\rm v=50km/s$. The resulting total column densities are: $\log$ N(FeII)=$14.48\pm0.004$, $\log$ N(SiII)=$15.0\pm0.01$ and $\log$ N(MgI)=$12.54\pm0.01$. In addition to these measures, the non-detection of CrII $\lambda$ 2056, MnII $\lambda$ 2594, NiII $\lambda$ 1751, and ZnII $\lambda$ 2062 is used to derive the following upper limits $\log  \rm N(CrII)<12.77$, $\log  \rm N(MnII)<12.13$, $\log \rm  N(NiII)<13.22$ and $\log \rm  N(ZnII)<12.18$. 

The high-ionization doublets of CIV $\lambda\lambda$ 1548 1550 and SiIV $\lambda\lambda$ 1393 and 1402 are detected in the spectrum albeit indicating strong saturation. A 7-component profile is used to fit these four transitions resulting in lower limits to the total column densities of $\log \rm N(SiIV)>14.55$ and $\log \rm  N(CIV)>15.13$. Interestingly, in this absorber, the velocity range of the high-ionization ion profile matches the one from the low-ionization ions extending to about $300$ km/s.

\begin{figure}
\begin{center}
\hspace*{-.8in}
\includegraphics[width=1.2\textwidth]{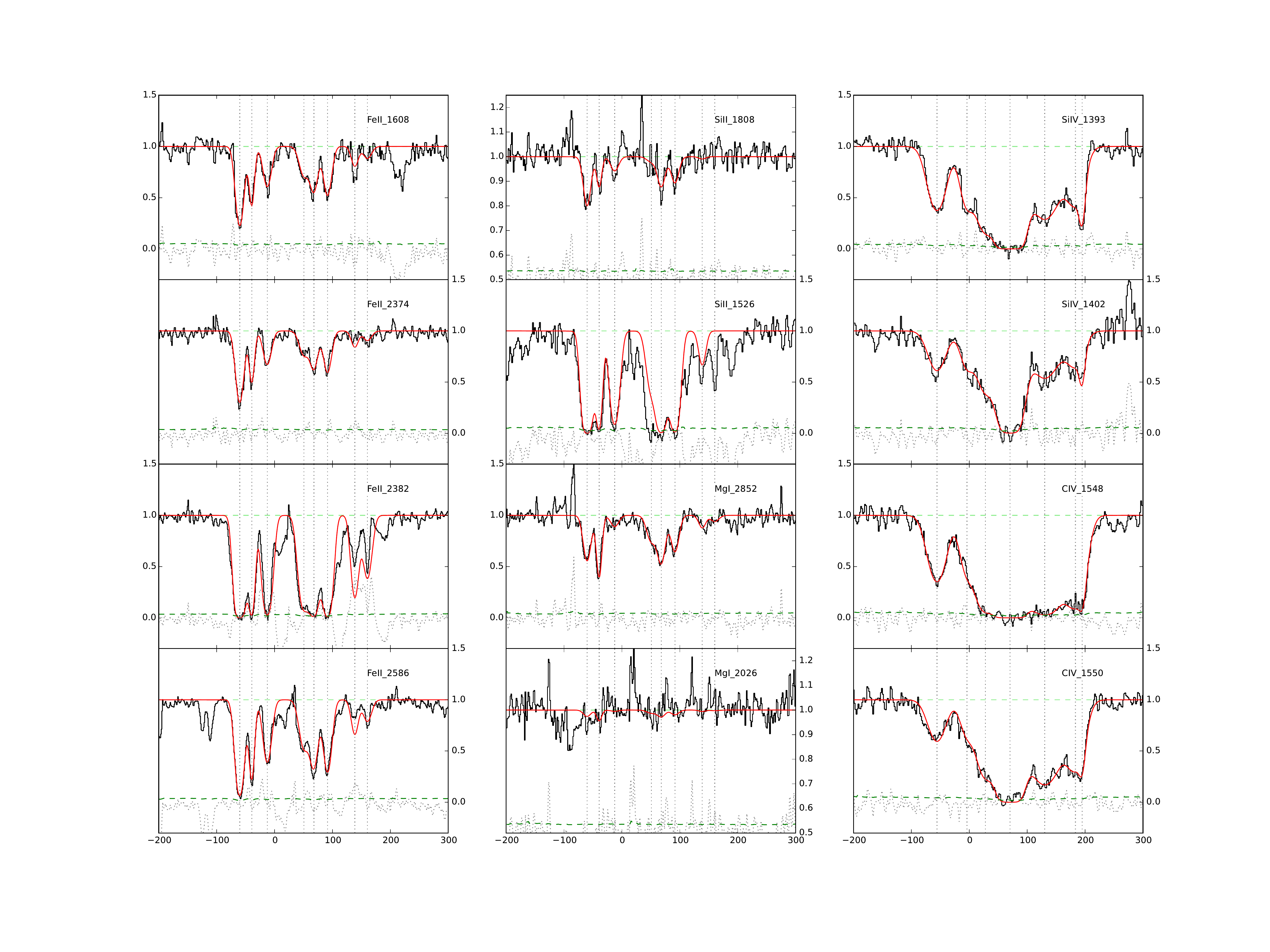}
\caption[QSOB0307-195B]{QSOB0307-195B}
\label{img:0307-195B}
\end{center}
\end{figure}

The resulting parameter fits for the low- and high-ionization profiles are listed in Table \ref{tab:0307-195B} and the corresponding Voigt profile fits are shown in Fig. \ref{img:0307-195B}.

\begin{table}
\begin{center}
\caption{Voigt profile fit parameters to the low- and high-ionization species for the z$_{\rm abs}$=1.788 log N(H\,I)=$19.00\pm0.10$ absorber towards QSO B0307-195B.}
\label{tab:0307-195B}
\begin{tabular}{ccccc}
\hline
\hline
Comp. & $z_{abs}$ & b & Ion & log $N$ \\
 & & km $s^{-1}$ & & cm$^{-2}$ \\
 \hline
1 & $1.78744$ & $7.8\pm0.1$ & FeII & $13.99\pm0.01$\\
   &   &   & SiII & $14.53\pm0.01$\\
   &   &   & MgI & $11.83\pm0.01$\\
2 & $1.78763$ & $4.3\pm0.1$ & FeII & $13.61\pm0.01$\\
   &   &   & SiII & $14.14\pm0.03$\\
   &   &   & MgI & $11.90\pm0.01$\\
3 & $1.78788$ & $8.6\pm0.4$ & FeII & $13.54\pm0.01$\\
   &   &   & SiII & $13.99\pm0.04$\\
   &   &   & MgI & $11.16\pm0.03$\\
4 & $1.78847$ & $9.9\pm0.1$ & FeII & $13.42\pm0.02$\\
   &   &   & SiII & $13.60\pm0.11$\\
   &   &   & MgI & $11.61\pm0.01$\\
5 & $1.78863$ & $8.2\pm0.2$ & FeII & $13.58\pm0.01$\\
   &   &   & SiII & $14.31\pm0.02$\\
   &   &   & MgI & $11.87\pm0.01$\\
6 & $1.78885$ & $8.2\pm0.2$ & FeII & $13.64\pm0.01$\\
   &   &   & SiII & $14.25\pm0.02$\\
   &   &   & MgI & $11.72\pm0.01$\\
7 & $1.78929$ & $7.0\pm0.7$ & FeII & $13.11\pm0.03$\\
   &   &   & SiII & $13.13\pm0.26$\\
   &   &   & MgI & $11.14\pm0.03$\\
8 & $1.78949$ & $8.9\pm2.0$ & FeII & $12.95\pm0.05$\\
   &   &   & SiII & $-$\\
   &   &   & MgI & $10.90\pm0.07$\\
   \hline
1 & $1.78748$ & $19.8\pm0.3$ & SiIV & $13.27\pm0.01$\\
   &   &   & CIV & $13.68\pm0.01$\\
2 & $1.78796$ & $15.1\pm0.6$ & SiIV & $13.10\pm0.03$\\
   &   &   & CIV & $13.46\pm0.03$\\
3 & $1.78826$ & $18.6\pm1.1$ & SiIV & $13.50\pm0.03$\\
   &   &   & CIV & $14.08\pm0.03$\\
4 & $1.78865$ & $19.8\pm0.5$ & SiIV & $14.37\pm0.02$\\
   &   &   & CIV & $14.89\pm0.02$\\
5 & $1.78921$ & $32.5\pm1.3$ & SiIV & $13.58\pm0.02$\\
   &   &   & CIV & $14.42\pm0.01$\\
6 & $1.78970$ & $18.9\pm0.6$ & SiIV & $13.15\pm0.02$\\
   &   &   & CIV & $13.97\pm0.02$\\
7 & $1.78981$ & $5.0\pm0.3$ & SiIV & $12.78\pm0.02$\\
   &   &   & CIV & $13.25\pm0.04$\\
   \hline
\end{tabular}

\end{center}
\end{table}

\onecolumn

\twocolumn

\subsection{QSO J0427-1302 z$_{\rm em}=2.166$, z$_{\rm abs}=1.562$, $\rm \log N(HI)=19.35\pm0.10$}
The low-ionization transitions in this low-redshift system are well fitted with two components, the redshift and Doppler parameter of which are fixed by a simultaneous fit of FeII $\lambda\lambda\lambda\lambda$ 2344, 2382, 2586, 2600 and AlII $\lambda$ 1670. The 2-component fit shows an interesting asymmetric distribution of FeII and AlII abundances: the blue component is stronger for FeII while the red component is stronger for AlII. The resulting column densities are $\log$ N(FeII)=$12.23\pm0.04$ and $\log$ N(AlII)=$11.78\pm0.1$. In addition, the non detection of CrII $\lambda$ 2056, MgI $\lambda$ 2026, MnII $\lambda$ 2576, NiII $\lambda$ 1741, and ZnII $\lambda$ 2026 leads to the following upper limits: $\rm \log N(CrII)<12.39$, $\rm \log N(MgI)<12.38$, $\rm \log N(MnII)<11.84$, $\rm \log N(NiII)<13.23$ and $\rm \log N(ZnII)<11.75$.

In addition to these low-ionization transitions, the EUADP spectrum covers several high-ionization species. The doublet of CIV $\lambda\lambda$ 1550, 1548 and SiIV $\lambda\lambda$ 1393 and 1402 expand a velocity range of about $250$ km/s. The CIV transitions are strongly saturated and SiIV $\lambda$ 1402 appears to be blended (greater absorption in spite of a lower oscillation factor than SiIV $\lambda$ 1393). Thus the fit is performed using the SiIV $\lambda$ 1393 line, considering a total of 12 components, with the redder one being associated with the low-ionization profile ($\rm v \sim$140 km/s). Many of the strongest components fitted for SiIV $\lambda$ 1393 match the velocity profiles of the CIV doublet. We obtain $\rm \log N(SiIV)=13.9\pm0.07$.

\begin{figure}
\begin{center}
\includegraphics[width=.5\textwidth]{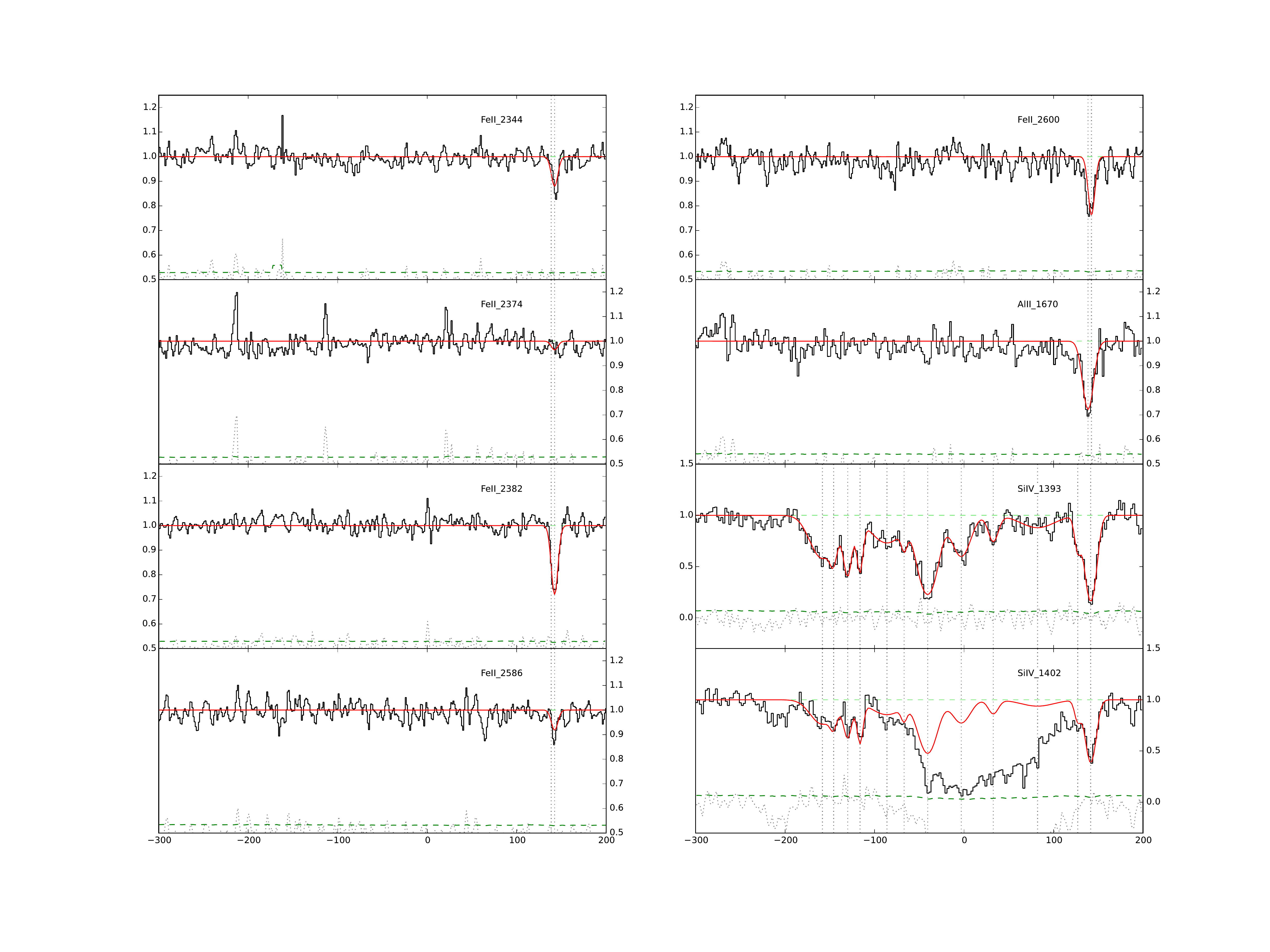}
\caption[QSOJ0427-1302]{QSOJ0427-1302}
\label{img:0427-1302}
\end{center}
\end{figure}

The parameter fits are listed in Table \ref{tab:0427-1302} and the corresponding Voigt profile fits are shown in Fig. \ref{img:0427-1302}.

\begin{table}
\begin{center}
\caption{Voigt profile fit parameters to the low- and high-ionization species for the z$_{\rm abs}$=1.562 log N(H\,I)=$19.35\pm0.10$ absorber towards QSO J0427-1302.}
\label{tab:0427-1302}
\begin{tabular}{ccccc}
\hline
\hline
Comp. & $z_{abs}$ & b & Ion & log $N$ \\
 & & km $s^{-1}$ & & cm$^{-2}$ \\
 \hline
1 & $1.5631845$ & $7.2\pm1.4$ & AlII & $11.78\pm0.10$\\
   &   &   & FeII & -\\
2 & $1.5632172$ & $2.9\pm0.6$ & AlII & -\\
   &   &   & FeII & $12.23\pm0.04$\\
 \hline
1 & $1.56065$ & $17.4\pm2.7$ & SiIV & $12.96\pm0.08$\\
2 & $1.56075$ & $3.1\pm2.4$ & SiIV & $12.31\pm0.22$\\
3 & $1.56089$ & $4.6\pm1.0$ & SiIV & $12.73\pm0.06$\\
4 & $1.56101$ & $1.5\pm0.7$ & SiIV & $12.93\pm0.62$\\
5 & $1.56126$ & $23.8\pm8.3$ & SiIV & $12.85\pm0.12$\\
6 & $1.56143$ & $1.2\pm2.0$ & SiIV & $12.13\pm0.39$\\
7 & $1.56165$ & $12.8\pm1.0$ & SiIV & $13.27\pm0.03$\\
8 & $1.56197$ & $12.7\pm1.8$ & SiIV & $12.81\pm0.04$\\
9 & $1.56228$ & $6.4\pm2.5$ & SiIV & $12.33\pm0.08$\\
10 & $1.56270$ & $25.4\pm5.2$ & SiIV & $12.49\pm0.08$\\
11 & $1.56309$ & $2.0\pm2.0$ & SiIV & $12.35\pm0.17$\\
12 & $1.56321$ & $6.4\pm0.6$ & SiIV & $13.15\pm0.03$\\
\end{tabular}

\end{center}
\end{table}

\subsection{QSO PKS0454-220 z$_{\rm em}=0.534$, z$_{\rm abs}=0.474$, $\rm \log N(HI)=19.45\pm0.03$}

This low-redshift absorption system contains a great number of transitions including three MnII lines: MnII $\lambda\lambda\lambda$ 2576, 2594, 2606, and seven FeII lines (four of which are saturated): FeII  $\lambda\lambda\lambda\lambda\lambda\lambda\lambda$ 2249, 2260, 2374, 2344, 2382, 2586 and 2600. It is interesting to notice the presence of a component in the blue part of the saturated lines which is not detected in the weaker transitions. Therefore, the fit is performed in two separate steps: on one hand the unsaturated lines are used to constrain the strongest components, on the other hand, this solution is applied to the saturated profiles to check its validity and to constrain the blue component. The absorption profile results in a total of seven components (five strong components as well as one blue and one red additional weaker components), spread in a velocity range of about 150 km/s. The column densities derived are $\log$ N(FeII)=$14.71\pm0.01$ and $\log$ N(MnII)=$12.58\pm0.01$. 

In this EUADP spectrum, no high-ionisation transitions are covered for this low-redshift absorber.

\begin{figure}
\begin{center}
\hspace*{-.6in}
\includegraphics[width=.6\textwidth, height=13cm]{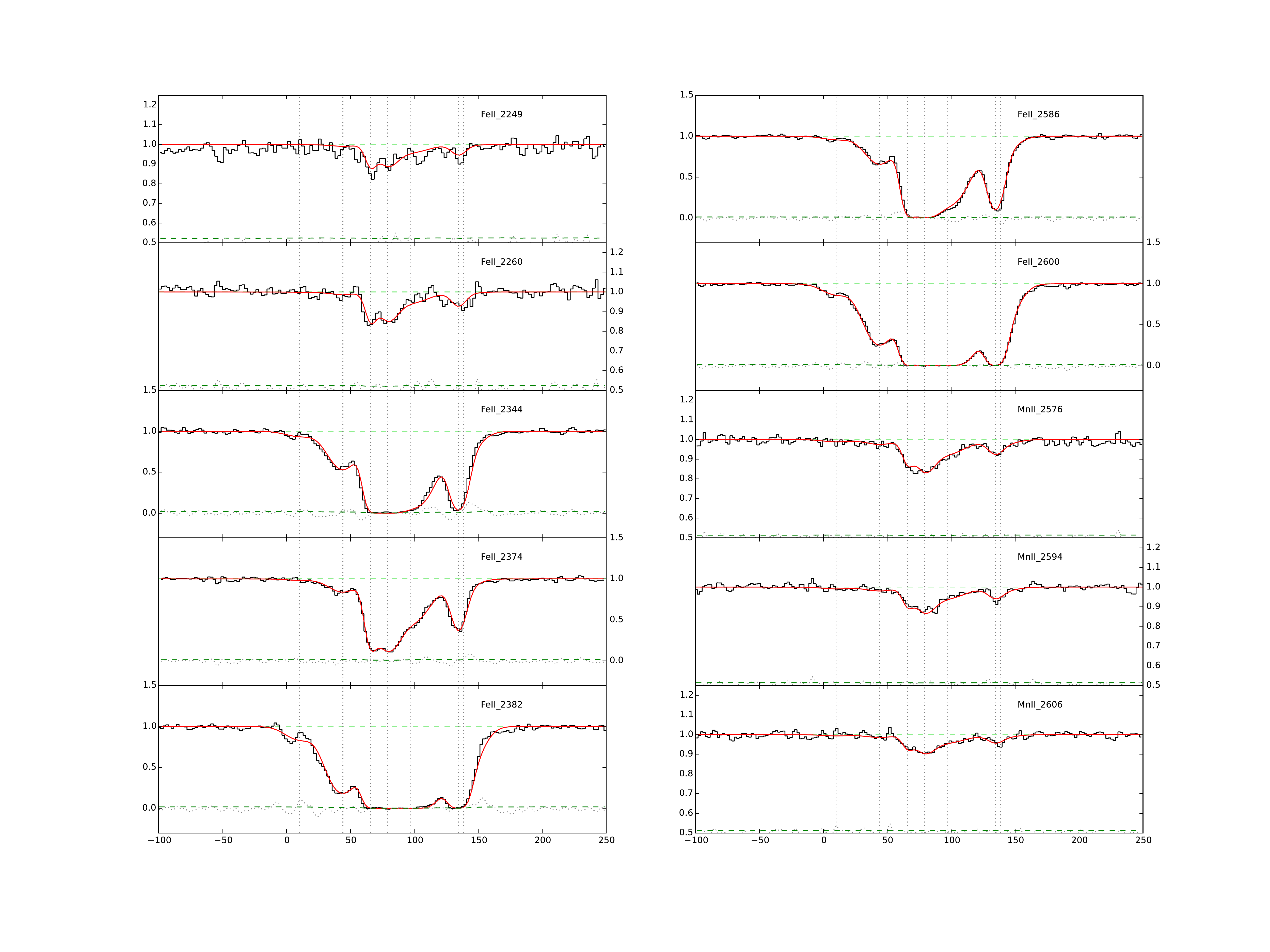}
\caption[QSOPKS0454-220]{QSOPKS0454-220}
\label{img:0454-220}
\end{center}
\end{figure}

The parameter fits are summarised in Table \ref{tab:0454-220} and Voigt profile fits are shown in Fig. \ref{img:0454-220}.

\begin{table}
\begin{center}
\caption{Voigt profile fit parameters to the low-ionization species for the z$_{\rm abs}$=0.474 log N(H\,I)=$19.45\pm0.03$ absorber towards QSO PKS 0454-220.}
\label{tab:0454-220}
\begin{tabular}{ccccc}
\hline
\hline
Comp. & $z_{abs}$ & b & Ion & log $N$ \\
 & & km $s^{-1}$ & & cm$^{-2}$ \\
 \hline
1 & $0.47405$ & $14.3\pm0.6$ & FeII & $12.39\pm0.02$\\
  &   &   & MnII & $11.12\pm0.07$\\
2 & $0.47422$ & $14.5\pm0.2$ & FeII & $13.38\pm0.01$\\
  &   &   & MnII & $11.46\pm0.03$\\
3 & $0.47432$ & $4.0\pm0.1$ & FeII & $14.00\pm0.02$\\
  &   &   & MnII & $11.65\pm0.02$\\
4 & $0.47439$ & $10.0\pm0.2$ & FeII & $14.26\pm0.01$\\
  &   &   & MnII & $12.09\pm0.01$\\
5 & $0.47448$ & $18.5\pm0.2$ & FeII & $14.12\pm0.01$\\
  &   &   & MnII & $12.03\pm0.01$\\
6 & $0.47466$ & $6.5\pm0.1$ & FeII & $13.81\pm0.01$\\
  &   &   & MnII & $11.48\pm0.04$\\
7 & $0.47468$ & $14.9\pm0.3$ & FeII & $13.11\pm0.04$\\
  &   &   & MnII & $11.50\pm0.06$\\
   \hline
\end{tabular}
\end{center}
\end{table}

\subsection{QSOJ0600-5040 z$_{\rm em}=3.13$, z$_{\rm abs}=2.149$, $\rm \log N(HI)=20.4\pm0.12$}
The EUADP spectrum for this absorber covers many low-ionization transitions including FeII $\lambda\lambda$ 1608, 1611, AlII $\lambda$ 1670, SiII $\lambda\lambda\lambda\lambda$ 1193, 1526, 1304, 1808, ZnII $\lambda$ 2062, CrII $\lambda\lambda$ 2062, 2026, NiII $\lambda\lambda\lambda$ 1709, 1741 and 1751. The absorption profile is clearly multi-component and covers a large velocity range of about $100$ km/s based on the strongest transitions (namely AlII $\lambda$ 1670, FeII $\lambda$ 1608, SiII $\lambda\lambda\lambda$ 1193, 1526 and 1304). The profile is well fitted with four components. This velocity profile is then applied to the weakest ZnII $\lambda$ 2026 line. It reveals a blend in the first two components related to the CrII $\lambda$ 2062 line which is therefore fitted simultaneously. The AlII $\lambda$ 1670 line is saturated, leading to a lower limit estimate in the column density of $\rm \log N(AlII)>14.33$ based on the four component profile (redshifts and Doppler parameters) described above. The AlIII $\lambda\lambda$ 1854 and 1862 profiles follow the low-ionization ions. However, a blend in AlIII $\lambda$ 1854 complicates the fit so that the velocity of the first component is fixed to the value derived above. The resulting column densities are: $\log$ N(FeII)=$14.84\pm0.03$, $\log$ N(SiII)=$15.08\pm0.01$, $\log$ N(NiII)=$13.62\pm0.02$, $\log$ N(CrII)=$13.10\pm0.01$, $\log$ N(ZnII)=$12.11\pm0.03$ and $\log$ N(AlIII)=$12.78\pm0.01$. 

In this EUADP spectrum, the high-ionization ions SiIV $\lambda\lambda$ 1393, 1402, and CIV $\lambda\lambda$ 1548 and 1550 are covered but are located in the forest, and hence suffer from important blending.

\begin{figure}
\begin{center}
\hspace*{-.8in}
\includegraphics[width=1.2\textwidth]{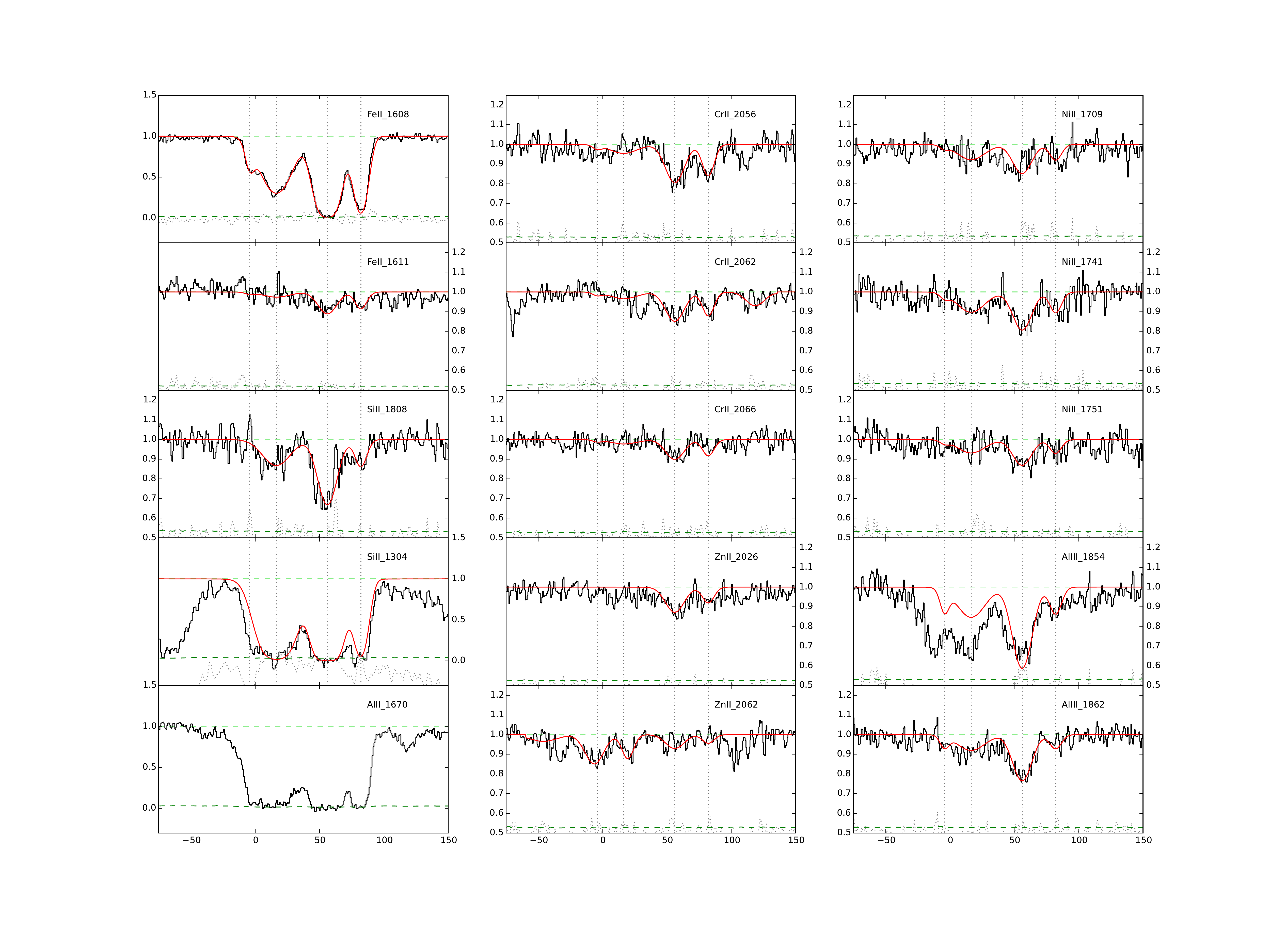}
\caption[QSOJ0600-5040]{QSOJ0600-5040}
\label{img:0600-5040}
\end{center}
\end{figure}

The parameter fits are summarized in Table \ref{tab:0600-5040} and Voigt profile fits are shown in Fig. \ref{img:0600-5040}.

\begin{table}
\begin{center}
\caption{Voigt profile fit parameters to the low- and intermediate-ionization species for the z$_{\rm abs}$=2.533 log N(H\,I)=$20.4\pm0.12$ absorber towards QSO J060008.1-504036.}
\label{tab:0600-5040}
\begin{tabular}{ccccc}
\hline
\hline
Comp. & $z_{abs}$ & b & Ion & log $N$ \\
 & & km $s^{-1}$ & & cm$^{-2}$ \\
 \hline
1 & $2.14895$ & $2.2\pm0.7$ & FeII & $13.22\pm0.04$\\
   &   &   & SiII & $-$\\
   &   &   & NiII & $12.06\pm0.18$\\
   &   &   & CrII & $11.50\pm0.22$\\
   &   &   & ZnII & $-$\\
   &   &   & AlIII & $11.60\pm0.02$\\
   &   &   & AlII & $>14.2$\\
2 & $2.14917$ & $14.1\pm0.4$ & FeII & $14.1\pm0.01$\\
   &   &   & SiII & $14.55\pm0.02$\\
   &   &   & NiII & $13.16\pm0.03$\\
   &   &   & CrII & $12.33\pm0.06$\\
   &   &   & ZnII & $-$\\
   &   &   & AlIII & $12.20\pm0.01$\\
   &   &   & AlII & $>13.1$\\
3 & $2.14959$ & $9.1\pm0.3$ & FeII & $14.58\pm0.01$\\
   &   &   & SiII & $14.84\pm0.01$\\
   &   &   & NiII & $13.29\pm0.02$\\
   &   &   & CrII & $12.83\pm0.01$\\
   &   &   & ZnII & $11.97\pm0.03$\\
   &   &   & AlIII & $12.54\pm0.01$\\
   &   &   & AlII & $>13.50$\\
4 & $2.14986$ & $4.7\pm0.3$ & FeII & $14.26\pm0.06$\\
   &   &   & SiII & $14.21\pm0.03$\\
   &   &   & NiII & $12.81\pm0.04$\\
   &   &   & CrII & $12.54\pm0.02$\\
   &   &   & ZnII & $11.56\pm0.06$\\
   &   &   & AlIII & $11.79\pm0.02$\\
   &   &   & AlII & $>13.10$\\
   \hline
\end{tabular}
\end{center}
\end{table}

\onecolumn

\twocolumn

\subsection{QSO B1036-2257 z$_{\rm em}=3.13$, z$_{\rm abs}=2.533$, $\rm \log N(HI)=19.3\pm0.10$}
A large number of low-ionisation elements are detected in the EUADP spectrum including SiII $\lambda\lambda\lambda\lambda$ 1190, 1193, 1260, 1526, CII $\lambda\lambda$ 1036, 1334, MgII $\lambda\lambda$ 2803 and 2796, AlII $\lambda$ 1670, AlIII $\lambda$ 1862 and FeII $\lambda\lambda$ 2382, 2600. Based on the SiII $\lambda\lambda$ 1193 and 1260 lines which are free from any saturation and blending, eleven components are used to fit the absorption profile. The low-ionization transitions for this system cover a large velocity range of about $200$ km/s. The strongest components of this profile are used to fit AlII $\lambda$ 1670, FeII $\lambda\lambda$ 2600, 2382 and AlIII $\lambda$ 1862. The resulting column densities are $\log$ N(SiII)=$13.64\pm0.01$, $\log$ N(AlII)=$12.52\pm0.01$, $\log$ N(AlIII)=$12.89\pm0.02$ and $\log$ N(FeII)=$12.93\pm0.01$. The profile estimated from the weakest transition is then used to fit the saturated CII $\lambda$ 1334 revealing a 12$^{\rm th}$ component around $v=0$ km/s. A manual fit using CII $\lambda\lambda$ 1036 and 1334 and the previous solution provides the following lower limit for CII $\rm \log N(CII)>15.98$. Nevertheless, both lines are contaminated with unrelated absorbers, which prevents us from deriving a robust estimate of the lower limit in CII.

A great number of high-ionization lines are detected in this EUADP spectra including OVI $\lambda\lambda$ 1031, 1037, NV $\lambda\lambda$ 1238, 1242, CIV $\lambda\lambda$ 1548, 1550, SiIV $\lambda\lambda$ 1393 and 1402. A number of these lines are located in the Lyman-$\alpha$ forest (SiIV doublet, NV $\lambda$1242 and OVI $\lambda$ 1031) and therefore appear to be blended. The CIV doublet is saturated in this case, such that only a lower limit is derived. Only the OVI $\lambda$ 1037 and NV $\lambda$ 1238 lines appear free from any saturation or blending, preventing from performing a reasonable fit. A three-component profile is used to fit SiIV $\lambda$ 1393. The resulting component velocities and Doppler parameters are then used to fit the other lines available. The resulting column densities are $\log$ N(SiIV)=$13.71\pm0.01$, $\log$ N(CIV)$>17.42$.

\begin{figure}
\begin{center}
\hspace*{-.8in}
\includegraphics[width=1.2\textwidth]{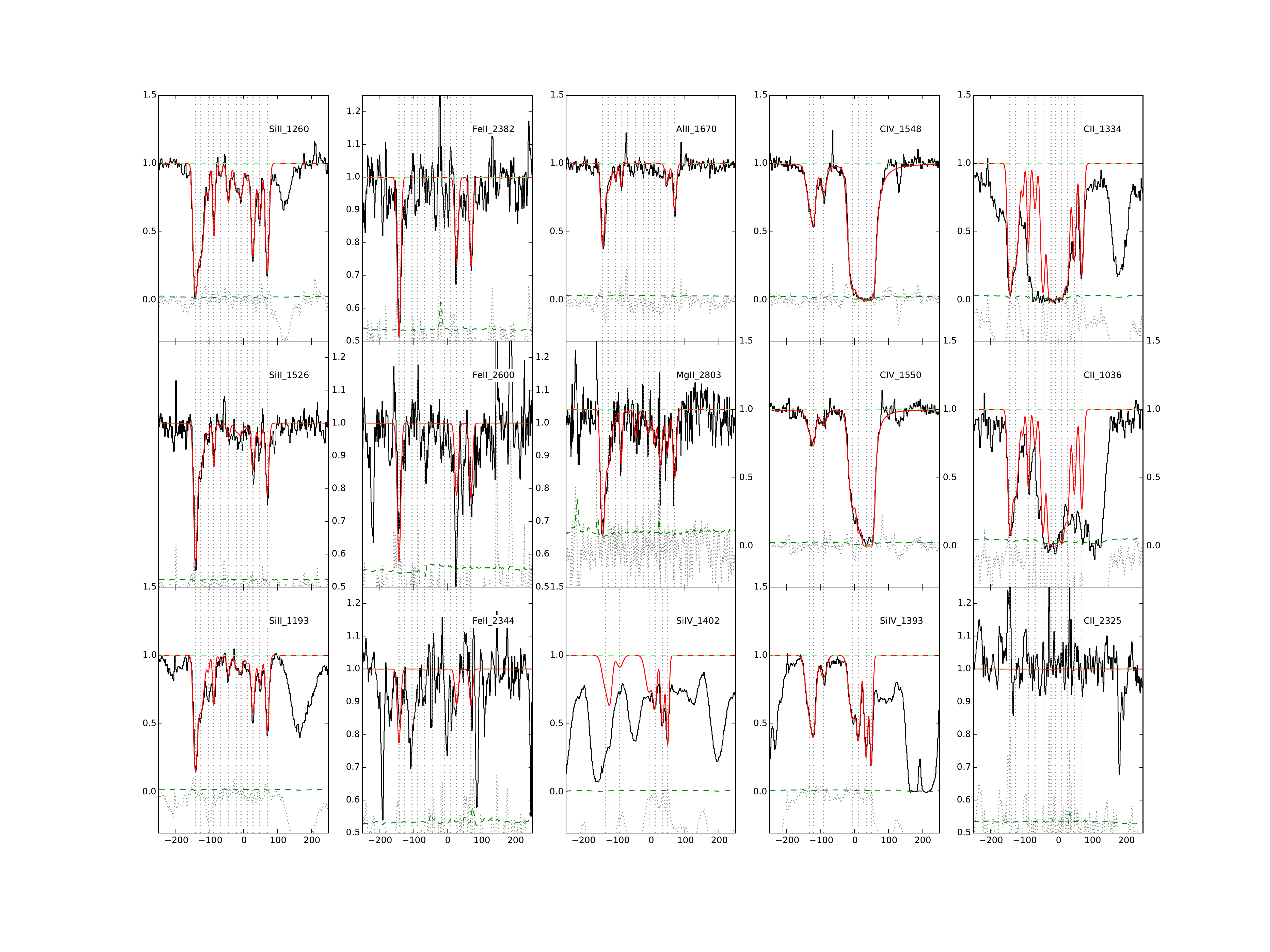}
\caption[QSOB1036-2257]{QSOB1036-2257}
\label{img:1036-2257}
\end{center}
\end{figure}

The parameter fits of the individual components are listed in Table \ref{tab:1036-2257} and the corresponding Voigt profile fits are shown in Fig. \ref{img:1036-2257}.

\twocolumn


\begin{table*}
\begin{center}
\caption{Voigt profile fit parameters to the low- and high-ionization species for the z$_{\rm abs}$=2.533 log N(H\,I)=$19.3\pm0.10$ absorber towards QSO B1036-2257}
\label{tab:1036-2257}
\begin{tabular}{ccccccccccc}
\hline
\hline
Comp. & $z_{abs}$ & b & Ion & log $N$ && Comp. & $z_{abs}$ & b & Ion & log $N$ \\
 & & km $s^{-1}$ & & cm$^{-2}$ &&  & & km $s^{-1}$ & & cm$^{-2}$ \\
1 & $2.53132$ & $5.4\pm0.1$ & SiII & $13.19\pm0.01$ && 1 & $2.53143$ & $12.9\pm0.6$ & CIV & $12.97\pm0.02$\\
   &   &   & FeII & $12.65\pm0.01$  &&    &   &   & SiIV & $12.83\pm0.02$\\
   &   &   & AlII & $12.16\pm0.01$  && 2 & $2.53158$ & $7.8\pm0.4$ & CIV & $12.98\pm0.02$\\
   &   &   & MgII & $13.20\pm0.04$  &&   &   &   & SiIV & $12.74\pm0.02$\\
   &   &   & CII & $14.00$ && 3 & $2.53192$ & $11.6\pm0.8$ & CIV & $12.83\pm0.02$\\
2 & $2.53152$ & $11.3\pm0.4$ & SiII & $12.87\pm0.03$ &&   &   &   & SiIV & $12.32\pm0.04$\\
   &   &   & FeII & $-$  && 4 & $2.53294$ & $13.2\pm2.5$ & CIV & $13.9\pm0.01$\\
   &   &   & AlII & $11.77\pm0.03$ &&   &   &   & SiIV & $12.91\pm0.01$\\
   &   &   & MgII & $12.71\pm0.03$  && 5 & $2.53314$ & $6.0\pm2.3$ & CIV & $>13.85$\\
   &   &   & CII & $13.81\pm0.01$   &&    &   &   & SiIV & $12.78\pm0.02$\\
3 & $2.53178$ & $2.2\pm0.2$ & SiII & $11.80\pm0.10$  && 6 & $2.53340$ & $5.4\pm4.4$ & CIV & $>16.57$\\
   &   &   & FeII & $-$  &&    &   &   & SiIV & $12.98\pm0.01$  \\
   &   &   & AlII & $11.14\pm0.03$  && 7 & $2.53358$ & $3.0\pm2.6$ & CIV & $>17.35$\\
   &   &   & MgII & $11.49\pm0.21$ &&    &   &   & SiIV & $13.15\pm0.01$\\
   &   &   & CII & $12.7$ && &&&&\\
4 & $2.53197$ & $2.1\pm0.2$ & SiII & $12.47\pm0.06$ && &&&&\\
   &   &   & FeII & $-$ && &&&&\\
   &   &   & AlII & $11.32\pm0.02$ && &&&&\\
   &   &   & MgII & $12.37\pm0.06$ && &&&&\\
   &   &   & CII & $13.60$ && &&&&\\
5 & $2.53220$ & $4.3\pm2.1$ & SiII & $11.49\pm0.13$ && &&&&\\
   &   &   & FeII & $-$ && &&&&\\
   &   &   & AlII & $-$ && &&&&\\
   &   &   & MgII & $11.19\pm0.43$ && &&&&\\
   &   &   & CII & $13.00$ && &&&&\\
6 & $2.53248$ & $5.3\pm0.9$ & SiII & $12.09\pm0.04$ && &&&&\\
   &   &   & FeII & $-$ && &&&&\\
   &   &   & AlII & $-$ && &&&&\\
   &   &   & MgII & $12.11\pm0.07$ && &&&&\\
   &   &   & CII & $14.00$ && &&&&\\
7 & $2.53275$ & $7.8\pm2.3$ & SiII & $11.95\pm0.05$ && &&&&\\
   &   &   & FeII & $-$ && &&&&\\
   &   &   & AlII & $-$ && &&&&\\
   &   &   & MgII & $11.27\pm0.48$ && &&&&\\
   &   &   & CII & $>15.90$ && &&&&\\
8 & $2.53291$ & $7.1\pm1.5$ & SiII & $12.09\pm0.04$ && &&&&\\
   &   &   & FeII & $-$ && &&&&\\
   &   &   & AlII & $-$ && &&&&\\
   &   &   & MgII & $12.11\pm0.07$ && &&&&\\
   &   &   & CII & $>15.00$ && &&&&\\
9 & $2.53313$ & $4.0\pm2.0$ & SiII & $11.60\pm0.04$ && &&&&\\
   &   &   & FeII & $-$ && &&&&\\
   &   &   & AlII & $-$ && &&&&\\
   &   &   & MgII & $12.14\pm0.06$ && &&&&\\
   &   &   & CII & $>14.30$ && &&&&\\
10 & $2.53332$ & $5.6\pm0.3$ & SiII & $12.65\pm0.01$ && &&&&\\
   &   &   & FeII & $12.31\pm0.01$ && &&&&\\
   &   &   & AlII & $-$ && &&&&\\
   &   &   & MgII & $12.52\pm0.04$ && &&&&\\
   &   &   & CII & $13.8$ && &&&&\\
11 & $2.53356$ & $5.9\pm0.7$ & SiII & $12.32\pm0.03$ && &&&&\\
   &   &   & FeII & $-$ && &&&&\\
   &   &   & AlII & $11.42\pm0.02$ && &&&&\\
   &   &   & MgII & $12.35\pm0.04$ && &&&&\\
   &   &   & CII & $13.55$ && &&&&\\
12 & $2.53382$ & $5.0\pm0.2$ & SiII & $12.82\pm0.01$ && &&&&\\
   &   &   & FeII & $12.30\pm0.01$ && &&&&\\
   &   &   & AlII & $11.80\pm0.01$ && &&&&\\
   &   &   & MgII & $12.58\pm0.03$ && &&&&\\
   &   &   & CII & $13.70$ && &&&&\\
   \hline
\end{tabular}
\end{center}
\end{table*}

\onecolumn

\twocolumn

\subsection{QSO J115538.6+053050 z$_{\rm em}=3.475$, z$_{\rm abs}=3.327$, $\rm \log N(HI)=21.0\pm0.10$}
Many ions are detected in this absorber, such as SiII $\lambda$ 1808, SII $\lambda\lambda\lambda$ 1259 1253 1250, AlIII $\lambda\lambda$ 1854 1862, NiII $\lambda\lambda$ 1370 1317, CIV $\lambda\lambda$ 1548 1550 and SiIV $\lambda\lambda$ 1393 and 1402. The low- and intermediate-ionization ions show a similar profile (strong absorption line near $\rm v=65$ km/s) and are therefore fitted together with a 7-component profile. Only the non-saturated lines, NiII $\lambda\lambda$ 1370 1317, SiII $\lambda$ 1808 and SII $\lambda$ 1253 are used in the derivation of the parameter. The ions AlII $\lambda$ 1670, CII $\lambda\lambda$ 1036 and 1334 are also detected but not fitted due to strong saturation. AlIII have not been fitted due to a blend in the AlIII $\lambda$ 1862 line. The resulting abundances for the low-ionization ions are $\log$ N(SiII)=$15.93\pm0.01$, $\log$ N(SII)=$15.31\pm0.01$ and $\log$ N(NiII)=$13.74\pm0.01$. A non-detection from MgI $\lambda$ 1827 gives the following upper limit: $\rm \log N(MgI)<13.33$.

The high-ionization ions CIV $\lambda\lambda$ 1548 1550 and SiIV $\lambda\lambda$ 1402 are fitted using a 3-component profile, extending to about 200 km/s. SiIV $\lambda$ 1393 is not considered for the fit as it is blended blueward the bluest component of the fit ($\rm v\sim150$ km/s). This gives the following abundances for the high-ionization transitions: $\log$ N(CIV)=$13.71\pm0.01$ and $\log$ N(SiIV)=$13.56\pm0.01$. We note that, although the reddest components for both low- and high-ionization profiles have velocities that differ by about 20 km/s, they both stand out from the bluer absorption profile.

\begin{figure}
\begin{center}
\hspace*{-.8in}
\includegraphics[width=1.2\textwidth]{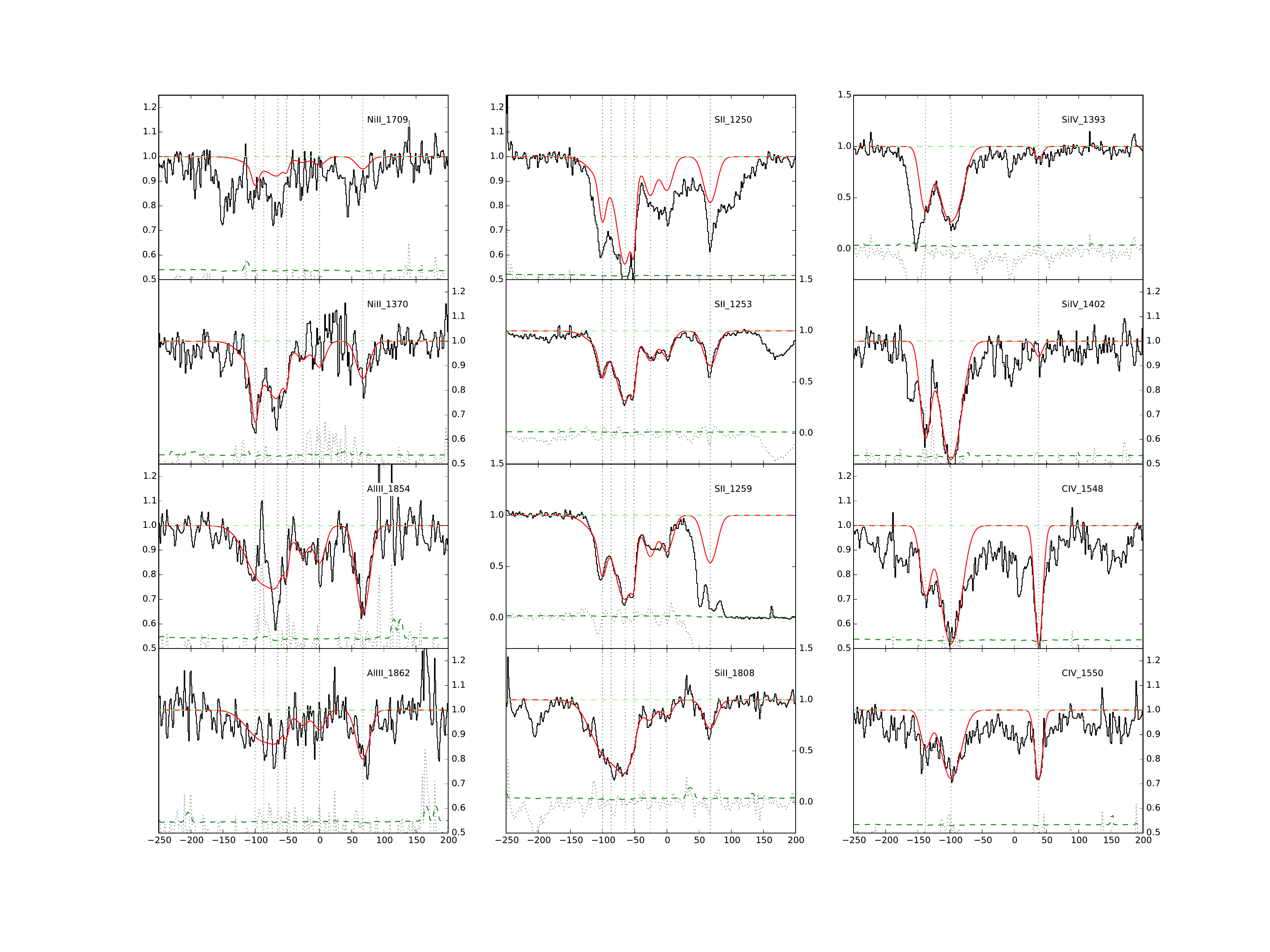}
\caption[QSOJ115538.6+053050 z=3.32]{QSOJ115538.6+053050 z=3.32}
\label{img:1155+0530}
\end{center}
\end{figure}

The parameter fits of the individual components are listed in Table \ref{tab:1155+0530} and the corresponding Voigt profile fits are shown in Fig. \ref{img:1155+0530}.

\begin{table}
\begin{center}
\caption{Voigt profile fit parameters to the low- and high-ionization species for the z$_{\rm abs}$=3.327 log N(H\,I)=$21.0\pm0.10$ absorber towards QSO J115538.6+053050.}
\label{tab:1155+0530}
\begin{tabular}{ccccc}
\hline
\hline
Comp. & $z_{abs}$ & b & Ion & log $N$ \\
 & & km $s^{-1}$ & & cm$^{-2}$ \\
 \hline
1 & $3.32555$ & $5.9\pm0.1$ & NiII & $12.84\pm0.01$\\
   &   &   & SII & $14.14\pm0.01$\\
   &   &   & SiII & $-$\\
   &   &   & AlIII & $-$\\
2 & $3.32575$ & $33.8\pm0.1$ & NiII & $13.35\pm0.01$\\
   &   &   & SII & $14.66\pm0.01$\\
   &   &   & SiII & $15.72\pm0.01$\\
   &   &   & AlIII & $12.77\pm0.01$\\
3 & $3.32606$ & $13.6\pm0.1$ & NiII & $12.88\pm0.01$\\
   &   &   & SII & $14.83\pm0.01$\\
   &   &   & SiII & $15.20\pm0.01$\\
   &   &   & AlIII & $11.96\pm0.01$\\
4 & $3.32626$ & $2.0\pm0.1$ & NiII & $12.27\pm0.02$\\
   &   &   & SII & $14.16\pm0.02$\\
   &   &   & SiII & $13.96\pm0.03$\\
   &   &   & AlIII & $11.55\pm0.04$\\
5 & $3.32663$ & $10.7\pm0.2$ & NiII & $12.48\pm0.01$\\
   &   &   & SII & $14.27\pm0.01$\\
   &   &   & SiII & $14.55\pm0.01$\\
   &   &   & AlIII & $11.94\pm0.02$\\
6 & $3.32700$ & $11.6\pm0.1$ & NiII & $12.71\pm0.01$\\
   &   &   & SII & $14.25\pm0.01$\\
   &   &   & SiII & $14.57\pm0.01$\\
   &   &   & AlIII & $12.12\pm0.01$\\
7 & $3.32797$ & $13.2\pm0.1$ & NiII & $12.93\pm0.01$\\
   &   &   & SII & $14.45\pm0.01$\\
   &   &   & SiII & $14.89\pm0.01$\\
   &   &   & AlIII & $12.60\pm0.01$\\
 \hline
1 & $3.32500$ & $10$ & CIV & $12.9\pm0.02$\\
   &   &   & SiIV & $13.00\pm0.01$\\
2 & $3.32558$ & $20$ & CIV & $13.48\pm0.02$\\
   &   &   & SiIV & $13.40\pm0.01$\\
3 & $3.32755$ & $7$ & CIV & $13.10\pm0.01$\\
   &   &   & SiIV & $12.00\pm0.02$\\
   \hline
\end{tabular}

\end{center}
\end{table}

\onecolumn

\twocolumn

\subsection{QSO LBQS 1232+0815 z$_{\rm em}=2.57$, z$_{\rm abs}=1.72$, $\rm \log N(HI)=19.48\pm0.13$}
The EUADP spectrum for this DLA absorber covers the following low- and intermediate-ionization transitions: FeII $\lambda\lambda\lambda\lambda$ 2382, 2374, 2344, 1608, AlIII $\lambda\lambda$ 1862, 1854, SiII $\lambda\lambda$ 1526 (blended) and 1808. It also covers CII $\lambda$ 1334, which is saturated. The low- and intermediate-ionization profiles are well fitted together with the transitions FeII $\lambda\lambda$ 2382, 2344, AlIII $\lambda\lambda$ 1854, 1862  and SiII $\lambda$ 1808 using 6 components spread over $\sim 200$ km/s, resulting in the following abundances $\log$ N(FeII)=$13.50\pm0.01$, $\log$ N(SiII) = $14.41\pm0.01$ and $\log$ N(AlIII)=$13.28\pm0.01$. We derived upper limits from non detection for SII $\lambda$ 1253, $\rm \log N(SII)<14.18$, CrII $\lambda$ 2056, $\rm \log N(CrII)<12.38$, MgI $\lambda$ 2026, $\rm \log N(MgI)<12.21$, NiII $\lambda$ 1751, $\rm \log N(NiII)<13.05$, and ZnII $\lambda$ 2026, $\rm \log N(ZnII)<11.58$.

The high-ionization ions detected in the spectrum are SiIV $\lambda\lambda$ 1393, 1402, CIV $\lambda\lambda$ 1548 and 1550.
The CIV transition lines are highly saturated and contaminated by an apparent blend, they are therefore not consider for the fit. From the less saturated SiIV transition lines, a 5-component profile provides the following lower limit $\rm \log N(SiIV)>14.67$. We notice that the low- and high-ionization ions seem to share the same components.

\begin{figure}
\begin{center}
\hspace*{-.8in}
\includegraphics[width=1.2\textwidth]{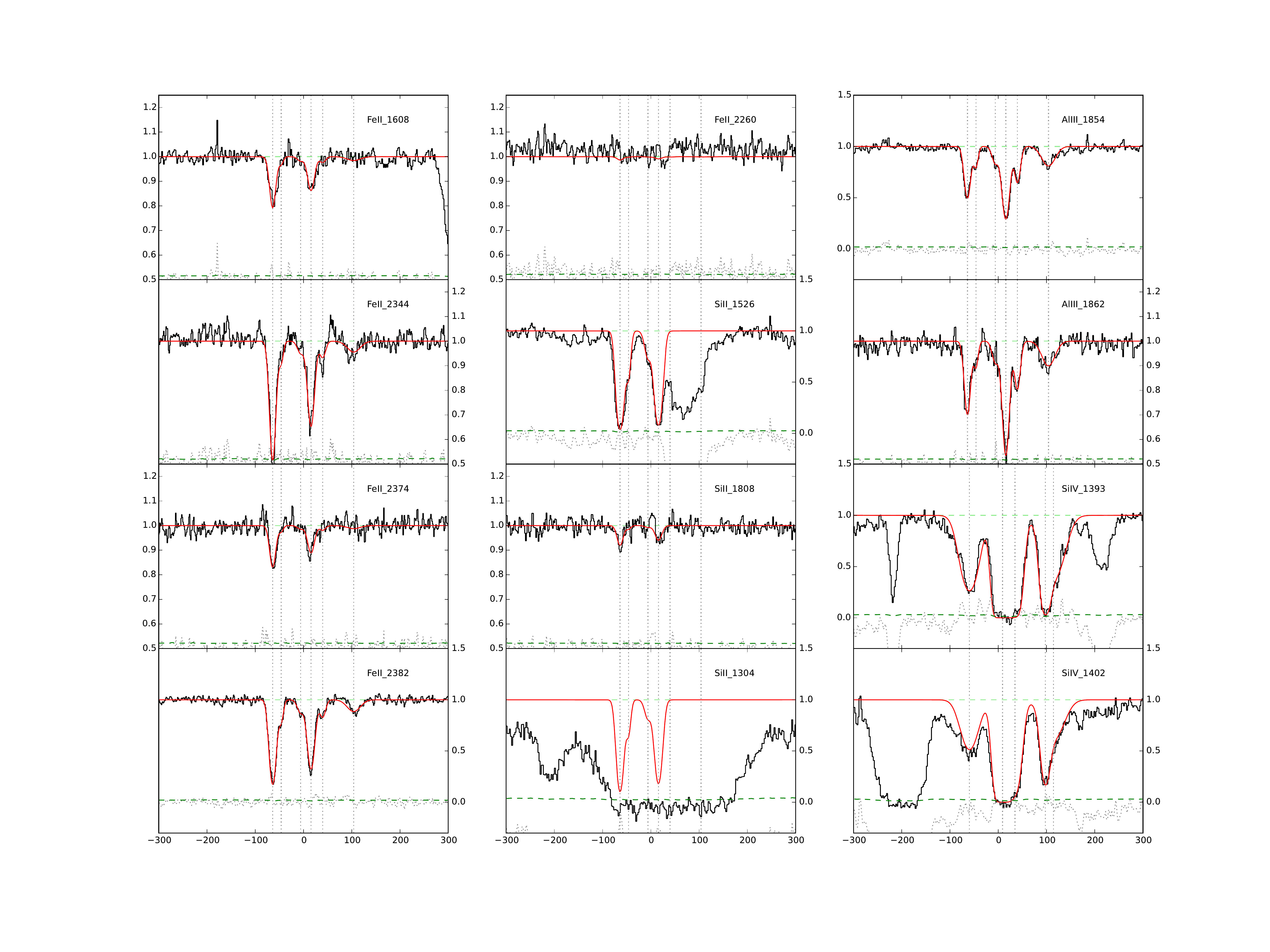}
\caption[QSOLBQS1232+0815]{QSOLBQS1232+0815}
\label{img:1232+0815}
\end{center}
\end{figure}

The parameter fits of the individual components are listed in Table \ref{tab:1232+0815} and the corresponding Voigt profile fits are shown in Fig. \ref{img:1232+0815}.

\begin{table}
\begin{center}
\caption{Voigt profile fit parameters to the low--ionization species for the z$_{\rm abs}$=1.72 log N(H\,I)=$19.48\pm0.13$ absorber towards  QSO LBQS 1232+0815.}
\label{tab:1232+0815}
\begin{tabular}{ccccc}
\hline
\hline
Comp. & $z_{abs}$ & b & Ion & log $N$ \\
 & & km $s^{-1}$ & & cm$^{-2}$ \\
 \hline
1 & $1.71942$ & $7.7\pm0.1$ & FeII & $13.17\pm0.01$\\
   &   &   & AlIII & $12.61\pm0.01$\\
   &   &   & SiII & $14.10\pm0.01$\\
2 & $1.71958$ & $4.6\pm0.4$ & FeII & $12.15\pm0.02$\\
   &   &   & AlIII & $11.96\pm0.02$\\
   &   &   & SiII & $13.20\pm0.02$\\
3 & $1.71995$ & $8.2\pm0.6$ & FeII & $12.07\pm0.03$\\
   &   &   & AlIII & $12.04\pm0.02$\\
   &   &   & SiII & $13.06\pm0.02$\\
4 & $1.72014$ & $8.8\pm0.1$ & FeII & $13.02\pm0.01$\\
   &   &   & AlIII & $12.90$\\
   &   &   & SiII & $14.01\pm0.01$\\
5 & $1.72036$ & $7.1\pm0.3$ & FeII & $12.16\pm0.02$\\
   &   &   & AlIII & $12.37\pm0.01$\\
   &   &   & SiII & $-$\\
6 & $1.72094$ & $18.3\pm0.6$ & FeII & $12.33\pm0.02$\\
   &   &   & AlIII & $12.41\pm0.01$\\
   &   &   & SiII & $-$\\


 \hline
1 & $1.71946$ & $23.3\pm0.3$ & SiIV & $13.47\pm0.01$\\
2 & $1.72008$ & $14.5\pm0.5$ & SiIV & $>14.50$\\
3 & $1.72032$ & $14.2\pm0.5$ & SiIV & $>13.82$\\
4 & $1.72089$ & $10.5\pm0.3$ & SiIV & $13.49\pm0.01$\\
5 & $1.72104$ & $27.8\pm0.6$ & SiIV & $13.41\pm0.01$\\
   \hline
\end{tabular}

\end{center}
\end{table}

\onecolumn

\twocolumn

\subsection{QSO J1330-2522 z$_{\rm em}=3.91$, z$_{\rm abs}=2.654$, $\rm \log N(HI)=19.56\pm0.13$}
This EUADP spectrum covers six ions, SiIV $\lambda\lambda$ 1393, 1402, SiII $\lambda$ 1526, AlII $\lambda$ 1670, AlIII $\lambda\lambda$ 1854 and 1862. Many of these transitions are blended and/or saturated, such that only AlII $\lambda$ 1670 and AlIII $\lambda$ 1854 have been fitted. The asymmetry of both lines suggests a 2-component profile, resulting in the following abundances: $\log$ N(AlIII)=$12.62\pm0.02$ and $\log$ N(AlII)=$12.18\pm0.2$. We derive an upper limit from non detection of NiII $\lambda$ 1741: $\log \rm N(NiII)<13.22$.

In this spectrum, the high-ionization ions SiIV $\lambda\lambda$ 1393, 1402 and CIV $\lambda\lambda$ 1548 and 1550 are covered but suffer from severe blending, such that no fit has been performed.

\begin{figure}
\begin{center}
\includegraphics[width=.45\textwidth]{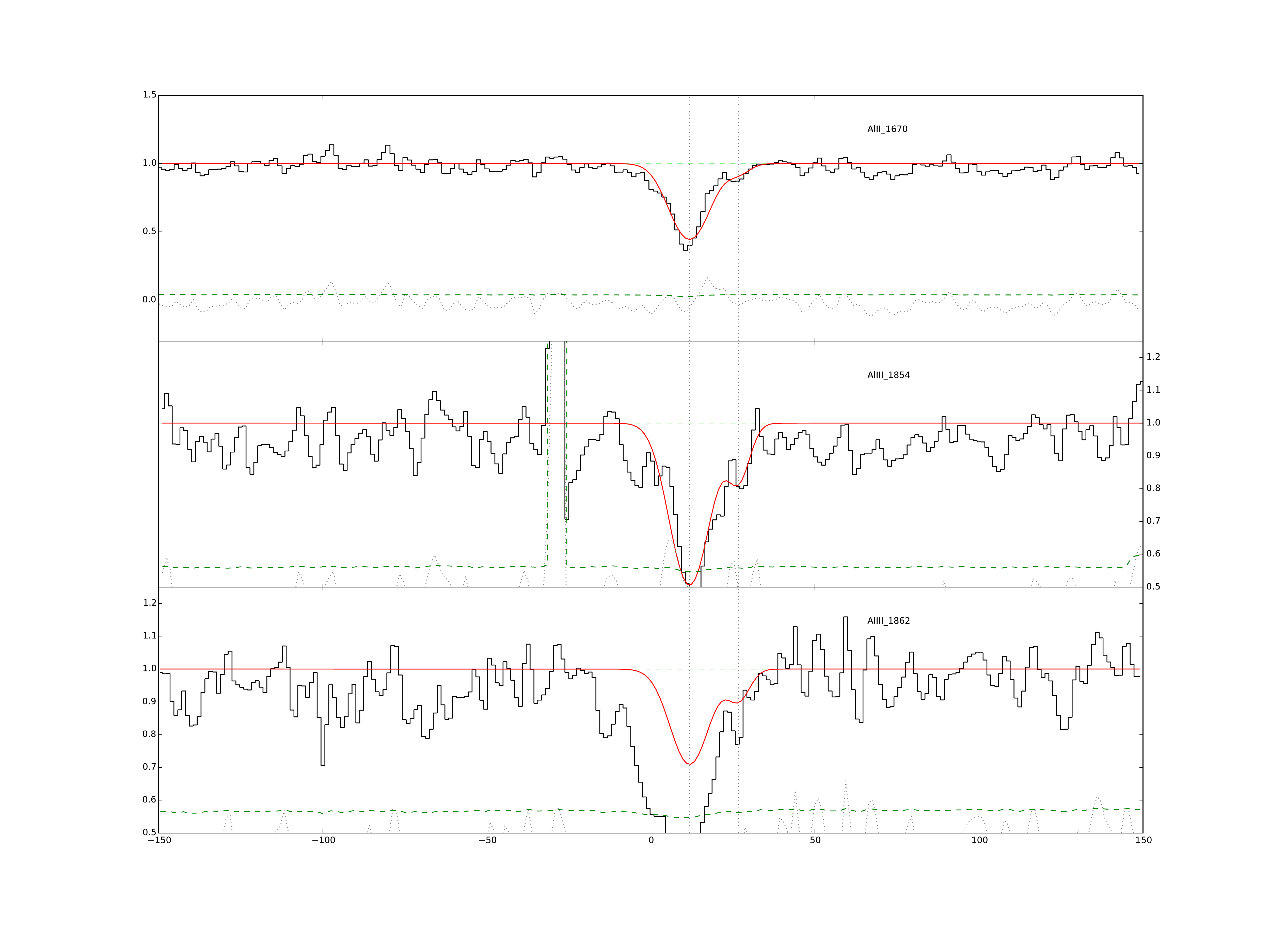}
\caption[QSOJ1330-2522 z=2.654]{QSOJ1330-2522 z=2.654}
\label{img:1330-2522}
\end{center}
\end{figure}

The parameter fits of the individual components are listed in Table \ref{tab:1330-2522} and the corresponding Voigt profile fits are shown in Fig. \ref{img:1330-2522}.

\begin{table}
\begin{center}
\caption{Voigt profile fit parameters to the low- and intermediate-ionization species for the z$_{\rm abs}$=2.654 log N(H\,I)=$19.56\pm0.13$ absorber towards QSO J1330-2522.}
\label{tab:1330-2522}
\begin{tabular}{ccccc}
\hline
\hline
Comp. & $z_{abs}$ & b & Ion & log $N$ \\
 & & km $s^{-1}$ & & cm$^{-2}$ \\
 \hline
1 & $2.65414$ & $6.7$ & AlII & $12.16\pm0.02$\\
   &   &   & AlIII & $12.55\pm0.02$\\
2 & $2.65433$ & $2.0$ & AlII & $10.90\pm0.21$\\
   &   &   & AlIII & $11.79\pm0.08$\\
   \hline
\end{tabular}

\end{center}
\end{table}

\subsection{QSO J1356-1101 z$_{\rm em}=3.006$, z$_{\rm abs}=2.397$, $\rm \log N(HI)=19.85\pm0.08$}
The spectrum covers four FeII lines and two SiIV lines associated with the absorber:  FeII $\lambda\lambda\lambda\lambda$ 2600 2344 2382 2586 and SiIV $\lambda\lambda$ 1393 and 1402. The low-ionization profile is well fitted with a 6-component profile in the red, and a single blue component isolated from the red group of components by about $\rm 250km/s$. The resulting column density is $\log \rm N(FeII)=13.44\pm0.01$. We derived upper limits from the non-detection of several transitions: CrII $\lambda$ 2056, $\rm \log N(CrII)<12.64$, MnII $\lambda$ 2576, $\rm \log N(MnII)<12.07$, NiII $\lambda$ 1317, $\rm \log N(NiII)<12.76$, and ZnII $\lambda$ 2062, $\rm \log N(ZnII)<12.38$.

The high-ionization ion SiIV $\lambda\lambda$ 1393 and 1402 is detected but not fitted because it is heavily saturated.

\begin{figure}
\begin{center}
\hspace*{-.5in}
\includegraphics[width=.6\textwidth,, height=12cm]{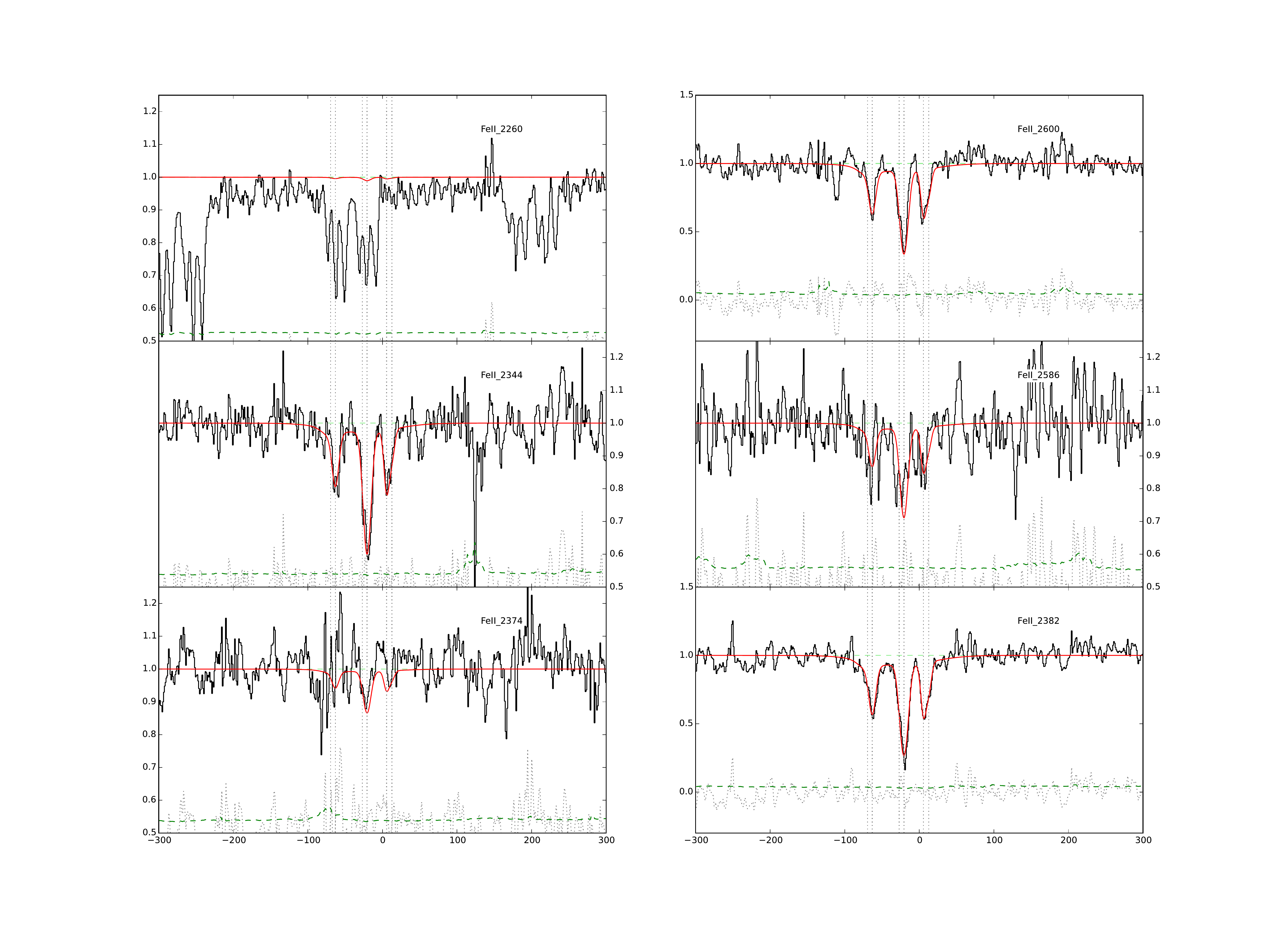}
\caption[QSOJ1356-1101]{QSOJ1356-1101}
\label{img:1356-1101}
\end{center}
\end{figure}

The parameter fits of the individual components are listed in Table \ref{tab:1356-1101} and the corresponding Voigt profile fits are shown in Fig. \ref{img:1356-1101}.

\begin{table}
\begin{center}
\caption{Voigt profile fit parameters to the low-ionization species for the z$_{\rm abs}$=2.397 log N(H\,I)=$19.85\pm0.08$ absorber towards QSO J1356-1101.}
\label{tab:1356-1101}
\begin{tabular}{ccccc}
\hline
\hline
Comp. & $z_{abs}$ & b & Ion & log $N$ \\
 & & km $s^{-1}$ & & cm$^{-2}$ \\
 \hline
1 & $2.39339$ & $11.0\pm0.3$ & FeII & $12.81\pm0.01$\\
2 & $2.39621$ & $14.8\pm1.4$ & FeII & $11.99\pm0.08$\\
3 & $2.39629$ & $4.5\pm0.3$ & FeII & $12.44\pm0.02$\\
4 & $2.39669$ & $61.1\pm5.4$ & FeII & $12.63\pm0.04$\\
5 & $2.39677$ & $5.8\pm0.1$ & FeII & $12.95\pm0.01$\\
6 & $2.39706$ & $2.4\pm0.3$ & FeII & $12.44\pm0.02$\\
7 & $2.39714$ & $2.6\pm0.6$ & FeII & $12.10\pm0.02$\\
\hline
\end{tabular}

\end{center}
\end{table}

\subsection{QSO J1621-0042 z$_{\rm em}=3.7$, z$_{\rm abs}=3.104$, $\rm \log N(HI)=19.7\pm0.20$}
The EUADP spectrum probing this high-redshift subDLA covers many transitions associated with the absorber such as FeII $\lambda\lambda$ 1608, 1611, CII $\lambda\lambda$ 1036, 1334, SII $\lambda\lambda\lambda$ 1250, 1253, 1259, SiII $\lambda\lambda\lambda\lambda\lambda$ 1190, 1193, 1260, 1304, 1526, SiIV $\lambda\lambda$ 1393, 1402 (partly), CIV $\lambda\lambda$ 1548 and 1550. Most of these lines are heavily blended (SiII $\lambda\lambda\lambda$ 1190, 1304, 1260) or saturated (CII $\lambda\lambda$ 1334 and 1036), but the wide coverage and SNR of the spectrum provide enough elements to derive the different parameters.

The low-ionization ions are fitted with a 12-component profile, with a broad velocity range (about 400 km/s). The component with the highest velocity ($\rm v\sim400$ km/s) is identified in three transitions: SiII $\lambda\lambda$ 1190 and CII $\lambda$ 1334. The low velocity components are less affected by the blending and are therefore well fitted, but the information about the group of component between $\rm v\sim 100$ km/s and $\rm v\sim 250$ km/s are only derived from FeII $\lambda$ 1608 and SiII $\lambda$ 1526. For CII in particular, this group of component is saturated so that a lower limit is derived based on the low velocity components $\log \rm N(CII)<14.41$. The SII line is most probably blended as its profile does not match the other low-ionization ions. The resulting abundances for the low-ionization ions are $\log \rm N(FeII)=13.30\pm0.04$ and $\log \rm N(SiII)=13.78\pm0.03$.

The high-ionization ion components are also detected with a broad velocity range (400km/s). The EUADP spectrum does not fully cover the SiIV $\lambda$ 1402 transition, thus preventing a proper fit. However, the blue part of SiIV $\lambda$ 1402 matches the blue parts of SiIV $\lambda$ 1393 and the CIV lines, confirming the detection of SiIV and CIV. It is interesting to note that in this case no satisfactory solutions could be found to fit simultaneously the SiIV and CIV doublets. To check the wavelength calibration of the spectrum, the SiII $\lambda\lambda$ 1526 and 1190 lines (which fall on two different arm of the spectrograph) are fitted independently. The redshifts determined for these transitions are consistent with each other ($\rm z=3.10408$ and $\rm z=3.10409$) thus indicating no systematic shift in the spectrum. Therefore, the SiIV and CIV transitions are fitted separately, with different Doppler parameters and velocities as seen in Table \ref{tab:1621-0042}. The 14-component fit results in $\log \rm N(SiIV)=14.24\pm0.03$ and $\log \rm N(CIV)=14.71\pm0.01$.

\begin{figure*}
\begin{center}
\hspace*{-.8in}
\includegraphics[width=1.2\textwidth]{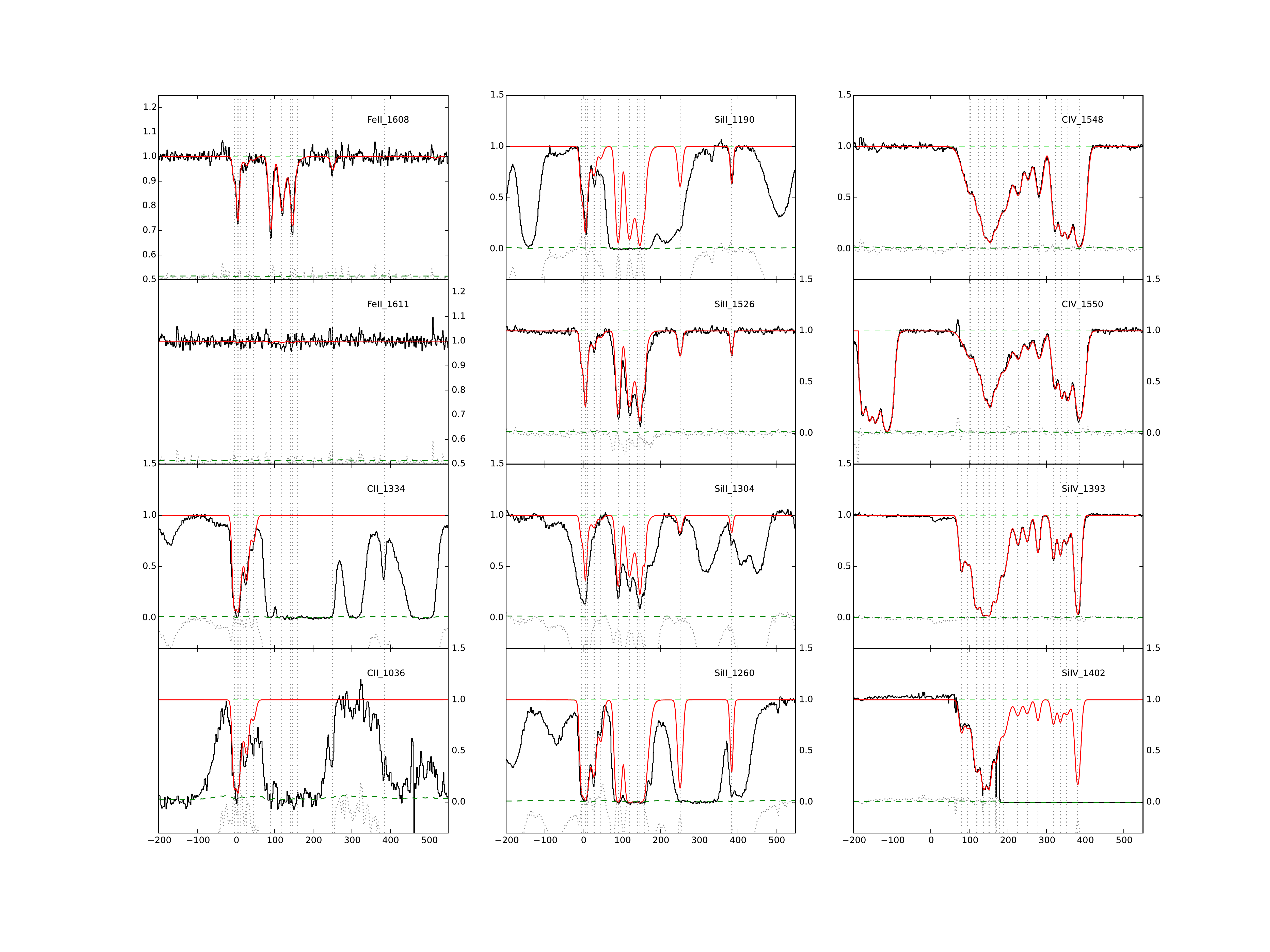}
\caption[QSOJ1621-0042]{QSOJ1621-0042}
\label{img:1621-0042}
\end{center}
\end{figure*}

The parameter fits of the individual components are listed in Table \ref{tab:1621-0042} and the corresponding Voigt profile fits are shown in Fig. \ref{img:1621-0042}.

\begin{table*}
\begin{center}
\caption{Voigt profile fit parameters to the low- and high-ionization species for the z$_{\rm abs}$=3.104 log N(H\,I)=$19.7\pm0.20$ absorber towards QSO J1621-0042.}
\label{tab:1621-0042}
\begin{tabular}{ccccccccccc}
\hline
\hline
Comp. & $z_{abs}$ & b & Ion & log $N$ && Comp. & $z_{abs}$ & b & Ion & log $N$ \\
 & & km $s^{-1}$ & & cm$^{-2}$ &&  & & km $s^{-1}$ & & cm$^{-2}$ \\
 \hline
1 & $3.10394$ & $3.7\pm0.7$ & FeII & $12.62\pm0.06$ && 1 & $3.10509$ & $7.0\pm1.0$ & SiIV & $12.77\pm0.07$\\
   &   &   & SiII & $12.95\pm0.03$  &&    & $3.10541$ & $24.6\pm1.0$ & CIV & $13.49\pm0.21$\\
   &   &   & CII & $13.90\pm0.03$  && 2 & $3.10530$ & $7.9\pm3$ & SiIV & $12.69\pm0.18$\\
2 & $3.10407$ & $2.7\pm0.5$ & FeII & $13.03\pm0.04$ &&    & $3.10539$ & $6.4\pm3.4$ & CIV & $12.13\pm0.21$\\
   &   &   & SiII & $13.55\pm0.03$ && 3 & $3.10564$ & $12.0\pm2.8$ & SiIV & $13.47\pm0.09$\\
   &   &   & CII & $14.00\pm0.17$ &&    & $3.10568$ & $8.5\pm1.0$ & CIV & $13.16\pm0.05$\\
3 & $3.10415$ & $8.4\pm2$ & FeII & $12.32\pm0.19$ && 4 & $3.10588$ & $6.1\pm1.9$ & SiIV & $13.40\pm0.12$\\
   &   &   & SiII & $12.93\pm0.15$  &&    & $3.10592$ & $9.0\pm0.6$ & CIV & $13.64\pm0.04$\\
   &   &   & CII & $13.57\pm0.02$&&5 & $3.10607$ & $7.5\pm1.7$ & SiIV & $13.52\pm0.09$\\
4 & $3.10438$ & $6.0\pm1.1$ & FeII & $12.28\pm0.11$&&   & $3.10612$ & $7.9\pm0.6$ & CIV & $13.69\pm0.05$\\
   &   &   & SiII & $12.73\pm0.06$&&6 & $3.10631$ & $8.1\pm1.5$ & SiIV & $13.16\pm0.10$\\
   &   &   & CII & $13.47\pm0.01$&&   & $3.10632$ & $8.6\pm0.5$ & CIV & $13.36\pm0.05$\\
5 & $3.10462$ & $7.3\pm1.4$ & FeII & $11.95\pm0.21$&&7 & $3.10657$ & $15.0\pm1.6$ & SiIV & $13.10\pm0.06$\\
   &   &   & SiII & $12.37\pm0.06$&&   & $3.10659$ & $21.5\pm0.7$ & CIV & $13.70\pm0.02$\\
   &   &   & CII & $13.00\pm0.01$&&8 & $3.10709$ & $9.3\pm1.2$ & SiIV & $12.51\pm0.04$\\   
6 & $3.10523$ & $6.1\pm0.13$ & FeII & $13.28\pm0.01$&&   & $3.10712$ & $12.1\pm0.4$ & CIV & $13.24\pm0.02$\\
   &   &   & SiII & $13.74\pm0.01$&&9 & $3.10742$ & $7.9\pm1.1$ & SiIV & $12.40\pm0.04$\\
   &   &   & CII & $-$&&   & $3.10746$ & $9.3\pm0.4$ & CIV & $12.95\pm0.02$\\
7 & $3.10563$ & $7.9\pm0.23$ & FeII & $13.16\pm0.02$&&10 & $3.10780$ & $6.0\pm0.6$ & SiIV & $12.50\pm0.02$\\
   &   &   & SiII & $13.73\pm0.03$&&   & $3.10785$ & $12.2\pm0.2$ & CIV & $13.27\pm0.01$\\
   &   &   & CII & $-$&&11 & $3.10836$ & $6.5\pm0.7$ & SiIV & $12.61\pm0.03$\\
8 & $3.10593$ & $20.0\pm1.08$ & FeII & $13.07\pm0.05$&&   & $3.10842$ & $10.4\pm0.2$ & CIV & $13.63\pm0.01$\\
   &   &   & SiII & $13.73\pm0.03$&&12 & $3.10859$ & $5.7\pm1.2$ & SiIV & $12.49\pm0.07$\\
   &   &   & CII & $-$&&   & $3.10865$ & $5.5\pm0.3$ & CIV & $13.43\pm0.02$\\
9 & $3.10601$ & $5.0$ & FeII & $13.10\pm0.02$&&13 & $3.10883$ & $10.6\pm2.1$ & SiIV & $12.53\pm0.07$\\
   &   &   & SiII & $13.66\pm0.02$&&   & $3.10887$ & $10.4\pm0.3$ & CIV & $13.76\pm0.01$\\
   &   &   & CII & $-$&&14 & $3.10922$ & $7.5\pm0.3$ & SiIV & $13.47\pm0.03$\\
10 & $3.10618$ & $0.9\pm0.2$ & FeII & $-$&&   & $3.10928$ & $14.1\pm0.1$ & CIV & $14.12\pm0.01$\\
   &   &   & SiII & $13.39\pm0.25$&&&&&&\\
   &   &   & CII & $-$&&&&&&\\
11 & $3.10743$ & $6.1\pm0.29$ & FeII & $12.47\pm0.06$&&&&&&\\
   &   &   & SiII & $12.93\pm0.01$&&&&&&\\
   &   &   & CII & $-$&&&&&&\\
12 & $3.10926$ & $2.5\pm0.25$ & FeII & $10.84\pm1.99$&&&&&&\\
   &   &   & SiII & $12.77\pm0.01$&&&&&&\\
   &   &   & CII & $-$&&&&&&\\
 \hline
\end{tabular}
\end{center}
\end{table*}

\subsection{QSO 4C12.59 z$_{\rm em}=1.792$, z$_{\rm abs}=0.531$, $\rm \log N(HI)=20.7\pm0.09$}
This very low-redshift DLA absorber presents a few absorption features in the EUADP spectrum partly due the limited wavelength coverage and an overall low SNR. The FeII ion is detected in the following transitions FeII $\lambda\lambda\lambda$ 2344, 2382 and 2374. The FeII $\lambda$ 2382 line is saturated. A satisfactory fit for the remaining transitions FeII $\lambda\lambda$ 2374 and 2344 is found with six components. The resulting column density is $\log \rm N(FeII)=14.26\pm0.08$. The CII $\lambda$ 2325 line is covered but not detected. The resulting upper limit is $\rm \log N(CII)<11.36$.

No high-ionization ions are covered in this EUADP spectrum.

\begin{figure}
\begin{center}
\includegraphics[width=.5\textwidth, height=14cm]{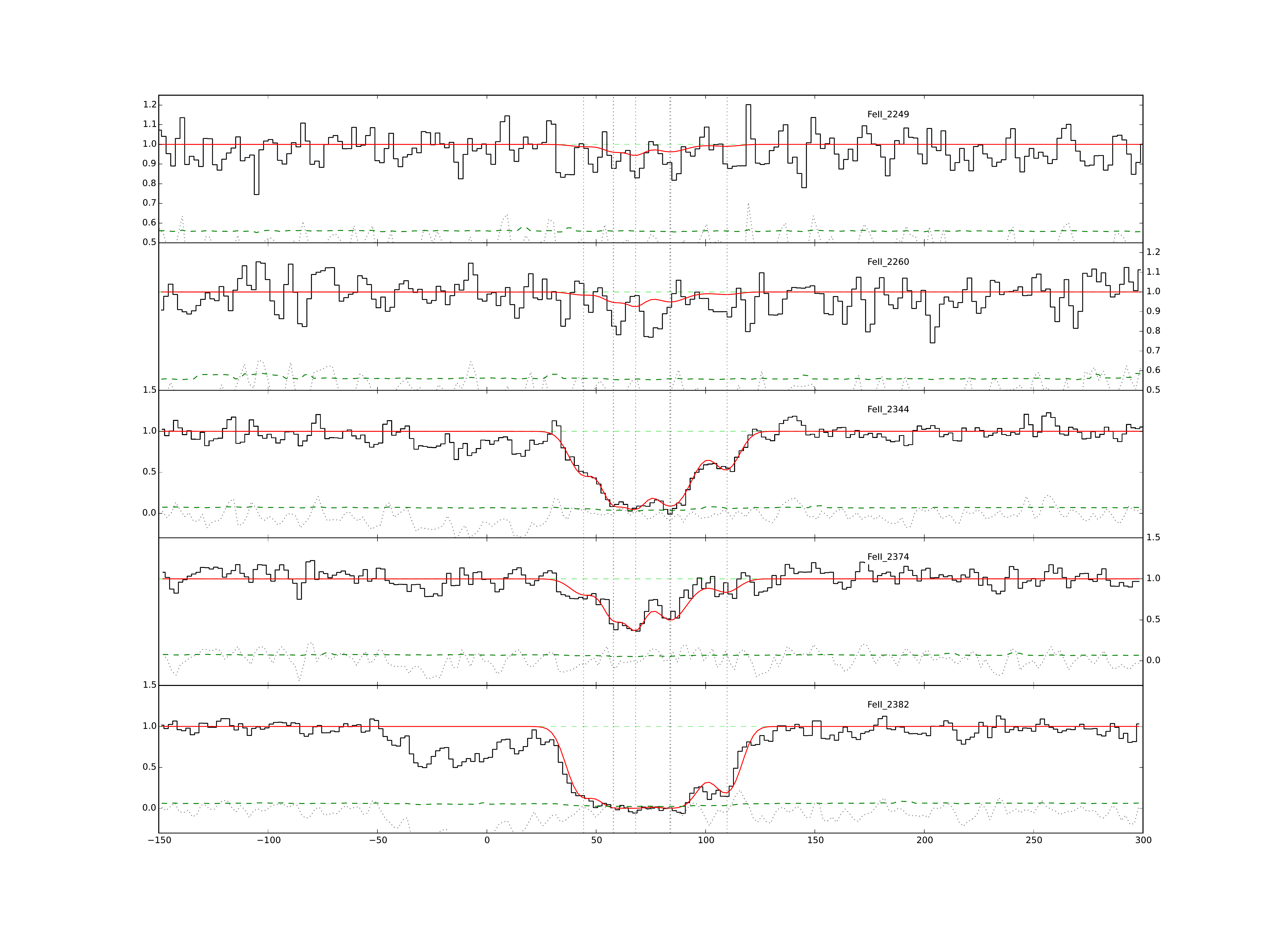}
\caption[QSO4C12.59]{QSO4C12.59}
\label{img:4C12-59}
\end{center}
\end{figure}

The parameter fits of the individual components are listed in Table \ref{tab:4C12-59} and the corresponding Voigt profile fits are shown in Fig. \ref{img:4C12-59}.

\begin{table}
\begin{center}
\caption{Voigt profile fit parameters to the low-ionization species for the z$_{\rm abs}$=0.531 log N(H\,I)=$20.7\pm0.09$ absorber towards QSO 4C 12.59.}
\label{tab:4C12-59}
\begin{tabular}{ccccc}
\hline
\hline
Comp. & $z_{abs}$ & b & Ion & log $N$ \\
 & & km $s^{-1}$ & & cm$^{-2}$ \\
 \hline
1 & $0.53123$ & $7.0\pm1.3$ & FeII & $13.20\pm0.08$\\
2 & $0.53130$ & $4.6\pm1.4$ & FeII & $13.58\pm0.08$\\
3 & $0.53135$ & $4.0\pm1.1$ & FeII & $13.68\pm0.10$\\
4 & $0.53143$ & $13.9\pm4.1$ & FeII & $13.59\pm0.26$\\
5 & $0.53143$ & $6.8\pm1.4$ & FeII & $13.47\pm0.28$\\
6 & $0.53156$ & $6.3\pm1.8$ & FeII & $13.06\pm0.12$\\
\hline
\end{tabular}
\end{center}
\end{table}

\subsection{QSO LBQS2114-4347 z$_{\rm em}=2.04$, z$_{\rm abs}=1.912$, $\rm \log N(HI)=19.5\pm0.10$}

The EUADP spectrum of this quasar covers the following low-ionization ions: SiII $\lambda\lambda$ 1304 1526, MgII $\lambda\lambda$ 2796 2803, FeII $\lambda\lambda\lambda\lambda\lambda\lambda$ 2600 1608 2260 2344 2374 2586, AlII $\lambda$ 1670 and CII $\lambda$ 1334. An 11-component fit is used to describe the lines free from saturations and blends, namely SiII $\lambda\lambda$ 1526 1304 and FeII $\lambda\lambda\lambda$ 1608 2374 and 2586.

This results in the following column densities: $\log \rm N(SiII)=14.39\pm0.02$, $\log \rm N(FeII)=14.02\pm0.01$, $\log \rm N(AlII)=13.00\pm0.01$ and $\log \rm N(MgII)=14.40\pm0.01$. In addition, the non-detections in the spectrum provide further upper limits as follows: AlIII $\lambda$ 1854, $\rm \log N(AlIII)<12.09$, CrII $\lambda$ 2056, $\rm \log N(CrII)<12.77$, MnII $\lambda$ 2576, $\rm \log N(MnII)<12.24$, NiII $\lambda$ 1317, $\rm \log N(NiII)<12.88$, SII $\lambda$ 1253, $\rm \log N(SII)<13.97$, and ZnII $\lambda$ 2026, $\rm \log N(ZnII)<12.17$.

A 4-component profile is used to fit the high-ionization ions: CIV $\lambda\lambda$ 1548 1550 and SiIV $\lambda\lambda$ 1393 and 1402. The resulting column densities are $\log \rm N(SiIV)=13.43\pm0.01$ and $\log \rm N(CIV)=14.39\pm0.01$.

\begin{figure}
\begin{center}
\hspace*{-.8in}
\includegraphics[width=1.2\textwidth]{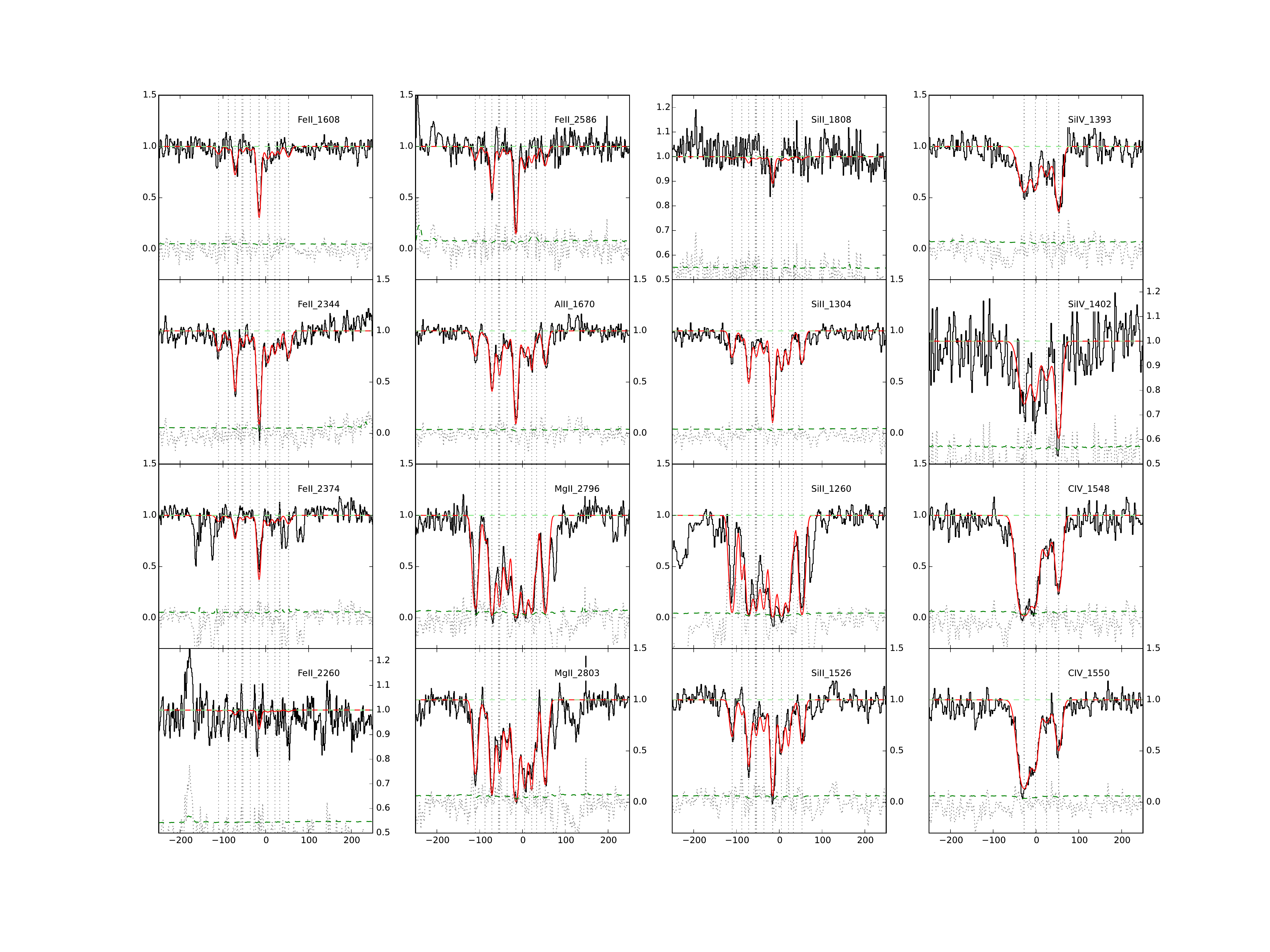}
\caption[QSOLBQS2114-4347]{QSOLBQS2114-4347}
\label{img:2114-4347}
\end{center}
\end{figure}

The parameter fits of the individual components are listed in Table \ref{tab:2114-4347} and the corresponding Voigt profile fits are shown in Fig. \ref{img:2114-4347}.

\begin{table}
\begin{center}
\caption{Voigt profile fit parameters to the low- and high-ionization species for the z$_{\rm abs}$=1.912 log N(H\,I)=$19.5\pm0.10$ absorber towards QSO LBQS 2114-4347.}
\label{tab:2114-4347}
\begin{tabular}{ccccc}
\hline
\hline
Comp. & $z_{abs}$ & b & Ion & log $N$ \\
 & & km $s^{-1}$ & & cm$^{-2}$ \\
 \hline
1 & $1.9109296$ & $6.7\pm0.1$ & SiII & $13.16\pm0.03$\\
   &   &   & FeII & $12.65\pm0.01$\\
   &   &   & AlII & $11.70\pm0.02$\\
   &   &   & MgII & $12.92\pm0.01$\\
2 & $1.9111506$ & $3.0\pm0.1$ & SiII & $12.53\pm0.01$\\
   &   &   & FeII & $12.11\pm0.01$\\
   &   &   & AlII & $10.84\pm0.01$\\
   &   &   & MgII & $11.76\pm0.01$\\
3 & $1.9113052$ & $4.6\pm0.1$ & SiII & $13.44\pm0.01$\\
   &   &   & FeII & $13.18\pm0.01$\\
   &   &   & AlII & $12.10\pm0.01$\\
   &   &   & MgII & $13.23\pm0.01$\\
4 & $1.91146$ & $12.3\pm0.1$ & SiII & $12.96\pm0.01$\\
   &   &   & FeII & $-$\\
   &   &   & AlII & $11.66\pm0.01$\\
   &   &   & MgII & $12.53\pm0.01$\\
5 & $1.91148$ & $3.7\pm0.1$ & SiII & $12.79\pm0.04$\\
   &   &   & FeII & $12.38\pm0.01$\\
   &   &   & AlII & $11.75\pm0.01$\\
   &   &   & MgII & $12.69\pm0.01$\\
6 & $1.91165$ & $5.5\pm0.1$ & SiII & $13.00\pm0.03$\\
   &   &   & FeII & $12.30\pm0.01$\\
   &   &   & AlII & $11.43\pm0.02$\\
   &   &   & MgII & $12.55\pm0.01$\\
7 & $1.91185$ & $4.1\pm0.1$ & SiII & $14.08\pm0.02$\\
   &   &   & FeII & $13.77\pm0.01$\\
   &   &   & AlII & $12.69\pm0.01$\\
   &   &   & MgII & $14.21\pm0.01$\\
8 & $1.91205$ & $6.7\pm0.1$ & SiII & $13.37\pm0.20$\\
   &   &   & FeII & $12.88\pm0.01$\\
   &   &   & AlII & $11.73\pm0.01$\\
   &   &   & MgII & $13.12\pm0.01$\\
9 & $1.91221$ & $4.0\pm1.0$ & SiII & $13.17\pm0.02$\\
   &   &   & FeII & $12.59\pm0.01$\\
   &   &   & AlII & $11.81\pm0.01$\\
   &   &   & MgII & $13.14\pm0.01$\\
10 & $1.91232$ & $0.9\pm0.1$ & SiII & $12.25\pm0.01$\\
   &   &   & FeII & $12.41\pm0.01$\\
   &   &   & AlII & $10.91\pm0.01$\\
   &   &   & MgII & $13.08\pm0.01$\\
11 & $1.91252$ & $6.3\pm1.0$ & SiII & $13.24\pm0.02$\\
   &   &   & FeII & $12.78\pm0.01$\\
   &   &   & AlII & $11.84\pm0.01$\\
   &   &   & MgII & $13.05\pm0.01$\\
 \hline
1 & $1.91173$ & $16.3\pm0.3$ & CIV & $14.19\pm0.01$\\
   &   &   & SiIV & $12.97\pm0.01$\\
2 & $1.91199$ & $8.5\pm0.3$ & CIV & $13.63\pm0.02$\\
   &   &   & SiIV & $12.65\pm0.02$\\
3 & $1.91224$ & $12.3\pm0.8$ & CIV & $13.18\pm0.03$\\
   &   &   & SiIV & $12.63\pm0.02$\\
4 & $1.91252$ & $8.8\pm0.2$ & CIV & $13.49\pm0.01$\\
   &   &   & SiIV & $12.97\pm0.01$\\
   \hline
\end{tabular}
\end{center}
\end{table}

\onecolumn

\twocolumn

\subsection{QSO B2126-15 z$_{\rm em}=3.268$, z$_{\rm abs}=2.638$, $\rm \log N(HI)=19.25\pm0.15$}
This EUADP spectrum covers the following transitions associated with the absorber: NiII $\lambda\lambda\lambda$ 1709 1741 1751, SiII $\lambda$ 1808, FeII $\lambda\lambda\lambda\lambda\lambda$ 2586 1608 2249 2600 2382 and AlIII $\lambda\lambda$I 1854 and 1862. The good quality of the spectrum and the presence of non blended lines enables a robust 3-component fit of the low-ionization ions with FeII $\lambda\lambda$ 1608 2586, SiII $\lambda$ 1808 and NiII $\lambda\lambda\lambda$ 1741 1751 and 1709. The apparent shift in FeII $\lambda$ 2586 is thought to originate from a poor continuum fit. The resulting abundances are $\log \rm N(FeII)=14.05\pm0.01$, $\log \rm N(SiII)=14.67\pm0.02$, $\log \rm N(NiII)=13.15\pm0.01$. We derive an upper limit from the non detection of ZnII $\lambda$ 2026: $\rm \log N(ZnII)<11.58$.

The intermediate-ionization transitions AlIII $\lambda\lambda$ 1854 and 1862 do not share the same absorption profile as the low-ionization ions (except for the strong narrow line in the middle of both profiles). These are therefore fitted separately with a 5-component profile. The resulting column density is $\log \rm N(AlIII)=13.24\pm0.02$.

The high-ionization transitions SiIV $\lambda\lambda$ 1398 and 1402 are clearly detected. However these lines are both blended and saturated, so that no fit are attempted.

\begin{figure*}
\begin{center}
\hspace*{-.8in}
\includegraphics[width=1.2\textwidth]{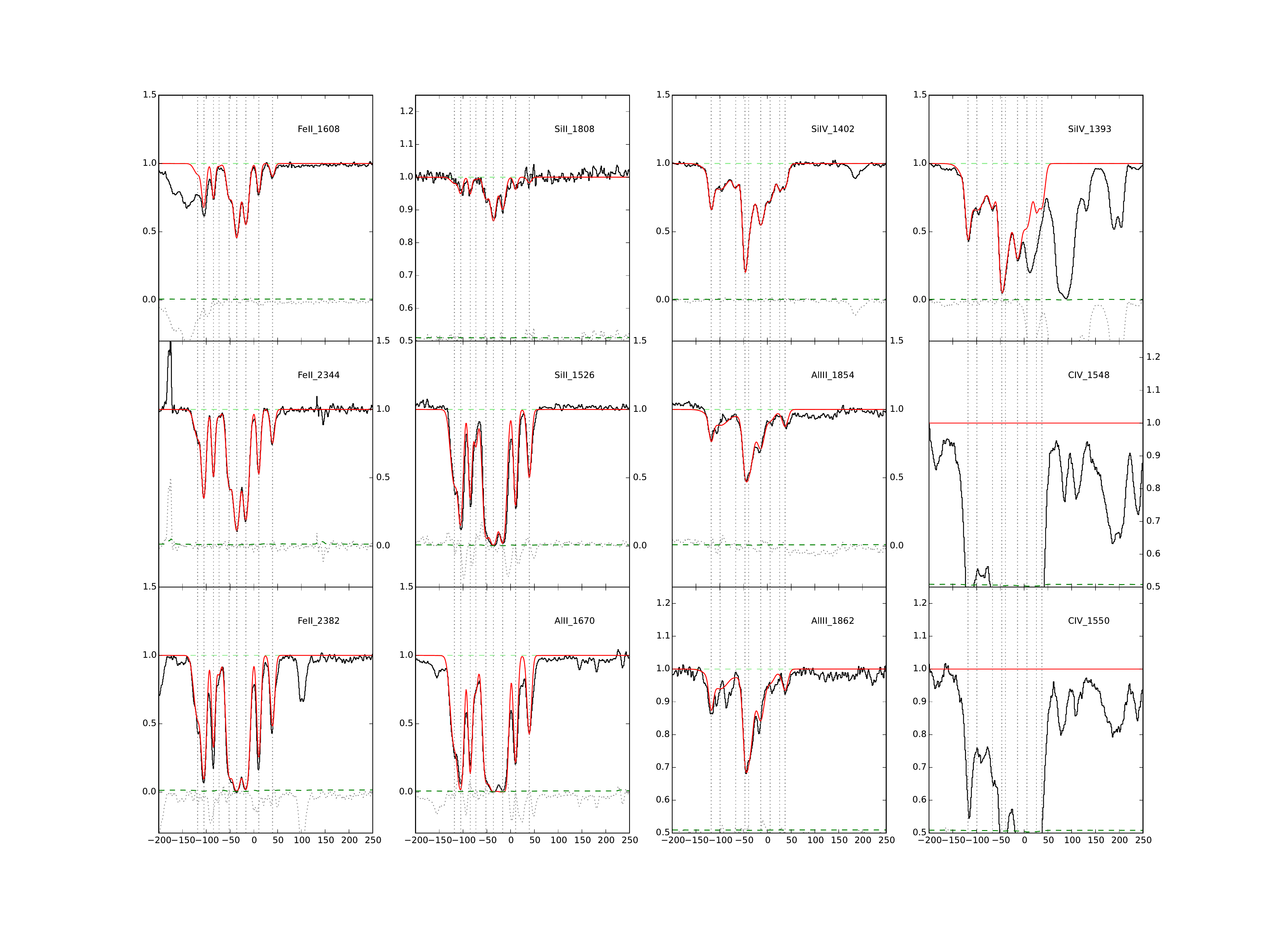}
\caption[QSOB2126-15 z=2.638]{QSOB2126-15 $\rm z_{abs}=2.638$}
\label{img:2126-15_2.638}
\end{center}
\end{figure*}

The parameter fits of the individual components are listed in Table \ref{tab:2126-15_2.638} and the corresponding Voigt profile fits are shown in Fig. \ref{img:2126-15_2.638}.

\begin{table}
\begin{center}
\caption{Voigt profile fit parameters to the low- and intermediate-ionization species for the z$_{\rm abs}$=2.638 log N(H\,I)=$19.25\pm0.15$ absorber towards QSO B2126-15.}
\label{tab:2126-15_2.638}
\begin{tabular}{ccccc}
\hline
\hline
Comp. & $z_{abs}$ & b & Ion & log $N$ \\
 & & km $s^{-1}$ & & cm$^{-2}$ \\
 \hline
1 & $2.63767$ & $17.8\pm0.1$ & FeII & $13.77\pm0.01$\\
   &   &   & SiII & $14.38\pm0.02$\\
   &   &   & NiII & $12.66\pm0.02$\\
2 & $2.63799$ & $4.1\pm0.1$ & FeII & $13.39\pm0.01$\\
   &   &   & SiII & $13.80\pm0.05$\\
   &   &   & NiII & $12.47\pm0.01$\\
3 & $2.63807$ & $19.3\pm0.3$ & FeII & $13.45\pm0.01$\\
   &   &   & SiII & $14.21\pm0.03$\\
   &   &   & NiII & $12.83\pm0.02$\\
\hline
1 & $2.63758$ & $9.9\pm0.5$ & AlIII & $12.38\pm0.02$\\
2 & $2.63784$ & $13.3\pm0.2$ & AlIII & $12.84\pm0.01$\\
3 & $2.63802$ & $4.2\pm0.5$ & AlIII & $12.34\pm0.05$\\
4 & $2.63818$ & $10.3\pm0.8$ & AlIII & $12.52\pm0.08$\\
5 & $2.63827$ & $11.75\pm1.8$ & AlIII & $12.43\pm0.11$\\
\hline
\end{tabular}
\end{center}
\end{table}

\subsection{QSO B2126-15 z$_{\rm em}=3.268$, z$_{\rm abs}=2.769$, $\rm \log N(HI)=19.2\pm0.15$}
Many lines associated to the absorber are detected in this EUADP spectrum: FeII $\lambda\lambda\lambda\lambda\lambda\lambda\lambda$ 2586 2374 2344 1608 1144 2382 2600, SiII $\lambda\lambda\lambda$ 1808 1526 1190, CII $\lambda\lambda$ 1334 1036, AlII $\lambda$ 1670, AlIII $\lambda\lambda$ 1862 1854, CIV $\lambda\lambda$ 1550 1548 and SiIV $\lambda\lambda$ 1402 and 1393.

The low-ionization ions are fitted considering the non blended lines, FeII $\lambda$ 2344, SiII $\lambda\lambda$ 1808, 1526 and AlII $\lambda$ 1670. This resulted in a 9-component profile with a $\rm 250km/s$ velocity range. The transition FeII $\lambda$ 2374 is not considered for the final fit due to the complexity in the continuum placement in this portion of the spectrum. The CII line is strongly saturated and therefore no fit could be performed on this transition. The AlII $\lambda$ 1670 transition is saturated, therefore resulting in a lower limit for the column density ($\rm \log N(AlII)<14.05$). The fit results in the following column densities: $\log \rm N(FeII)=14.17\pm0.01$ and $\log \rm N(SiII)=14.79\pm0.01$. In addition, the non-detections led to the following upper limits: CrII $\lambda$ 2056, $\rm \log N(CrII)<12.40$, MnII $\lambda$ 2606, $\rm \log N(MnII)<12.28$, and ZnII $\lambda$ 2062, $\rm \log N(ZnII)<11.95$.

The intermediate- and high-ionization ions AlIII $\lambda\lambda$ 1854 1862 and SiIV $\lambda$ 1402 are fitted together given the similarities in their absorption profile. The SiIV $\lambda$ 1393 transition is not considered because of an unidentified blending in the red part of the profile. The blending does not match any transition from the absorber at redshift $\rm z_{abs}=2.638$. The 9-component profile of about $\rm 200km/s$ and results in the following column densities $\log \rm N(SiIV)=13.84\pm0.13$ and $\log \rm N(AlIII)=13.11\pm0.01$. The CIV doublet is also covered by the data but the CIV $\lambda$ 1548 line is saturated and both profiles show evidence for the presence of blending. It is interesting to note that the CIV doublet presents some components far in the red (up to $\rm \sim240km/s$) which are not seen in SiIV. The OVI $\lambda\lambda$ 1031 and 1037 lines are also detected, but they are blended as often the case in the Ly$\alpha$ forest.

\begin{figure}
\begin{center}
\hspace*{-.8in}
\includegraphics[width=1.2\textwidth]{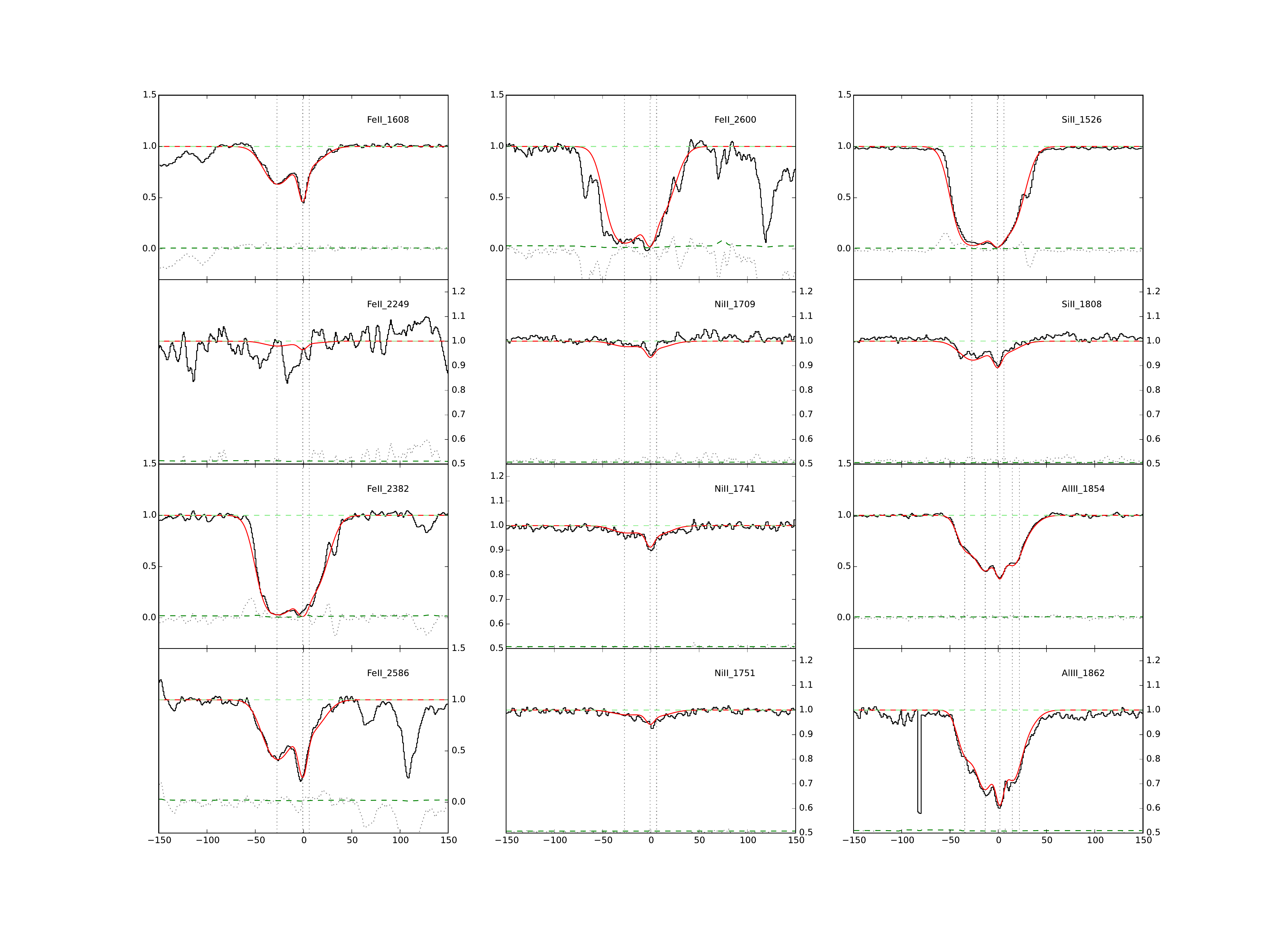}
\caption[QSOB2126-15 z=2.769]{QSOB2126-15 z=2.769}
\label{img:2126-15 z=2.769}
\end{center}
\end{figure}

The parameter fits of the individual components are listed in Table \ref{tab:2126-15 z=2.769} and the corresponding Voigt profile fits are shown in Fig. \ref{img:2126-15 z=2.769}.

\begin{table}
\begin{center}
\caption{Voigt profile fit parameters to the low- and high- and intermediate-ionization species for the z$_{\rm abs}$=2.769 log N(H\,I)=$19.20\pm0.15$ absorber towards QSO B2126-15.}
\label{tab:2126-15 z=2.769}
\begin{tabular}{ccccc}
\hline
\hline
Comp. & $z_{abs}$ & b & Ion & log $N$ \\
 & & km $s^{-1}$ & & cm$^{-2}$ \\
 \hline
1 & $2.76751$ & $9.4\pm0.8$ & FeII & $12.78\pm0.01$\\
   &   &   & SiII & $13.52\pm0.01$\\
   &   &   & AlII & $12.40\pm0.01$\\
2 & $2.76768$ & $4.4\pm0.1$ & FeII & $13.23\pm0.01$\\
   &   &   & SiII & $13.70\pm0.01$\\
   &   &   & AlII & $13.07\pm0.01$\\
3 & $2.76793$ & $2.1\pm0.2$ & FeII & $13.06\pm0.01$\\
   &   &   & SiII & $13.64\pm0.07$\\
   &   &   & AlII & $13.42\pm0.01$\\
4 & $2.76808$ & $3.7\pm1.4$ & FeII & $11.91\pm0.05$\\
   &   &   & SiII & $12.87\pm0.01$\\
   &   &   & AlII & $11.58\pm0.01$\\
5 & $2.76835$ & $6.2\pm0.2$ & FeII & $13.19\pm0.01$\\
   &   &   & SiII & $13.94\pm0.05$\\
   &   &   & AlII & $12.65\pm0.01$\\
6 & $2.76855$ & $7\pm0.2$ & FeII & $13.67\pm0.01$\\
   &   &   & SiII & $14.30\pm0.02$\\
   &   &   & AlII & $13.13\pm0.01$\\
7 & $2.76879$ & $6.8\pm0.1$ & FeII & $13.54\pm0.01$\\
   &   &   & SiII & $14.15\pm0.03$\\
   &   &   & AlII & $13.70\pm0.01$\\
8 & $2.76913$ & $3.2\pm0.2$ & FeII & $12.99\pm0.01$\\
   &   &   & SiII & $13.55\pm0.09$\\
   &   &   & AlII & $12.45\pm0.01$\\
9 & $2.76949$ & $4.4\pm0.3$ & FeII & $12.65\pm0.01$\\
   &   &   & SiII & $13.26\pm0.01$\\
   &   &   & AlII & $12.10\pm0.01$\\
 \hline
1 & $2.76751$ & $4.9\pm0.9$ & SiIV & $12.55\pm0.06$\\
 &   &   & AlIII & $11.91\pm0.02$\\
2 & $2.76775$ & $26.3\pm2.5$ & SiIV & $13.02\pm0.04$\\
  &   &  & AlIII & $12.33\pm0.01$\\
3 & $2.76816$ & $7.3\pm6.9$ & SiIV & $12.37\pm0.47$\\
  &   &   & AlIII & $11.02\pm0.11$\\
4 & $2.76841$ & $4.6\pm2.9$ & SiIV & $13.17\pm0.38$\\
  &   &   & AlIII & $12.01\pm0.02$\\
5 & $2.76851$ & $12.0\pm6.2$ & SiIV & $13.18\pm0.46$\\
  &   &   & AlIII & $12.72\pm0.01$\\
6 & $2.76882$ & $9.1\pm1.7$ & SiIV & $13.02\pm0.10$\\
  &  &   & AlIII & $12.32\pm0.01$\\
7 & $2.76907$ & $12.4\pm4.3$ & SiIV & $12.88\pm0.15$\\
  &   &   & AlIII & $11.87\pm0.02$\\
8 & $2.76933$ & $4.7\pm3.9$ & SiIV & $12.30\pm0.36$\\
  &  &  & AlIII & $10.70\pm0.21$\\
9 & $2.76947$ & $7.0\pm2.6$ & SiIV & $12.47\pm0.18$\\
  &  &  & AlIII & $11.85\pm0.02$\\
\hline
\end{tabular}
\end{center}
\end{table}

\onecolumn

\twocolumn

\subsection{QSO LBQS2132-4321 z$_{\rm em}=2.42$, z$_{\rm abs}=1.916$, $\rm \log N(HI)=20.74\pm0.09$}
This EUADP spectrum provides the coverage of many ions associated to the absorber, often free from blending and with a good SNR.
The low-ionization ions detected are SiII $\lambda\lambda\lambda\lambda\lambda\lambda\lambda$ 1190 1304 1808 1253 1250 1260 1193, FeII $\lambda\lambda\lambda$ 1608 2260 2249, CrII $\lambda$ 2056, NiII $\lambda\lambda$ 1741 1709, ZnII $\lambda$ 2026 and SII $\lambda\lambda$ 1250 and 1253. The profile presents two distinct groups of components spread over about $300$ km/s: the weak components, only detected in large oscillator strength ions and the strong components, detected in all low-ionization ions and saturated otherwise. Therefore, the column densities of the strongest components are measured from the small oscillator strength ions only. 
We fit SiII $\lambda$ 1808, FeII $\lambda$ 2260, CrII $\lambda\lambda$ 2056 2026, NiII $\lambda\lambda$ 1741 1709 and ZnII $\lambda$ 2026 with four components. The intermediate ion AlIII $\lambda$ 1854 which presents the same components (among others) is added to the fit. The remaining weak group is fitted with 3 components for SiII, AlIII and FeII. In this latter fit, some parameters such as the Doppler-parameter are fixed during the process because of a rather low SNR and/or blending. We detected a contamination in the third component of SiII $\lambda$ 1808, based on SiII $\lambda\lambda$ 1526 and 1304, we therefore fitted the first three red components fixing the doppler parameter and column density of the third component of SiII. The fit results in the following column densities: 
$\log \rm N(SiII)=15.75\pm0.02$, %
$\log \rm N(FeII)=15.06\pm0.04$, %
$\log \rm N(CrII)=13.38\pm0.03$, %
$\log \rm N(NiII)=13.80\pm0.03$, %
$\log \rm N(ZnII)=12.69\pm0.02$ and %
$\log \rm N(AlIII)=13.18\pm0.03$. %
For SII $\lambda\lambda$ 1250 and 1253, a blend in the blue part of the absorption prevent from fitting the full profile. However, a lower limit is derived from the fit of the strong component in the red based on the other elements. The resulting column density is $\rm \log N(SII)>14.90$.

The spectrum cover several high-ionization ions: CIV $\lambda\lambda$ 1550, 1548 and SiIV $\lambda\lambda$ 1393 and 1402. However, no satisfactory fit could be found for the CIV doublet because of the low SNR in this portion of the spectrum. On the contrary, a 4-component profile is derived for the SiIV $\lambda$ 1402. It is also interesting to note that the velocity components are quite similar to the low-ionization ions components. The SiIV $\lambda$ 1393 line suffers from a blend and again a limited SNR, so that the fit is performed with fix parameters but for the second Doppler-parameter. This fit results in $\log \rm N(SiIV)=14.2\pm0.01$.

\begin{figure*}
\begin{center}
\hspace*{-.8in}
\includegraphics[width=1.2\textwidth, height=18cm]{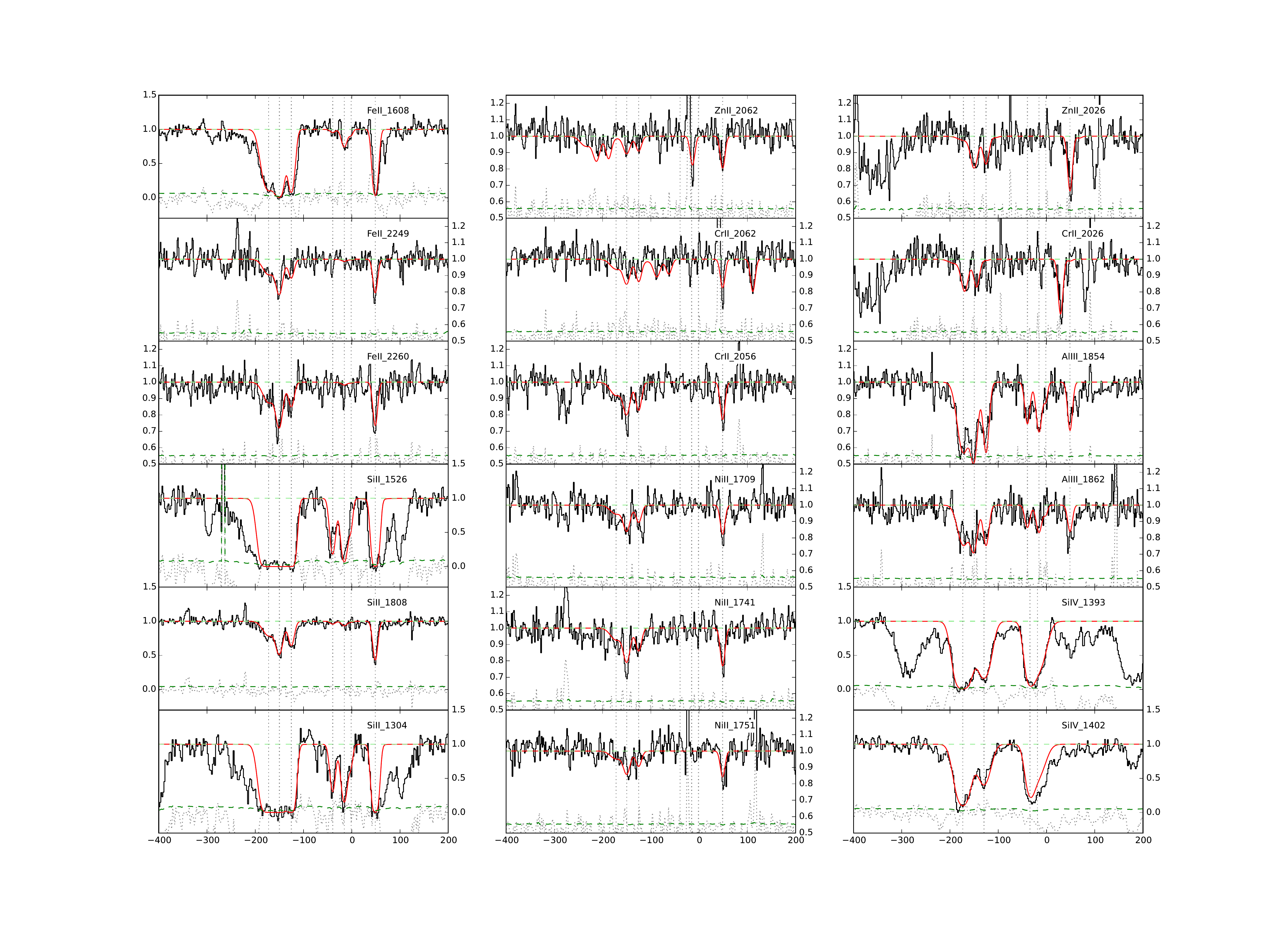}
\caption[QSOLBQS2132-4321]{QSOLBQS2132-4321}
\label{img:2132-4321}
\end{center}
\end{figure*}

The parameter fits of the individual components are listed in Table \ref{tab:2132-4321} and the corresponding Voigt profile fits are shown in Fig. \ref{img:2132-4321}.

\begin{table}
\begin{center}
\caption{Voigt profile fit parameters to the low- and high-ionization species for the z$_{\rm abs}$=1.916 log N(H\,I)=$20.74\pm0.09$ absorber towards QSO LBQS 2132-432.}
\label{tab:2132-4321}
\begin{tabular}{ccccc}
\hline
\hline
Comp. & $z_{abs}$ & b & Ion & log $N$ \\
 & & km $s^{-1}$ & & cm$^{-2}$ \\
 \hline
1 & $1.91433$ & $15.60\pm1.2$ & SiII & $14.82\pm0.04$\\
   &   &   & FeII & $14.42\pm0.06$\\
   &   &   & CrII & $12.63\pm0.05$\\
   &   &   & NiII & $13.04\pm0.08$\\
   &   &   & ZnII & $11.43\pm0.17$\\
   &   &   & AlIII & $12.77\pm0.02$\\
2 & $1.91454$ & $8.9\pm0.4$ & SiII & $15.03\pm0.02$\\
   &   &   & FeII & $14.57\pm0.03$\\
   &   &   & CrII & $12.81\pm0.03$\\
   &   &   & NiII & $13.30\pm0.03$\\
   &   &   & ZnII & $12.15\pm0.03$\\
   &   &   & AlIII & $12.58\pm0.02$\\
3 & $1.91478$ & $7.00$ & SiII & $14.80$\\
   &   &   & FeII & $14.22\pm0.06$\\
   &   &   & CrII & $12.69\pm0.03$\\
   &   &   & NiII & $13.05\pm0.05$\\
   &   &   & ZnII & $12.00\pm0.04$\\
   &   &   & AlIII & $12.48\pm0.02$\\
4 & $1.91562$ & $6.0$ & SiII & $13.74\pm0.02$\\
   &   &   & FeII & $12.40\pm0.32$\\
   &   &   & CrII & $-$\\
   &   &   & NiII & $-$\\
   &   &   & ZnII & $-$\\
   &   &   & AlIII & $12.15\pm0.03$\\
5 & $1.91585$ & $7.0$ & SiII & $13.9\pm0.02$\\
   &   &   & FeII & $13.26\pm0.05$\\
   &   &   & CrII & $-$\\
   &   &   & NiII & $-$\\
   &   &   & ZnII & $-$\\
   &   &   & AlIII & $12.29\pm0.03$\\
6 & $1.91599$ & $3.0$ & SiII & $13.07\pm0.06$\\
   &   &   & FeII & $12.52\pm0.22$\\
   &   &   & CrII & $-$\\
   &   &   & NiII & $-$\\
   &   &   & ZnII & $-$\\
   &   &   & AlIII & $11.60\pm0.10$\\
7 & $1.91647$ & $5.2\pm0.3$ & SiII & $15.01\pm0.02$\\
   &   &   & FeII & $14.41\pm0.05$\\
   &   &   & CrII & $12.74\pm0.04$\\
   &   &   & NiII & $13.21\pm0.04$\\
   &   &   & ZnII & $12.28\pm0.02$\\
   &   &   & AlIII & $12.19\pm0.03$\\
   \hline
1 & $1.91431$ & $17.8\pm0.3$ & SiV & $13.90\pm0.01$\\
2 & $1.91474$ & $16.5\pm0.5$ & SiV & $13.46\pm0.01$\\
3 & $1.91567$ & $11.1\pm0.8$ & SiV & $13.40\pm0.07$\\
4 & $1.91582$ & $19.4\pm1.1$ & SiV & $13.40\pm0.05$\\
   \hline
\end{tabular}
\end{center}
\end{table}

\subsection{QSO B2318-1107 z$_{\rm em}=2.96$, z$_{\rm abs}=1.629$, $\rm \log N(HI)=20.52\pm0.14$}
A broad variety of ions are covered in the EUADP spectrum of this low-redshift absorber.
The low-ionization ions detected are FeII $\lambda\lambda\lambda$ 2260, 2249 2374, MnII $\lambda$ 2576, AlII $\lambda$ 1670, CII $\lambda$ 1334, SiII $\lambda\lambda\lambda$ 1260, 1304 and 1526. The profile of the CII line appears to be saturated and blended from a comparison with other low-ionisation ions. This line is therefore not considered for a fit. Likewise, the SiII lines are saturated and are not fitted. The AlII $\lambda$ 1670 line appears to be blended (strong absorption on the blue side of the line), and therefore only an upper limit is available: $\rm \log N(AlII)<14.93$.
FeII and MnII are well fitted with 2 components (from the asymmetry of FeII $\lambda$ 2249): resulting in %
$\log \rm N(FeII)=14.14\pm0.02$ and %
$\log \rm N(MnII)=11.78\pm0.04$. 
Also, we derive upper limits from non detection of SII $\lambda$ 1250, $\rm \log N(SII)<14.54$, CrII $\lambda$ 2062, $\rm \log N(CrII)<12.47$, ZnII $\lambda$ 2026, $\rm \log N(ZnII)<11.74$, MgI $\lambda$ 2026 and $\rm \log N(MgI)<12.37$.

The detected intermediate-ionization transitions are AlIII $\lambda\lambda$ 1854 and 1862. The absorption profiles differ significantly from the high- or the low-ionization ions (red component stronger than the blue one). The fit is thus performed  separately with two components. The resulting column density is $\log \rm N(AlIII)=12.17\pm0.02$.

The high-ionization ions detected are SiIV $\lambda\lambda$ 1393, 1402 and CIV $\lambda\lambda$ 1548 (saturated) and 1550. 
The position of the SiIV lines (on the red wing of the Lya absorber for $\lambda$ 1393 and blended from the forest for $\lambda$ 1402) prevents a robust fit for the SiIV lines. 
Similarly to the low-ionization ions, the asymmetry of CIV $\lambda$ 1550 suggests a 2-component profile. We derive an upper limit for CIV from a 2-component fit of CIV $\lambda$ 1550, as the saturated CIV $\lambda$ 1548 brings no constraints on possible contamination of CIV $\lambda$ 1550. We obtain $\log \rm N(CIV)<14.10$.

\begin{figure*}
\begin{center}
\hspace*{-.8in}
\includegraphics[width=1.2\textwidth]{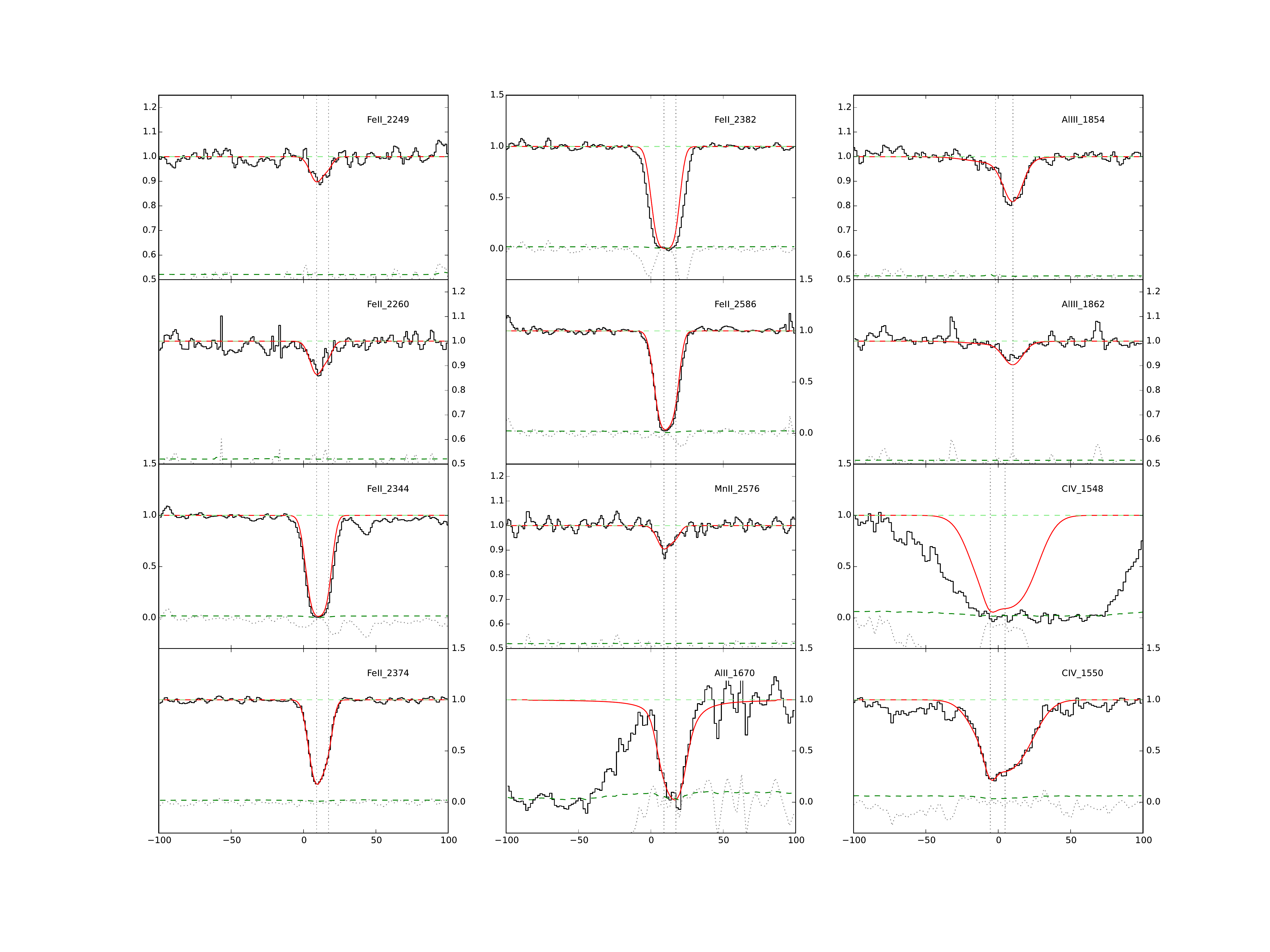}
\caption[QSOB2318-1107]{QSOB2318-1107}
\label{img:2318-1107}
\end{center}
\end{figure*}

The parameter fits of the individual components are listed in Table \ref{tab:2318-1107} and the corresponding Voigt profile fits are shown in Fig. \ref{img:2318-1107}.

\begin{table}
\begin{center}
\caption{Voigt profile fit parameters to the low- and high-ionization species for the z$_{\rm abs}$=1.629 log N(H\,I)=$20.52\pm0.14$ absorber towards QSO B2318-1107.}
\label{tab:2318-1107}
\begin{tabular}{ccccc}
\hline
\hline
Comp. & $z_{abs}$ & b & Ion & log $N$ \\
 & & km $s^{-1}$ & & cm$^{-2}$ \\
 \hline
1 & $1.62908$ & $4.7\pm0.3$ & FeII & $14.05\pm0.01$\\
   &   &   & MnII & $11.67\pm0.04$\\
   &   &   & AlII & $12.06\pm0.10$\\
2 & $1.62915$ & $1.4\pm0.4$ & FeII & $13.42\pm0.08$\\
   &   &   & MnII & $11.13\pm0.12$\\
   &   &   & AlII & $14.93\pm0.40$\\
 \hline
1 & $1.62898$ & $24.2\pm15.0$ & AlIII & $11.60\pm0.35$\\
2 & $1.62909$ & $8.0\pm1.4$ & AlIII & $12.03\pm0.11$\\
   \hline
1 & $1.62895$ & $2.7\pm3.2$ & CIV & $13.21\pm0.17$\\
2 & $1.62904$ & $19.8\pm1.1$ & CIV & $14.04\pm0.03$\\
   \hline
\end{tabular}

\end{center}
\end{table}

\end{document}